\documentclass[11pt,thmsa]{report}%
\usepackage{amssymb}
\usepackage{makeidx}
\usepackage{sw20mitt}
\usepackage{amsmath}
\usepackage{amsfonts}
\usepackage{graphicx}%
\providecommand{\U}[1]{\protect\rule{.1in}{.1in}}
\begin{document}
\ifx\href\undefined\else\hypersetup{linktocpage=true}\fi 

%\title{Anelastic spectroscopy studies of high-T$_c$ superconductors}
%\author{Francesco Cordero}
%\thisdegree{Doctor of Philosophy in Engineering}
%\university{University of Tsukuba}
%\degreemonth{October}
%\degreeyear{2005}
%\date{}
%\chairmanname{}
%\chairmantitle{}
%\super{}
%\supertitle[Research Head]{}
%\super{}
%\supertitle{}

%\maketitle
\vspace{3 cm}

\begin{center}
{\LARGE Anelastic spectroscopy studies of high-Tc superconductors}

\vspace{0.5cm}

{\Large Francesco Cordero }
\end{center}

{\Large \vspace{1 cm} }

\begin{center}
{\Large Submitted to the Graduate School of}

{\Large Pure and Applied Sciences}

{\Large in Partial Fulfillment of the Requirements}

{\Large for the Degree of Doctor of Philosophy in}

{\Large Engineering}

{\Large at the}

{\Large University of Tsukuba}

October 2005 
\end{center}

%\begin{abstract}
%Replace with the thesis abstract
%\end{abstract}

\newpage

{\center {\textbf {\large Abstract}}}

Two families of cuprates exhibiting high-$T_{c}$ superconductivity, YBa$_{2}%
$Cu$_{3}$O$_{6+x}$ (YBCO) and La$_{2{}x}$Sr$_{x}$CuO$_{4+\delta}$ (LSCO), have
been extensively studied by anelastic spectroscopy, by measuring the complex
dynamic Young's modulus $E\left(  \omega,T\right)  =E^{\prime}+iE^{\prime
\prime}$ of ceramic samples at frequencies of 0.5-20~kHz between 1 and 900~K.
Results are also presented on oxygen vacancies in the ruthenocuprate
RuSr$_{2}$GdCu$_{2}$O$_{8-\delta}$. The elastic energy loss curves as a
function of temperature, $Q^{-1}\left(  \omega,T\right)  =$ $E^{\prime\prime
}/E^{\prime}$, contain peaks at the temperatures $T_{m}$ such that $\omega
\tau(T_{m})~\simeq$ 1, where $\omega/2\pi$ is the measuring frequency and
$\tau$ is the relaxation time of any microscopic process coupled to strain,
like hopping of O atoms, tilting of O\ octahedra or fluctuations of the hole
stripes. By measuring such anelastic spectra at different frequencies, it is
therefore possible to selectively probe the dynamics of the various relaxation
processes, precisely determining their characteristic times $\tau\left(
T\right)  $. The reliability of the anelastic experiments and of the
assignments to various microscopic mechanisms is also discussed.

The richest anelastic spectra are those of LSCO, and have been studied in the
whole range of Sr doping ($0<x<0.2$) and at few Ba doping levels. As in the
parent perovskite compounds, LSCO is made of O octahedra unstable against
tilting, which give rise to low symmetry phases. The layered coordination of
such octahedra and the possibility of achieving a low density of pinning
defects (interstitial O atoms), makes it possible to observe solitonic tilt
waves and fast local motion of the octahedra among the several minima of the
local tilt potential far below the structural transition temperature. The fast
local motion is driven by tunneling of the O atoms and is enormously enhanced
and accelerated by even small doping, demonstrating direct coupling between
the tilt modes of the octahedra and the hole excitations.

In addition, it has been possible to probe the dynamics of the stripes into
which the holes segregate, at liquid He temperature, when they act as walls
between domains of antiferromagnetically correlated spins (cluster spin
glass), and also at higher temperature, when they can overcome by thermal
activation the pinning barriers provided by Sr$^{2+}$ dopants. A picture is
proposed in which the high temperature motion of the stripes involves the
formation of kink pairs, while at lower temperature only the motion of the
existing kinks can occur.

In YBCO the anelastic spectra are dominated by the motion of the O atoms in
the CuO$_{x}$ planes, which are responsible for the doping of the charge
carriers (holes) into the superconducting CuO$_{2}$ planes. The doping level,
and therefore the superconducting properties, are determined not only by the
content $x$ of highly mobile non-stoichiometric O, but also by its ordering.
The various processes involved in ordering and diffusion of O have different
characteristic times $\tau$, and therefore produce distinct peaks in the
anelastic spectra at the temperatures for which $\omega\tau(T_{m})=$ 1; they
have been studied in the whole stoichiometry ($0<x<1$) and temperature
(50~K~$<T<800$~K) ranges. It is shown that there are three types of O jumps
with different rates, depending whether they involve:  \textit{i)}  ordered
Cu-O chains in the orthorhombic O-I phase ($x$~$\sim$ 1); \textit{ii)} sparser
chains fragments in the O-II and tetragonal phases; \textit{iii)} isolated O
atoms. The first two types of jumps occur over a barrier of $\sim$1.0~eV,
whereas the latter has a barrier of only 0.11~eV. It is discussed how such
widely different barriers for O hopping are possible and how the
extraordinarily high mobility of isolated O atoms is compatible with the slow
times for O ordering. In addition to the diffusive jumps, hopping of O between
off-center positions within the Cu-O chains is proposed to occur. Finally, an
anelastic process has been observed, whose intensity increases steeply in the
overdoped state, $x>0.9$, where all the other physical properties remain
practically constant. Such a process is attributed to the reorientation of
small bipolarons on orbitals that do not contribute to the electrical
conduction, and can be useful for characterizing materials with non-optimal O
content, like thin films.

RuSr$_{2}$GdCu$_{2}$O$_{8-\delta}$ is isostructural with YBCO, except for
RuO$_{2-\delta}$ planes with $\delta$ below few percents instead of the widely
nonstoichiometric CuO$_{x}$ planes. In this case, the situation of O\ mobility
and ordering is simpler than in YBCO, and it is shown that it can be
quantitatively explained in terms of hopping of O vacancies whose elastic
quadrupoles are weakly interacting and start becoming parallel to each other
below $T_{\mathrm{C}}\sim470$~K.
\newpage

\tableofcontents
{\bf Bibliography}

{\bf List of acronyms and symbols}

{\bf Acknlowledgments}

\newpage
\chapter{Introduction}

The cuprates exhibiting high-$T_{\text{c}}$ superconductivity (HTS) have been
receiving enormous attention in the scientific literature, due to the great
number of interesting physical effects they exhibit and to their technological
applications such as Superconducting Quantum Interference Devices (SQUID)\ for
highly sensitive measurements of magnetic fields, filters for the
transmissions with mobile phones, or electric power applications \cite{LGF01}.
Among the most studied and not yet completely understood issues are those
connected with nonstoichiometric oxygen, its ordering and role in doping, and
those connected with spin and charge inhomogeneities on the scale of
nanometers, generally called stripes (see Ref. \cite{KBF03} for a recent review).

In HTS cuprates, superconductivity sets in mainly in CuO$_{2}$ planes doped
with holes (or electrons in the case of Nd$_{2-x}$Ce$_{x}$CuO$_{4+\delta}$),
and doping is due to the charge unbalance from aliovalent substitutional
cations (\textit{e.g.} Sr$^{2+}$ in La$_{2-x}$Sr$_{x}$CuO$_{4}$) and from
nonstoichiometric oxygen (generally excess O$^{2-}$). The oxygen stoichiometry
may be varied over relatively wide ranges, but the amount of charge doping
depends also on the ordering of nonstoichiometric O atoms, which are the most
mobile atomic species. For this reason, the detailed knowledge of the
diffusion and ordering mechanisms of oxygen in the cuprates are of great
interest, especially in materials of the YBa$_{2}$Cu$_{3}$O$_{6+x}$ (YBCO)
family where doping is totally due to oxygen. A host of studies have been
carried out on the oxygen\ mobility and ordering, especially with diffraction
and permeation from gas phase methods, revealing complex phase diagrams and
non trivial effects. Anelastic spectroscopy is one of the most powerful
methods to study these complex phenomena, thanks to its ability of selectively
measuring different types of hopping rates, from isolated or differently
aggregated O atoms, which produce different elastic energy loss peaks in the
temperature scale.

The topic of the intrinsic charge and spin inhomogeneities in HTS cuprates is
also of great interest; in fact, not only the observation that the conducting
holes may segregate into fluctuating stripes is difficult from the
experimental point of view and counterintuitive, but it is also debated
whether it is a phenomenon competing against superconductivity \cite{Ric97} or
instead is at the basis of HTS \cite{BSR96,KFE98,Zaa00}. Rather unexpectedly,
the anelastic measurements on cuprates of the LSCO family reveal also features
attributable to the slow collective dynamics of charge stripes and
antiferromagnetic domains in the CuO$_{2}$ planes, and, to my knowledge, they
are the only experiments where some of these dynamic processes are observable
at acoustic frequencies. In fact, ac magnetic susceptibility, and dielectric,
NMR, $\mu$SR spectroscopies are dominated by the single charge or spin
fluctuations, while anelastic spectroscopy is insensitive to them and may
probe the collective charge and spin motions through their weak coupling to strain.

Two families of HTS cuprates, YBa$_{2}$Cu$_{3}$O$_{6+x}$ (YBCO) and La$_{2-x}
$Sr$_{x}$CuO$_{4}$ (LSCO), have been extensively studied by anelastic
spectroscopy, by measuring the complex dynamic Young's modulus of ceramic
samples at frequencies of 0.5-20~kHz between 1 and 900~K. In addition, some
results are presented on the ruthenocuprate compund RuSr$_{2}$GdCu$_{2}$%
O$_{8}$ (Ru-1212). All the samples have been obtained from a collaboration
with the Department of Chemistry and Industrial Chemistry of the University of
Genova, Italy (M. Ferretti), while several anelastic experiments have been
done in collaboration with the Physics Department of the University of Rome
''La Sapienza'' (G. Cannelli, R. Cantelli, A. Paolone, F. Trequattrini).

Several phenomena have been found and studied, among which diffusive hopping,
ordering and off-center dynamics of O atoms, collective and local tilt
dynamics of the octahedra in LSCO, and charge stripe fluctuations and
depinning. Unless otherwise specified, all the results presented here were the
first studies of such phenomena by anelastic spectroscopy.

The Thesis is organized as follows. First is an introduction to anelasticity,
limited to the concepts that are necessary for interpreting the phenomena
discussed later, and emphasizing those concepts that are treated little in
books on anelasticity; in particular, relaxation between energetically
inequivalent states, interaction between elastic dipoles in the mean field
approximation, and the relationship between anelastic and other
spectroscopies. A brief description follows of the method of measurement and
of the samples treatments. A chapter is devoted to LSCO, with a short
description of its structural and magnetic phase diagram, with the unstable
tilt modes of the oxygen\ octahedra, and of the charge and spin stripes. The
anelastic measurements are presented starting with the phase transformations,
then interstitial O, followed by the newly found relaxational dynamics of the
unstable tilts of the octahedra, and finally the observations of the thermally
activated depinning dynamics of the hole stripes and their motion in the
cluster spin glass phase, identifiable with the motion of pinned domain walls
between antiferromagnetic domains. The following chapter is devoted to YBCO,
starting with a presentation of its complex structural phase diagram due to
various types of ordering of O in the CuO$_{x}$ planes, and of the main
results in literature on the mobility of this oxygen species. Section
\ref{sect YBCO AR O diff} is devoted to the diffusive jumps of oxygen and
starts with an overview of the anelastic spectra at different stoichiometries,
followed by a discussion of what kind of effects one might expect from the
interaction among oxygen atoms, at least in the simple Bragg-Williams
approximation. Then, the three distinct elastic energy loss peaks due to the
oxygen diffusive jumps are discussed, with emphasis on the extremely fast
hopping rate of the isolated O atoms in the semiconducting state, and on the
role of the charge transfer between Cu-O chains and CuO$_{2}$ planes in
determining various types of O jumps; also the ordering dynamics of oxygen is
discussed. The chapter on YBCO terminates with a relaxation process identified
with the hopping of O between off-center position in the Cu-O chains, which
diffraction studies suggest to be slightly zig-zag instead of straight, and
with a peak which develops for $x>0.85$, a doping range where all the physical
properties are practically constant, and attributed to reorientation of pairs
of holes (bipolarons) in the apical O atoms.

The anelastic spectra of LSCO and YBCO reflect the great complexity of the
structural, magnetic and charge phenomena occurring in these cuprates, and
their interpretation is not always straightforward; therefore, at the end of
the chapters devoted to LSCO and YBCO, a summary of the main results is
provided, together with brief explanations of how the various anelastic
effects have been assigned to specific mechanisms.

Finally, a chapter is devoted to the diffusive hopping of oxygen in the
ruthenocuprate RuSr$_{2}$GdCu$_{2}$O$_{8-\delta}$, where it is more
appropriate to talk of oxygen vacancies, since in the RuO$_{2-\delta}$ planes
the oxygen stoichiometry is rather stable and close to the maximum. In this
case, therefore, there is no complex phase diagram for the oxygen ordering,
and the hopping dynamics may be described in terms of interaction among the
elastic dipoles in the mean-field approximation.

\chapter{Anelasticity\label{sect anel}}

The anelastic spectroscopy consists in the measurement of the complex dynamic
compliance, or its reciprocal, the elastic stiffness, generally as a function
of temperature at fixed frequencies. It is the mechanical analogue of the
dielectric spectroscopy or ac magnetic susceptibility. Comprehensive
treatments of the theory of anelasticity and of the application of the
anelastic spectroscopy to the study of solids can be found in the seminal book
of C. Zener \cite{Zen48}, in the classic book by Nowick and Berry \cite{NB72}
and in the most recent book edited by Schaller, Fantozzi and Gremaud
\cite{SFG01}. The use of tensors for describing the elastic properties of
solids can be found in books like Ref. \cite{Nye57,LB1,SS82}. In the present
chapter I will only mention what is strictly necessary for defining the
notation and for the comprehension of the results discussed in the Thesis,
with emphasis on the issues that are not treated in the above texts.

\subsection{Elastic dipole and thermodynamics of the
relaxation\label{sect thermod}}

\begin{figure}[tbh]
\begin{center}
\includegraphics[
%natheight=312.250000pt,
%natwidth=710.625000pt,
%height=6.6426cm,
width=15.0447cm
]{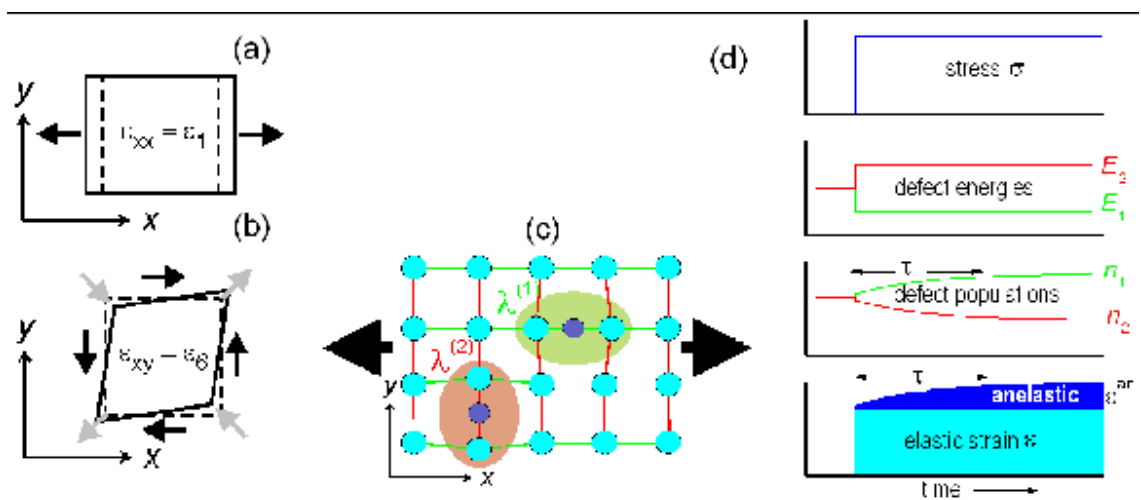}
\end{center}
\caption{(a) Uniaxial and (b) shear strains. (c) Interstitial atoms and
corresponding elastic dipoles under the application of a uniaxial $\sigma_{1}$
stress and (d)\ effect of $\sigma_{1}$ on the defects and strain.}%
\label{fig anelasre}%
\end{figure}

For a perfectly elastic solid, Hooke's law can be written (in matrix instead
of tensor notation) as
\begin{equation}
\varepsilon_{i}=s_{ij}\sigma_{j} \label{Hooke s}%
\end{equation}
where $\varepsilon_{i}$ and $\sigma_{i}$ are components of the strain and
stress tensors and $s_{ij}$ the elastic \textbf{compliance} matrix, and
summation over repeated indices is understood, or equivalently
\begin{equation}
\sigma_{i}=c_{ij}\varepsilon_{j}\, \label{Hooke c}%
\end{equation}
where $c_{ij}$ is the \textbf{stiffness} matrix. In matrix notation, index
$i=1,2,3$ denote uniaxial strains along $x,y$ and $z$ respectively, while
$i=4,5,6$ denote shears of type $yz,xz$ and $xy$ (see Fig. \ref{fig anelasre}%
a,b; note that a shear strain is equivalent to two perpendicular uniaxial
strains at 45$^{\mathrm{o}}$ with different sign and equal magnitude, as shown
by the gray arrows). Anelasticity results from the response to the application
of a stress from defects or excitations that can change their state and also
their contribution to the overall strain, with a characteristic relaxation
time $\tau$; then, the elastic response according to the above equations is
accompanied by a retarded anelastic response.

For simplicity, I will refer to a molar concentration $c$ of point defects
uniformly distributed over the solid, and having at least two possible states;
\textit{e.g.} interstitial atoms that visit sites of type 1 and 2, according
whether the first neighboring lattice atoms along the $x$ or $y$ directions
(Fig. \ref{fig anelasre}c).

One then defines the concentrations $c_{1}$ and $c_{2}$ of defects
in states 1 and 2, with $c_{1}+c_{2}=c$ and their specific contributions to
strain
\begin{equation}
\lambda_{ij}^{(\alpha)}=\frac{\partial\varepsilon_{ij}}{\partial c_{\alpha}}
\label{lambda}%
\end{equation}
where $\lambda_{ij}^{(\alpha)}$ is the \textbf{elastic dipole} of the defects
of type $\alpha$. The definition ''elastic dipole''\ derives from the analogy
with the electric or magnetic dipoles of polar or magnetic defects
\cite{NB72}, but, being a centrosymmetric strain tensor of the 2nd rank and
not a vector, it is actually a \textbf{quadrupole} \cite{LB1} representable as
an ellipsoid with the principal axes defined by the tensor eigenvalues. This
is shown in Fig. \ref{fig anelasre}c for interstitial atoms causing greater
lattice expansion in the direction of the nearest neighbor atoms. From Eq.
(\ref{lambda}) it follows that the anelastic strain is
\begin{equation}
\varepsilon_{ij}^{\text{an}}=\sum_{\alpha}c_{\alpha}\lambda_{ij}^{(\alpha
)}\;\,. \label{e_an}%
\end{equation}
From now on I will drop the tensor indices unless necessary, and assume that a
pure type of stress is applied, \textit{e.g.} uniaxial, and the corresponding
components of strain and compliance are probed.

It is easy to show with a thermodynamic argument that the elastic dipole is
also minus the rate of change of the elastic energy of a defect on application
of stress. In fact, the differential of the Gibbs free energy $g$ per unit
volume (all the extensive variables are expressed per unit volume) is:
\begin{equation}
dg=-\varepsilon d\sigma-sdT+\frac{1}{v_{0}}\sum_{\alpha}E_{\alpha}dc_{\alpha}%
\end{equation}
where $v_{0}$ is the molecular volume and $s$ here is the entropy per unit
volume; differentiating twice one obtains that
\begin{equation}
\lambda^{\left(  \alpha\right)  }=\frac{\partial\varepsilon}{\partial
c_{\alpha}}=-\frac{\partial^{2}g}{\partial c_{\alpha}\partial\sigma}=-\frac
{1}{v_{0}}\frac{\partial E_{\alpha}}{\partial\sigma}. \label{l=-dE/ds}%
\end{equation}
and therefore the application of a stress $\sigma$ changes the \textbf{defect
elastic energy} as
\begin{equation}
E_{\alpha}\left(  \sigma\right)  =E_{\alpha}\left(  0\right)  -v_{0}%
\lambda^{\alpha}\sigma\,. \label{elastic energy}%
\end{equation}

The defects occupy the possible states (or energetic levels) according to some
distribution function and, for the sake of simplicity, let us considered
diluted defects whose populations obey the Boltzmann distribution function;
then
\begin{equation}
c_{\alpha}=c\frac{e^{-E_{\alpha}/k_{\text{B}}T}}{Z}\,,\quad Z=\sum_{\alpha
}e^{-E_{\alpha}/k_{\text{B}}T}. \label{c Boltzmann}%
\end{equation}
The application of a stress $\sigma$ (see Fig. \ref{fig anelasre}c and
d)\ changes the energies $E_{\alpha}\left(  \sigma\right)  $ of
\begin{equation}
\delta E_{\alpha}=\sigma\frac{\partial E_{a}}{\partial\sigma}=-v_{0}%
\lambda^{\left(  \alpha\right)  }\sigma\label{dE(s)}%
\end{equation}
and the new thermal equilibrium requires a repopulation of the states (with a
relaxation time $\tau$) such that%

\begin{equation}
\delta c_{\alpha}=\sum_{\beta}\frac{\partial c_{\alpha}}{\partial E_{\beta}%
}\frac{\partial E_{\beta}}{\partial\sigma}\sigma=-\sum_{\beta}\frac{\partial
c_{\alpha}}{\partial E_{\beta}}\,v_{0}\lambda^{\left(  \beta\right)  }%
\sigma\,,
\end{equation}
which results in the anelastic strain%

\begin{equation}
\varepsilon^{\text{an}}=\sum_{\alpha}\lambda^{\left(  \alpha\right)  }\,\delta
c_{\alpha}=-v_{0}\sum_{\alpha}\sum_{\beta}\frac{\partial c_{\alpha}}{\partial
E_{\beta}}\lambda^{\left(  \alpha\right)  }\lambda^{\left(  \beta\right)
}\,\sigma
\end{equation}
It can be shown \cite{33} that (the case of only two states is trivial)%

\begin{equation}
\varepsilon^{\text{an}}=v_{0}\sigma\sum_{\text{pairs }\alpha<\beta}%
c\frac{n_{\alpha}n_{\beta}}{k_{\text{B}}T}\left(  \lambda^{\left(  a\right)
}-\lambda^{\left(  \beta\right)  }\right)  ^{2}%
\end{equation}
where $n_{\alpha}=c_{\alpha}/c$, which can also be extended to non-Boltzmann
statistics \cite{33}. The \textbf{relaxation strength} is defined as%

\begin{equation}
\Delta=\frac{\varepsilon^{\text{an}}}{\varepsilon^{\text{el}}}=\frac{v_{0}}%
{s}\sum_{\text{pairs }\alpha<\beta}c\frac{n_{\alpha}n_{\beta}}{k_{\text{B}}%
T}\left(  \lambda^{\left(  a\right)  }-\lambda^{\left(  \beta\right)
}\right)  ^{2}=\sum_{\text{pairs }\alpha<\beta}\Delta_{a\beta} \label{relstr0}%
\end{equation}
and this equation tells us that \textit{i)} the anelastic response is the sum
of the partial contributions $\Delta_{\alpha\beta}$ from the repopulation of
all the pairs of defect states; \textit{ii)} $\Delta_{\alpha\beta}$ is
proportional to the square of the anisotropy of the elastic dipole $\left(
\lambda^{\left(  \alpha\right)  }-\lambda^{\left(  \beta\right)  }\right)
^{2}$ and therefore elastically equivalent states ($\lambda^{\left(
\alpha\right)  }=\lambda^{\left(  \beta\right)  }$) are not repopulated with
respect to each other and do not cause relaxation; \textit{iii)}
$\Delta_{\alpha\beta}$ is proportional to the defect concentration but also to
the \textbf{depopulation factor} $n_{\alpha}n_{\beta}$, meaning that if the
two states $\alpha$ and $\beta$ are energetically inequivalent also in the
absence of stress, then one of the two is less populated and this limits the
stress-induced repopulation and the relaxation intensity. Note that in case of
high density of defects \cite{33}, the term $n_{\alpha}n_{\beta}$ becomes of
the type $n_{\alpha}\left(  1-n_{\alpha}\right)  n_{\beta}\left(  1-n_{\beta
}\right)  $ since \textit{e.g.} the jump of a defect from a site $\alpha$ to a
site $\beta$ is proportional to $n_{\alpha}$ but also requires that site
$\beta$ is empty, hence the term $\left(  1-n_{\beta}\right)  $, and
analogously for the $\beta\rightarrow\alpha$ transitions. Such expressions are
also symmetric in $n_{\alpha}$ and $1-n_{\alpha}$, meaning that, \textit{e.g.
}when dealing with the jumps of the O atoms in the CuO$_{2c}$ or RuO$_{2c}$
planes of YBCO or Ru-1212 ($0<c<1$) for $c\rightarrow0$ the O atoms are the
defects, but for $c\rightarrow1$ the O vacancies can rather be considered as
the defects. Finally, the $1/k_{\text{B}}T$ term comes from $\frac{\partial
n_{\alpha}}{\partial E_{\beta}}\propto\left(  k_{\text{B}}T\right)  ^{-1}$,
which is a consequence of the fact that with increasing temperature all the
defect states tend to become equiprobable.

For the case of relaxation between only two states with
\begin{equation}
\Delta\lambda=\lambda_{2}-\lambda_{1}\,,\quad\Delta E=E_{2}-E_{1}
\label{dl dE}%
\end{equation}
one obtains
\begin{equation}
\Delta\left(  T\right)  =\frac{v_{0}c\left(  \Delta\lambda\right)  ^{2}%
}{4s\,k_{\text{B}}T\,\cosh^{2}\left(  \Delta E/2k_{\text{B}}T\right)  }
\label{relstr}%
\end{equation}
which further reduces to the well known expression
\begin{equation}
\Delta\left(  T\right)  =\frac{v_{0}c\left(  \Delta\lambda\right)  ^{2}%
}{4s\,k_{\text{B}}T\,} \label{relstr1}%
\end{equation}
for equivalent states with $\Delta E=0$. The term $\cosh^{-2}\left(  \Delta
E/2kT\right)  $, coming from the depopulation factor $n_{1}n_{2}$ in Eq.
(\ref{relstr0}) is generally overlooked, but becomes very important in all
situations in which $\Delta E\sim k_{\text{B}}T$, since it produces a maximum
in $\Delta\left(  T\right)  $ at $T=0.65\Delta E/k_{\text{B}}$ and then falls
off as $\exp\left(  -\Delta E/k_{\text{B}}T\right)  $ for $T\rightarrow0$. For
example, when dealing with jumps of the O atoms in the CuO$_{x}$ planes of
YBCO, oxygen\ can pass from the isolated to the aggregated state and
\textit{vice versa}, which might well differ in energy by several tenths of
electronvolt, which means thousands of kelvin in the temperature scale (I will
often measure the energy in Kelvin, by considering $E/k_{\text{B}}$ instead of
$E;$ the conversion is 1~eV\thinspace= 11600~K); the $\cosh^{-2}\left(  \Delta
E/2kT\right)  $ term would cause a reduction of $\Delta\left(  T=500~\text{K}%
\right)  $ by a factor 0.04 for $\Delta E\sim0.2$~eV and a factor $0.004$ for
$\Delta E\sim0.3$~eV, making certain types of jumps completely unobservable in
the anelastic relaxation.

The depopulation factor becomes important at low enough temperature even for
relaxation between states that ideally are energetically equivalent; in fact,
unless dealing with extremely low concentrations of impurities in crystals of
high perfection, there will always be long range elastic interactions among
defects that cause random shifts $\Delta E$ to the defect energies. Such
shifts have been estimated for the case of O-H\ pairs in Nb to be of the order
of 100~K for impurity concentrations of $\sim1$~at\% \cite{WN84}. This means
that, especially for disordered solids like the HTS, relaxation processes
below 100~K are very likely affected by the depopulation factor, whose effect
is of changing the temperature dependence of the relaxation intensity from
$T^{-1}$ to a nearly constant or increasing function of $T$. In such cases, I
will include the depopulation factor in the analysis.

\subsection{Relaxation kinetics\label{sect kin}}

There is no general treatment for the relaxation kinetics and I\ will limit to
the relaxation between two states from the start. In this case the rate
equations for the defect populations are
\begin{equation}
\left\{
\begin{array}
[c]{c}%
\dot{n}_{1}=-\nu_{21}n_{1}+\nu_{12}n_{2}\\
\dot{n}_{2}=+\nu_{21}n_{1}-\nu_{12}n_{2}%
\end{array}
\right.  \,
\end{equation}
with $\nu_{ij}$ the rate for passing from $j$ to $i$; thanks to the condition
$n_{1}+n_{2}=1$ there is actually one independent $n_{i}$. Setting $\dot
{n}_{i}=0$ the equilibrium populations are found as%

\begin{equation}
\overline{n}_{i}=\frac{\nu_{ij}}{\nu_{12}+\nu_{21}}\,,
\end{equation}
which agree with the thermodynamic result if $\nu_{12}$ e $\nu_{21}$ satisfy
the detailed balance principle
\begin{equation}
\frac{\overline{n}_{1}}{\overline{n}_{2}}=\frac{\nu_{12}}{\nu_{21}}=e^{\Delta
E/k_{\text{B}}T}\text{\thinspace.} \label{detailb}%
\end{equation}
Incidentally, Eq. (\ref{detailb}) requires that, for hopping over a saddle
point $E_{\text{s}}$ according to the \textbf{Arrhenius law},
\begin{equation}
\nu_{ij}=\frac{1}{2}\tau_{0}^{-1}\exp\left(  \frac{E_{\text{s}}-E_{j}%
}{k_{\text{B}}T}\right)  \label{Arrh}%
\end{equation}
with the same $\tau_{0}$ for both states. Defining
\begin{equation}
\Delta n=n_{2}-n_{1}\,,
\end{equation}
the previous equations can be put in the form
\begin{subequations}
\begin{align}
\Delta\dot{n}  &  =-\frac{\Delta n-\Delta\overline{n}}{\tau}\label{rate_eqn}\\
\tau^{-1}  &  =\left(  \nu_{12}+\nu_{21}\right)  =\tau_{0}^{-1}\exp\left(
-E/k_{\text{B}}T\right)  \cosh\left(  \Delta E/2k_{\text{B}}T\right)
\label{tau asym}%
\end{align}
which says that the rate for $\Delta n$ reaching the equilibrium value
$\Delta\overline{n}$ is proportional to its deviation from $\Delta\overline
{n}$ through the relaxation rate $\tau^{-1}$; the latter has been written in
terms of the mean activation energy $E=E_{\text{s}}-\frac{1}{2}\left(
E_{1}+E_{2}\right)  $ and therefore contains a factor $\cosh\left(  \Delta
E/2k_{\text{B}}T\right)  $ that should be taken into account when dealing with
asymmetric states.

On application of a time dependent stress $\sigma\left(  t\right)  $, there
will be an instantaneous elastic response
\end{subequations}
\begin{equation}
\varepsilon^{\text{el}}\left(  t\right)  =s_{\text{U}}\sigma\left(  t\right)
\label{e_el}%
\end{equation}
where subscript ''U'' stands for unrelaxed, and in addition an anelastic
strain $\varepsilon^{\text{an}}=$ $c\left(  n_{1}\lambda_{1}+n_{2}\lambda
_{2}\right)  =$ $c\left[  \frac{\lambda_{1}+\lambda_{2}}{2}+\frac{1}{2}%
\Delta\lambda\Delta n\right]  $ where the first term is constant and the time
dependent anelastic response is
\begin{equation}
\varepsilon^{\text{an}}\left(  t\right)  =\frac{c}{2}\Delta\lambda\Delta
n\left(  t\right)  \label{e_an(t)}%
\end{equation}
with $\Delta n$ determined by Eq. (\ref{rate_eqn}).

\subsection{Dynamic compliance\label{sect dyn s}}

Let us calculate the time dependent anelastic response Eq. (\ref{e_an(t)}) on
application of a periodic stress $\sigma=\sigma_{0}e^{i\omega t}$ that
modulates $\Delta E$. It is convenient to refer $\Delta n$ to the equilibrium
values in the absence of stress, defining:
\begin{equation}
\Delta n^{\prime}=\Delta n-\Delta\overline{n}\left(  \sigma=0\right)
\end{equation}
and for the instantaneous equilibrium $\Delta\overline{n}=-\tanh\left(  \Delta
E/2k_{\text{B}}T\right)  $%
\begin{equation}
\Delta\overline{n}^{\prime}=\Delta\overline{n}-\Delta\overline{n}\left(
\sigma=0\right)  =\frac{d\Delta\overline{n}}{d\Delta E}\frac{d\Delta
E}{d\sigma}\sigma=\frac{v_{0}\Delta\lambda}{2k_{\text{B}}T\cosh^{2}\left(
\Delta E/2k_{\text{B}}T\right)  }\sigma
\end{equation}
By substituting into Eq. (\ref{rate_eqn})%

\begin{equation}
\Delta n^{\prime}=\frac{v_{0}\Delta\lambda\sigma}{2k_{\text{B}}T\cosh
^{2}\left(  \Delta E/2k_{\text{B}}T\right)  }\frac{1}{1+i\omega\tau}%
\end{equation}
and using Eqs. (\ref{e_el},\ref{e_an(t)}) the dynamic compliance can be
written as
\begin{equation}
s\left(  \omega\right)  =s^{\prime}-is^{\prime\prime}=\frac{\varepsilon
^{\text{el}}+\varepsilon^{\text{an}}}{\sigma}=s_{\text{U}}\left[
1+\frac{\Delta\left(  T\right)  }{1+i\omega\tau}\right]
\end{equation}
where $\Delta\left(  T\right)  $ is the same as given by Eq. (\ref{relstr}).
The real and imaginary parts of $s\left(  \omega\right)  $ are therefore
\begin{align}
s^{\prime}\left(  \omega,T\right)   &  =s_{\text{U}}\left[  1+\frac
{\Delta\left(  T\right)  }{1+\left(  \omega\tau\right)  ^{2}}\right]
\label{s Debye}\\
s^{\prime\prime}\left(  \omega,T\right)   &  =s_{\text{U}}\Delta\left(
T\right)  \frac{\omega\tau}{1+\left(  \omega\tau\right)  ^{2}}\nonumber
\end{align}
where the $s^{\prime\prime}$ presents the well known \textbf{Debye peak} of
amplitude $\frac{1}{2}\Delta\left(  T\right)  $ at the\textbf{\ condition for
maximum relaxation}
\begin{equation}
\omega\tau=1\,, \label{wt=1}%
\end{equation}
while $s^{\prime}$ a step of amplitude $\Delta\left(  T\right)  $. In the
$\omega\tau\gg1$ limit the defects are too slow to follow the periodic stress
and corresponds to the elastic limit without defects ($s^{\prime}=s_{\text{U}%
}$, $s^{\prime\prime}=0$). In the $\omega\tau\ll1$ limit the defect relaxation
is so fast that instantaneously complies to the periodic stress,
$s^{\prime\prime}=0$, and the compliance is totally relaxed, $s^{\prime
}=s_{\text{U}}\left(  1+\Delta\right)  $.

It is possible to make an analogous derivation for the \textbf{dynamic
stiffness}
\begin{equation}
c\left(  \omega,T\right)  =c^{\prime}+ic^{\prime\prime}\,,
\end{equation}
finding
\begin{align}
c^{\prime}  &  =c_{\text{U}}\left(  1-\frac{\Delta\left(  T\right)
}{1+\left(  \omega\tau\right)  ^{2}}\right) \label{M}\\
c^{\prime\prime}  &  =c_{\text{U}}\Delta\left(  T\right)  \frac{\omega\tau
}{1+\left(  \omega\tau\right)  ^{2}}\nonumber
\end{align}

\section{Elastic energy loss and anelastic spectrum}

The fact that the dynamic compliance $s\left(  \omega,T\right)  $ or stiffness
$c\left(  \omega,T\right)  $ is complex means that, due to the retarded
anelastic response, strain is out-of-phase with respect to stress by the loss
angle
\begin{equation}
\tan\phi=\frac{s^{\prime\prime}}{s^{\prime}}=\frac{c^{\prime\prime}}%
{c^{\prime}}\,.
\end{equation}
Generally, it is $\Delta\left(  T\right)  \ll1$ so that
\begin{equation}
\tan\phi\simeq\frac{s^{\prime\prime}}{s_{\text{U}}}=\Delta\left(  T\right)
\frac{\omega\tau}{1+\left(  \omega\tau\right)  ^{2}}\,. \label{Debye peak}%
\end{equation}
The tangent of the loss angle can be measured from the dissipation of elastic
energy or acoustic absorption; in fact, if $\sigma=\sigma_{0}\cos\left(
\omega t\right)  $ and $\varepsilon=\varepsilon_{0}\cos\left(  \omega
t-\phi\right)  $, the elastic energy dissipated in one vibration cycle is
\begin{equation}
\Delta W=\int_{\omega t=0}^{\omega t=2\pi}\sigma d\varepsilon=\pi
\varepsilon_{0}\sigma_{0}\sin\phi\nonumber
\end{equation}
while the maximum elastic energy stored is $W=\int_{\omega t=0}^{\omega
t=\pi/2}\sigma d\varepsilon^{\prime}=$ $\frac{1}{2}\varepsilon_{0}\sigma
_{0}\cos\phi$, where $\varepsilon^{\prime}=\varepsilon_{0}\cos\phi\cos\left(
\omega t\right)  $ is the strain component in phase with $\sigma$. One defines
the \textbf{elastic energy loss coefficient} as
\begin{equation}
Q^{-1}=\frac{1}{2\pi}\frac{\Delta W}{W}=\tan\phi\,\text{,} \label{Q-1}%
\end{equation}
which coincides with the reciprocal of the mechanical $Q$ of the sample.

\begin{figure}[tbh]
\begin{center}
\includegraphics[
%natheight=324.687500pt,
%natwidth=577.687500pt,
%height=8.0484cm,
width=14.2693cm
]{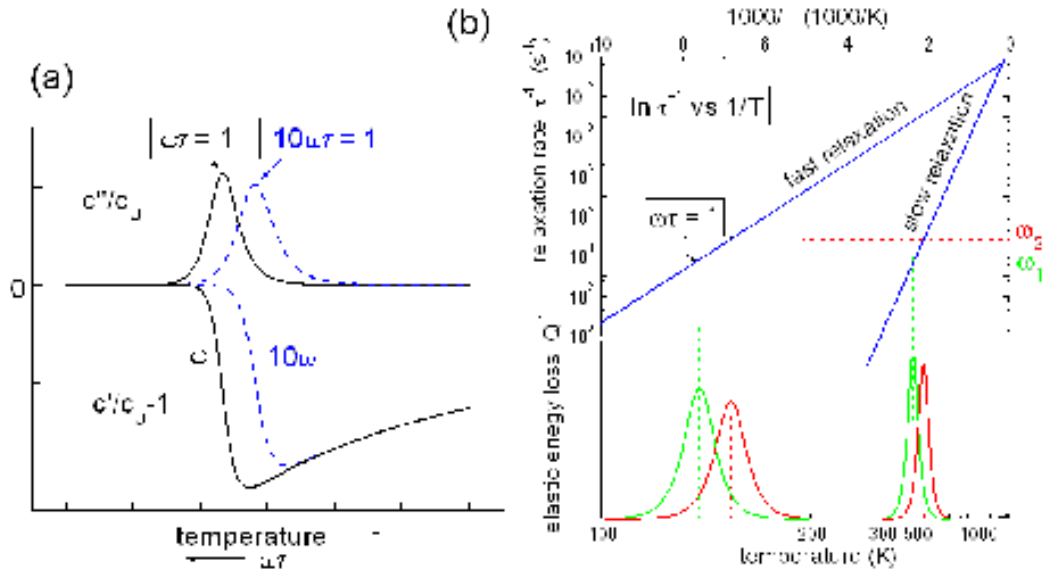}
\end{center}
\caption{(a) Contribution of a Debye relaxation process to the imaginary and
real parts of the dynamic stiffness at two measuring frequencies. (b)
Anelastic spectrum with two processes having the relaxation rates plotted in
the upper part; measured at the two frequencies $\omega_{1}$ and $\omega_{2}%
$.}%
\label{fig ArrDebye}%
\end{figure}

Generally, the anelastic measurements are made sweeping temperature at nearly
fixed frequency $\omega$, and the resulting spectrum contains absorption peaks
in correspondence with the temperatures $T_{\text{max}}$ for which the
condition of maximum relaxation $\omega\tau\left(  T_{\text{max}}\right)  =1$
is verified, while $c^{\prime}$ ($s^{\prime}$) contain negative (positive)
steps. This is shown in Fig. \ref{fig ArrDebye}, assuming that the relaxation
rates follow the Arrhenius law, $\tau^{-1}=\tau_{0}^{-1}\exp\left(
E/k_{\text{B}}T\right)  $, between states with the same energy, so that
$\Delta\left(  T\right)  \propto1/T$. Figure \ref{fig ArrDebye}b shows how
slower processes are peaked at higher temperature, and how it is possible to
estimate the energy barrier $E$ by measuring at different frequencies,
exploiting the condition $\tau\left(  T_{\text{max}}\right)  =\omega^{-1}$ at
the peak temperatures.

It is more convenient to analyze the $Q^{-1}\left(  \omega,T\right)  $ curves
rather than the \textbf{real parts} $s^{\prime}\left(  \omega,T\right)  $ or
$c^{\prime}\left(  \omega,T\right)  $, especially if $\Delta$ is very small,
because the absorption peaks stand out of a usually small background (at least
for $\omega<10$~MHz, when anharmonic effects and sound wave diffraction at the
boundaries are small); instead, even for a perfectly elastic solid, the real
part $s_{\text{U}}\ $or $c_{\text{U}}$ has a temperature dependence due to
\textbf{anharmonic effects} that should be subtracted in order to analyze the
anelastic contribution. In addition, the elastic moduli are affected by the
sample \textbf{porosity, by microcracks, and internal stresses due to
anisotropic thermal expansion} building up during thermal cycling, which may
result in anomalies and hystereses on varying temperature.

\begin{figure}[tbh]
\begin{center}
\includegraphics[
%natheight=432.875000pt,
%natwidth=866.500000pt,
%height=6.1418cm,
width=12.2396cm
]{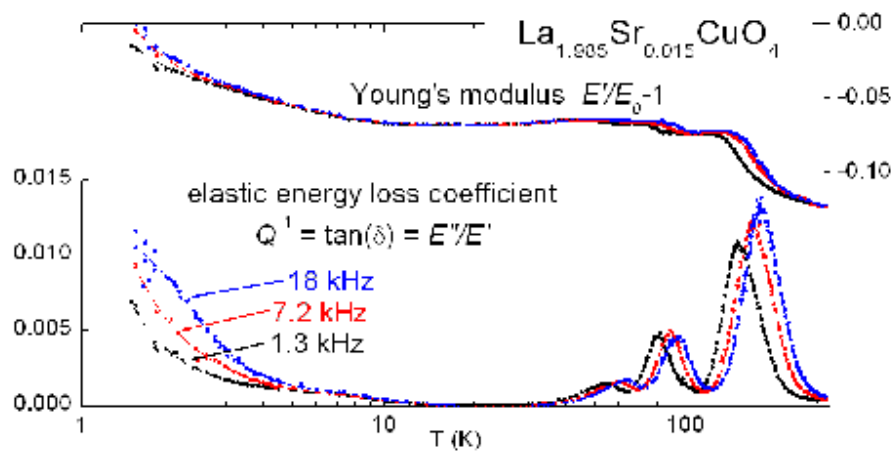}
\end{center}
\caption{Example of anelastic spectrum (real and absorption parts) of LSCO
measured at three vibration frequencies during the same run.}%
\label{fig spectrum}%
\end{figure}

Figure \ref{fig spectrum} presents an example of anelastic spectrum measured
on LSCO at three resonance frequencies during the same run; the stiffness
which is measured is the Young's modulus $E$. Note that the scale of the real
part is almost 10 times larger than that of the absorption. Note also that the
amplitudes of the steps in the real part are larger than $2\Delta$ deduced
from the amplitudes $\Delta$ of the corresponding absorption peaks, as one
expects from Eq. (\ref{s Debye}). This is due to the fact that the relaxation
processes are broadened by distributions of relaxation times; the elementary
peaks are shifted with respect to each other, so that the resulting peak
%height is smaller than the sum of the elementary amplitudes, while the
resulting step in $E^{\prime}$ is the sum of the elementary steps.

The pre-exponential factor $\tau_{0}^{-1}$ in the Arrhenius law is the
relaxation rate extrapolated to infinite temperature, and its value deduced
from experiments at low temperature should not be taken too seriously, since
one should also take into account the temperature dependence of all the
quantities that affect the jump rate, including the vibration entropies.
Still, as a rule of thumb, $\tau_{0}\sim10^{-13}$~s, of the order of magnitude
of the local vibrations promoting the atomic jump, is indicative of point
defect relaxation, while $\tau_{0}>10^{-12}$~s is indicative of extended
defects or collective motions.

\section{Elastic energy loss and spectral density of degrees of freedom
coupled to strain\label{sect FDT}}

The dynamic susceptibility $\chi$ is defined as the ratio between a response
$\partial r$ and the excitation force $\partial f$: $\chi=\frac{\partial
r}{\partial f}$; in the elastic case, the compliance $s=\frac{\partial
\varepsilon}{\partial\sigma}$ is the elastic susceptibility with $f=\sigma$
and $r=\varepsilon$. The fluctuation-dissipation theorem \cite{LL5} correlates
the imaginary part of a susceptibility $\chi^{\prime\prime}$ with the spectral
density $J\left(  \omega,T\right)  $ (Fourier transform of the autocorrelation
function) of the spontaneous fluctuations of $r$; for the elastic case and in
the classical limit it can be written \cite{LL5,WK96}:
\begin{equation}
s^{\prime\prime}\left(  \omega,T\right)  =\frac{\omega V}{k_{\text{B}}T}\int
dt\,e^{i\omega t}\left\langle \varepsilon\left(  t\right)  \varepsilon\left(
0\right)  \right\rangle =\frac{\omega V}{k_{\text{B}}T}J\left(  \omega
,T\right)  \label{FDT}%
\end{equation}
where $\left\langle ...\right\rangle $ denotes the thermal average and $V$ is
the sample volume. This important equation tells us that the elastic energy
absorption at angular frequency $\omega$ is proportional to the corresponding
Fourier component of the spontaneous strain fluctuations, which are due to any
motion or excitation coupled to strain $\varepsilon$ (as \textit{e.g.} in Eq.
(\ref{e_an})). For a process with relaxation time $\tau$, also called of the
diffusive or pseudodiffusive type, the strain autocorrelation function is
$\left\langle \varepsilon\left(  t\right)  \varepsilon\left(  0\right)
\right\rangle =<\delta\varepsilon^{2}>\exp\left(  -\left|  t\right|
/\tau\right)  $, and its Fourier transform is
\begin{equation}
J\left(  \omega\right)  =\text{ }<\delta\varepsilon^{2}>\frac{\tau}{1+\left(
\omega\tau\right)  ^{2}} \label{cp}%
\end{equation}
which, introduced into Eq. (\ref{FDT}) yields the usual Debye formula
(\ref{s Debye}) with the $T^{-1}$ thermodynamic factor (\ref{relstr1}). This
formulation of $s^{\prime\prime}\left(  \omega,T\right)  $ and therefore of
$Q^{-1}\left(  \omega,T\right)  $ will be useful when comparing anelastic and
NQR\ data in Sec. \ref{sect tiltwav}, and in general when comparing an
anelastic spectrum with those from other spectroscopies (\textit{e.g.}
neutrons or infrared). In fact, the intensities of the transmitted or
diffracted radiations are proportional to the spectral densities of the
fluctuations of the physical quantities $x$ coupled to those radiations
(nuclei positions for neutron scattering, electric polarization for light,
\textit{etc.}). A $J\left(  \omega\right)  $ of the form of Eq. (\ref{cp}) is
also called a ''central peak'' for the following reason. Motions $x\left(
t\right)  $ with resonance frequency $\omega_{0}$ and mean lifetime $\tau$
(vibrations, excitations with energy $\hbar\omega_{0}$, tunneling) have a
correlation function $\left\langle x\left(  t\right)  x\left(  0\right)
\right\rangle \propto$ $\cos\omega_{0}t\exp\left(  -\left|  t\right|
/\tau\right)  $, and their spectral density is a lorentzian peak $J\left(
\omega\right)  =\frac{\tau^{-1}}{\left(  \omega-\omega_{0}\right)  ^{2}%
+\tau^{-2}} $ centered in $\omega_{0}$ and with width $\tau^{-1}$; Eq.
(\ref{cp}) can be obtained by setting $\omega_{0}=0$, and therefore is a peak
centered at the origin of the frequency or energy scale.

\begin{figure}[tbh]
\begin{center}
\includegraphics[
%natheight=241.000000pt,
%natwidth=361.375000pt,
%height=8.5119cm,
width=12.736cm
]{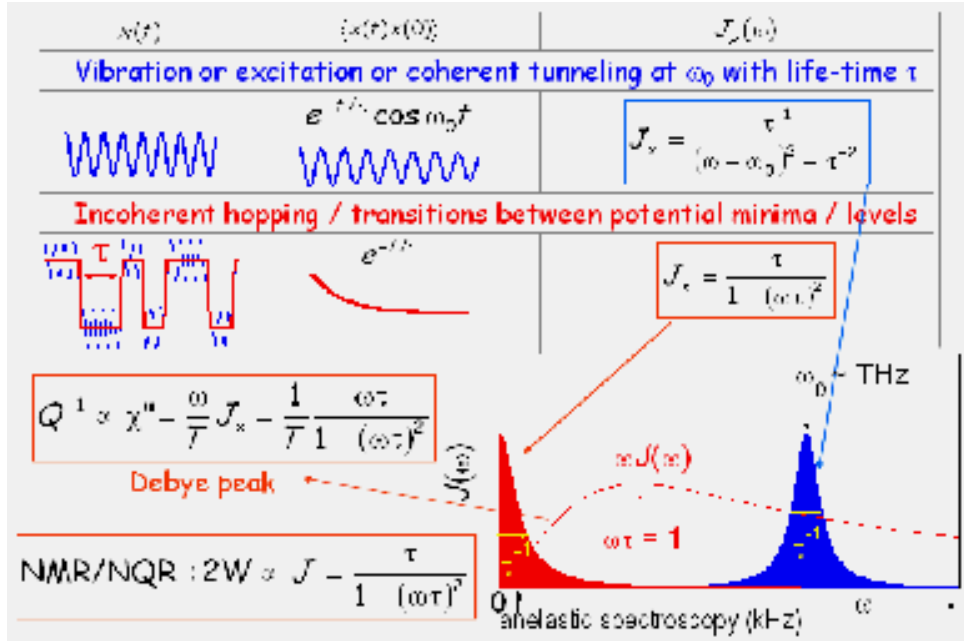}
\end{center}
\caption{Spectral densities $J\left(  \omega\right)  $ of $x\left(  t\right)
$ for resonant (blue)\ and relaxational or diffusive (red)\ processes. The
imaginary elastic compliance or energy loss is related to $J\left(
\omega\right)  $ through the fluctuation-dissipation theorem.}%
\label{fig spectral}%
\end{figure}

This is schematically shown in Fig. \ref{fig spectral}, where the physical
quantity $x\left(  t\right)  $, \textit{e.g.} atomic positions coupled to
strain $\varepsilon$ and to the electric field gradient producing NQR
relaxation (see Sec. \ref{sect tiltwav}), displays resonant and
pseudodiffusive types of motion. The spectral density of the fluctuations of
$x$ contains both the peak centered at $\omega_{0}>10^{12}$~s$^{-1}$ (local
vibration in a potential minimum) and the central peak of width $\tau^{-1}$
(hopping between two potential minima with rate $\tau^{-1}$). At acoustic
frequencies the resonant process has negligible spectral weight and only the
central peak is observed; $s^{\prime\prime}$ can be obtained from Eq.
(\ref{FDT}), while the NQR relaxation rate is given directly by $J\left(
\omega\right)  $ \cite{RBC98}.

\section{Distributions of relaxation times}

Up to now we considered anelastic relaxation from equivalent non-interacting
elastic dipoles, all having the same relaxation time $\tau$, as is the case of
very diluted solid solutions. More common are the situations with
distributions of relaxation times $g\left(  \tau\right)  $, due to static
disorder or interactions among the dipoles, which result in broader peaks in
$s^{\prime\prime}\left(  \omega,T\right)  $.

In what follows we are mainly concerned with the imaginary or absorption part
of the compliance or modulus, because it is easier to analyze than the real
part. In fact, the imaginary part $s^{\prime\prime}\left(  \omega,T\right)  $
associated with a relaxation process with characteristic time $\tau$ is peaked
around the temperature at which $\omega\tau\left(  T\right)  =1$ holds, and is
superimposed over a background contribution $s_{\text{bkg}}^{\prime\prime
}\left(  \omega,T\right)  $ which is usually small and slowly varying over the
temperature range of the peak in $s^{\prime\prime}\left(  \omega,T\right)  $.
Instead, the contribution to $s^{\prime}\left(  \omega,T\right)  $ extends
well above the temperature for which $\omega\tau\left(  T\right)  =1$, and is
superimposed to elastic compliance $s_{\text{el}}^{\prime}\left(  T\right)  $,
which is large and temperature dependent. This is at variance with the
dielectric case, where the equivalent of $s_{\text{el}}^{\prime}\left(
T\right)  $ is close to the vacuum permittivity $\varepsilon_{0}$.

\subsection{Uniform distribution of activation energies or $\ln\tau$}

The simplest distribution of relaxation times with a clear physical meaning is
a uniform distribution in the activation energy $E$: $g\left(  E\right)
=\left(  E_{2}-E_{1}\right)  ^{-1}$ for $E_{1}\le E\le E_{2}$. Setting
$\ln\tau=\ln\tau_{0}+E/T$, this is equivalent to a uniform distribution in
$\ln\tau$: $\ln\tau_{0}+E_{1}/T\le$ $\ln\tau\le$ $\ln\tau_{0}+E_{2}/T$:
\begin{equation}
g\left(  E\right)  dE=\frac{dE}{E_{2}-E_{1}}=\frac{d\left(  \ln\tau\right)
}{\ln\tau_{2}-\ln\tau_{1}}=g\left(  \ln\tau\right)  d\left(  \ln\tau\right)
\end{equation}

The imaginary part of the compliance can be easily integrated over such a
distribution, where we include a relaxation strength $\propto1/T$ [see Eq.
(\ref{relstr1})]:
\begin{align}
s^{\prime\prime}\left(  \omega,T\right)   &  =\int_{E_{1}}^{E_{2}}\frac
{dE}{\left(  E_{2}-E_{1}\right)  }\frac{\Delta s}{T}\frac{2\,\omega\tau
_{0}e^{E/T}}{1+\left(  \omega\tau_{0}e^{E/T}\right)  ^{2}}=\nonumber\\
&  =\frac{2\Delta s}{\left(  E_{2}-E_{1}\right)  }\left[  \arctan\left(
\omega\tau_{0}e^{E_{2}/T}\right)  -\arctan\left(  \omega\tau_{0}e^{E_{1}%
/T}\right)  \right]  \,.
\end{align}

This distribution, however, usually provides a poor fit to very broadened
processes, since it develops a plateau at the maximum. It has been used to
describe glassy processes, by letting $\tau_{2}\left(  T\right)  $ diverge at
the freezing temperature, and assuming a constant relaxation strength. The
omission of the $1/T$\ term in the relaxation strength transforms the broad
plateau into a linear temperature increase from the freezing temperature up to
the maximum at $\omega\tau_{2}\left(  T_{m}\right)  \simeq1$, which describes
reasonably well, for example, the dielectric maximum of some relaxor
ferroelectrics \cite{KBP00}, possibly by smoothing the cutoff at $\tau_{2}$
\cite{116}.

\subsection{Fuoss-Kirkwood distribution}

Fuoss and Kirkwood \cite{FK41} showed that, when the imaginary part of a
susceptibility $\chi^{\prime\prime}$ may be written as%

\begin{equation}
\chi^{\prime\prime}\left(  \omega\right)  =\int_{0}^{\infty}d\left(  \ln
\tau\right)  \,g\left(  \ln\tau\right)  \frac{\omega\tau}{1+\left(  \omega
\tau\right)  ^{2}}%
\end{equation}
then the distribution function $g$ of the relaxation times may be expressed by
analytical continuation in terms of $\chi^{\prime\prime}$:%

\begin{equation}
g\left(  \ln\tau\right)  =\frac{1}{\pi}\left[  \chi^{\prime\prime}\left(
-\ln\tau/\tau_{m}+i\frac{\pi}{2}\right)  +\chi^{\prime\prime}\left(  -\ln
\tau/\tau_{m}-i\frac{\pi}{2}\right)  \right]  \,, \label{gFK}%
\end{equation}
where $\tau_{m}=\omega^{-1}$ at the maximum of $\chi^{\prime\prime}$ is the
mean value of $\tau$, when $g\left(  \ln\tau\right)  $ is even in $\ln
\tau/\tau_{m}=$ $\ln\tau-\ln\tau_{m}$ and therefore $\chi^{\prime\prime}$ is
even in $\ln\omega\tau_{m}$. In this manner, it is possible to associate a
distribution function $g\left(  \ln\tau\right)  $ to any analytical form of
broadened peak in $s^{\prime\prime}\left(  \omega,T\right)  $. A commonly used
expression is
\begin{equation}
s^{\prime\prime}\left(  \omega\right)  =\frac{\alpha\left(  \omega\tau\right)
^{\alpha}}{1+\left(  \omega\tau\right)  ^{2\alpha}}=\frac{\alpha}%
{2\cosh\left[  \alpha\ln\left(  \omega\tau\right)  \right]  } \label{pFK}%
\end{equation}
with $0<\alpha\le1$; $\alpha=1$ corresponds to a Debye peak, whereas
$\alpha<1$ broadens the peak by a factor $\alpha^{-1}$ in the $\ln\tau$ and
therefore $T^{-1}$ scale (and lowers its amplitude by $\alpha$); the
corresponding distribution, called Fuoss-Kirkwood distribution \cite{NB72},
can be calculated through eq. (\ref{gFK}):
\begin{align}
g_{\text{FK}}\left(  \ln\tau\right)   &  =\frac{1}{\pi}\left[  \frac{\alpha
}{2\cosh\left[  -\alpha\ln\left(  \tau/\tau_{m}\right)  +i\alpha\frac{\pi}%
{2}\right]  }{\small +}\frac{\alpha}{2\cosh\left[  -\alpha\ln\left(  \tau
/\tau_{m}\right)  -i\alpha\frac{\pi}{2}\right]  }\right]  =\nonumber\\
&  =\frac{\alpha}{\pi}\frac{\cosh\left[  \alpha\ln\tau/\tau_{m}\right]
\cos\frac{\alpha\pi}{2}}{\cos^{2}\frac{\alpha\pi}{2}+\sinh^{2}\left[
\alpha\ln\tau/\tau_{m}\right]  }%
\end{align}
and is normalized to 1; $g_{\text{FK}}\left(  x\right)  $ can also be well
approximated with a Lorentzian
\begin{equation}
g_{\text{FK}}\left(  x\right)  \simeq\frac{\alpha}{\pi\cos\frac{\alpha\pi}{2}%
}\frac{\left(  w/2\right)  ^{2}}{x^{2}+\left(  w/2\right)  ^{2}}%
\,\,\text{with}\quad w=3.38\times\left(  \alpha^{-1}-1\right)
\label{LorentzFK}%
\end{equation}
for $\alpha$ down to $0.4$. If $\ln\tau=\ln\tau_{0}\exp\left(  E/k_{\text{B}%
}T\right)  $, and the effect of a distribution in the values of $\tau_{0}$ is
neglected in comparison to those of $E$, then $g_{\text{FK}}\left(  \ln
\tau\right)  $ can be attributed to a distribution in $E$, roughly Lorentzian
with a temperature dependent full width at half maximum $w_{E}=3.38\times
\left(  \alpha^{-1}-1\right)  k_{\text{B}}T$.

\subsection{Cole-Cole distribution}

The simple expression
\begin{equation}
\chi=\frac{1}{1+\left(  i\omega\tau\right)  ^{\alpha}}\,,\quad\chi
^{\prime\prime}=\frac{\left(  \omega\tau\right)  ^{\alpha}\sin\left(
\frac{\pi}{2}\alpha\right)  }{1+\left(  \omega\tau\right)  ^{2\alpha}+2\left(
\omega\tau\right)  ^{\alpha}\cos\left(  \frac{\pi}{2}\alpha\right)  }=\frac
{1}{2}\frac{\sin\left(  \frac{\pi}{2}\alpha\right)  }{\cosh\left[  \alpha
\ln\left(  \omega\tau\right)  \right]  +\cos\left(  \frac{\pi}{2}%
\alpha\right)  } \label{pCC}%
\end{equation}
is very popular in the analysis of the dielectric susceptibility. The
corresponding distribution function is
\begin{equation}
g_{\text{CC}}\left(  \ln\tau\right)  =\frac{1}{2\pi}\frac{\sin\left(
\pi\alpha\right)  }{\cosh\left[  \alpha\ln\left(  \tau/\tau_{m}\right)
\right]  +\cos\left(  \pi\alpha\right)  }\,. \label{gCC}%
\end{equation}

\subsection{Other expressions}

Other expressions for $\chi\left(  \omega\right)  $ and corresponding
distributions in $g\left(  \ln\tau\right)  $ are used, especially in the
dielectric literature, for example the Havriliak-Negami one,
\begin{equation}
\chi=\frac{1}{\left[  1+\left(  i\omega\tau\right)  ^{\alpha}\right]
^{\gamma}}\,. \label{pHN}%
\end{equation}
To my knowledge there are no particular physical reasons for preferring one
over the others, except the empirical fact that different processes may be
better interpolated by different expressions. For example, a process whose
$\chi^{\prime\prime}\left(  \omega,T\right)  $ is not even in $1/T$ will be
better interpolated by an expression of the Havriliak-Negami type, which is
not even in $\ln\omega\tau$, but also a simpler expression, like
\begin{equation}
\chi^{\prime\prime}\left(  \omega\right)  =\frac{1}{\left(  \omega\tau\right)
^{\alpha}+\left(  \omega\tau\right)  ^{-\beta}} \label{pFK2}%
\end{equation}
may be used \cite{48}. The latter is a generalization of the Fuoss-Kirkwood
expression, where rescaling in $1/T$ by $\alpha$ is adopted for the
low-$T$/high-$\tau$ region and rescaling by $\beta$ in the high-$T$ region. An
even more flexible expression was proposed by Jonscher \cite{Jon75},
\begin{equation}
\chi^{\prime\prime}\left(  \omega\right)  =\frac{1}{\left(  \omega\tau
_{1}\right)  ^{\alpha}+\left(  \omega\tau_{2}\right)  ^{-\beta}}
\label{Jonscher}%
\end{equation}
where $\tau_{1}\left(  T\right)  $ and $\tau_{2}\left(  T\right)  $ are two
relaxation times, possibly following the Arrhenius law, which describe the
relaxation of the system at long and short times, respectively.

\begin{figure}[h]
\begin{center}
\includegraphics[
%natheight=288.312500pt,
%natwidth=432.875000pt,
%height=2.661in,
width=3.9885in
]{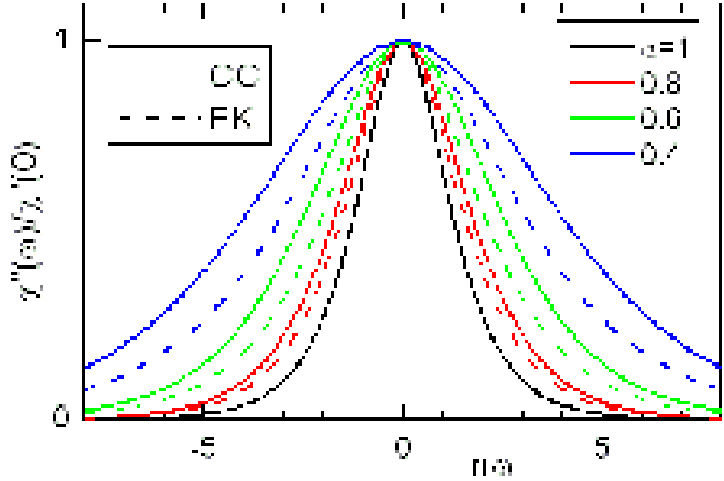}
\end{center}
\caption{Normalized $\chi^{\prime\prime}$ vs $\ln\omega$ according to the
Fuoss-Kirkwood and Cole-Cole expressions for different values of $\alpha$.}%
\label{figFKCC}%
\end{figure}

\section{Interacting elastic dipoles\label{sect interact}}

The treatment of interacting elastic dipoles is of particular importance in
the case of oxygen in the CuO$_{x}$ planes of YBCO, where the concentration of
dipoles $0.1<x<1$ is by no means small, and their interaction is so strong to
give rise to a complicated phase diagram. Under such conditions, any attempt
at describing the anelastic relaxation from the hopping of the O atoms should
somehow take into account their mutual interactions, which are both of
electronic and elastic origin. The description of the structural phase diagram
of YBCO, however, requires rather sophisticated models with asymmetric
interactions (different along the $a$ and $b$ axes) at least up to the
next-nearest neighbors (so-called ASYNNNI models \cite{FCA90}). Such models
can be solved only with Monte Carlo techniques, and are generally adopted to
reproduce the YBCO structural phase diagram, in only few cases to evaluate the
oxygen\ diffusion coefficient \cite{AP95,SPA99}, and no attempt exists to
evaluate the dynamic elastic compliance due to oxygen\ hopping. The numerical
results on the tracer diffusion coefficient \cite{AP95,SPA99} are of little
help in analyzing the anelastic data.

Wipf and coworkers \cite{BSW94} carried out an analysis of the high
temperature anelastic measurements of YBCO where the interaction between the O
atoms is treated in the Bragg-Williams or mean-field approximation; in that
manner, the complexity of the YBCO phase diagram cannot be obtained, since
only a single ordering phase transformation is reproduced, identified with the
tetragonal to orthorhombic one, but useful expressions of the dynamic
compliance can be obtained. This treatment is particularly appropriate to the
analysis of anelastic relaxation from oxygen jumps in the RuO$_{2-\delta}$
planes of Ru-1212, which can indeed be considered as a diluted solution of O
vacancies with long range elastic interactions (see Sect \ref{sect Ru V_O}).

A\ treatment of the dynamics of interacting elastic dipoles had also been
carried out previously by Dattagupta \cite{Dat82,DBR82} in the mean field
approximation, assuming that the actual elastic field that a particular dipole
senses is substituted with a mean stress field $\sigma_{\text{MF}}$ due to all
the other dipoles, the same for all dipoles. Such a treatment has also been
reviewed in connection with the Snoek relaxation of O interstitial atoms in
\textit{bcc} metals \cite{HWD92}. Dattagupta's model does not explicitly take
into account the concentration of relaxing dipoles, since it assumes that
there is one dipole per cell, which is able to change between three possible
orientations; it is the elastic analogous of the Curie-Weiss theory of
interacting magnetic dipoles, and the result is essentially the same. If the
elastic interaction $E=\lambda\sigma_{\text{MF}}$ among the dipoles on the
average favors parallel orientations of the major axis of $\lambda$, $\Delta
E=E_{\perp}-E_{\parallel}>0$,\ a cooperative alignment of the dipoles occurs,
which results in an increase of both the relaxation strength $\Delta\left(
T\right)  $ and relaxation time $\tau$ by a factor $\left(  1-T_{\mathrm{C}%
}/T\right)  ^{-1}$. The effect of the cooperative motion can be roughly viewed
as the coordinated motion of several dipoles instead of independent dipoles,
which results in a larger effective elastic dipole, but also in a slower
reorientation time. When the anisotropic component of the interaction energy,
$\Delta E$, is of the order of the thermal energy $k_{\text{B}}T$, the dipoles
start freezing into a fixed orientation, resulting in a ferroelastic ordering
transition, similar to the ferromagnetic or ferroelectric one. The expression
of $s^{\prime\prime}$ above $T_{\mathrm{C}}$ becomes
\begin{align}
\frac{s^{\prime\prime}}{s^{\prime}}  &  =\tilde{\Delta}\frac{\omega\tilde
{\tau}}{1+\left(  \omega\tilde{\tau}\right)  ^{2}}\label{CW}\\
\tilde{\Delta}  &  =\frac{\Delta\left(  T\right)  }{1-T_{\mathrm{C}}/T}%
\propto\frac{T_{\mathrm{C}}}{T-T_{\mathrm{C}}}\,,\quad\tilde{\tau}=\frac
{\tau\left(  T\right)  }{1-T_{\mathrm{C}}/T}%
\end{align}
and, assuming $\Delta\left(  T\right)  =\Delta_{0}/T$ can be put in the form
\begin{equation}
\frac{s^{\prime\prime}}{s^{\prime}}=\frac{\Delta_{0}}{T}\frac{\omega\tau
}{\left(  1-T_{\text{c}}/T\right)  ^{2}+\left(  \omega\tau\right)  ^{2}}%
\end{equation}
where $\tau$ is the relaxation time for non-interacting dipoles. Such a
treatment might be adapted to only two orientations, as is the case of oxygen
in YBCO, and the expressions of the elastic compliance and relaxation time
might be extended below $T_{\mathrm{C}}$; the dependence on the dipole
concentration $c$ might be taken into account as an interaction energy $\Delta
E\simeq$ $\lambda\sigma_{\text{MF}}\propto$ $\left(  \Delta\lambda\right)
^{2}c$, giving rise to $k_{\text{B}}T_{\mathrm{C}}\simeq$ $\Delta E\propto c$,
as also suggested in \cite{Dat82,DBR82}. However, the result would be correct
only in the limit of low concentration, and in what follows I\ will adopt
Wipf's approach \cite{BSW94}.

\subsection{Ordering transition}

Let us consider the CuO$_{x}$ plane of YBCO, where, like the interstitial
atoms in Fig. \ref{fig anelasre}c, the O atoms can occupy sites of type 1 and
2, corresponding to the positions generally called O(1)\ and O(5) (see Fig.
\ref{fig YBCO6_7}); in the completely ordered phase only sites of type 1 are
occupied. The starting point of Wipf's analysis \cite{BSW94} is the
equilibrium condition for the O\ atoms between the sites of type 1 and 2:
\begin{equation}
\mu_{1}\left(  c_{1},c,T\right)  =\mu_{2}\left(  c_{2},c,T\right)
\end{equation}
where $\mu_{\alpha}$ is the chemical potential of the O\ atoms in the
sublattice $\alpha$ and $c=\frac{x}{2}$ is the fraction of sites populated in
the CuO$_{x}$ plane ($0\le c\le0.5$ for YBCO) with
\begin{equation}
c_{1}+c_{2}=c\,.
\end{equation}
The chemical potentials are evaluated from the free energy $F=U-TS$, as
\begin{equation}
\mu_{\alpha}=\frac{\partial F}{\partial c_{\alpha}}%
\end{equation}
and in the Bragg-Williams approximation is $F=\overline{E}-k_{\text{B}}%
T\ln\mathcal{N}$, with $\overline{E}$ the average energy of the dipoles and
$\mathcal{N}$ the number of ways a state with that energy may be obtained,
disregarding the interactions among dipoles. Therefore, if the crystal
contains $N$ cells, $\mathcal{N}$ is the number of ways $N_{1}$ O atoms can be
put in $N$ sites of type 1 and $N_{2}$ O\ atoms in the other $N\;$sites of
type 2:
\begin{equation}
\mathcal{N}=\frac{N\left(  N-1\right)  ...\left(  N-N_{1}+1\right)  }{N_{1}%
!}\frac{N\left(  N-1\right)  ...\left(  N-N_{2}+1\right)  }{N_{2}!}%
=\prod_{\alpha=1}^{2}\frac{N!}{N_{\alpha}!\left(  N-N_{\alpha}\right)  !}
\label{Omega}%
\end{equation}
and using Stirling's formula $\ln N!\simeq N\ln N$ and $c_{\alpha}=N_{\alpha
}/N$
\begin{equation}
-\ln\mathcal{N}=N\sum_{\alpha=1}^{2}\left[  c_{\alpha}\ln c_{\alpha}+\left(
1-c_{\alpha}\right)  \ln\left(  1-c_{\alpha}\right)  \right]
\end{equation}
The mean energy is
\begin{equation}
\overline{E}=\sum_{\alpha,\beta}\sum_{i,j<i=1}^{N}E_{ij}^{\alpha\beta
}c_{\alpha}c_{\beta}%
\end{equation}
where in the mean-field approximation \cite{DBR82} the interaction energy
between two dipoles of type $\alpha$ and $\beta$ in the cells $i$ and $j$ is
set
\begin{equation}
E_{ij}^{\alpha\beta}=\left\{
\begin{array}
[c]{c}%
E_{\parallel}\quad\text{if }\alpha=\beta\\
E_{\perp}\quad\text{if }\alpha\neq\beta
\end{array}
\right.
\end{equation}
yielding a mean energy per unit cell
\begin{equation}
\overline{E}=E_{\parallel}\sum_{\alpha}c_{\alpha}^{2}+E_{\perp}\sum_{\alpha
}c_{\alpha}\left(  c-c_{\alpha}\right)  =\Delta E\sum_{\alpha}c_{\alpha}%
^{2}+E_{\perp}c^{2}%
\end{equation}
where
\begin{equation}
\Delta E=E_{\parallel}-E_{\perp}%
\end{equation}
and $\Delta E<0$ favors parallel orientations. The free energy per unit cell
can then be written as
\begin{equation}
F=\overline{E}-k_{\text{B}}T\ln\mathcal{N}=\Delta E\sum_{\alpha}c_{\alpha}%
^{2}+E_{\perp}c^{2}+k_{\text{B}}T\sum_{\alpha=1}^{2}\left[  c_{\alpha}\ln
c_{\alpha}+\left(  1-c_{\alpha}\right)  \ln\left(  1-c_{\alpha}\right)
\right]
\end{equation}
and the chemical potentials
\begin{equation}
\mu_{\alpha}=\frac{\partial F}{\partial c_{\alpha}}=2\Delta Ec_{\alpha
}+k_{\text{B}}T\ln\left(  \frac{c_{\alpha}}{1-c_{\alpha}}\right)
\end{equation}
or, adopting the notation of \cite{BSW94},
\begin{equation}
\mu_{\alpha}=\frac{\partial F}{\partial c_{\alpha}}=-\alpha\left(  c_{\alpha
}-\frac{c}{2}\right)  +k_{\text{B}}T\ln\left(  \frac{c_{\alpha}}{1-c_{\alpha}%
}\right)  \label{mu}%
\end{equation}
where the energy is referred to that of the completely disordered state,
$c_{\alpha}=\frac{c}{2}$, and $\alpha=-2\Delta E>0$ is the difference in
energy between states 1 and 2, supposed to be weakly dependent on temperature.
By equating $\mu_{1}$ and $\mu_{2}$ we obtain
\begin{equation}
\frac{\alpha}{k_{\text{B}}T}\left(  c_{1}-c_{2}\right)  =\ln\left[
\frac{c_{1}\left(  1-c_{2}\right)  }{c_{2}\left(  1-c_{1}\right)  }\right]
\label{c1}%
\end{equation}
having solutions different from the trivial one $c_{1}=c_{2}=\frac{1}{2}$ only
below the critical temperature
\begin{equation}
k_{\text{B}}T_{\mathrm{C}}=\alpha\frac{c}{2}\left(  1-\frac{c}{2}\right)
\,;\; \label{TC1}%
\end{equation}
below $T_{\mathrm{C}}$, the equation can be solved numerically for $c_{1}$, as
shown in Appendix A.

\subsection{Coupling to stress and relaxation strength\label{sect intr s}}

On application of a stress $\sigma_{kl}$, the elastic energy of each O atom of
type $\alpha$ changes by
\begin{equation}
-v_{0}\lambda_{kl}^{(\alpha)}\sigma_{kl}\,,
\end{equation}
where for tetragonal elastic dipoles $\lambda_{xx}^{(1)}=\lambda_{yy}%
^{(2)}=\lambda_{1}\,,\quad\lambda_{yy}^{(1)}=\lambda_{xx}^{(2)}=\lambda_{2}$,
while the $z$ component has no interest since it is the same for both types of
sites, so that the anelastic strain due to the oxygen dipoles is (compare with
Eq. (\ref{e_an(t)}))
\begin{equation}
\varepsilon_{xx}=-\varepsilon_{yy}=\frac{1}{2}\left(  c_{1}-c_{2}\right)
\left(  \lambda_{1}-\lambda_{2}\right)  \,.
\end{equation}
The change in elastic energy, given by Eq. (\ref{dE(s)}), results in a change
of the populations from $c_{\alpha}\ $to $c_{\alpha}^{\prime}\left(
\sigma\right)  $ such that
\begin{equation}
\mu_{1}\left(  c_{1}^{\prime},c,T\right)  -v_{0}\lambda_{kl}^{(1)}\sigma
_{kl}=\mu_{2}\left(  c_{2}^{\prime},c,T\right)  -v_{0}\lambda_{kl}^{(2)}%
\sigma_{kl}%
\end{equation}
and expanding to first order in $c_{\alpha}^{\prime}-c_{\alpha}$ and using the
fact that $\left(  c_{1}^{\prime}-c_{1}\right)  =-\left(  c_{2}^{\prime}%
-c_{2}\right)  $,
\begin{equation}
c_{1}^{\prime}-c_{1}=\frac{v_{0}\left(  \lambda_{kl}^{(1)}-\lambda_{kl}%
^{(2)}\right)  \sigma_{kl}}{\frac{\partial\mu_{1}}{\partial c_{1}}%
+\frac{\partial\mu_{2}}{\partial c_{2}}}=\frac{v_{0}\left(  \lambda
_{1}-\lambda_{2}\right)  \left(  \sigma_{xx}-\sigma_{yy}\right)  }%
{\frac{\partial\mu_{1}}{\partial c_{1}}+\frac{\partial\mu_{2}}{\partial c_{2}%
}}%
\end{equation}
which results in an anelastic strain
\begin{equation}
\varepsilon_{xx}^{\text{an}}=-\varepsilon_{yy}^{\text{an}}=\frac{1}{2} \left[
\left(  c_{1}^{\prime}-c_{2}^{\prime}\right)  -\left(  c_{1}-c_{2}\right)
\right]  \left(  \lambda_{1}-\lambda_{2}\right)  =\frac{v_{0}\left(
\lambda_{1}-\lambda_{2}\right)  ^{2}\left(  \sigma_{xx}-\sigma_{yy}\right)
}{\frac{\partial\mu_{1}}{\partial c_{1}}+\frac{\partial\mu_{2}}{\partial
c_{2}}}%
\end{equation}
or relaxation of the compliances $\delta s_{ijkl}=\varepsilon_{ij}^{\text{an}%
}/\sigma_{kl}$:
\begin{equation}
\delta s_{11}=\delta s_{22}=-\delta s_{12}=\frac{v_{0}\left(  \lambda
_{1}-\lambda_{2}\right)  ^{2}}{\frac{\partial\mu_{1}}{\partial c_{1}}%
+\frac{\partial\mu_{2}}{\partial c_{2}}}=F_{1}\left(  T\right)  \frac
{v_{0}\left(  \lambda_{1}-\lambda_{2}\right)  ^{2}}{k_{\text{B}}T}
\label{s BW}%
\end{equation}
where the factor $F_{1}\left(  T\right)  $ is computed from eq. (\ref{mu}):
\begin{equation}
F_{1}\left(  T\right)  =\left[  \frac{1}{8}\left[  \frac{c\left(  2-c\right)
}{c_{1}\left(  1-c_{1}\right)  }+\frac{c\left(  2-c\right)  }{c_{2}\left(
1-c_{2}\right)  }\right]  -\frac{T_{\mathrm{C}}}{T}\right]  ^{-1}%
\end{equation}
For $T>T_{\mathrm{C}}$ it is $c_{1}=c_{2}=c/2$ and therefore
\begin{equation}
F_{1}\left(  T\right)  =\frac{T}{T-T_{\mathrm{C}}}\quad\left(  T>T_{\mathrm{C}%
}\right)
\end{equation}
and $\delta s$ becomes the Curie-Weiss susceptibility; below $T_{\mathrm{C}}$,
$F_{1}\left(  T\right)  $ falls off to zero, as shown in Fig. \ref{fig F1F2}.

\subsection{Relaxation rate}

Following \cite{BSW94}, the rate equation for the instantaneous
out-of-equilibrium concentrations $c_{1}^{\prime\prime}$ and $c_{2}%
^{\prime\prime}=c-c_{1}^{\prime\prime}$ is
\begin{align}
\frac{dc_{1}^{\prime\prime}}{dt}  &  =\frac{1}{2}\tau_{0}^{-1}\exp\left(
-E_{0}/k_{\text{B}}T\right)  \left[  c_{2}^{\prime\prime}\left(
1-c_{1}^{\prime\prime}\right)  e^{\Delta E/2k_{\text{B}}T}-c_{1}^{\prime
\prime}\left(  1-c_{2}^{\prime\prime}\right)  e^{-\Delta E/2k_{\text{B}}%
T}\right] \label{rateqn}\\
&  =\frac{1}{2\tau}\left[  c_{2}^{\prime\prime}\left(  1-c_{1}^{\prime\prime
}\right)  e^{\alpha\left(  c_{1}^{\prime\prime}-\overline{c}\right)
/k_{\text{B}}T}-c_{1}^{\prime\prime}\left(  1-c_{2}^{\prime\prime}\right)
e^{-\alpha\left(  c_{1}^{\prime\prime}-\overline{c}\right)  /k_{\text{B}}%
T}\right] \nonumber
\end{align}
where $\tau^{-1}=\tau_{0}^{-1}\exp\left(  -E_{0}/k_{\text{B}}T\right)  $ is
the relaxation rate when sites 1 and 2 are equivalent (absence of stress and
disordered state), the two terms are the rates from 1 to 2 and \textit{vice
versa}, taking into account the probability that the neighboring site is
empty, while the $e^{\pm\Delta E/2k_{\text{B}}T}$ factors account for the
non-equivalence introduced by the ordering transition, where the difference in
the site energies is given by the difference in chemical potential, without
counting the entropic term:
\begin{equation}
\Delta E=E_{2}-E_{1}=\alpha\left(  c_{1}^{\prime\prime}-c_{2}^{\prime\prime
}\right)  =2\alpha\left(  c_{1}^{\prime\prime}-\overline{c}\right)  \,.
\end{equation}
On setting $dc_{1}/dt=0$ one finds again eq. (\ref{c1}), which demonstrates
the consistence between the thermodynamic-statistical and the kinetic analyses.

\begin{figure}[tbh]
\begin{center}
\includegraphics[
%natheight=397.125000pt,
%natwidth=516.562500pt,
%height=4.9424cm,
width=7.8551cm
]{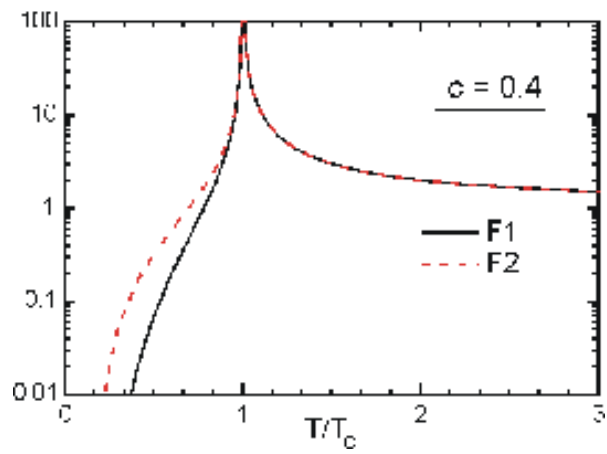}
\end{center}
\caption{$F_{1}$ and $F_{2}$ versus reduced temperature for $c=0.4$.}%
\label{fig F1F2}%
\end{figure}

In order to find the dynamic response to a small applied stress, we set
$c_{1}^{\prime\prime}=c_{1}+\delta c$ and $c_{2}^{\prime\prime}=c_{2}-\delta
c$ into the rate equation, where $\delta c\left(  t\right)  $ is the small
response to the applied stress, and keep terms to the first order in $\delta
c$; after some algebra one obtains
\begin{equation}
\frac{d\delta c}{dt}=-\frac{\delta c}{\tau_{\text{eff}}}%
\end{equation}
with
\begin{equation}
\tau_{\text{eff}}=\tau\frac{\frac{c}{2}\left(  1-\frac{c}{2}\right)  }%
{\sqrt{c_{1}c_{2}\left(  1-c_{2}\right)  \left(  1-c_{1}\right)  }}%
F_{1}\left(  T\right)  =\tau\,F_{2}\left(  T\right)  \,. \label{tau_eff}%
\end{equation}
For $T>T_{\mathrm{C}}$ the factor before $F_{1}$ becomes unity, and one finds
that
\begin{equation}
\tau_{\text{eff}}=\tau\frac{T}{T-T_{\mathrm{C}}}\quad\left(  T>T_{\mathrm{C}%
}\right)
\end{equation}
which represents the critical slowing down on approaching the phase
transformation. The functions $F_{1}$ and $F_{2}$ that enhance the relaxation
strength and time near $T_{\mathrm{C}}$ are shown in Fig. \ref{fig F1F2}.

\chapter{Experimental}

\section{The samples\label{sect samples}}

The YBa$_{2}$Cu$_{3}$O$_{6+x}$ (YBCO), La$_{2-x}$Sr$_{x}$CuO$_{4}$ (LSCO) and
Nd$_{2}$CuO$_{4+\delta}$ (NCO) samples have been prepared by M. Ferretti at
the Department of Chemistry and Industrial Chemistry of the University of
Genova, Italy. The powders of the starting oxides were mixed in appropriate
amounts and calcined, typically 1233~K in air for 6~h for YBCO, 1323~K for
18~h for LSCO and 90~h for NCO. The powders were then pressed into bars
$5\times5\times50$~mm$^{3}$ and sintered in fluent O$_{2} $ (typically YBCO at
1253~K for 22~h and LSCO at 1323~K for 18 h), and further oxygenated in fluent
O$_{2}$ at lower temperature for the case of YBCO. The powders were checked
with X-ray diffraction for impurity phases from decomposition or incomplete
solid state reaction, and sometimes intermediate grinding and sintering
treatments were made. Especially RuSr$_{2}$GdCu$_{2}$O$_{8}$ (Ru-1212)
required a long processing route \cite{ACC02} before the final oxygenation at
1343~K in fluent O$_{2}$ for one week, followed by cooling at 50~K/h.

The ingots were finally cut with a circular diamond saw to the dimensions
suitable for the anelastic measurements: length of $40-50$~mm, thickness of
$0.2-1.5$~mm, width of $4-5$~mm. The samples were rather porous (from 10\% up
to 50\% for Ru-1212) and some of the oil used during cutting remains in the
pores and it is almost impossible to completely remove it with solvents
(toluene and acetone) in a ultrasound bath. The complete removal of the oil
can be accomplished by burning it, heating the sample at high temperature. The
residual oil, which can also be adsorbed from the pumping system \cite{KGG91},
has a freezing transition around 220~K giving rise to an absorption peak and
anomalies in the elastic moduli (see also Sect \ref{sect 240K}).

Further characterization of the samples was made by measuring the resistivity
and superconducting transition with a four-probe technique with a closed cycle
cryocooler from 300 down to 12~K. Checks were made that resistivity $\rho$ did
not depend on the amplitude and direction of the applied voltage. The
oxygen\ stoichiometry in YBa$_{2}$Cu$_{3}$O$_{6+x}$ was estimated by combined
analysis of the $\rho\left(  T\right)  $ curves (compared with literature
data) and of the anelastic spectra; in some cases series of samples have been
prepared with different stoichiometries, checked by iodometric titration and
from the cell parameters from X-ray diffraction.

\section{Anelastic measurements with resonating samples}

The equation of the vibration of a solid near one of its resonances is that of
a damped harmonic oscillator with resonance frequency \cite{NB72}
$f_{0}=\alpha\sqrt{M/\rho}=$ $\frac{^{\omega_{0}}}{2\pi}$ where $M$ is an
effective elastic modulus and $\alpha$ a geometrical factor depending on the
vibration mode. Then, the equation of motion for strain under application of a
stress $\sigma_{0}e^{i\omega t}$ with $\omega\simeq\omega_{0}$ is%

\begin{equation}
\sigma_{0}e^{i\omega t}=\alpha^{2}\left(  M^{\prime}+iM^{\prime\prime}\right)
\,\varepsilon+\rho\ddot{\varepsilon}=\rho\omega_{0}^{2}\left(  1+i\tan
\phi\right)  \varepsilon-\rho\omega^{2}\varepsilon\label{eq motion}%
\end{equation}
where the real and imaginary parts of the modulus have been introduced.

\begin{figure}[tbh]
\begin{center}
\includegraphics[
%natheight=210.062500pt,
%natwidth=524.312500pt,
%height=2.9501cm,
width=7.4224cm
]{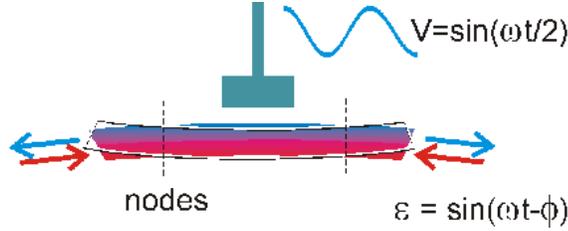}
\end{center}
\caption{Electrostatic excitation of the flexural vibrations of a thin bar.
Strain is inhomogeneous along the sample thickness: from a uniaxial expansion
(blue) to a compression (red).}%
\label{fig flexure}%
\end{figure}

All measurements presented here have been made by exciting the samples on
their 1st, 3rd and 5th flexural modes. Thanks to the sample dimensions, with
length $30-800$ times larger than the thickness, and to the low vibration
amplitude, always within the linear regime, the approximation of simple
bending of thin bars is a good one. Under such approximation the modulus
involved in the flexural vibrations is the \textbf{Young's modulus }$E$, that
relates uniaxial stress and strain of a long sample of uniform cross section.
In the case of pure extensional vibrations, where stress and strain obey
$\sigma=E\varepsilon$ all over the sample, the resonance frequency of the
$n$-th mode is given by $f_{n}=\left(  n/2l\right)  \sqrt{E/\rho}$ where $l$
and $\rho$ are the sample length and density (neglecting a correction factor
that takes into account the finite size of the cross-section with respect to
the length and proportional to the square of Poisson's ratio \cite{74}). In
the case of flexure, strain is inhomogeneous along the sample thickness,
passing from extension on the convex side to compression on the concave one,
as shown in Fig. \ref{fig flexure}. The effective modulus (the real part) is
therefore $E$, but reduced of a geometrical factor taking into account the
sample cross section; for the fundamental mode of a thin bar of thickness $h$,
it is \cite{NB72}
\begin{equation}
f=1.03\frac{h}{l^{2}}\sqrt{E/\rho}\,; \label{f flex}%
\end{equation}
the frequencies of third and fifth modes are respectively 5.40 and 13.3 times larger.

We never tried to estimate the absolute value of $E$ because of irregularities
in the samples shapes and above all due to the large sample porosity (from 10
to 50\%), but it is possible to measure the variation of $E\left(  T\right)  $
with respect to a reference temperature $T_{0}$ through
\begin{equation}
\frac{E\left(  T\right)  }{E\left(  T_{0}\right)  }=\frac{f^{2}\left(
T\right)  }{f^{2}\left(  T_{0}\right)  }\,,
\end{equation}
where the temperature dependence of $\rho\left(  T\right)  $ is neglected in
comparison to that of $E\left(  T\right)  $.

The \textbf{elastic energy loss} can be measured with two methods. In the
\textbf{free decay} method, the excitation is switched off and, according to
Eq. (\ref{eq motion}) with $\sigma_{0}=0$, the vibration amplitude falls off
as
\begin{equation}
\varepsilon\left(  t\right)  =\varepsilon_{0}e^{i\omega_{0}\sqrt{1+i\tan\phi
}\,t}\simeq\varepsilon_{0}e^{i\omega_{0}t}e^{-\phi\,\omega_{0}t/2}.
\end{equation}
Therefore, interpolating with a straight line the logarithm of the vibration
amplitude $A\left(  t\right)  =\varepsilon_{0}\exp\left(  -\phi\,\omega
_{0}t/2\right)  $, $\log\left[  A\left(  t\right)  /A\left(  0\right)
\right]  =-\frac{1}{2}Q^{-1}\omega_{0}t$, it is possible to deduce $\tan
\phi=Q^{-1}$. This method is useful when the decay constant $Q/(\pi f_{0})$ is
long enough to let the amplitude decay to be recorded.

For larger dissipations or frequencies the \textbf{forced oscillation} method
can be used, where the vibration amplitude is measured on sweeping frequency
near resonance. From Eq. (\ref{eq motion}), the vibration amplitude is
\begin{equation}
\left|  A\right|  ^{2}=\frac{\left(  \sigma_{0}/\rho\right)  ^{2}}{\left(
\omega_{0}^{2}-\omega^{2}\right)  ^{2}+\omega_{0}^{4}\tan^{2}\phi}%
\end{equation}
and by least square fitting of $A^{-2}\left(  \omega^{2}\right)  $ one obtains
the resonance frequency $\omega_{0}/2\pi$ and the elastic energy loss
coefficient $Q^{-1}$.

\section{The anelastic relaxation setup\label{sect AS setup}}

The anelastic measurements were based on resonant sample techniques developed
in the Laboratory of Solid State Acoustics at the Istituto di Acustica ''O.~M.
Corbino'' of CNR, starting from Prof. P.~G. Bordoni up to Profs. G. Cannelli
and R. Cantelli. The sample is suspended on thin thermocouple wires in
correspondence with the nodal lines for the flexural vibrations; generally the
nodes at $0.225l$ from the sample ends ($l=$ sample length) are used, since
they are very close to nodes of both the 1st and 5th flexural modes, whose
resonance frequencies are in the ratio 1:13.3. It is therefore possible to
measure these two frequencies during the same run.

As schematically shown in Fig. \ref{fig flexure}, an electrode is kept very
close the sample surface and the sample, if not conducting, is covered with
silver paint in correspondence with the electrode and suspension wires, in
order to have electrical continuity and proper grounding (through one of the
thermocouple wires). An excitation ac voltage $V_{\text{exc}}$ is applied to
the electrode with frequency $f/2$ and amplitude up to 200~V; this induces a
potential of opposite sign on the surface of the sample and the resulting
electrostatic force $\propto V_{\text{exc}}^{2}$ excites vibrations of the
sample at frequency $f$ (100~Hz-100~kHz depending on the sample shape and
vibration mode). These vibrations modulate the sample/electrode capacitance,
which is part of a high frequency ($\sim10$~MHz) circuit, and therefore
modulate its resonant frequency, which is in turn demodulated with a coupled
resonating circuit with tunable capacitance. This revelator, called
vibrometer, was constructed in the Solid State Acoustics Laboratory by Profs.
Nuovo and Bordoni. I improved the data acquisition by introducing a lock-in
amplifier that measures the signal amplitude and can detect low vibration
levels at high noise level, and developed the acquisition software. Now it is
routine to perform measurements at several frequencies during the same run
sometimes in semiautomatic manner. This is very important whenever the status
of the sample is modified by the measurement itself, for example due to oxygen
loss in vacuum at high temperature, and a subsequent run at a different
frequency could not be compared with the previous one for making an analysis
at different frequencies. If the sample is properly mounted and its intrinsic
dissipation is $Q^{-1}\sim10^{-4}$ or larger, it is possible to measure
several flexural and torsional modes with the same mounting; Fig.
\ref{fig spectrum} is an example with 1st, 3rd and 5th flexural modes
vibrating at 1.3, 7.2 and 18~kHz respectively.

There are two separate inserts for low (1-300~K) and high temperature
(100-950~K) measurements, evacuated by a diffusion pump; He exchange gas is
used for thermalization, and an adsorption pump is used at low temperature, in
order to minimize the adsorption by the sample of residual gases when the
diffusion pump is closed (being cooled with liquid N$_{2}$, the adsorption
pump does not condense the He exchange gas).

The low temperature insert has a cylindrical brass cell that can be closed
both with a conical coupling with vacuum grease to the flange or with In wire
for measurements below 2~K. The cell is equipped with heater and silicon diode
for temperature control and a wound copper tube for the flow of LHe. The whole
insert may be closed with a vacuum tight glass dewar and works as a LN bath
cryostat and LHe flow or bath cryostat: below 5~K the dewar is filled with LHe
and may be pumped in order to lower temperature down to 1.1~K. The
thermocouple is Au-0.003Fe versus chromel.

The high temperature insert is made of stainless steel and ceramic insulating
parts and is closed with a quartz tube. Temperature is regulated with a
tubular furnace or dewar with LN.

\section{Reliability of the anelastic spectroscopy measurements}

Sometimes, there is scepticism on interpretations of anelastic relaxation
processes outside mechanisms based on atomic defects, dislocations, domain
boundaries and grain boundaries; in addition, it is often felt that the
anelastic spectroscopy should be particularly sensitive to the polycrystalline
nature of ceramic samples and to grain boundaries, making the interpretation
of the anelastic spectra very unreliable. In this section few arguments are
put forward demonstrating that such doubts are generally unjustified.

Regarding the effect of impurity phases, generally found at the grain
boundaries, it should be remarked that anelastic spectroscopy probes the whole
volume of the sample; therefore, if defects motions or phase transformations
occur in an impurity phase, their effect will be weighted with the molar
fraction of the phase (and another factor of order unity taking into account
the different moduli of bulk and impurity phase \cite{97}). Therefore,
spurious phases may affect the anelastic spectra exactly as they affect other
bulk measurements like specific heat or neutron spectroscopy, and much less
than techniques sensitive to the grain surfaces, like resistivity, dielectric
susceptibility or optical spectroscopies. In the samples considered here the
molar fraction of impurity phases did never exceed few percents, as checked by
X-ray diffraction, which means that their possible contribution to the
anelastic spectrum was rather small. Of course, if the impurity phase presents
a phase transformation, even small amount of it may produce noticeable effects
on the anelastic spectrum, as for example occurs with La$_{1-x}$Ca$_{x}%
$MnO$_{3}$ samples containing Mn$_{3}$O$_{4}$ below the detection limit with
standard powder X-ray diffraction \cite{97}. The only cases considered here of
little characterized anomalies looking like phase transformations are those in
YBCO around 240~K (Sec. \ref{sect 240K}) and 120-160~K (Sec. \ref{sect AFE}),
but even in these cases, their peculiar dependence on oxygen\ doping strongly
suggest mechanisms involving the CuO$_{x}$ planes of YBCO.

\begin{figure}[tbh]
\begin{center}
\includegraphics[
%natheight=568.062500pt,
%natwidth=425.625000pt,
%height=4.3457in,
width=3.2621in
]{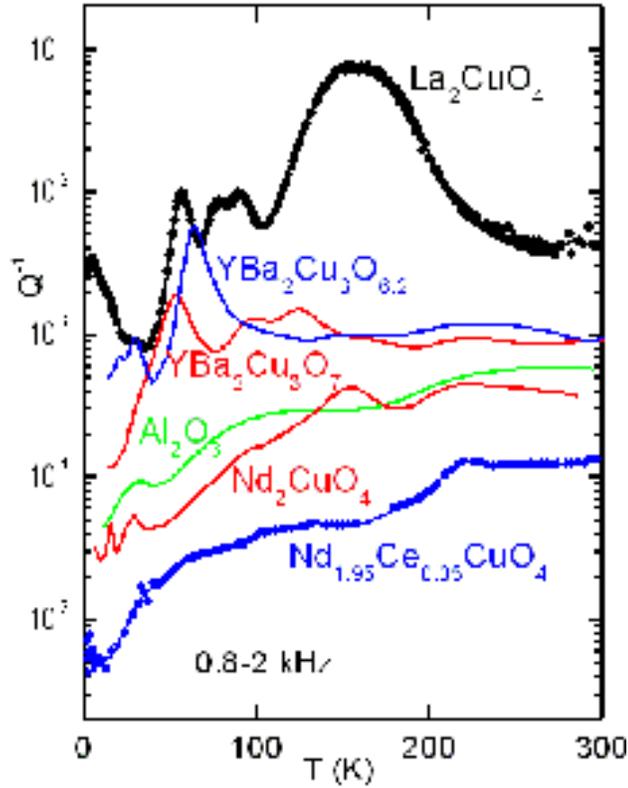}
\end{center}
\caption{Elastic energy loss versus tempertaure of ceramic samples of LSCO,
YBCO, NCCO at various dopings and of porous alumina, measured in the same
conditions at frequencies between 0.8 and 2 kHz.}%
\label{fig ceramics}%
\end{figure}

\begin{figure}[tbh]
\begin{center}
\includegraphics[
%natheight=3.327800in,
%natwidth=4.746900in,
%height=3.3278in,
width=4.7469in
]{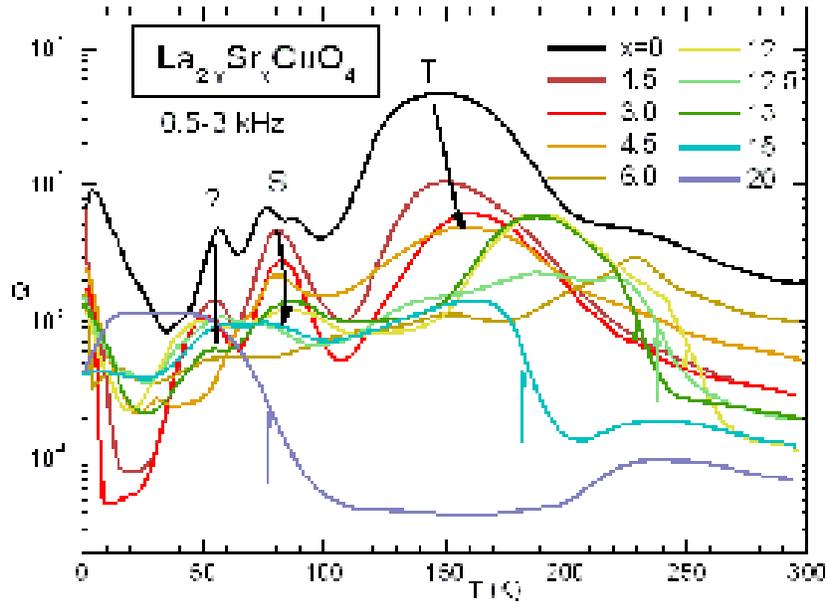}
\end{center}
\caption{Anelastic spectra of La$_{2-x}$Sr$_{x}$CuO$_{4}$ at various doping
levels $x$ in \%; the vertical coloured arrows indicate the structural
transitions from cubic to tetragonal.}%
\label{fig LSCO spectra}%
\end{figure}

A convincing indication against the contribution of impurity phases and also
against instrumental contributions to the spectra examined in the present
study is the strong dependence of these spectra on doping and the fact that
they are completely different for different families of cuprates. This is
shown in Fig. \ref{fig ceramics}, comparing the $Q^{-1}\left(  T\right)  $
curves of ceramic samples of La$_{2-x}$Sr$_{x}$CuO$_{4}$, YBa$_{2}$Cu$_{3}%
$O$_{6+x}$ and Nd$_{2-x}$Ce$_{x}$CuO$_{4+\delta}$ at various dopings, measured
in the same conditions exciting the mode at lower frequency, between 0.8 and 2
kHz. All the samples were prepared in the same laboratory by M. Ferretti and
coworkers, with the same equipments and procedures, except for sintering times
and temperatures, but the spectra are completely different from each other.
The fact that there is no single feature that is shared among all the spectra
excludes the influence of systematic instrumental effects. The only possible
instrumental contribution is from freezing of adsorbed oils around 220~K, as
discussed in Sec. \ref{sect 240K}, but the curve of Nd$_{1.95}$Ce$_{0.05}%
$CuO$_{4}$ sets the upper limit of this possible contribution to a value too
small to affect the spectra of LSCO and YBCO; in addition, a sample of
Al$_{2}$O$_{3}$ with 50\%\ porosity, also shown in Fig. \ref{fig ceramics},
does not present any anomaly at that temperature, in spite of being much more
susceptible to oil uptake. The influence of gases adsorbed by the samples is
discussed in Sec. \ref{sect AFE}. The lack of peaks common to all the curves
in Fig. \ref{fig ceramics} excludes the influence of a spurious phase like
copper oxides, certainly absent in the almost flat and very low $Q^{-1}\left(
T\right)  $ curve of Nd$_{1.95}$Ce$_{0.05}$CuO$_{4}$. The same observation is
made by varying doping within the same cuprate family, as shown in Fig.
\ref{fig LSCO spectra} for LSCO. In this figure, there is a progressive change
of the spectrum with increasing $x$, indicating that all the features depend
on doping of La$_{2-x}$Sr$_{x}$CuO$_{4}$ and therefore not on spurious phases.
If they contributed to one of the peaks labeled T or S or to the maximum at
LHe temperature, they should exist also at $x=0.2$, where instead the elastic
energy loss is $2-3$ orders of magnitude smaller (except for the step at the
structural transformation from cubic to tetragonal, shifted down to 80~K at
such doping).

In Fig. \ref{fig LSCO spectra}, the only instance in which there
is no clear smooth evolution of the curves with doping is the double peak at
80~K for $x=0$, which should not be identified with peak S in the Sr-doped
samples (see Sec. \ref{sect depin}). The curve with $x=0$ was measured after a
rather strong outgassing treatment that possibly introduced O vacancies in the
CuO$_{2}$ planes (see Sec. \ref{sect VOinLSCO}), and the double peak might be
related to such vacancies. The peak at 50~K also appears as some doping
dependent relaxation in La$_{2-x}$Sr$_{x}$CuO$_{4}$, but it is labeled with a
question mark since we were not able to find a plausible mechanism for it, and
it will not be discussed in this Thesis.

Finally, the possibility that the motion of grain boundaries itself may
contribute to the present measurements can be excluded; this type of
relaxation, in fact, requires the diffusion of the cations, which occurs at
temperatures comparable to the sintering temperatures, above 1200~K. Instead,
all the thermally activated relaxation processes observed here at high
temperature are clearly related to diffusion of nonstoichiometric oxygen.

\section{The UHV system for sample treatments\label{sect UHV}}

The oxygenation and some of the outgassing treatments were made in a Ultra
High Vacuum (UHV) system realized by Ing. Dalla Bella at RIAL Vacuum (Parma,
Italy) after my project. It consists of a main spherical chamber with
Bayard-Alpert head and Residual Gas Analyzer, pumped by turbomolecular, ionic
and Ti sublimation pumps. Two CF63 flanges on opposite sides connect on one
side, through an all-metal gate valve, another chamber connected with the gas
inlet line, capacitive heads and a quartz tube flanged CF63 where the sample
is put for the treatments. On the other side, the spherical chamber is
connected to a chamber for introducing the sample without breaking vacuum in
the rest of the system, equipped with a magnetically coupled rotary-linear
feedthrough with a tray for depositing the sample. The trail has the
possibility of some lateral movement and is surrounded by stainless steel
wire, so that with a rotary movement it is possible to retrieve also fragile
samples, generally wrapped with Pt wire for protection and for avoiding direct
contact with the quartz tube. The sample is heated up to 1100~$^{\mathrm{o}}$C
by a horizontal tubular furnace mounted on wheels.

The gas inlet line consists of a multiway valve connecting to various pure gas
bottles (O$_{2}$, H$_{2}$, D$_{2}$, N$_{2}$), and a bakeable all metal section
with inlet and outlet needle valves, a small (17.3~cm$^{3}$) and a large
(200~cm$^{3}$) calibrated volumes and capacitance head, in order to be able to
admit a known amount of gas.

The base vacuum before any treatment was in the $10^{-9}$~mbar range, but
could be improved with baking to the $10^{-10}$~mbar range in particular cases.

\chapter{LSCO}

La$_{2-x}$(Sr/Ba)$_{x}$CuO$_{4}$ (LSCO or LBCO)\ is the first
high-$T_{\mathrm{C}}$ superconductor discovered by Bednorz and M\"{u}ller.
Having $T_{\mathrm{C}}\le40$~K, it is not particularly attractive for
applications, but has the simplest structure among the superconducting
cuprates and is probably the best characterized. Doping may be achieved both
through excess oxygen in La$_{2}$CuO$_{4+\delta}$ and by partial substitution
of La$^{3+}$ with Sr$^{2+}$ or Ba$^{2+}$; in the latter case, if excess oxygen
is completely removed, one does not have the complications due to the ordering
of the nonstoichiometric oxygen, which characterize the other HTS cuprates.

\section{Structural phase diagram\label{sect LSCOstrc}}

\begin{figure}[tbh]
\begin{center}
\includegraphics[
%natheight=373.062500pt,
%natwidth=544.000000pt,
%height=6.9172cm,
width=10.0583cm
]{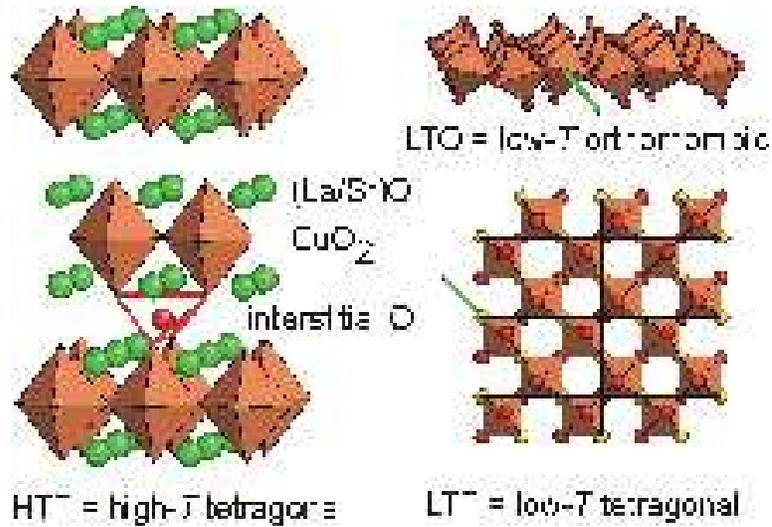}
\end{center}
\caption{Structure of LSCO. Left panel: ideal HTT structure; the O atoms (not
shown) are at the vertices of the octahedra; also shown is an interstitial O
atom and its tetrahedric coordination with the nearest neighbor apical O
atoms. Right panel: the two more stable tilt patterns of the octahedra, giving
rise to the LTO and LTT structures; in the latter case, adjacent layers of the
octahedra are rotated by 90$^{\mathrm{o}}$, so that the overall structure is
tetragonal.}%
\label{fig LSCOstrc}%
\end{figure}

LSCO is formed by layers of CuO$_{6}$ octahedra intercalated by (La/Sr) atoms,
as shown in Fig. \ref{fig LSCOstrc}. The substitution $x$ trivalent La ions
with divalent Sr or Ba introduces $p=x$ holes/unit cell in the CuO$_{2}$
layers, which form a conducting band. Interstitial oxygen (O$_{\text{i}}$) can
also be introduced in tetrahedral coordination with four apical O atoms and
four La atoms; each O$^{2-}$ provides two conducting holes. A small amount of
excess O ($\delta\sim10^{-3}$) is present in as-prepared La$_{2}%
$CuO$_{4+\delta}$ and can be increased to few percent by equilibrating in
O$_{2}$ at moderate temperatures \cite{63}; up to 12\% O$_{\text{i}}$ can be
introduced by electrochemical oxidation \cite{CCJ92,RJS93}. By increasing the
Sr doping, the equilibrium concentration of O$_{\text{i}}$ decreases, and
electrochemical oxidation is necessary to obtain La$_{2-x}$Sr$_{x}%
$CuO$_{4+\delta}$.

\subsection{Tolerance factor and Low-Temperature Orthorhombic
(LTO)\ phase\label{sect LSCO t}}

When decreasing temperature, the equilibrium bond lengths in the (La/Sr)O
layers decrease faster than the equilibrium CuO bond length within the
CuO$_{2}$ planes, and the resulting lattice mismatch is relieved by a buckling
of the CuO$_{2}$ planes \cite{ZCG94} below a temperature $T_{t}(x)$. This
phenomenon is typical of the perovskite structure, formed by a
three-dimensional network of BO$_{6}$ octahedra intercalated by A cations. In
the ideal cubic case, which is the high temperature structure of most
perovskites, the ratio of the A-O and B-O bond lengths is $l_{\text{A-O}%
}/l_{\text{B-O}}=\sqrt{2}$, and it is usual to define a \textbf{tolerance or
Goldschmidt factor}
\begin{equation}
t=\frac{l_{\text{A-O}}}{\sqrt{2}l_{\text{B-O}}}\,, \label{tolerance}%
\end{equation}
which is 1 in the ideal cubic case, and, when becomes $<1$, indicates that the
size of the BO$_{6}$ octahedra is too large for the equilibrium A-O bond
length, resulting in rotations of the octahedra. In many cases, the
transitions of perovskites from cubic to lower symmetries involving rotations
of the octahedra are understood in terms of $t$ decreasing with temperature.

\begin{figure}[ptb]
\begin{center}
\includegraphics[
%natheight=330.000000pt,
%natwidth=560.687500pt,
%height=4.1143cm,
width=6.9545cm
]{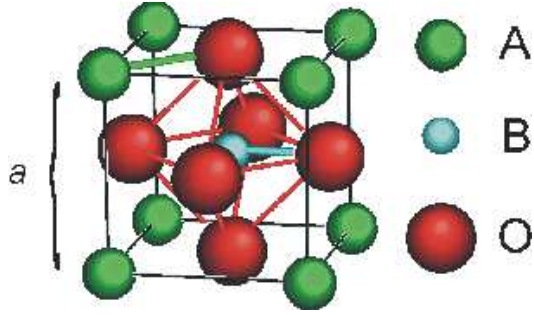}
\end{center}
\caption{Perovskite structure with evidenced the A-O and B-O\ bonds, whose
lengths in the ideal cubic structure are in the ratio $\sqrt{2}$. }%
\label{fig perovsk}%
\end{figure}The same reasoning can be applied to the perovskite layers of
La$_{2-x}$Sr$_{x}$CuO$_{4}$, defining the tolerance factor \cite{ZCG94}
\begin{equation}
t=\frac{l_{\text{La-O}}}{\sqrt{2}l_{\text{Cu-O}}}\,.
\end{equation}
Since the octahedra are relatively rigid units, the buckling results in a
collective tilting of the octahedra, and the structure changes from
high-temperature tetragonal (HTT) to low-temperature orthorhombic (LTO, see
Figs. \ref{fig LSCOstrc} and \ref{fig TiltPot}a).

\begin{figure}[tbh]
\begin{center}
\includegraphics[
%natheight=324.687500pt,
%natwidth=577.687500pt,
%height=6.9062cm,
width=12.2396cm
]{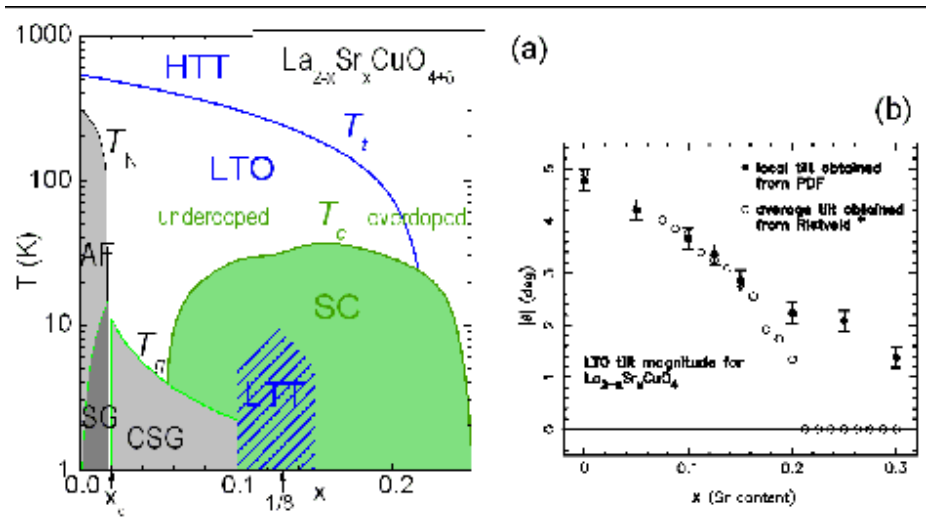}
\end{center}
\caption{(a) Structural (blue), magnetic (gray) and electric (green) phase
diagram of LSCO. AF = antiferoomagnetic, SG = spin glass, CSG = cluster spin
glass, SC = superconducting. (b) average and local tilt angle of LTO phase
versus doping (Ref. \cite{BBK99}).}%
\label{fig LSCOpd}%
\end{figure}Doping reduces the mismatch between LaO blocks and CuO$_{2}$
planes in two ways \textit{i)} substitution of La$^{3+}$ with larger Sr$^{2+}$
or insertion of interstitial O expand the lattice, and therefore relieve the
compressive stress on the CuO$_{2}$ planes; \textit{ii)} doping holes in the
CuO$_{2}$ planes removes charge from the CuO antibonds, therefore shortening
them \cite{CCJ92,RMA95}. Therefore, the HTT phase is stabilized by doping, and
$T_{t}\left(  x\right)  $ is an almost linearly decreasing function of $x$, as
appears from the LSCO phase diagram in Fig. \ref{fig LSCOpd}. In the same
figure is also reported the average tilt angle $\theta$ in the LTO phase from
diffraction measurements, again a decreasing function of doping; the decrease
seems to be much more regular at a local level, as found by extracting the
pair-distribution functions (PDF) from neutron diffraction \cite{BBK99}, and
prosecutes into the HTT phase, which therefore should consist of disordered
tilted instead of untilted octahedra.

\subsection{Other tilt patterns and the Low-Temperature tetragonal (LTT)
phase\label{sect LSCO LTT}}

\begin{figure}[tbh]
\begin{center}
\includegraphics[
%natheight=211.500000pt,
%natwidth=578.187500pt,
%height=5.6343cm,
width=15.2973cm
]{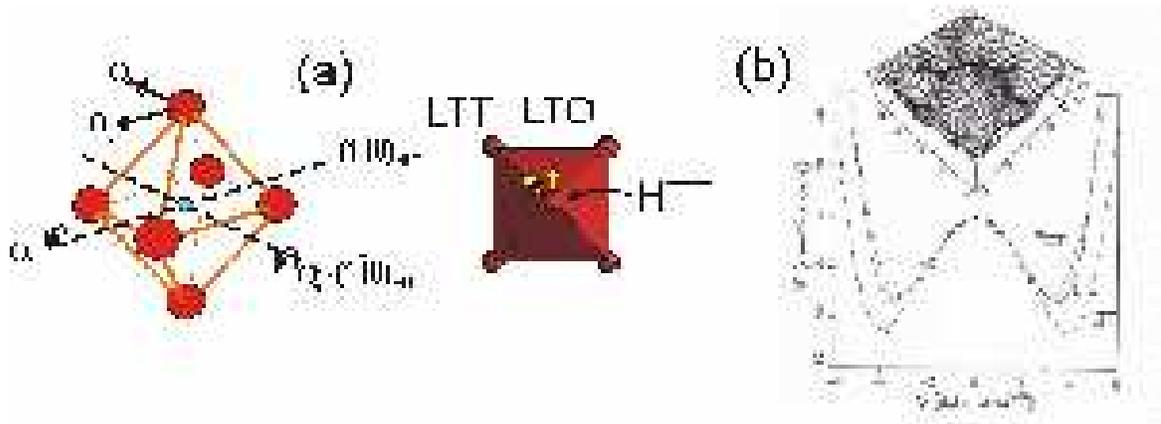}
\end{center}
\caption{(a) The possible tilts of the CuO$_{6}$ octahedra in LSCO; (b)
calculated local potential for an octahedron in LBCO, with eight minima
corresponding to LTT and LTO tilts (Ref. \cite{PCK91}).}%
\label{fig TiltPot}%
\end{figure}

Actually, other tilt patterns besides the LTO one are possible, all
describable in terms of a rotation axis of the octahedra within the $ab$
plane, and therefore in terms of two rotation angles $Q_{1}$ and $Q_{2}$ about
two orthogonal axes within $ab$. Such axes are generally chosen at
45$^{\mathrm{o}}$ with the Cu-O bond directions (the direction of the $a$ axes
of the LTO cell, while the $a$ and $b$ axes of the HTT cell is parallel to the
Cu-O bonds) \cite{AMH89,CHC91}, as shown in Fig. \ref{fig TiltPot}a.

The two variants of the LTO phase are then described by $\left(
Q_{1},0\right)  $ and $\left(  0,Q_{2}\right)  $, while the LTT phase by
$\left|  Q_{1}\right|  =\left|  Q_{2}\right|  $. The intermediate cases
$Q_{1}\neq$ $Q_{2}\neq$ 0 are also possible, and produce intermediate phases,
generally precursors to the LTT one \cite{CHC91,NSM93,KWI95}. Tilt patterns
intermediate between LTT and LTO are present also within the twin boundaries
in the LTO phase. Such twins walls have been observed to be nucleation sites
for the LTT phase \cite{MWZ98}.

The LTT structure is stable only at low temperature, near the doping $x=1/8$,
and if there is sufficient disorder in the ionic sizes in the La sublattice.
The latter is obtained by substituting La with Ba instead of Sr, which has a
still larger radius \cite{TR91} (the radii of 12-fold coordinated Ba$^{2+}$,
Sr$^{2+}$ and La$^{3+}$\emph{\ }are 1.61, 1,44 and 1.36~\AA ), or by doping
with Sr$^{2+}$ and substituting part of La$^{3+}$ with the larger Nd$^{3+}$.

\begin{figure}[tbh]
\begin{center}
\includegraphics[
%natheight=288.562500pt,
%natwidth=649.937500pt,
%height=1.9519in,
width=4.382in
]{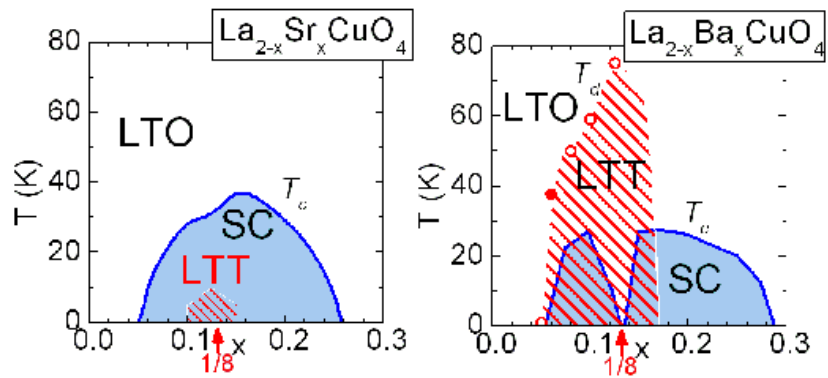}
\end{center}
\caption{Low-temperature region of the phase diagram of LSCO and LBCO, with
the different regions of stability of the LTT structure below $T_{t}$. The
open circles are $T_{t}$ from Ref. \cite{AMH89}, while the closed one from the
measurement of Fig. \ref{fig LSBCOTt} .}%
\label{figLSCO+LBCO}%
\end{figure}Figure \ref{figLSCO+LBCO} shows the region of stability of the
LTT\ phase for La$_{2-x}M_{x}$CuO$_{4}$ with $M=$ Sr, Ba, deduced from various
types of measurements \cite{AMH89,FNH90}. It appears also that the LTT phase
is associated with a depression of the superconducting $T_{\mathrm{c}}$.

It is important to note that in the LTO phase all the O atoms in the CuO$_{2}
$ planes are equivalent, while in the LTT phase one can distinguish between
the O atoms on the rotation axes, and therefore remaining on the plane, and
those shifted out of the plane; this provides a \textbf{modulation of the
potential felt by the charge carriers}, which results in a mutual
stabilization of the LTT modulation and of the static hole stripes (see Sec.
\ref{sect stripes}), whose spacing becomes commensurate with the lattice
spacing at $x=\frac{1}{8}$.

The \textbf{local potential} felt by each octahedron free to tilt has been
theoretically found to have \textbf{eight minima} in correspondence of the
four possible LTO and four LTT tilts separated by barriers $E/k_{\text{B}}%
\sim300-500~$K \cite{PCK91,BMF91} (see Fig. \ref{fig TiltPot}b).

\section{Electric phase diagram}

The electric and magnetic phase diagram of LSCO can be considered as
representative of the other cuprates, except for complications arising from
the O nonstoichiometry in the latter.

As anticipated above, the concentration of holes doped in the CuO$_{2}$ planes
of La$_{2-x}$Sr$_{x}$CuO$_{4+\delta}$ is $p=x+2\delta$, neglecting clustering
of O$_{\text{i}}$ (see Sec. \ref{sect Oi}). The transport properties
(conductivity, Hall coefficient, dielectric permittivity) of lightly doped
La$_{2-x}$Sr$_{x}$CuO$_{4+\delta}$ can be understood in terms of conventional
semiconductor physics \cite{TII89,CBC93,LG98b}: La$_{2}$CuO$_{4}$ has static
dielectric constant $\varepsilon=310$, and the hole effective mass is
$m_{h}=2m_{e}$. The holes are thermally ionized from the acceptors, to which
are bound with an energy $E_{b}$(Sr) $=$ $10-20$~meV for the case of Sr
dopants and $E_{b}$(O) $\sim$ 31~meV for O$_{\text{i}}$. Conduction is of
band-type at high temperature, namely $\sigma\propto$ $p\propto$ $\exp
(-E_{b}/kT)$ at $T>70~$K, and variable-range-hopping below 50~K ($\sigma
\propto$ $\exp[-(T_{0}/T)^{1/4}]$).

For $p>0.05$ the planes start to superconduct below $T_{\mathrm{c}}\left(
p\right)  $, which has a maximum versus doping at $p_{\text{opt}}\simeq0.15$
(see Figs. \ref{fig LSCOpd} and \ref{figLSCO+LBCO}); the system is called
underdoped, optimally doped and overdoped, depending on the value of $p$ with
respect to $p_{\text{opt}}$. Underdoped cuprates exhibit various anomalies,
partly interpreted in terms of opening of pseudogaps in the charge or spin
excitations and partly in terms of charge stripes; the latter will be dealt
with in some detail in Sect \ref{sect stripes}. In overdoped cuprates,
instead, a uniform metallic state sets in and becomes so stable that
superconductivity eventually disappears.

In cuprates of the LSCO family, a depression of superconductivity occurs in
correspondence with the formation of the LTT phase, and this is understood in
terms of locking of the charge stripes to the LTT lattice modulation, as
explained in Sect \ref{sect stripes}.

\section{Magnetic phase diagram\label{sect magn pd}}

The study of the superconducting and magnetic phase diagram of the CuO$_{2}$
planes of the superconducting cuprates is a complex and fascinating subject
(for a review see \textit{e.g.} \cite{RBC98}), but I will mention only the
issues relevant to the anelastic measurements. In the absence of doping, the
CuO$_{2}$ planes are semiconducting, with Cu in the Cu$^{2+}$ oxidation state
having spin $s=\frac{1}{2}$; these spins order \textbf{antiferromagnetically
(AF)} below the N\'{e}el temperature $T_{\text{N}}=315$~K, with the staggered
magnetization within the $ab$ plane, and mainly parallel to $b$. There is also
a small component of the spins pointing out of the planes, due to the small
tilt of the Cu-O bases of the octahedra; this weak canting produces a
ferromagnetic component that dominates the low frequency magnetic
susceptibility \cite{LAK01}. Doping holes causes some Cu atoms to pass into
the Cu$^{3+}$ state with $s=0$, and this disturbs the AF order; in La$_{2-x}%
$Sr$_{x}$CuO$_{4}$ $T_{\text{N}}\left(  x\right)  $ drops to 0~K already at
the \textbf{critical doping }$x_{c}\simeq\,0.02$ (see Fig. \ref{fig LSCOpd}).

A\ model of how the holes disturb the AF order has been proposed by Gooding
\textit{et al. }\cite{GSB97}, starting from the hypothesis that at low
temperature and low doping the holes are localized near the Sr dopants. The
ground state for one isolated hole would be doubly degenerate with the hole
circulating either clockwise or anticlockwise over the four Cu atoms nearest
neighbors to the Sr atom. The hole motion couples to the transverse
fluctuations of the Cu spins and produces a spiraling distortion of the AF
order within the $ab$ plane; at this point it should be noted that the model
assumes the staggered magnetization of the hole-free plane along $c$, while in
fact it is along $b$, but the model may help in focusing some mechanisms
responsible for the magnetic phase diagram of the HTS\ cuprates. The ground
states with disordered distributions of Sr impurities would consist of AF
correlated domains delimited by the Sr atoms and with the in-plane AF order
parameter randomly oriented, resulting in a \textbf{(cluster) spin-glass}
state. The domains would be separated by narrow domain walls with disordered
spins and ferromagnetic character which connect the Sr atoms. The hole
mobility is much higher in FM rather than AFM regions (the hopping of a hole
in a AF domain requires also a spin flip, if destruction of AF order has to be
avoided), and therefore with increasing temperature and doping the holes move
along these \textbf{domain walls}, which would therefore correspond to the
\textbf{charge stripes} of the next Section.

Whatever model is chosen, experiments probing the local spin fluctuations,
like NQR \cite{RBC98,CBC93} and $\mu$SR \cite{NBB98}, indicate that the spin
degrees of freedom associated with the doped holes are different from the
in-plane Cu$^{2+}$\ spin degrees of freedom that order themselves below
$T_{\text{N}}$, and the localization of the doped holes allows the associated
spins to progressively slow down and freeze \cite{RBC98,CBC93}. For $x<x_{c}$
one has long range AF order below $T_{\text{N}}$, and the doped spins freeze
into a \textbf{spin glass (SG)} state below $T_{f}\left(  x\right)
\ $linearly increasing with $x$ (see Fig. \ref{fig LSCOpd}). For $x>x_{c}%
$\ there is no long range AF order and approaching $T_{g}\simeq0.2$~K$/x$\ AF
correlations develop within domains separated by hole-rich walls, and with
easy axes uncorrelated between different clusters, giving rise to a
\textbf{cluster spin-glass (CSG)} state (see Fig. \ref{fig LSCOpd}).

More recent neutron scattering experiments \cite{MFY02} of the magnetic
correlations in La$_{2-x}$Sr$_{x}$CuO$_{4}$\ for $x<x_{c}$ suggests a
different picture of the spin glass phase, with the 3D AF ordered phase
coexisting below $\sim30$\ K with domains of the stripe phase observed for
$x>x_{c}$ (see also next Section). It has been proposed that the hole
localization starting around 150~K involves an \textbf{electronic phase
separation} into regions with $x_{1}\sim0$\ and $x_{2}\sim0.02$, and the
volume fraction of the $x_{2}=0.02$\ phase changes as a function of the Sr
doping, in order to achieve the average $x$.

\section{Hole stripes\label{sect stripes}}

A phenomenon that seems to be common to many superconducting cuprates, and has
been attracting enormous interest, is the \textbf{segregation of the
conducting holes into stripes}, while maintaining very good conductivity or
even superconductivity \cite{TSA95,TIU99,WSE99,WBK00,RBC98,KBF03}. The
literature on the subject is vast, and I will deal only with those aspects
related to the anelasticity. On the theoretical side, it is debated whether
these charge stripes compete against superconductivity \cite{Ric97} or on the
contrary they are an essential ingredient of HTS \cite{KFE98,Zaa00}. In both
cases, an important issue is the dynamics of the transverse fluctuations of
these stripes, which has been modeled in terms of collective pinning from the
doping impurities by C. Morais Smith \cite{MDH98}. To my knowledge, the only
experimental results on the low frequency stripe fluctuations are the
anelastic measurements presented here \cite{86,88,93,94,99}.

\begin{figure}[tbh]
\begin{center}
\includegraphics[
%natheight=212.250000pt,
%natwidth=483.062500pt,
%height=3.787cm,
width=8.5493cm
]{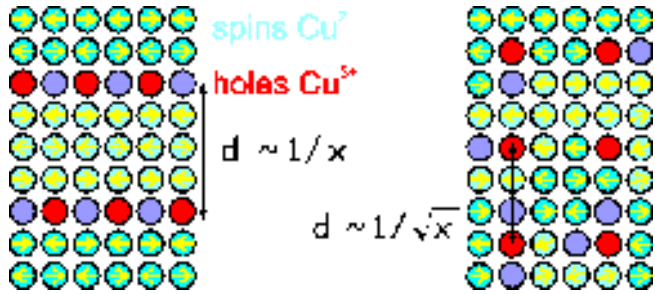}
\end{center}
\caption{Left: ordered stripes with periodicity $d$ commensurate with the
lattice at $x=\frac{1}{8}$. The holes (Cu$^{3+}$ with spin $s=0$) are in red,
while the violet sites along the hole stripes do not have a definite spin
direction, since they are also walls between different AF domains. Right:
possible situation if the holes were uniformly distributed.}%
\label{fig stripes}%
\end{figure}

The first indications of segregation of the charge carriers into stripes came
from measurements of the correlation length $\xi\left(  x\right)  $ for the AF
order in LSCO, deduced from $T_{\text{N}}\left(  x\right)  $ \cite{CCJ93} and
from NQR experiments \cite{RBC98}: the observation is that the increase of
$\xi$ on cooling is limited to a length $l\sim a/x$ ($a$ is the lattice
constant). Since the AF correlations can develop only over regions free holes
(which have $s=0$ instead of $\frac{1}{2}$), $l$ should represent the size of
domains free of holes and, if the holes were uniformly distributed over the
CuO$_{2}$ plane, their separation should scale as $l\sim a/\sqrt{x}$, while
$l\sim a/x$ indicates that the holes are in one-dimensional walls of fixed
hole density separating hole-free domains or stripes (see also Fig.
\ref{fig stripes}). Note that these charge stripes are not charge-density
waves, having a much sharper modulation and allowing the Cu$^{2+}$ spins to
form AF domains between in the charge-poor regions.

Several other indirect indications of the existence of the charge stripes have
been found \cite{RBC98,Mar97}, including the observation of inhomogeneous Cu-O
bond lengths in the CuO$_{2}$ planes with probes of the local structure like
EXAFS \cite{BSR96} and pair-distribution functions (PDF) from neutron
diffraction \cite{BD02}. The strongest evidence of the existence of parallel
magnetic and charge stripes comes from magnetic inelastic neutron diffraction,
which reveals one-dimensional dynamic charge and magnetic correlations with a
spacing $d$ incommensurate with the lattice parameter and decreasing with
doping as $d\propto1/x$ \cite{YLE97}. In addition, the direction of the
modulation changes from diagonal to parallel with respect to the Cu-O bonds
when increasing doping above $x=0.055$, which also separates the
semiconducting from the superconducting region \cite{WBK00,FYH02}. These
correlations are observable also as static structural modulations in the
region of the phase diagram $x\simeq\frac{1}{8}$ with the LTT\ phase
stabilized by partial substitution of La with Nd \cite{TSA95,TIU99}. In fact,
at $x=\frac{1}{8}$ the stripe spacing becomes commensurate with the lattice,
and the LTT tilt pattern provides a modulation to which the stripes are
locked. The situation for $x=\frac{1}{8}$ deduced from neutron diffraction is
represented in the left hand of Fig. \ref{fig stripes}. The hole stripes act
as antiphase boundaries between regions where the Cu$^{2+}$ spins have AF
correlation; varying doping does not modify the hole density within a charge
stripe, which remains 0.5, but only the stripe separation (hence $d\propto
1/x$). It is also found that, on cooling, the formation of the \textbf{charge
stripes precedes} that of the \textbf{AF spin stripes} \cite{TSA95}.

\section{Nd$_{2-x}$Ce$_{x}$CuO$_{4+\delta}$}

\begin{figure}[tbh]
\begin{center}
\includegraphics[
%natheight=313.437500pt,
%natwidth=373.500000pt,
%height=6.6667cm,
width=7.9342cm
]{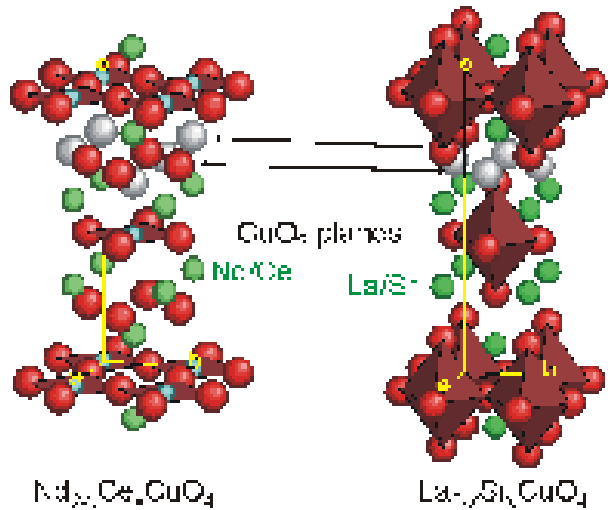}
\end{center}
\caption{Comparison between the T (LSCO) and T ' (NCCO) structures. The arrows
indicate the correspondence between the O interstitial positions (white) in
one structure and O in the NdO or LaO blocks in the other one.}%
\label{fig LCONCOstr}%
\end{figure}

The structure of La$_{2}$CuO$_{4}$, also called T structure \cite{TTU89}, is
one of three possible structures of A$_{2}$BO$_{4}$. Nd$_{2-x}$Ce$_{x}%
$CuO$_{4}$, instead, presents the so-called T' structure, where the cations
and the O atoms of the CuO$_{2}$ planes are in the same positions as in the T
structures, while the O atoms in the (Nd/Ce)O blocks correspond to the
interstitial sites of La$_{2}$CuO$_{4+\delta}$; on the other hand, the
interstitial positions in the T' structure correspond to the apical O atoms in
the T structure. The correspondence between the oxygen positions in the two
structures is shown by the arrows in Fig. \ref{fig LCONCOstr}. The result of
this shift in the oxygen positions is that there are no short Cu-O bonds along
the $c$ axis and therefore no CuO$_{6}$ octahedra, but only flat CuO$_{2}$ planes.

It should be mentioned that Nd$_{2}$CuO$_{4}$ supports only electron doping
(substituting Nd$^{3+}$ with Ce$^{4+}$), at variance with all the other
superconducting cuprates.

\section{The anelastic spectrum}

\begin{figure}[tbh]
\begin{center}
\includegraphics[
%natheight=340.375000pt,
%natwidth=681.500000pt,
%height=6.0385cm,
width=12.0331cm
]{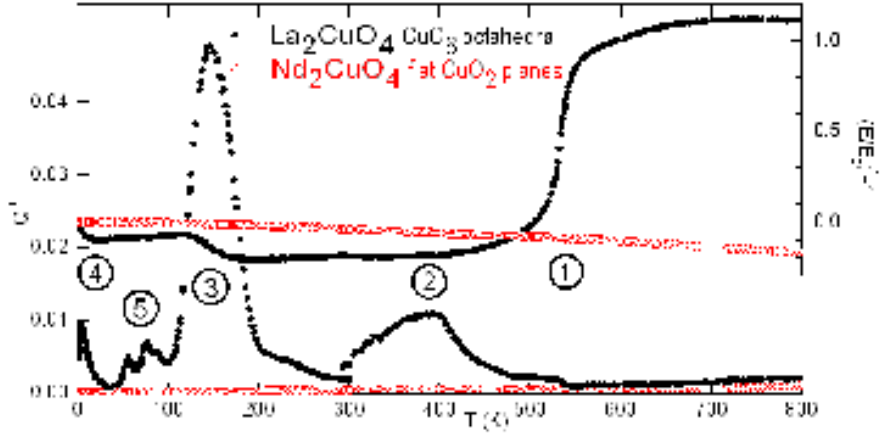}
\end{center}
\caption{Elastic energy loss and Young's modulus of La$_{2}$CuO$_{4}$ (500 Hz)
and Nd$_{2}$CuO$_{4}$ (800 Hz) from 1 to 800~K.}%
\label{fig LCONCO_AS}%
\end{figure}

The anelastic spectrum of LSCO contains several relaxation processes whose
intensity and appearance strongly depend on the type and level of doping.
Figure \ref{fig LCONCO_AS}\ present the elastic energy loss and Young's
modulus between 1 and 800~K of stoichiometric semiconducting La$_{2}$CuO$_{4}
$ and Nd$_{2}$CuO$_{4}$. It should be stressed that as-prepared La$_{2}%
$CuO$_{4+\delta}$ has $\delta\sim0.003$ that drastically modifies the
anelastic spectrum, and the result of Fig. \ref{fig LCONCO_AS} is obtained
after accurate outgassing of the excess oxygen. Nd$_{2}$CuO$_{4}$ appears like
a normal solid without defects or excitations:\ the absorption is low and the
elastic modulus decreases regularly by less than 20\% between 0 and 800~K.
Stoichiometric La$_{2}$CuO$_{4}$ is also free of defects, in principle, but
its anelastic spectrum presents extremely intense anomalies. The main
difference between the two compounds is that Nd$_{2}$CuO$_{4}$ has flat
CuO$_{2}$ planes while La$_{2}$CuO$_{4}$ has CuO$_{6}$ octahedra unstable
against tilting (see Sec. \ref{sect LSCOstrc} and Fig. \ref{fig LCONCOstr}).
In fact, almost all the anelastic processes in La$_{2}$CuO$_{4}$ are due to
some type of motion of the octahedra. Starting from high temperature we find
(the numbers are those in Fig. \ref{fig LCONCOstr}): 1) the transformation
from HTT to LTO structure; 2) motion of the domain walls between the two LTO
variants (orthorhombic $b$ axis along $x$ or $y$); 3) octahedra tilt waves of
solitonic type with thermally activated dynamics; 4) local tilts with dynamics
governed by tunneling and interaction with the charge excitations; 5)
thermally activated fluctuations of the charge stripes whose interaction with
the lattice is mainly mediated by the octahedral tilts.

Apart from the well known structural HTT/LTO transformation, all
the other processes have been revealed by the anelastic experiments presented
in the next Sections.

\section{Structural phase transitions}

\subsection{HTT/LTO transformation and determination of the Sr content from
$T_{t}$\label{sect HTT/LTO}}

The acoustic anomalies connected with this transformation have been studied by
other authors \cite{LLN90,SMM94} and, in view of the polycrystalline nature of
our samples, I did not attempt any quantitative analysis of the huge elastic
anomaly and the accompanying rise in acoustic losses (\#1 in Fig.
\ref{fig LCONCO_AS}). I only observe, together with Lee, Lew and Nowick
\cite{LLN90}, that only the region at $T>T_{t}$ may be analyzed in terms of
Landau free energy, as in Ref. \cite{SMM94}, because below $T_{t}$ the motion
of the walls between the two possible LTO variants is predominant both in the
imaginary and real moduli. The motion of the walls is very sensitive to
defects like O$_{\text{i}}$ and Sr, and is difficult to be modeled; their
effect is to mask in most cases the cusps or kinks that would otherwise be
observed in the moduli, making difficult even to establish where $T_{t}$
exactly is.

In the present Thesis the main interest in analyzing the HTT/LTO transition is
for an accurate determination of the actual concentration of Sr and an
estimate of its homogeneity. In fact, the transition temperature $T_{t}$ is
strongly dependent on doping \cite{Joh97}
\begin{equation}
T_{t}\left(  x\right)  =\left(  1-x/0.217\right)  \times535~\text{K ,}
\label{Tt}%
\end{equation}
and the width of the modulus step provides an upper limit to the possible
inhomogeneities in $x$ over a sample, or, at least, by comparing widths of
different $E\left(  T\right)  $ curves it is possible to tell whether a
particular sample appears to be more inhomogeneous than the others.

\begin{figure}[tbh]
\begin{center}
\includegraphics[
%natheight=266.937500pt,
%natwidth=722.187500pt,
%height=6.1549cm,
width=16.556cm
]{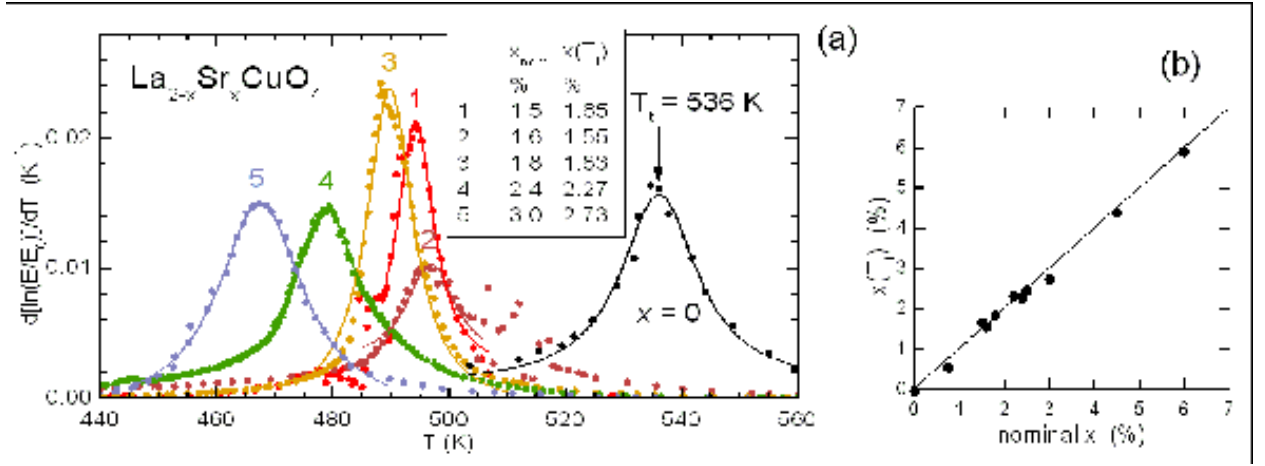}
\end{center}
\caption{(a) Derivative of the relative variation of the Young's modulus of
La$_{2-x}$Sr$_{x}$CuO$_{4}$ at the HTT/LTO transition at various values of
$x$. (b) Sr content deduced from $T_{t}$ versus nominal $x$.}%
\label{fig LSCO xTt}%
\end{figure}

As mentioned above, there is no model available for fitting the $Q^{-1}\left(
T\right)  $ and $E\left(  T\right)  $ curve; as a uniform criterion to extract
$T_{t}\,$from the anelastic spectra, I chose to identify $T_{t}$ with the
inflection point of $E\left(  T\right)  $. In practice I fitted the derivative
of the relative variation of the Young's modulus, $d\ln\left[  E/E\left(
0\right)  \right]  /dT$, with lorentzian peaks, as in Fig. \ref{fig LSCO xTt}%
a, obtaining $T_{t}$ from the temperature of the maximum, and a transition
width $\Delta T_{t}$ from the peak width. The method seems to work, since the
curve for $x=0$ has a maximum exactly at $535$~K and the fitting Lorentzian at
536~K, as expected from Eq. (\ref{Tt}) (it should be mentioned that sometimes
distorted shapes of $E\left(  T\right)  $ have been measured on La$_{2}%
$CuO$_{4+\delta}$, possibly connected with the presence of O$_{\text{i}}$).
The Sr concentrations estimated in this manner for the samples investigated
here are plotted against the nominal $x$ in Fig. \ref{fig LSCO xTt}b, and show
a good agreement. The relationship between transition width and inhomogeneity
of the Sr concentration cannot be simply deduced by translating the peak width
$\Delta T$ into $\Delta x$, since there is a considerable intrinsic width of
the transition (critical softening above $T_{t}$ and domain wall motion
below). This is particularly evident by observing that the curve for $x=0$,
where $\Delta x=0$, is about twice broader than those for $x\sim0.015$. It is
seen, however, that transition \#2 is broader than the others at similar
doping, suggesting homogeneity problems with that sample.

\subsection{LTO/LTT\ transformation\label{sect LTT AS}}

\begin{figure}[tbh]
\begin{center}
\includegraphics[
%natheight=577.437500pt,
%natwidth=577.437500pt,
%height=8.1758cm,
width=8.1758cm
]{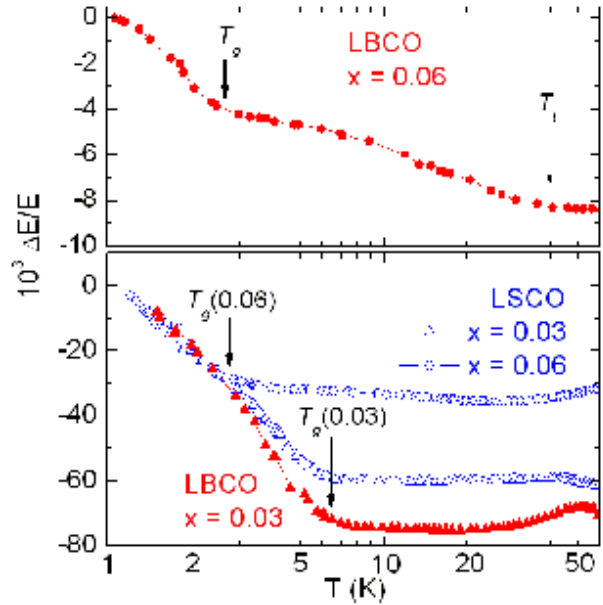}
\end{center}
\caption{Young's modulus of LSCO and LBCO at $x=0.03$ and 0.06. $T_{g}$
indicates the CSG freezing temperature, and $T_{t}$ the onset of the
LTO/LTT\ transition.}%
\label{fig LSBCOTt}%
\end{figure}

As anticipated in Sec. \ref{sect LSCO LTT}, extended domains of LTT phase can
be found only in LBCO or Nd-substituted LSCO near $x=\frac{1}{8}$. The
signature of the LTO/LTT transformation in the anelastic spectrum has been
identified as a small acoustic absorption peak and stiffening below the
transition temperature $T_{t}$ \cite{FNH90}. Figure \ref{fig LSBCOTt} presents
the Young's modulus curves normalized to the value extrapolated to $T=0$~K for
samples where La is substituted with 3\% and 6\% Sr and Ba.

The only clear indication of formation of LTT\ phase is for the
6\%~Ba sample, below $T_{t}$ $\simeq40$~K; such a stiffening is compatible
with the ultrasonic measurements on LBCO with $0.10\leq x\leq0.16$%
\ \cite{FNH90} and with the available phase diagram \cite{AMH89,PR91}, as
shown in Fig. \ref{figLSCO+LBCO}.

Also LSCO at $x=0.06$ exhibits similar weak anomalies, which suggest a
tendency to the formation of LTT domains also in LSCO, in accordance with
high-resolution diffraction experiments \cite{MC00}. In LSCO, however, such
domains should be either confined to the twin boundaries or fluctuating,
extremely small, and without long range correlation; instead, in LBCO and LSCO
co-doped with Nd the diffraction experiments reveal a stable phase with long
range order, which can also provide a pinning potential for the stripes. The
anelastic experiments do not provide any direct information on the extension
or topology of the LTT domains, and the elastic anomalies are likely connected
with the domain boundaries. It is therefore possible that narrow domains of
minority LTT phase in LSCO, with a high perimeter to area ratio, and extended
LTT domains in LBCO produce elastic anomalies of comparable amplitude.

\section{Interstitial oxygen\label{sect Oi}}

In literature extensive reports can be found on La$_{2}$CuO$_{4+\delta}$ that
is heavily doped by electrochemical methods, generally enough to become
superconducting. The interest in these researches is on electrical and
structural phase separation and various types of ordering of O$_{\text{i}}$
that occur under those conditions. My research is instead focused on the low
concentration limit. Under normal conditions, the amount of excess oxygen in
La$_{2}$CuO$_{4+\delta}$ is very small: by electrochemical reduction it has
been estimated $\delta\sim0.005$ in the as-prepared state \cite{BBC98,FLH95};
many estimates of $\delta\sim0.01-0.02$ have also been reported, but, in view
of the results presented in Sec. \ref{sect pxT}, I think that $\delta
\lesssim0.005$ is a more reliable evaluation.

\begin{figure}[tbh]
\begin{center}
\includegraphics[
%natheight=283.312500pt,
%natwidth=738.562500pt,
%height=5.5333cm,
width=14.3352cm
]{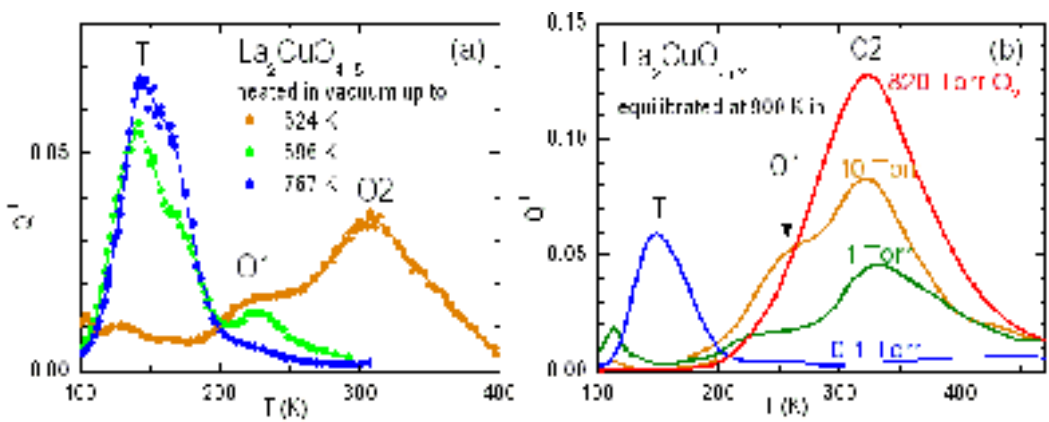}
\end{center}
\caption{Dependence of the anelastic spectrum of La$_{2}$CuO$_{4+\delta}$ on
$\delta$ (a) outgassing an as prepared sample at progressively higher
temperatures in vacuum (300 Hz); (b) equilibrating at 900 K at different O
pressures (800 Hz).}%
\label{fig LCO O1O2T}%
\end{figure}

Figure \ref{fig LCO O1O2T}a presents the effect of outgassing La$_{2}%
$CuO$_{4+\delta}$ at progressively higher temperatures in vacuum starting from
the as prepared state. Initially, two peaks labelled O1 and O2 are present at
225 and 300~K (at 300~Hz). Outgassing in vacuum reduces the intensity of these
peaks in favor of another peak, labeled T, at 150~K. It is obvious to assign
O1 and O2 to interstitial O, which is known to be present in small quantities
in the sites indicated in Fig. \ref{fig LSCOstrc}, and is lost in vacuum at
high temperature \cite{59}. The hypothesis is confirmed by following the
O$_{2}$ partial pressure during heating at 6~K/min in dynamic vacuum in a UHV
system \cite{63}; $p_{\text{O}_{2}}$ starts increasing above 550~K, indicating
the loss of interstitial O, as also explained in Sec. \ref{sect pxT}.

The phenomenology is confirmed in Fig. \ref{fig LCO O1O2T}b with another
sample equilibrated at 900~K in partial pressures of O$_{2}$ ranging from 0.1
to 820 torr \cite{61}. Each oxygenation was preceded by an outgassing at
700-750~$^{\mathrm{o}}$C, in order to obtain an initial reference state; the
sample was then equilibrated at 620~$^{\mathrm{o}}$C for 90-120~min in a
static atmosphere of pure O$_{2}$\ and rapidly cooled.

\begin{figure}[tbh]
\begin{center}
\includegraphics[
%natheight=454.812500pt,
%natwidth=568.500000pt,
%height=5.6519cm,
width=7.0512cm
]{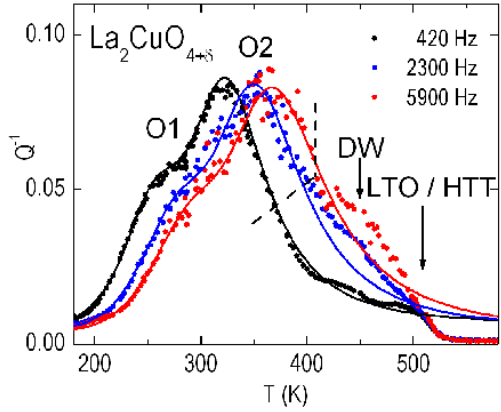}
\end{center}
\caption{Anelastic spectrum of La$_{2}$CuO$_{4+\delta}$ in the region of the
peaks due to hopping of isolated (O1) and paired (O2) interstitial O atoms.
The curves are a fit of the data at the left of the dashed lines.}%
\label{fig LSCO Oi}%
\end{figure}

Figure \ref{fig LSCO Oi} presents a fit to the $Q^{-1}\left(  T\right)  $
curves after equilibration in 10~torr with two Fuoss-Kirkwood peaks, Eq.
(\ref{pFK}), excluding the region at the right of the dashed lines, where
contributions from the structural transition and the DW motion are important.
The fitting parameters are $\tau_{0}=$ $2\times10^{-14}$~s , $E_{\text{O1}%
}/k_{\text{B}}=6000$~K and $\alpha=0.4$ for peak O1 and $\tau_{0}=$
$8\times10^{-14}$~s , $E_{\text{O2}}/k_{\text{B}}=7000$~K and $\alpha=0.5$ for
peak O2.

The most likely explanation for peak O1 and O2 is that \textbf{O1 is due to
hopping of isolated O}$_{\text{i}}^{2-}$\textbf{\ and O2 to pairs of
O}$_{\text{i}}$, possibly peroxyde complexes with lower oxidation state, as
suggested by the dependence of $\delta$ on the oxygen pressure (see Sec.
\ref{sect pxT}). This hypothesis explains the fact that peak O1, expected to
increase linearly with $\delta$, is rapidly overwhelmed by O2, which should
increase with $\delta^{2}$. The parameter $\tau_{0}^{-1}$ of peak O1 is of the
order of magnitude of the local phonon modes promoting the atomic jumps, and
the resulting rate $\tau^{-1}\left(  T\right)  $ agrees with the observation
that the excess oxygen becomes immobile below 140-150~K \cite{CJ96}: in that
temperature range the estimated mean time $\tau$ between subsequent jumps
passes from 5~min to more than 1~hour. It should be mentioned that experiments
based on the analysis of the time and temperature dependence of the
superconducting properties, which in turn depend on the ordering of
O$_{\text{i}}$ \cite{CJ96}, provide activation enthalpies different from
$E_{\text{O1}}$. Those experiments, however, determine the characteristic time
of a complex process, like the aggregation of oxygen into domains and the
possible ordering within the domains, while peak O1directly probes the hopping
rate for single O$_{\text{i}}$ defects.

Two features of peaks O1 and O2 seem at first at odds with a mechanism based
on jumps of O$_{\text{i}}$: the fact that they are very broad and that they
shift in temperature with increasing doping. The broad width can be justified
by the \textbf{strong interaction between excess oxygen and the tilts of the
surrounding octahedra}, as discussed in the next Section. The network of these
octahedra with their easy tilting modes provides a disordered environment for
excess oxygen and mediates the interaction between the interstitial atoms much
more effectively than in a regular lattice. In fact, at lower values of
$\delta$ peak O1 is narrower (\textit{e.g.} $\alpha=0.7$ for the intermediate
curve of Fig. \ref{fig LCO O1O2T}a \cite{61}). This could also explain the
shift of peak O1 to higher temperature with increasing $\delta$ in terms of
correlated and therefore slower dynamics of O$_{\text{i}}$.

These results can be compared with similar measurements made by Kappesser
\textit{et al.} \cite{KWK96}, where only one thermally activated peak was
observed, which was attributed to oxygen\ diffusion. Since that peak was found
after electrochemical doping to $\delta=0.013$ and $\delta=0.035$, it should
correspond to peak O2, with O1 possibly masked as in the highest curve of Fig.
\ref{fig LCO O1O2T}b. The peak temperature is close to that of O2 at a similar
frequency, but the attempt frequency and activation enthalpy were estimated to
be rather low: $\tau_{0}^{-1}=6.5\times10^{8}~$s and $H/k_{\text{B}}=3600$~K.

\subsection{The structure of the interstitial oxygen defect\label{sect Oi str}%
}

\begin{figure}[tbh]
\begin{center}
\includegraphics[
%natheight=275.500000pt,
%natwidth=549.062500pt,
%height=5.8694cm,
width=11.6377cm
]{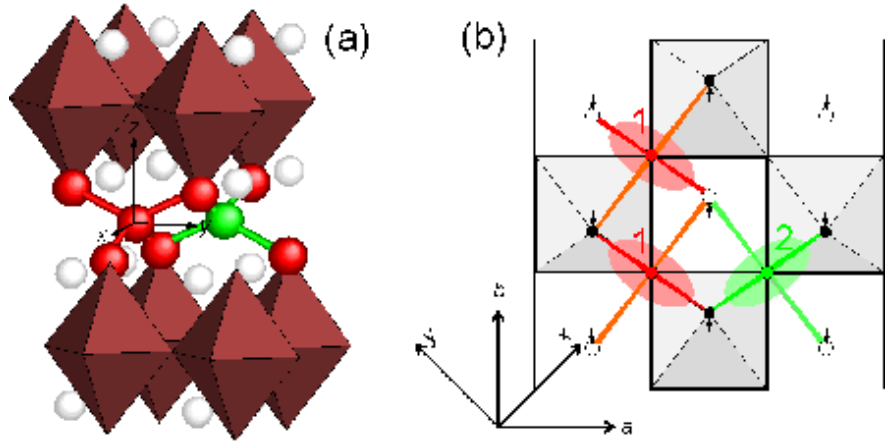}
\end{center}
\caption{(a) two interstitial O atoms in the HTT\ structure; b) interstitial
sites for O in the LTO structure labeled as 1 or 2 depending whether the
shorter distance with the surrounding apical O atoms is along $\left(
1\overline{1}0\right)  $ or $\left(  110\right)  $. The white (black) circles
are the apical O(a) atoms above (below) the plane of the O(i) sites; the
arrows show their shift with respect to the ideal tetragonal HTT structure.
The octahedra below O(i) are shadowed. The elastic dipoles due to the
interstitial O atom are represented as ellipsoids with major axis along the
shorter O(i)-O(a) distances.}%
\label{fig O_int}%
\end{figure}

The \textbf{interstitial sites O(i)}\ for O$_{\text{i}}$ have been determined
by neutron diffraction \cite{JDP89,CCF89,RJS93} and are shown in Figs.
\ref{fig LSCOstrc} and \ref{fig O_int}a. In the HTT structure and in the
absence of ordering into superstructures they are at the center of regular
tetrahedra formed by the neighboring \textbf{apical oxygen atoms, O(a)}, with
O(a)-O(i) distances in the range $1.6\div2.2$~\AA , and larger regular
tetrahedra formed by La atoms. Inspection of Fig. \ref{fig O_int}a shows that
all the interstitial sites for oxygen\ in the ideal HTT structure are
equivalent from the elastic point of view. In fact, by symmetry the elastic
dipole of O$_{\text{i}}$ in the red site must have principal axes $z\parallel
c$ and $x$ and $y$ within the $ab$ plane, parallel to the horizontal edges of
the tetrahedron; the tetrahedron corresponding to the neighboring green site
is rotated by 90$^{\mathrm{o}}$ about the $c$ axis with respect to the red
one, but it may be also be obtained from it by inversion. This means that also
the elastic dipole of O$_{\text{i}}$ in the green sites may be obtained by
inversion of the red one, and therefore the principal axes $x$ and $y$ must be
equivalent; then, the strain due to O$_{\text{i}}$ is isotropic within the
$ab$ plane and there is no change of the elastic dipole for jumps of
O$_{\text{i}}$ in the ideal HTT\ structure.

In the LTO structure, where the anelastic relaxation peaks are observed, the
octahedra are tilted about the $a$ axis (of the LTO representation). Figure
\ref{fig O_int}b shows a top view of three interstitial sites (red and green
circles) and the surrounding apical O atoms, with their displacements with
respect to the tetragonal structure. The black circles are O(a) below the
plane of O(i), with the corresponding octahedra shaded in gray, while the
white circles are O(a) above it. The sites are labeled as 1 (red) or 2 (green)
depending whether the O(i)-O(a) distances are shorter along $\left(
1\overline{1}0\right)  $ (or $x$) or along $\left(  110\right)  $ (or $y$).
After the insertion of an O atom, the lattice will expand in different ways
along the directions of the shorter bonds with respect to the longer bonds. In
the figure is represented the case in which the O(a) which are closer are more
displaced outwards (but in the case of the formation of a short peroxide bond
the opposite can be true): the elastic dipole, represented as an ellipse, has
the major axis along the direction of the shorter bonds and reorients by
90$^{\text{o}}$ when the O atom jumps to a site of different type, modulating
the $\varepsilon_{6}$ strain and giving rise to anelastic relaxation of the
$c_{66}$ elastic constant.

\textbf{The anisotropy of the elastic dipole in the }$ab$\textbf{\ plane
arises from the deviation of the LTO structure with respect to the HTT one},
and therefore is expected to depend on temperature and doping. In fact, the
tilt angles of the octahedra decrease continuously with doping, at least in
La$_{2-x}$Sr$_{x}$CuO$_{4+\delta}$ \cite{HSH96}; in addition, the
orthorhombicity of La$_{2}$CuO$_{4+\delta}$, defined as $\left(  b-a\right)
/a$, varies from 0.0165 at $\delta=0$ to 0.0119 in the oxygen-rich phase
\cite{RJS93,CHC95}.

From Fig. \ref{fig O_int}b it also appears that only jumps along $a$
contribute to anelastic relaxation, because they involve a rotation by
90$^{\mathrm{o}}$ of the anisotropic elastic dipole, while jumps in the $b$
direction do not. The hopping rates in the two directions may also be
different, as discussed in more detail in Ref. \cite{61}, but up to now there
are no experimental indications in this sense.

\subsection{Dependence of $\delta$ on the O$_{2}$ pressure at high temperature
\label{sect pxT}}

Under normal conditions, the amount of excess oxygen in La$_{2}$%
CuO$_{4+\delta}$ is very small: by electrochemical reduction it has been
estimated $\delta\sim0.005$ in the as-prepared state \cite{BBC98,FLH95}, but
no systematic data exist of the equilibrium values of $\delta$ as a function
of temperature and pressure (the $p\delta T$ phase diagram). For this reason,
I studied the absorption of interstitial O and the formation of O\ vacancies
in La$_{2}$CuO$_{4+\delta}$ in equilibrium with O$_{2}$ pressures between
$10^{-4}$ and $10^{3}$ torr or in vacuum at high temperature, in the UHV
system described in Sec. \ref{sect UHV} \cite{63}. Such a study could be done
in a more straightforward manner with a thermobalance; therefore I will omit
the complications arising from measuring the amount of O absorbed or lost in a
closed volume \cite{63} and will concentrate on the results and the
implications on the aggregation and valence state of excess oxygen in La$_{2}%
$CuO$_{4+\delta}$.

Two different temperature regions are found: \textbf{below about
600}$~^{\text{o}}$\textbf{C} the equilibrium down to very low pressures of
O$_{2}$ involves only \textbf{interstitial oxygen}\ atoms with negligible
influence of O vacancies (V$_{\text{O}}$), while \textbf{above 700}%
$~^{\text{o}}$\textbf{C} large amount of \textbf{vacancies} are created.

I determined the equilibrium values of $\delta$ as a function of pressure at
550$~^{\text{o}}$C, and in view of the very low amount of oxygen absorbed
($\delta<0.01$), it seems appropriate to initially suppose that O$_{\text{i}}$
dissolves as O$^{2-}$ ions, with the charge balance guaranteed by the change
of the valence of Cu from Cu$^{2+}$ to Cu$^{3+}$ (La$^{3+}$ does not support
mixed valence). The reaction involved in the equilibrium between gaseous
O$_{2}$ and interstitial O$_{\text{i}}^{2-}$ can then be written as
\begin{equation}
\frac{1}{2}\text{O}_{2}+2\text{Cu}^{2+}\longleftrightarrow\text{O}_{\text{i}%
}^{2-}+2\text{Cu}^{3+} \label{O2 - O2+}%
\end{equation}
and at equilibrium the concentrations must satisfy the mass action law
\begin{equation}
K=\frac{\left[  \text{O}_{\text{i}}^{2-}\right]  \,\left[  \text{Cu}%
^{3+}\right]  ^{2}}{p^{1/2}\,\left[  \text{Cu}^{2+}\right]  ^{2}}\,,
\label{eq O2-}%
\end{equation}
where $K\left(  T\right)  $ is the reaction constant and the square brackets
represent the molar concentrations, assumed to be $\ll1$. Introducing the
neutrality condition, $\left[  \text{Cu}^{3+}\right]  =2\left[  \text{O}%
_{\text{i}}^{2-}\right]  =2\delta$, and the fact that for $\delta\ll1$ it is
$\left[  \text{Cu}^{2+}\right]  \simeq1$, one obtains
\begin{equation}
\delta\propto p^{1/6}\text{.} \label{p 1/6}%
\end{equation}

\begin{figure}[tbh]
\begin{center}
\includegraphics[
%natheight=432.875000pt,
%natwidth=432.875000pt,
%height=6.9018cm,
width=6.9018cm
]{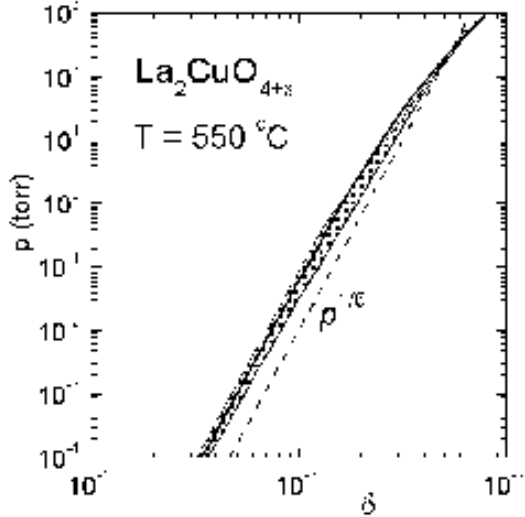}
\end{center}
\caption{The experimental $\delta$ after equilibration at 550~$^{\mathrm{o}}$C
in a pressure $p$ of O$_{2}$ is within the gray region; the continuous line is
a fit with Eq. ( \ref{delta pairs} ) and $a=0.003$, $b=0.02$.}%
\label{fig LCO pvsd}%
\end{figure}

The experimental $\delta\left(  p\right)  $ relationship is the gray region in
Fig. \ref{fig LCO pvsd} and shows deviation from $\delta\propto p^{1/6}$
already at $\delta\sim0.002$. A transition to $p^{1/n}$ laws with $n=4,3,2$
could result from the formation of \textbf{peroxide species} like
O$_{\text{i}}^{-}$ or neutral interstitial O; for example, the formation of
neutral O$_{\text{i}}$ involves the reaction $\frac{1}{2}$O$_{2}%
\longleftrightarrow$O$_{\text{i}}$ with the equilibrium condition $K=$
$\delta/p^{1/2}$ instead of Eq. \ref{eq O2-}

Peroxide species are indeed found in La$_{2}$CuO$_{4+\delta}$
\cite{CCJ92,MC97}, but at much higher doping levels; on the other hand, the
anelastic spectra of Fig. \ref{fig LCO O1O2T} clearly show that interstitial O
forms two distinct species already at $\delta\ll0.01$; there are also
indications that the second species consists of pairs of interstitial O atoms
stabilized by partially covalent bonds with one of the neighboring apical O
atoms. These peroxide complexes can be indicated as O$_{\text{2i}}^{2-}$ in
the equations for the chemical equilibrium, since involve two interstitial O
atoms, and the reaction for their formation is
\begin{equation}
\text{O}_{2}+2\text{Cu}^{2+}\longleftrightarrow\text{O}_{\text{2i}}%
^{2-}+2\text{Cu}^{3+}\text{.} \label{O2 - 2O2-}%
\end{equation}
We introduce the notation $c_{1}=\left[  \text{O}_{\text{i}}^{2-}\right]  $,
$c_{2}=2\left[  \text{O}_{\text{2i}}^{2-}\right]  $, $h=\left[  \text{Cu}%
^{3+}\right]  $, which yield $\delta=c_{1}+c_{2}$ and the neutrality
condition
\begin{equation}
h=2c_{1}+c_{2}\text{.} \label{neutr}%
\end{equation}
The equilibria between gaseous O$_{2}$ and O$_{\text{i}}^{2-}$ (Eq.
\ref{eq O2-}) and between gaseous O$_{2}$ and O$_{\text{2i}}^{2-}$ can be
written as
\begin{align}
K  &  =\frac{c_{1}h^{2}}{p^{1/2}}\label{equil}\\
K^{\prime}  &  =\frac{\left(  c_{2}/2\right)  h^{2}}{p}\,.\nonumber
\end{align}
which yield
\begin{align*}
c_{1}  &  =K^{1/3}\frac{p^{1/6}}{\left(  2+\frac{2K^{\prime}}{K}%
p^{1/2}\right)  ^{2/3}}\\
c_{2}  &  =\frac{2K^{\prime}}{K}p^{1/2}c_{1}%
\end{align*}
and
\begin{equation}
\delta\left(  p\right)  =c_{1}+c_{2}=ap^{1/6}\frac{\left(  1+bp^{1/2}\right)
}{\left(  2+bp^{1/2}\right)  ^{2/3}} \label{delta pairs}%
\end{equation}
with $a=K^{1/3}\,$and $b=\frac{2K^{\prime}}{K}$. At low pressure, O$_{2}$
dissolves only as isolated O$_{\text{i}}^{2-}$ ions, with $\delta
\simeq2^{-2/3}ap^{1/6}$, while at $p>b^{-2}$ also the peroxide pairs are
formed and $\delta$ tends to $\left(  2K^{\prime}\right)  ^{2/3}p^{1/3}$. The
case without formation of the peroxide pairs is reobtained setting $b=0$ (or
$K^{\prime}=0$). The continuous line in Fig. \ref{fig LCO pvsd} is Eq.
(\ref{delta pairs}) with $a=3.0\times10^{-3}$~torr$^{-1/6}$, $b=0.02^{-3}%
$~torr$^{-1/2}$, which yield a perfect fit except between 0.2 and 0.4~torr,
where it is higher than the experiment. It can be concluded that around
550$~^{\text{o}}$C the equilibrium content of excess oxygen in La$_{2}%
$CuO$_{4+\delta}$ depends on the O$_{2}$ pressure as expected from the
\textbf{dissolution as interstitial divalent O ions }($\delta\sim p^{1/6}%
$)\textbf{, but with the formation of peroxide complexes already at }%
$\delta\sim2\times10^{-3}$; the concentration of such complexes presumably
increases with decreasing temperature, in agreement with the anelasticity results.

\subsection{Coexistence of interstitial and vacancy
defects\label{sect VOinLSCO}}

When heating La$_{2}$CuO$_{4+\delta}$ above 700~$^{\mathrm{o}}$C in a closed
volume, the evolution of O$_{2}$ becomes massive, indicating that O vacancies
(V$_{\text{O}}$) start forming (up to $\delta=-0.08$ for $T=750$%
~$^{\mathrm{o}}$C). Such vacancies can be reversibly filled by increasing
$p_{\text{O}_{2}}$, and no sample decomposition occurs: after 9 cycles of
outgassing followed by oxygenation, a sample was still superconducting at 34
K, with a resistivity at 5 K lower than 1~$\mu$Ohm~cm, and the anelastic
spectrum was still reproducible \cite{63}.

The kinetics for the formation and filling of V$_{\text{O}}$ has been measured
by cycling in temperature and pressure and recording the $p\left(  t\right)  $
curves \cite{63}, and is much slower than that of O$_{\text{i}}$. If
reoxygenation is done at the relatively low temperature of 200~$^{\mathrm{o}}%
$C, a small and fast decrease of $p\left(  t\right)  $ in a closed volume
indicates that $\delta\sim0.002$ oxygen enters as O$_{\text{i}}$, but the
filling of the V$_{\text{O}}$ requires equilibration times of years \cite{63}.
The first consequence is that O$_{\text{i}}$ and V$_{\text{O}}$ may coexist;
the second is that they must be in the CuO$_{2}$ planes, otherwise they would
be immediately filled by the much more mobile O$_{\text{i}}$. The coexistence
of V$_{\text{O}}$ and O$_{\text{i}}$ is confirmed by the fact that traces of
the relaxation peaks due to hopping of O$_{\text{i}}$ are seen both in the
anelastic and NQR spectra \cite{59,62} of samples which have been outgassed
above 700$~^{\text{o}}$C and cooled in vacuum; such samples should have up to
8\% of O vacancies in the CuO$_{2}$ plane. This would also be in accordance
with the observation by neutron diffraction that vacancies are formed in the
CuO$_{2}$ plane and not at the apical positions \cite{KIA90}.

It can be concluded that if one tries to obtain perfectly stoichiometric
La$_{2}$CuO$_{4}$ with a long anneal at temperatures exceeding 700$~^{\text{o}%
}$C in flowing inert gas, then a large amount of O\ vacancies is introduced in
the CuO$_{2}$ plane, unless the inert gas contains O$_{2}$ as an impurity,
with a partial pressure larger than 0.1~torr.

The absorption kinetics at room temperature could not be followed, but it
seems that \textbf{interstitial oxygen can enter in a nearly stoichiometric
sample even at room temperature}. In fact, a sample that was carefully reduced
in vacuum and H$_{2}$ atmosphere displayed intense peaks due to interstitial
oxygen hopping after the sample had been kept for 10 months in a glass tube
closed with a rubber bung and containing silica gel for absorbing humidity.

\section{Collective tilt motion of the oxygen octahedra\label{sect tiltwav}}

\begin{figure}[tbh]
\begin{center}
\includegraphics[
%natheight=230.812500pt,
%natwidth=577.687500pt,
%height=6.1418cm,
width=15.2863cm
]{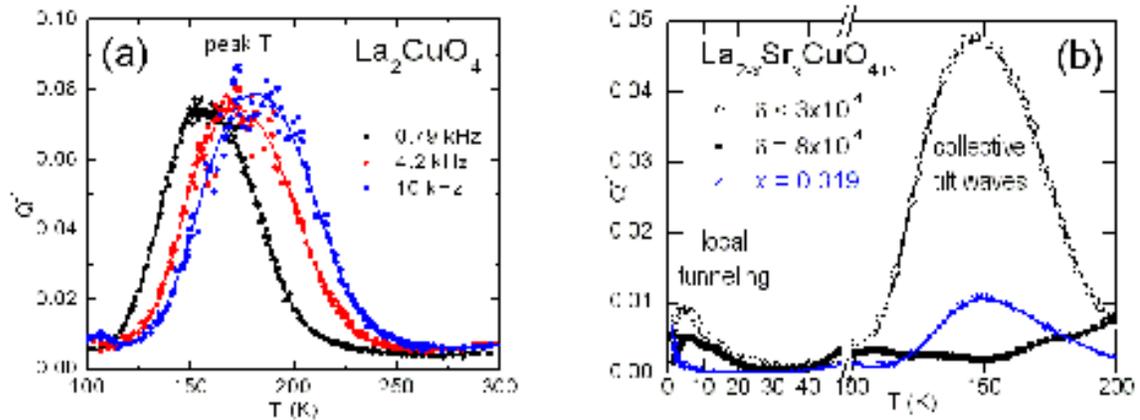}
\end{center}
\caption{(a) peak T attributed to the tilt waves of the octahedra in
stoichiometric La$_{2}$CuO$_{4}$ measured at three frequencies. (b) Blocking
effect of $8\times10^{-4}$ excess O and 0.019 substitutional Sr on peak T and
on the relaxation at LHe temperatures.}%
\label{fig LSCO block}%
\end{figure}

In this section I\ will deal with the peak around 150~K labeled as (3) in Fig.
\ref{fig LCONCO_AS} and T in Fig. \ref{fig LCO O1O2T}. This relaxation process
is thermally activated, as shown in Fig. \ref{fig LSCO block}a, and has the
particularity of presenting the \textbf{maximum intensity in the perfectly
stoichiometric material}. This fact is demonstrated in Fig.
\ref{fig LSCO block}b, showing the effect of O$_{\text{i}}$ and substitutional
Sr. The relaxation must be assigned to some intrinsic lattice mechanism, in
view of the extremely high intensity that it may reach in a stoichiometric and
defect-free sample; the amplitude of the process may be better appreciated by
noting that the maximum $Q^{-1}$ is only 0.08 in Fig. \ref{fig LSCO block}a
because of the peak broadening, but the corresponding relative modulus defect
is $\Delta E/E=0.25$. An intrinsic thermally activated relaxation in La$_{2}%
$CuO$_{4}$ is certainly connected with the \textbf{unstable tilt modes of the
octahedra} (see Sec. \ref{sect LSCOstrc}) \cite{59}, but it must also be
different from the motion of twin walls between LTO variants \cite{LLN90},
which should be responsible for the broad relaxation maximum below the
structural transformation at $T_{t}$ , marked ''DW'' in Fig. \ref{fig LSCO Oi}
and ''2'' in Fig. \ref{fig LCONCO_AS}. It must also be a \textbf{collective
motion of the octahedra}, since it is completely blocked by $\delta<0.001$ of
O$_{\text{i}}$: peak T\ in the curve $\delta=8\times10^{-4}$ of
\ref{fig LSCO block}b) is completely absent (the $Q^{-1}$ rise above 150~K is
due to peak O1 from O$_{\text{i}}$ hopping). Substitutional Sr is instead much
less effective than O$_{\text{i}} $ in blocking this collective tilt motion,
since peak T\ is reduced of only $\frac{1}{5}$ by $x=0.019$, a concentration
20 times larger than the concentration of O$_{\text{i}}$ that completely
inhibits the peak; the evolution of peak T\ can be followed for $x$ up to at
least 0.08 (see Fig. \ref{fig LSCOp80K}). This fact can be easily explained by
looking at Fig. \ref{fig LSCOstrc} or \ref{fig O_int}a and considering that an
O$_{\text{i}}^{2-}$ pushes outwards 4 apical O atoms and therefore firmly
blocks the tilts of the corresponding octahedra, which in turn may block
several surrounding octahedra. Instead, substitutional Sr$^{2+}$ constitutes a
much weaker disturbance, than interstitial O$^{2-}$, both from the steric
point of view and because it brings only one instead of two holes.

\begin{figure}[tbh]
\begin{center}
\includegraphics[
%natheight=268.250000pt,
%natwidth=557.750000pt,
%height=5.7156cm,
width=11.82cm
]{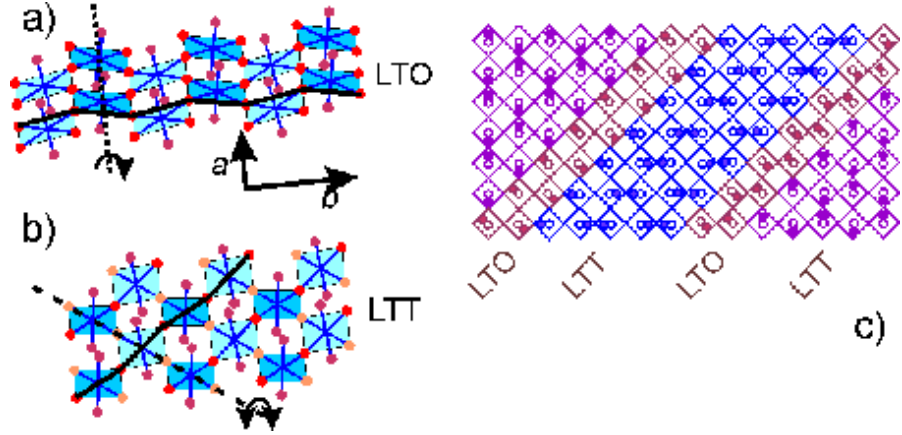}
\end{center}
\caption{Tilt patterns of the LTO (a)\ and LTT (b)\ phases; (c) possible
representation of tilt waves, where only the apical atoms (filled circles) and
their average positions (empty circles) are represented.}%
\label{fig tiltwave}%
\end{figure}

A possible picture of the excitations involving the planes of CuO$_{6}$
octahedra has been provided by Markiewicz \cite{Mar93b}. In a model of rigid
octahedra free to tilt as in the LTO\ structure (Fig. \ref{fig LSCOstrc}), the
tilt of an octahedron propagates all over the plane (Fig. \ref{fig tiltwave}%
a), but if the rotation axes pass through the CuO bonds as in the LTT case,
the correlations are only in rows of octahedra along directions perpendicular
to the tilt axes, since they share the O atoms which are displaced from the
CuO plane, while adjacent rows are little correlated, since the share O\ atoms
remaining in the CuO plane (Fig. \ref{fig tiltwave}b). Markiewicz therefore
observed that a system with \textbf{LTT ground state is reduced to a
one-dimensional array of rows of octahedra}, where all the octahedra belonging
to the same row are tilted according to the same pattern. The resulting
equations of motion are non-linear and admit \textbf{solitonic solutions},
which correspond to one-dimensional \textbf{propagating walls between domains
with different tilt patterns} \cite{Mar93b}, as schematically represented in
Fig. \ref{fig tiltwave}c). Although extended X-ray absorption fine structure
\cite{HSH96} (EXAFS) and atomic pair distribution function \cite{BBK99} (PDF)
measurements of La$_{2-x}$Sr$_{x}$CuO$_{4}$ exclude a prevalent LTT local tilt
at small $x$, the model by Markiewicz \cite{Mar93b} of a LTO\ structure
arising from a LTT ground state (dynamic Jahn-Teller phase) seems to provide a
good framework for analyzing the dynamics of the connected octahedra without
charge doping.

\begin{figure}[tbh]
\begin{center}
\includegraphics[
%natheight=396.750000pt,
%natwidth=938.750000pt,
%height=5.6365cm,
width=13.2544cm
]{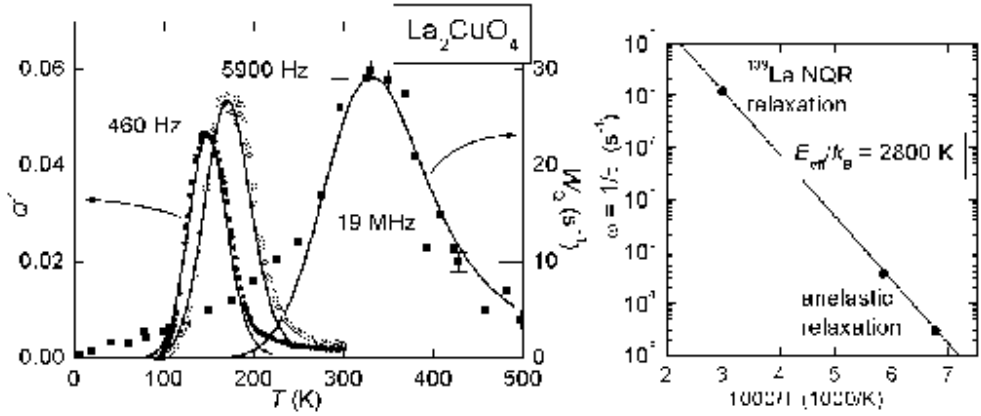}
\end{center}
\caption{Peak T after subtraction of the background measured in the anelastic
$Q^{-1}$ and $^{139}$La NQR relaxation rate. Also shown is the Arrhenius plot
$\tau^{-1}$ deduced from the three maxima. }%
\label{fig LCO ASNQR}%
\end{figure}

Additional evidence that peak T\ is connected with the oxygen octahedra and
more insight on the nature of motion involved came from the collaboration with
Rigamonti and Corti, who made \textbf{Nuclear Quadrupolar Resonance} (NQR)
experiments on the same samples measured by us \cite{64}. Details on the NQR
principles and technique can be found in Refs. \cite{64,RBC98} and will be
omitted here for reasons of space. I will only mention that in those
experiments the spins of the $^{139}$La nuclei can be driven out of
equilibrium by radio frequency pulses, and the effective relaxation rate
$W_{\text{Q}}$ for reaching equilibrium can be measured. Such a rate is
determined in the present case by the fluctuations of electric field gradient
(EFG) at the La nuclei, and in particular by the \textbf{fluctuating
displacements of the surrounding apical O atoms}. Under such conditions, there
is correspondence between $W_{\text{Q}}\left(  \omega,T\right)  $ and
$Q^{-1}\left(  \omega,T\right)  $, since \cite{RBC98}
\begin{equation}
W_{\text{Q}}\left(  \omega,T\right)  \propto\int dt\,e^{-i\omega
t}\left\langle x\left(  t\right)  x\left(  0\right)  \right\rangle =J\left(
\omega,T\right)  \label{WQ}%
\end{equation}
where $J$ is the spectral density of the displacements $x$ of the apical O
atoms. On the other hand, as shown in Sec. \ref{sect FDT}, also $Q^{-1}\left(
\omega,T\right)  \propto J$, since the same displacements, for nearly rigid
octahedra, are coupled to the in-plane shear. Therefore, if such displacements
produce peak T in the elastic energy loss, they must also produce a peak in
$W_{\text{Q}}\left(  \omega,T\right)  .$ This has been indeed observed
\cite{64}, as shown in Fig. \ref{fig LCO ASNQR}.

The relaxation rate $\tau^{-1}$, determined from the condition
$\omega\tau=1$ at the maxima of the $Q^{-1}\left(  T\right)  $ and
$W_{\text{Q}}\left(  T\right)  $ curves, is also plotted in logarithmic scale
against the reciprocal of temperature. The three points can be closely fitted
by a straight line $\tau^{-1}=\tau_{0}^{-1}\exp\left(  -E_{\text{eff}%
}/k_{\text{B}}T\right)  $, $\tau_{0}=1.9\times10^{-12}$~s and $E_{\text{eff}%
}/k_{\text{B}}=2800~$K, clearly indicating that the same process is observed
by both techniques. Both sets of curves have been fitted (continuous
lines)\ with $\tau_{0}=$ $1.7\times10^{-12}~$s integrating over a gaussian
distribution of $E_{\text{eff}}/k_{\text{B}}$ with mean value $2800~$K and
width of $\sim300~$K \cite{64}.

In order to gain some insight on the nature of the barrier $E_{\text{eff}}$,
one should develop the statistical mechanics of the one-dimensional model of
correlated tilts of Markiewicz \cite{Mar93b}, which is a rather difficult
task. However, one can apply the one-dimensional models of non-linear lattice
dynamics \cite{Aub75}, which have been used to describe the correlated
dynamics of off-centre atoms in perovskites \cite{TR87}. In these models the
$i$-th atom moves in a double-potential of the form $V=\sum_{i}-ax_{i}%
^{2}+\,bx_{i}^{4}+\,\overline{c}x_{i}x_{i+1}$, with minima separated by an
energy barrier $E=(a^{2}/4b)$ and with a bilinear coupling to the neighbors.
The coupling constant $\overline{c}$ takes into account a cluster average over
configurations. The resulting equation of motion has solitonic solutions
similar to those found for the octahedra in La$_{2}$CuO$_{4}$,\cite{Mar93b}
and it has been shown \cite{Aub75,TR87} that the spectral density $J\left(
\omega\right)  $ of the displacements $x$ contains a resonant peak at the
frequency of vibration in each minimum, and a \textbf{central peak due to the
jumps to the other minimum with mean hopping frequency }$\tau^{-1}=\tau
_{0}^{-1}\exp\left(  E_{\text{eff}}/T\right)  $ with effective barrier
$E_{\text{eff}}\approx1.75E\sqrt{2\overline{c}/a}$, increased with respect to
the local barrier $E$ by the interaction between the octahedra, and $\tau
_{0}=\frac{d}{v}\sqrt{\overline{c}/a}$, where $v$ is the average velocity of
propagation of the soliton-like excitation through the atoms spaced by $d$.

The central peak (see Sec. \ref{sect FDT}) is therefore responsible for the
peak in the NQR relaxation rate and in the elastic energy loss; the effective
barrier $E_{\text{eff}}/k_{\text{B}}=2800$~K is compatible with the
theoretically estimated \cite{PCK91,BMF91} local barrier of $\sim500~$K (see
also Sec. \ref{sect LSCO LTT}) assuming a mean coupling constant $\overline
{c}_{0}\simeq10a$. The width of the gaussian distribution of $\bar{c}$\ is
$\sigma=0.18\overline{c}_{0}$\ for the NQR and $0.25\overline{c}_{0} $\ for
the anelastic data and is justified by the distribution in size and shape of
the regions where the octahedra clusters build up the cooperative dynamics;
such regions are delimited by O$_{\text{i}}$, which blocks the relaxational
dynamics. An asymmetry energy $\Delta E=280~$K, 10 times smaller than
$E_{\text{eff}}$, has been assumed to reproduce the increase of the intensity
of the anelastic peak T, and may also be justified as follows.\ For nuclear
quadrupolar relaxation to occur it is sufficient that a solitonic front passes
near a $^{139}$La nucleus producing a fluctuation of the atomic positions; for
anelastic relaxation to occur, instead, it is necessary that a strain
fluctuation occurs, and the simple propagation of a tilt wave is not
sufficient: referring to Fig. \ref{fig tiltwave}c, for example, it is
necessary that also the size of an LTT region changes with respect to the
neighboring LTO\ region, since LTT and LTO regions have different strain (in
other words it must be $\Delta\lambda\neq0$ in Eq. (\ref{relstr})); such a
process involves energetically inequivalent states and therefore $\Delta
E\neq0$.

The proposed physical picture is therefore that tilt waves can arise,
propagate with a velocity determined by the coupling $\overline{c}$ between
rows of octahedra, and disappear within regions free of defects that can block
the tilt degrees of freedom (O$_{\text{i}}$ in the first place, substitutional
Sr and twin boundaries). The creation and annihilation of such excitations
would result in strain fluctuations and therefore anelastic relaxation. On the
other hand, such tilt waves could extend over more than few tens of lattice
spacings only in accurately outgassed and defect-free crystals and would have
little or no correlation along the $c$ axis, making difficult their
observation by diffraction techniques. \textbf{The lack of correlation among
different planes would be the main difference between these tilt solitons and
the usual twin boundaries} and, to my knowledge, La$_{2}$CuO$_{4}$ is a unique
case exhibiting this type of lattice excitations. The \textbf{conditions
necessary for their existence} are: \textit{i)} the existence of
\textbf{unstable lattice modes describable by a one-dimensional non-linear
equation of motion} (therefore having solitons as possible solutions);
\textit{ii)} a \textbf{low density of pinning centers}, in order to allow the
formation and propagation of such excitations to occur. La$_{2}$CuO$_{4}$
satisfies both conditions, since it has planar arrays of octahedra with little
coupling between different planes, and with the further possibility of
decoupling each plane into rows \cite{Mar93b}; in addition, the level of
defects and impurities can be lowered enough to let these solitonic tilt waves
to develop. I tried to observe similar effects in the isostructural La$_{2}%
$NiO$_{4+\delta}$, but in that case it is impossible to lower the content of
O$_{\text{i}}$ below acceptable levels. Another isostructural material with
planes of octahedra close to an instability is Sr$_{2}$RuO$_{4+\delta}$, but
it might present problems with excess oxygen \cite{BMN97}, and above all the
instability is rotational about the $c$ axis \cite{BRN98}.\ In that case the
rotation of an octahedron is strongly coupled to the rotations of all the
octahedra in the plane, excluding the possibility of a one-dimensional
equation of motion; it is also intuitive that when a rotation of a single
octahedron determines the rotations of the neighboring octahedra in both $x$
and $y$ directions, excitations like these cannot arise (see also Fig.
\ref{fig Ru octa}). Indeed, such rotations would be coupled to the $c_{66}$
elastic constant, which however stiffens normally below 220~K \cite{MYY98}.
Also in RuSr$_{2}$GdCu$_{2}$O$_{8}$ the octahedra are tilted about the $c$
axis, and in fact its low temperature anelastic spectrum does not show any
anomaly of this type \cite{111}.

\section{Tunneling driven local motion of the oxygen octahedra}

\begin{figure}[tbh]
\begin{center}
\includegraphics[
%natheight=324.687500pt,
%natwidth=577.687500pt,
%height=7.2554cm,
width=12.8634cm
]{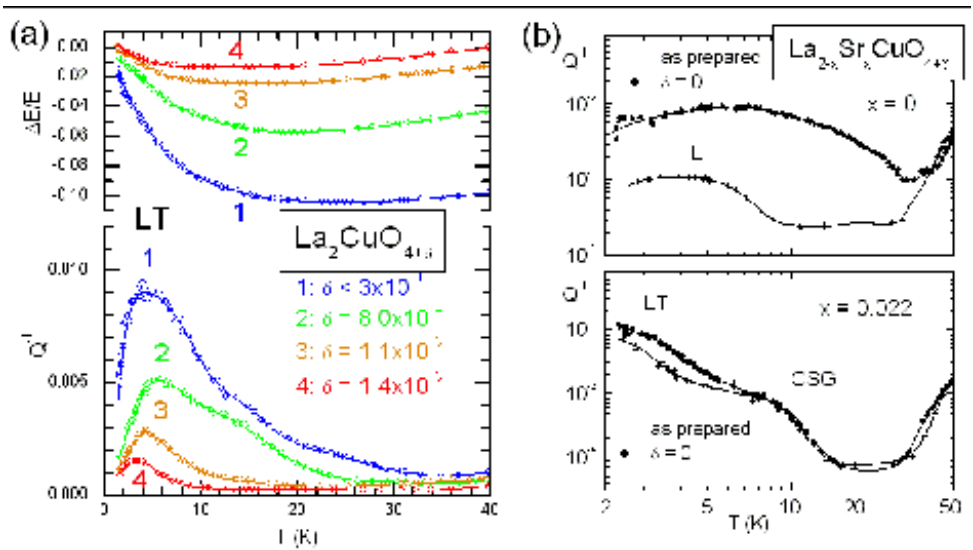}
\end{center}
\caption{Effect of excess O on the relaxation processes LT attributed to
O\ tunneling (a) in a Sr-free sample and (b) in Sr-doped LSCO, where peak
LT\ is accompanied by the relaxation of domain walls in the CSG phase (CSG).}%
\label{fig LSCO tunn1}%
\end{figure}

This section is devoted to peak \#4 in Fig. \ref{fig LCONCO_AS}, also shown in
Fig. \ref{fig LSCO block}b and indicated as ''local tunneling'', and studied
in Refs. \cite{59,72}. For this relaxation process, which I label here as LT,
it is possible to make considerations similar to those made in the previous
section for peak T. Its intensity, again better appreciated from a modulus
defect up to 10\% (Fig. \ref{fig LSCO tunn1}a), is extraordinarily high if one
considers that at 10~K only small atomic displacements that driven by
tunneling may occur. Therefore, a great fraction of the lattice must
participate in the motion, and, again, the unstable \textbf{tilt modes of the
octahedra} are the first candidates, but \textbf{on a local scale} (single
octahedra or even oxygen atoms).

A confirmation from this hypothesis comes from the \textbf{effect of
O}$_{\text{i}}$: with increasing $\delta$ the relaxation amplitude is
depressed (Fig. \ref{fig LSCO tunn1}a), since the tilts of the octahedra
surrounding O$_{\text{i}}$ are fixed to accommodate the interstitial. Still,
the effect is much less marked than for the collective motion, as can be
appreciated from Fig. \ref{fig LSCO block}b, where in the curve for
$\delta=8\times10^{-4}$ peak T is completely absent, but peak LT is only
halved in intensity. It also appears that the peak shifts to lower temperature
with increasing doping, and I will show now that this effect has to be
attributed to the holes doped by O$_{\text{i}}$.

The effect of \textbf{hole doping} is better studied by substitution of La
with Sr, which we have seen disturbs the lattice much less than O$_{\text{i}}
$, and reveals extremely interesting effects of lattice-charge coupling, also
providing evidence that such a local motion is dominated by atomic tunneling.
The introduction of holes above $\sim0.01$, however, introduces another
relaxation process attributed to the motion of the domain walls (hole stripes)
in the cluster spin glass (CSG) phase, as discussed in Sec. \ref{sect CSG},
which has to be distinguished from peak LT. Figure \ref{fig LSCO tunn1}b shows
the low temperature anelastic spectra of La$_{2-x}$Sr$_{x} $CuO$_{4}$ at $x=0$
and $x=0.022$, where the Sr-doped sample presents an additional step around
10~K in correspondence with the freezing into the CSG phase. The different
nature between the tunneling process, LT, and the CSG absorption is put in
evidence by the fact that excess O depresses peak LT (in the as-prepared
state) in both samples, but does not affect the CSG step at all. This fact
allows the two processes to be distinguished from each other.

\begin{figure}[tbh]
\begin{center}
\includegraphics[
%natheight=324.437500pt,
%natwidth=808.687500pt,
%height=5.7617cm,
width=14.2671cm
]{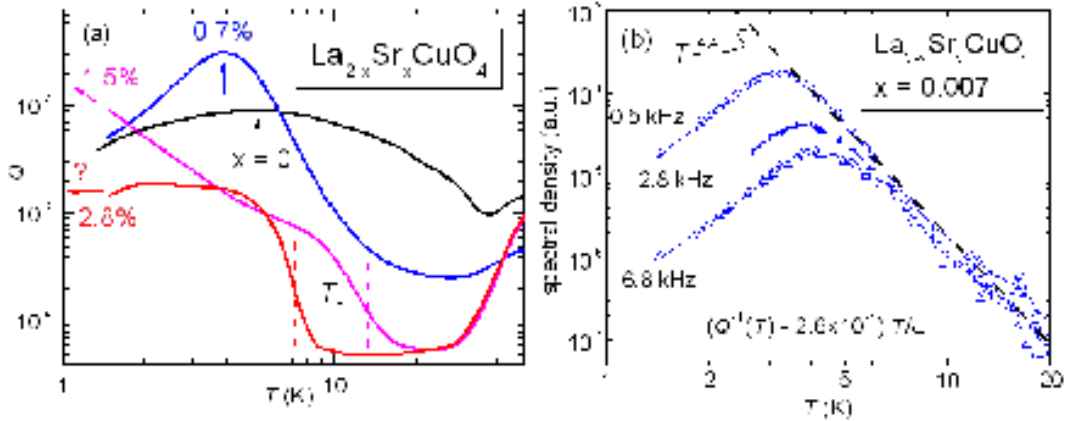}
\end{center}
\caption{Left: elastic energy loss coefficient of slightly doped LSCO at LHe
temperature. The arrows indicate peak LT; the steps in correspondence with the
vertical bars at $T_{g}\left(  x\right)  $ are due to the onset of the CSG
phase. Right: peak LT for $x=0.007 $ measured at three frequencie, after
subtraction of a constant backgroung.}%
\label{fig LSCO tunn}%
\end{figure}

Figure \ref{fig LSCO tunn} shows the $Q^{-1}\left(  T\right)  $ curves of
outgassed LSCO at four doping levels between $0<x<0.028$. What appears from
the first three curves at $x=0$, 0.007 and 0.015 is that Sr doping has two
extremely marked effects on peak LT: \textit{i)} a \textbf{shift to lower
temperature, } \textit{ii)} a narrowing and \textit{iii)} an \textbf{increase
of the intensity}. Already at a doping level as low as $x=0.015$ the maximum
of peak LT is below our experimental temperature window, and therefore it is
impossible to say whether the intensity is still higher than that, already
considerable, of the curve for $x=0.007$. For $x=0.028$ even the tail of the
relaxation is no more visible. The effect of shift to lower temperature is
also observable with doping with O$_{\text{i}}$ (Fig. \ref{fig LSCO tunn1}b),
but is obscured by the concomitant depression due to blocking.

With Sr doping the blocking effect is reduced and it appears that
\textbf{doping has a very strong effect on the low temperature relaxation}%
\textit{: }it \textbf{increases its intensity} and shifts it to lower
temperature, which implies an \textbf{acceleration of the pseudodiffusive
dynamics of the octahedra}. Unfortunately the relaxation could be followed
only up to the very small doping level of $x=0.015$, and it is impossible to
extrapolate to the superconducting region of the phase diagram.

The peak for $x=0.007$ is sufficiently well defined to be analyzed, and Fig.
\ref{fig LSCO tunn}b shows the spectral density $J\left(  \omega,T\right)
=\frac{T}{\omega}Q^{-1}\left(  T\right)  $ measured on the three vibration
modes in double logarithmic scale, after subtraction of a common constant
background. It appears that all the curves collapse together in the high
temperature region, as expected from a process that is a superposition of
Debye relaxations with some distributions $g\left(  \tau\right)  $ and
$h\left(  \Delta\right)  $ of relaxation times and strengths. In fact, at high
temperature one can discard $\left(  \omega\tau\right)  ^{2}$ in the
denominator of Eq. (\ref{Debye peak}) and set $J\left(  \omega,T\right)
=\frac{T}{\omega}Q^{-1}\left(  T\right)  \simeq T\int d\tau\,g\left(
\tau\right)  \int d\Delta\,h\left(  \Delta\right)  \,\tau$ which is
independent of $\omega$. The $J\left(  \omega,T\right)  $ obtained for $x=0$
does not satisfy the same condition, because it is described by a broad
distribution of $\tau\left(  T\right)  $ which vary with temperature at a
slower rate than in the case $x>0$, so that the condition $\omega\tau\ll1$ is
not satisfied in the high temperature side of the peak for all the elementary
relaxations. Therefore, from the temperature position and shape of peak LT we
deduce that \textbf{doping changes the mean relaxation rate }$\tau^{-1}%
$\textbf{\ from a slowly varying function of temperature in the undoped case
to a function which increases faster with temperature}. The high temperature
side of $J\left(  \omega,T\right)  $ indicates that, for $x=0.007$, the rate
$\tau^{-1}$ increases faster than $T^{4}$ above 6$~$K.

The above analysis demonstrates that peak LT is due to atomic displacements
changing at an average rate $\tau^{-1}$, and both the low temperature of the
maximum and the fact that $\tau^{-1}\propto T^{n}$ indicate that the rate is
not due to overbarrier hopping with the Arrhenius law, but is determined by
quantum effects. In the standard \textbf{tunneling model}, the changes of
average atomic positions with rate $\tau^{-1}$ are associated with transitions
of the tunnel system (TS) between its eigenstates \cite{CCC98}. The
eigenstates arise from the overlap of the atomic wave functions (of a single
oxygen\ or possibly of a whole octahedron in our case) over two or more minima
of the multiwell potential, like that of Fig. \ref{fig TiltPot}b; the
transitions between the eigenstates are promoted by the interaction between
the TS and the various excitations of the solid, generally consisting of
emission and absorption of phonons and scattering of the conduction electrons
\cite{CCC98}. An important difference between the relaxation process LT and
those due to TS in glasses is that the latter are characterized by an
extremely broad distribution of parameters, mainly the tunneling energy $t$
and the asymmetry between the minima of the double-well potential; instead, in
La$_{2-x}$Sr$_{x}$CuO$_{4}$ the geometry of the tunnel systems is much better
defined, being some particularly unstable configurations of the octahedra. As
a consequence, the TLS relaxation in glasses produces a plateau in the
acoustic absorption and a linear term in the specific heat as a function of
temperature, whereas in La$_{2-x}$Sr$_{x}$CuO$_{4}$ we observe a well defined
peak in the absorption, and no linear contribution to the specific heat has
been reported.

\subsection{Interaction between the tilts of the octahedra and the hole
excitations}

I argue now that the marked acceleration of the local fluctuations with doping
(narrowing and shift to lower temperature of peak LT) is the manifestation of
a \textbf{direct coupling between the tilts of the octahedra and the holes},
similarly to the TS in metals, whose dynamics is dominated by the interaction
with the conduction electrons \cite{CCC98}. The transition rate of a TS is
generally of the form
\begin{equation}
\tau^{-1}\left(  T\right)  \propto t^{2}\left[  f_{\text{ph}}\left(  T\right)
+f_{\text{el}}\left(  T\right)  \right]  \,, \label{tau tunn}%
\end{equation}
where $t$ is the tunneling matrix element and $f_{\text{ph}}\left(  T\right)
$ and $f_{\text{el}}\left(  T\right)  $ contain the interaction between the TS
and the phonons and electrons, whose excitation spectra depend on temperature
and hole density. As shown in Sec. \ref{sect LSCO t} and Fig. \ref{fig LSCOpd}%
b, doping reduces the tilt instability and tilt angle of the octahedra, and
hence makes the minima of the multiwell potential closer to each other and
reduces the potential barriers between them, resulting in a larger $t$
\cite{61}. Such changes are however gradual, and a doping as low as $x=0.007$
can hardly account for the drastic rise of $\tau^{-1}$ only through an
increase of $t$. The greater contribution has to be attributed to the
increased interaction with the doped holes, corresponding to the term
$f_{\text{el}}\left(  T\right)  ,$which is null in the perfectly undoped
material, but immediately becomes $\gg f_{\text{ph}}$. Models for the
interaction between tunneling systems and electrons have been developed and
successfully adopted, but the electronic excitation spectrum of the cuprate
superconductors is certainly different from that of metals, and new models for
the interaction between octahedra and holes are needed. In fact, the
asymptotic behavior of $J\left(  \omega,T\right)  $ at high and low
temperature for $x=0.007$ indicate that $f_{\text{el}}\left(  T\right)  \sim
T^{n}$ with $n\sim3-5$, completely different from TS in standard metals, where
$f_{\text{el}}\left(  T\right)  $ is less than linear \cite{CCC98}.

Unfortunately, with the present experiments it has been possible to follow
peak LT only up to $x=0.015$, but \textbf{it cannot be excluded that} the
pseudo-diffusive lattice modes giving rise to peak LT \textbf{modify the
electron-phonon coupling at higher doping}. In fact, the characteristic
frequencies of these modes are far too low for having any influence on the
electron dynamics for $x\le0.02$, but increase dramatically with doping. In
terms of the multiwell lattice potential, it is possible that at higher doping
the barriers between the minima become small enough to give rise to an
enhancement of the electron-phonon coupling predicted by some models with
anharmonic potentials \cite{BBB91}.

\subsection{Static and dynamic tilt disorder}

The observation of an increased intensity of peak LT at $x>0$ with respect to
$x=0$ (Fig. \ref{fig LSCO tunn}a) provides an explanation for the apparent
discrepancy between the tilt disorder observed by anelastic experiments on
La$_{2}$CuO$_{4+\delta}$ on one side and EXAFS \cite{HSH96} and atomic PDF
\cite{BBK99} measurements on La$_{2-x}$Sr$_{x}$CuO$_{4}$ on the other side.
The anelastic spectra of Fig. \ref{fig LSCO tunn1} indicates a dynamic tilt
disorder increasing with reducing $\delta$ (and therefore doping), while the
latter techniques see the opposite effect of an increasing tilt disorder with
increasing doping through $x$. Actually, the EXAFS spectra and atomic PDF are
sensitive to both static and dynamic disorder; at $x=0$ the instantaneous
fraction of octahedra swept by the tilt waves is relatively small; according
to a crude estimate \cite{64} a few percents of octahedra instantaneously
involved in the tilt waves account for the intensity of NQR relaxation (peak
T) and their effect on the PDF or EXAFS spectra would be undetectable. At
higher $x$, the instantaneous tilt disorder will increase due to both the
disorder in the La/Sr sublattice and the lattice fluctuations. Instead, the
anelastic spectroscopy is sensitive only to the dynamic disorder with
characteristic frequency comparable to the measurement frequency; the
introduction of interstitial O certainly increases the static tilt disorder
but also inhibits the dynamic one, resulting in the depression of peaks LT and
T observed in Figs. \ref{fig LSCO tunn1} and \ref{fig LCO O1O2T}. By
introducing substitutional Sr, which disturbs the lattice much less than
interstitial O, it is possible to observe that \textbf{doping actually
increases the fraction of fluctuating octahedra, and not only the static
disorder}.

\section{Charge and spin inhomogeneities on nanometer scale}

\subsection{The cluster spin glass phase: fluctuations of the hole stripes
between the pinning points\label{sect CSG}}

\begin{figure}[tbh]
\begin{center}
\includegraphics[
%natheight=255.125000pt,
%natwidth=312.250000pt,
%height=5.4388cm,
width=6.6426cm
]{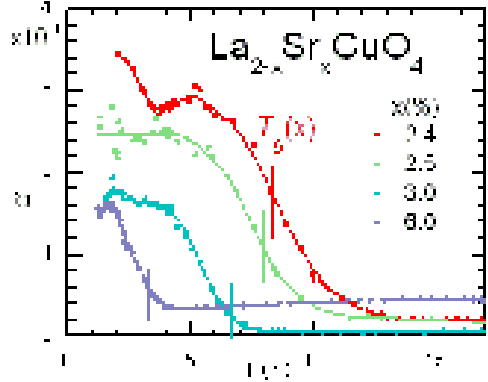}
\end{center}
\caption{Step in the $Q^{-1}$ below the temperature $T_{g}$ for freezing into
the cluster spin glass state at different dopings.}%
\label{fig LSCO CSG}%
\end{figure}

The low-temperature anelastic spectrum of LSCO presents a step-like increase
of the absorption below a doping-dependent temperature $T_{g}\left(  x\right)
$, as already noted in Figs. \ref{fig LSCO tunn1}b and \ref{fig LSCO tunn}a.
Figure \ref{fig LSCO CSG} clearly shows that the temperature at which the
increase of $Q^{-1}$ occurs is in close agreement with the temperature
$T_{g}\left(  x\right)  $ for freezing into the cluster-spin glass
(CSG)\ state (Sec. \ref{sect magn pd} and Fig. \ref{fig LSCOpd}). As also
discussed in Refs. \cite{WUE00,88}, $T_{g}$ is not easy to be determined, and
is often identified with the temperature $T_{f}\left(  \omega\right)  $ at
which the rate of the spin fluctuations falls below the frequency $\omega$
probed by the experiment, $>10^{11}$~s$^{-1}$ for neutron scattering
\cite{WSE99}, $10^{8}-10^{9}$~s$^{-1}$ for NMR/NQR \cite{JCR01}, $10^{6}%
$~s$^{-1}$ for muon spin relaxation \cite{NBB98} $10^{3}-10^{5}$~s$^{-1}$ for
anelastic spectroscopy and $<0.1$~s$^{-1}$ for magnetization measurements
\cite{WUE00}. A system with glassy dynamics is generally describable with a
broad distribution of relaxation times $\tau$, and the maximum relaxation time
$\tau_{\text{M}}$ of such a distribution diverges at $T_{g}$, often following
the Vogel-Fulcher law $\tau_{\text{M}}=\tau_{0}\exp\left[  \frac
{E}{k_{\text{B}}\left(  T-T_{g}\right)  }\right]  $, resulting in a maximum of
the dynamic susceptibilities at $\omega\tau_{\text{M}}\simeq1$. By adopting
the above criterion, one measures the freezing temperatures $T_{f}\left(
\omega\right)  $ such that $\omega^{-1}\sim\tau_{0}\exp\left[  \frac
{E}{k_{\text{B}}\left(  T_{f}-T_{g}\right)  }\right]  $, and the lower is
$\omega$ the closer $T_{f}$ approximates $T_{g}$. For the case of freezing
into the CSG state in LSCO, it seems that $\tau_{\text{M}}$ diverges fast
enough that $T_{f}$ measured by magnetization is a good estimate of $T_{g}$
and can be well approximated by the expression \cite{Joh97}
\begin{equation}
T_{g}\left(  x\right)  \simeq\left(  0.2~\text{K}\right)  /x\,. \label{Tg}%
\end{equation}
The vertical bars in Fig. \ref{fig LSCO CSG} are obtained from this expression
of $T_{g}\left(  x\right)  $ and have an excellent correlation with the rise
of $Q^{-1}\left(  T\right)  $ \cite{75,86,88}, clearly indicating that
\textbf{the acoustic absorption is correlated to the CSG state}. A strong
indication that the absorption step in the CSG state is of \textbf{magnetic
origin within the CuO}$_{2}$\textbf{\ planes} is its \textbf{complete
insensitiveness to interstitial O}, as shown in Fig. \ref{fig LSCO tunn1}b,
whereas the rest of the anelastic spectrum, involving tilts of the octahedra,
is drastically affected by the presence of even minute amounts of
O$_{\text{i}}$.

Regarding the mechanism producing dissipation, it might be similar to the well
known \textbf{stress-induced movement of domain walls} (DW) in ferromagnetic
materials \cite{NB72,SFG01}. In this case, the spin order is antiferromagnetic
(AF), and the \textbf{anisotropic strain might be correlated with the easy
axis }$\mathbf{\hat{m}}$\textbf{\ for the staggered magnetization} \cite{75}.
The elastic energy of domains with different orientations of $\mathbf{\hat{m}%
}$ would be differently affected by a shear stress, and the domains with lower
energy would grow at the expenses of those with higher energy. According to
the prevalent interpretation presented in Sec. \ref{sect stripes}, \textbf{the
walls between the AF clusters should be the charge stripes} where the holes
are segregated, and therefore \textbf{the acoustic absorption below }$T_{g}%
$\textbf{\ would be due to the fluctuations of the hole stripes}. According to
this view, the rise of $Q^{-1}\left(  T\right)  $ would be determined by the
freezing of the spins into AF domains with some disorder in the directions of
the staggered magnetization $\mathbf{\hat{m}}$. According to neutron
diffraction and magnetic susceptibility, in\ La$_{2}$CuO$_{4}$ or slightly
doped La$_{2-x}$Sr$_{x}$CuO$_{4}$ ($x<0.03$) the direction of $\mathbf{\hat
{m}}$ is parallel to the $b$ axis \cite{LAK01}, but we suppose that some
\textbf{distortion of the spin order} must be present, like \textit{e.g.} in
Gooding's model \cite{GSB97}, otherwise the anisotropic strain would be
parallel to $b$ in all domains and there would be no anelastic relaxation.
Note that the fact that the hole stripes must have fluctuations and
topological defects at all temperatures is predicted by several authors
\cite{HCM99,BS01,BSZ01}, but what is needed here is disorder in the average
direction of $\mathbf{\hat{m}} $ from domain to domain. Within this picture,
the diffraction experiments see the ordered fraction of stripes, with regular
spacing and separating antiphase AF domains, while \textbf{anelastic
relaxation is due to the disordered fraction of the spin structure}. The
relaxation of DW pinned by impurities, the Sr dopants in our case, is
generally characterized by a broad distribution of relaxation times, and this
would explain why a plateau rather than a peak is observed. The $Q^{-1}\left(
T\right)  $ step at $T_{g}$ would than be analogous to the $Q^{-1}\left(
T\right)  $ step occurring at $T_{t}$ in correspondence with the structural
HTT/LTO transformation with consequent formation and motion of twin walls (see
the step \#1 followed by the broad peak \#2 in Fig. \ref{fig LCONCO_AS}).

A possible alternative or concomitant mechanism of relaxation within the CSG
phase is the \textbf{motion of kinks} on the otherwise straight stripes, in
analogy with the motion of kinks in dislocations \cite{SFG01}. The dynamics of
such kinks in the hole stripes has been theoretically analyzed \cite{MDH98},
also in the presence of pinning, and seems to correspond to another anelastic
relaxation process attributed to collective stripe depinning, as described in
Sec. \ref{sect depin}.

The above results indicate that the anelastic relaxation below $T_{g}$ is due
to the stripe fluctuations, when they act as walls between frozen AF domains.
On the other hand, it has been established that \textbf{the stripes can be
locked by the LTT structure}, when they are parallel to the LTT modulation
(for $x>0.055$ \cite{WBK00,FYH02}) and commensurate with it ($x=\frac{1}{8}$
\cite{TSA95,TIU99}). The time scale of the neutron scattering experiments that
established this phenomenology is extremely short, and what appears static in
those experiments may be fast for low frequency anelastic spectroscopy;
nonetheless, in case of real locking of the stripes by the LTT\ structure, the
absorption step at the CSG transition should be depressed of the fraction of
stripes that become static. For this reason, we extended the anelastic
measurements to \textbf{La}$_{2-x}$\textbf{Ba}$_{x}$\textbf{CuO}$_{4}$ (LBCO)
\cite{86}, that has a greater tendency to form the LTT phase (see Fig.
\ref{figLSCO+LBCO}).

\begin{figure}[tbh]
\begin{center}
\includegraphics[
%natheight=433.125000pt,
%natwidth=801.750000pt,
%height=6.9106cm,
width=12.7404cm
]{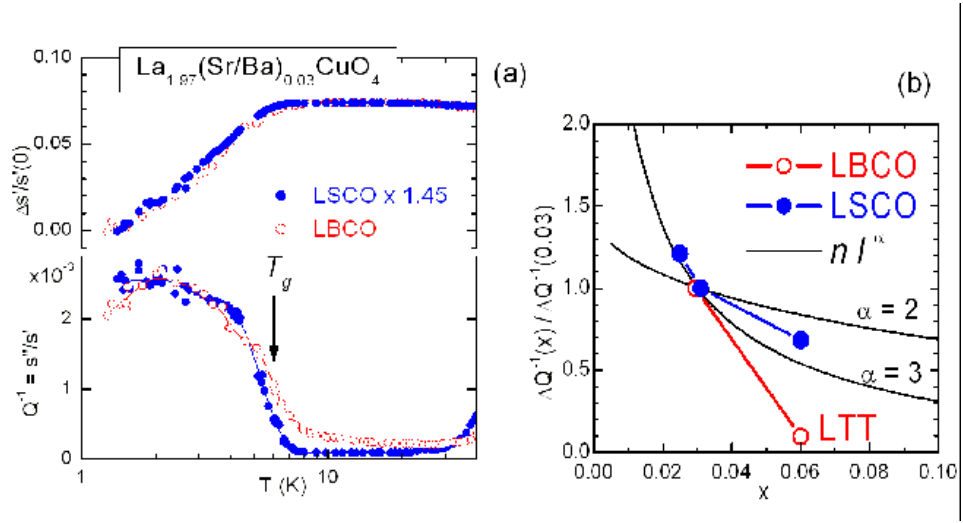}
\end{center}
\caption{(a) Compliance of LBCO and LSCO (rescaled to overlap with that of
LBCO) with 3\% Sr. (b) Doping dependence of the intensity of the $Q^{-1}$ step
at $T_{g}$ for LSCO and LBCO, compared with the relaxation amplitude expected
from dislocations ($\alpha=3$) or ferromagnetic domain walls ($\alpha=2$).}%
\label{fig LSCOLBCO}%
\end{figure}Figure \ref{fig LSCOLBCO}a shows both the absorption and real
parts of the compliance of LSCO and LBCO at $x=0.03$. The data of LSCO
practically overlap with those of LBCO if multiplied by 1.45, and the onset of
the decrease of the compliance coincides with $T_{g}$, indicating that it is
determined by the same mechanism producing the absorption, \textit{i.e.} the
freezing of the stripe fluctuations. The factor $\sim1.5$ can be ascribed to a
greater magnetoelastic coupling in the Ba compound, possibly resulting from
the slightly different atomic sizes and distances.

Figure \ref{fig LSCOLBCO}b presents the amplitude of the $Q^{-1}$ step at
$T_{g}$ for LSCO at $x=0.025$, 0.03 and 0.06, and of LBCO at $x=$ 0.03 and
0.06 normalized to the value assumed at $x=0.03$. These data are compared with
the relaxation strengths expected from dislocations, $\Delta_{\text{d}}\left(
x\right)  $, and from ferromagnetic DW, $\Delta_{\text{DM}}\left(  x\right)
$, strongly pinned by a concentration $x$ of defects in a bidimensional
geometry. In both cases a power law $\Delta\propto n\left(  x\right)
l^{\alpha}\left(  x\right)  $ is expected, with $l$ the mean length between
pinning points, $n$ the line density and with $\alpha=3$, in the case of
dislocations \cite{NB72} or $\alpha=2$ in the case of FM DW \cite{NSV90}. The
$\Delta\left(  x\right)  $ curves have been calculated assuming that the
stripes are fixed at the pinning points \cite{86}, obtaining $\Delta
\propto2x\left(  \frac{1}{\sqrt{x}}-1\right)  ^{\alpha}$, as explained in
Appendix C. In this manner, the experimental data of LSCO are reasonably well
reproduced with $\alpha\lesssim3$, intermediate between the dislocation and DW
cases. It should be remarked that the condition of \textbf{strong pinning},
namely that the walls are really pinned at the dopants and can move only
between them \cite{NSV90}, is probably not verified. In fact, as discussed in
Sec. \ref{sect depin}, Morais Smith \textit{et al.} \cite{MDH98} estimate that
the stripes in LSCO are in the \textbf{weak pinning} regime, where the stripe
may accommodate into the pinning potential over a \textbf{collective length
}$L_{c}$ enclosing several pinning centers. Nonetheless, the dependence of
$\Delta Q^{-1}$ for LSCO on $x$ is intermediate between that typical of
ferromagnetic DW and that of dislocations, and this appears as an indication
that indeed \textbf{the relaxation process below }$T_{g}$\textbf{\ and
therefore the stripe fluctuations in the CSG phase can be assimilated to the
motion of line defects pinned by the Sr dopants}.

Comparing now the data on LSCO with those on LBCO in Fig. \ref{fig LSCOLBCO}b,
it turns out that when La is substituted with 6\% Ba instead of 6\% Sr, then
$\Delta Q^{-1}$ is $\sim7$ times smaller than expected. This depression has to
be attributed to the transformation into the \textbf{LTT structure} below
$T_{d}\simeq40$~K, as indicated by the upward inflection in the Young's
modulus observed only for La$_{2.94}$Sr$_{0.06}$CuO$_{4}$ (Fig.
\ref{fig LSBCOTt}), and is a clear indication of pinning of the DW motion
within the LTT phase, which causes a modulation of the potential felt by the
holes, as explained in Sec. \ref{sect LSCO LTT}. As explained in Sec.
\ref{sect LTT AS}, there are indications of the formation of LTT domains also
in LSCO, but only in LBCO and LSCO co-doped with Nd a stable LTT phase with
long range order sets in, which can provide a pinning potential for the
stripes (see Sec. \ref{sect LSCO LTT}). Only the stripes parallel or nearly
parallel to this modulation should be clamped, and therefore from the
reduction factor $\sim7$ we deduce that about 87\% of the DW are parallel to
the direction of the Cu-O bonds in the LTT phase (or even more if the
transition to the LTT structure is incomplete); this stripe orientation is in
agreement with the direction of the magnetic modulation observed by neutron
scattering for $x\ge0.055$ \cite{WBK00}.

The pinning of the DW of the CSG phase within the LTT structure is well
documented for $x\simeq\frac{1}{8}$, when the stripe spacing is commensurate
with the lattice spacing \cite{TSA95,TIU99}. Figure \ref{fig LSCOLBCO}
demonstrates that \textbf{almost complete pinning of the stripes can occur
also} at $x=0.06$, \textbf{far from the condition of commensurability with the
lattice}; in addition, the anelastic spectroscopy provides evidence of pinning
\textbf{over a time scale }$\tau>$\textbf{\ }$\omega^{-1}\sim10^{-3}%
$\textbf{~s}, much longer than the time scale of neutron diffraction,
demonstrating that the pinned stripes are really static.

\subsection{Electronic phase separation at $x<x_{c}$\label{sect phasep}}

\begin{figure}[tbh]
\begin{center}
\includegraphics[
%natheight=141.625000pt,
%natwidth=170.062500pt,
%height=5.0281cm,
width=6.0231cm
]{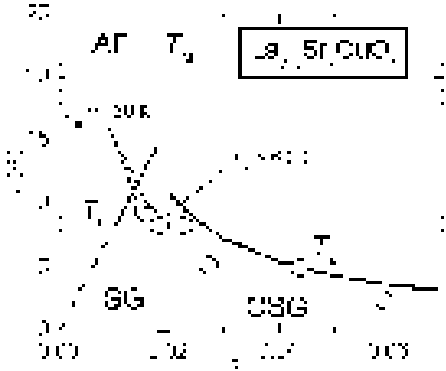}
\end{center}
\caption{Low-$T$ and low-$x$ region of the phase diagram of LSCO. The open
circles are $T_{g}$ deduced from the step in the $Q^{-1}\left(  T\right)  $
curves.}%
\label{fig CSG ps}%
\end{figure}

The study of the low-temperature anelastic spectrum of LSCO has been
concentrated around the critical doping $x_{c}\simeq0.02$ above which no more
long ranged AF order exists (Fig. \ref{fig LSCOpd}), on samples with nominal
doping $x=0.015$, 0.016, 0.018, 0.024, 0.03. The prevalent view is that for
$x<x_{c}$ a spin glass (SG) state still occurs, but different from the CSG one
both in the nature (long range AF order with uncorrelated spin distortions
near the dopants instead of AF clusters) and doping dependence of the freezing
temperature $T_{f}$ ($T_{f}\propto x$ instead of $T_{g}\propto1/x$). This
differences should appear also in the anelastic spectra, since in this picture
there are no walls in the SG state; therefore, below $x_{c}$ there should be
no step-like rise of the acoustic absorption at $T_{f}$, or possibly one
expects a sharp peak corresponding to those appearing in the NQR and $\mu$SR
relaxation rates, in correspondence with the crossover for spin fluctuations
becoming slower than the experimental frequency \cite{RBC98} (the condition
$\omega\tau\simeq1$). On the contrary, \textbf{the anelastic spectra at
}$x<x_{c}$\textbf{\ present the same }$Q^{-1}\left(  T\right)  $\textbf{\ step
as for }$x\simeq0.02$\textbf{, at nearly the same temperature but with smaller
amplitude }\cite{75,93}. The temperatures of the $Q^{-1}\left(  T\right)  $
steps found for both $x>x_{c}$ and $x<x_{c}$ are plotted as open circles in
the region of low-$T$ and low-$x$ of the phase diagram of LSCO\ in Fig.
\ref{fig CSG ps}, where also the canonical $T_{f}\left(  x\right)  $ and
$T_{g}\left(  x\right)  $ lines are indicated, and there is no hint of a
crossover from $T_{f}\left(  x\right)  $ to $T_{g}\left(  x\right)  $.

In view of the difficulties in preparing samples with $x$ determined to better
than 0.1\%\ and considering the possibility of inhomogeneities in the Sr
distribution, we carried out a rather accurate characterization of our
samples. The Sr content $x$ was checked from the temperature $T_{t}\left(
x\right)  $ of the HTT/LTO transformation and the homogeneity was checked from
the width of the transformation appearing in the\ $Q^{-1}\left(  T\right)  $
and $E\left(  T\right)  $ curves, as explained in Sec. \ref{sect HTT/LTO}. The
deviations of the resulting $x$ from the nominal value were not important, as
shown Fig. \ref{fig LSCO xTt}b, and no inhomogeneity could be detected, to
justify regions with $x>0.02$ in samples with $x\le0.018$. The magnetic state
of the samples was characterized by means of NQR and SQUID\ measurements
performed by M. Corti and A. Rigamonti in Pavia: $T_{\text{N}}$ of the samples
with $x=$ $0.015$ and 0.016 was $\sim150$ K, while for $x=$ $0.018$ it was
only possible to obtain an upper limit $T_{\text{N}}<60$~K, due to the overlap
with the intense maximum in the $^{139}$La relaxation rate usually attributed
to spin freezing.

It can be concluded that the similarity of the anelastic spectra for $x<0.02$
with those for $x>0.02$ is a real effect and not due to poor samples. The
anelastic data are then in perfect agreement with the neutron scattering
experiment by Matsuda \textit{et al.}\cite{MFY02} on the magnetic correlations
in La$_{2-x}$Sr$_{x}$CuO$_{4}$ for $x<x_{c}$, where it is found that also at
$x<x_{c}$ the 3D AF ordered phase coexists below $\sim30$ K with domains of
''diagonal'' stripe phase (with the hole stripes at $45^{\text{o}}$ with
respect to the Cu-O bonds). According to these authors, the hole localization
starting around 150~K involves an \textbf{electronic phase separation into
regions with }$x_{1}\sim0$\textbf{\ and }$x_{2}\sim0.02$, and variations in
doping affect only the volume fractions of the two phases. The $Q^{-1}$ step
in samples with $x<0.02$ \cite{75,93} is therefore due to the fraction with
$x_{2}\sim0.02$, so explaining the fact that the onset of the absorption
occurs at the same temperature appropriate for the local doping $x=x_{2}$,
while the decrease of the absorption amplitude reflects the decrease in the
phase with $x=x_{2}$.

\subsection{Thermally activated depinning of the charge
stripes\label{sect depin}}

\begin{figure}[tbh]
\begin{center}
\includegraphics[
%natheight=5.441000cm,
%natwidth=15.642200cm,
%height=5.441cm,
width=15.6422cm
]{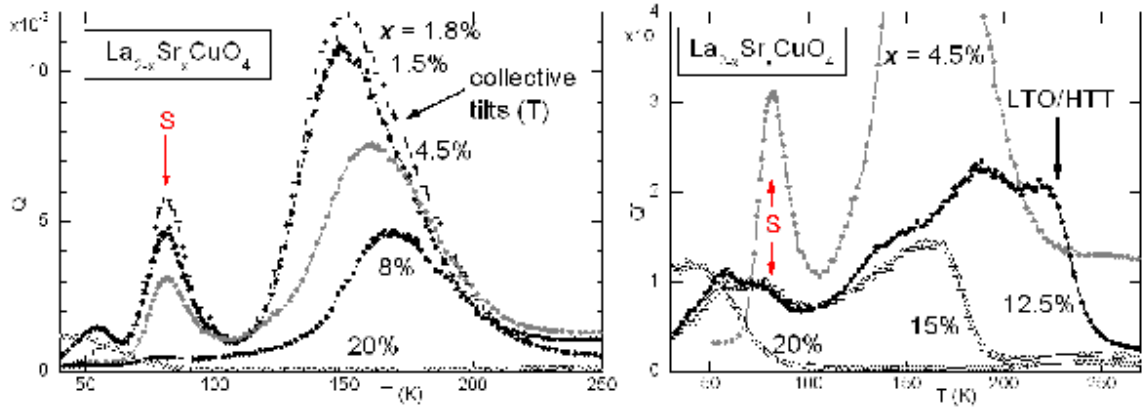}
\end{center}
\caption{Anelastic spectra of LSCO between 50 and 250~K from the underdoped to
the overdoped state, measured at $\sim1$~kHz. The red arrow indicates the peak
attributed to the stripe depinning.}%
\label{fig LSCOp80K}%
\end{figure}

The elastic energy loss rise in correspondence with the freezing into the CSG
phase discussed in the previous section, although connected with the stripe
fluctuation, exists only below $T_{g}$, and for this reason is attributed to
magnetoelastic coupling with the AF domains rather than to the stripes
themselves. The charge stripes, however, do exist also at higher temperature,
\textit{e.g. }in LSCO substituted with Nd and 12\% Sr the superlattice peaks
of charge order appear below 70~K \cite{TSA95}, and various types of phase
diagrams with charge stripes at temperature well above that of the spin glass
state are extensively discussed \cite{KFE98,KBF03}. Carrying on the analogy
between the charge stripes and other linear defects, like dislocations or
domain walls, we tried to identify a possible anelastic signature of the
stripes that overcome the pinning potential by thermal activation, and found
that a relaxation process with maximum at $\sim80$~K for $\sim1$~kHz, here
labeled peak S, has all the expected characteristics \cite{99}; in Ref.
\cite{99} it is also discussed why alternative explanations should be very
unlikely. The peak is evidenced with a red arrow in Fig. \ref{fig LSCOp80K},
where the evolution of the anelastic spectrum of La$_{2-x}$Sr$_{x}$CuO$_{4}$
from $x=0.018$ to $x=0.20$ is shown. The peak well visible up to $x=0.045$ is
peak T\ discussed in Sec. \ref{sect tiltwav} and attributed to the solitonic
tilt waves of the octahedra; it is progressively suppressed by the disorder
introduced by Sr.

\begin{figure}[tbh]
\begin{center}
\includegraphics[
%natheight=288.562500pt,
%natwidth=577.687500pt,
%height=8.1714cm,
width=16.299cm
]{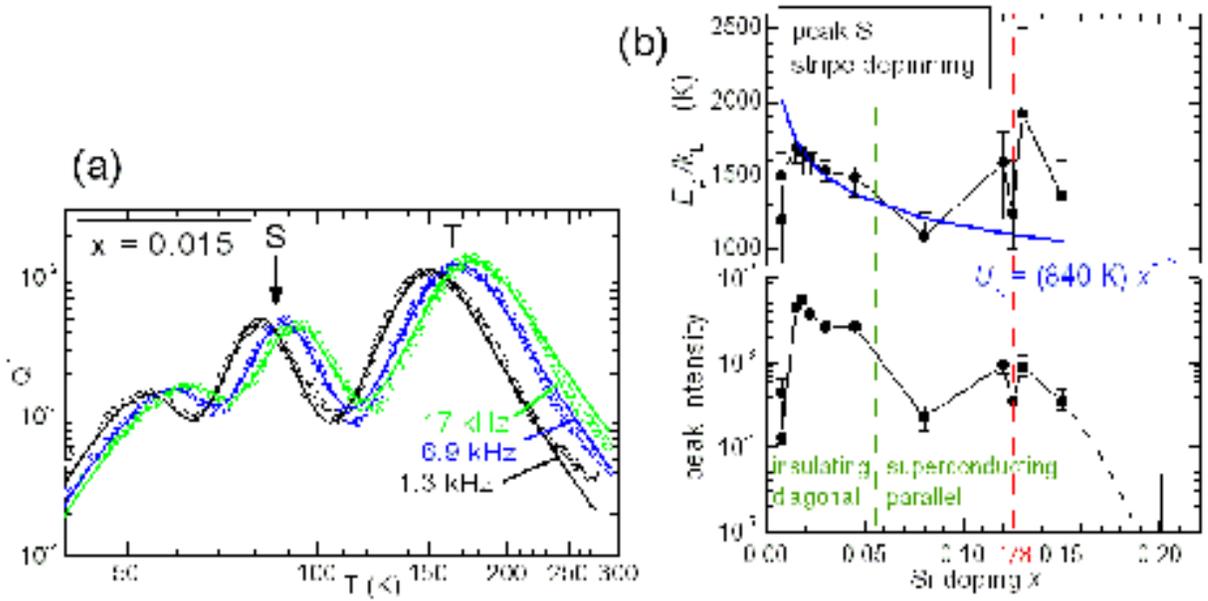}
\end{center}
\caption{(a)\ Example of fit of the anelastic spectrum of LSCO with $x=0.015$,
in order to deduce the activation energy and intensity of the peak attributed
to stripe depinning. (b) Doping dependence of the intensity and activation
energy of the peak at 80~K (for $\sim1$~kHz) attributed to the stripe
depinning. The blue line is the predicted collective pinning barrier,
according to \cite{MDH98}.}%
\label{fig peak S}%
\end{figure}

For $x\ge0.125$ the temperature $T_{t}$ of the HTT/LTO
transformation becomes lower than 250~K and is easily identifiable as a sharp
rise of $Q^{-1}\left(  T\right)  $ and softening of the Young's modulus (the
latter not shown in Fig. \ref{fig LSCOp80K}). Peak S has an intensity with
non-monotonic dependence on $x$ and is visible up to $x\le0.15$; in the
overdoped state, $x=0.20$, there is no trace of the peak, and the only feature
of the anelastic spectrum is the HTT/LTO transformation at 70~K. The
$Q_{\text{max}}^{-1}\left(  x\right)  $ data for peak S are reported in the
lower part Fig. \ref{fig peak S}b, and it is apparent that \textbf{the
intensity of peak S has exactly the doping dependence expected from a
polaronic relaxation due to the hole stripes}: \textit{i)} it \textbf{vanishes
at }$x\rightarrow0$, where there are no holes and therefore no stripes;
\textit{ii)} it \textbf{vanishes in the overdoped region}, $x>0.15$, where the
holes are uniformly distributed as in a normal metallic state; \textit{iii)}
it presents an \textbf{abrupt drop} (note the logarithmic scale of Fig.
\ref{fig peak S}) between at $x=0.045$ and 0.08, in correspondence with the
change of the \textbf{stripe order from parallel to diagonal with respect to
the LTO lattice modulation} \cite{WBK00,FYH02}; \textit{iv) }it has a
\textbf{local maximum around }$x\sim\frac{1}{8}$, where the
\textbf{commensurability between stripe spacing and lattice enhances the
stripe-lattice coupling}, and the \textbf{sharp minimum exactly at }%
$x=\frac{1}{8}$ is most likely due to the fact that the locking between LTT
lattice fluctuations and stripes is so strong that the fluctuations are
depressed. The agreement between the $Q_{\text{max}}^{-1}\left(  x\right)  $
data and what is expected from a relaxation process due to fluctuations of
pinned hole stripes is certainly remarkable, but there is more.

Figure \ref{fig peak S}a shows the anelastic spectrum for $x=0.015$ from 40 to
300~K measured at three vibration frequencies, which allows peak S together
with the neighboring peaks to be reliably fitted, in order to extract the
various parameters of the relaxation S and particularly its activation energy,
that can be identified with the \textbf{effective pinning barrier }$E_{p}$
(the origin of the peak around 60~K is not yet identified). The continuous
curves are a simultaneous fit at all frequencies with Eq. (\ref{pFK2}) for all
three peaks. Peak S is rather broad and symmetric, with $\alpha=\beta=0.7$,
$\tau_{0}=7\times10^{-14}$~s and $E_{p}/k_{\text{B}}=1690$~K. Fits like this
can be done up to $x=0.045$, while for higher doping there is a larger error
in extracting the parameters of peak S, since it is smaller and masked by the
other processes. Nevertheless, the doping dependence of $\tau_{0}$ and
$E_{p}/k_{\text{B}}$ parameters could be followed over the complete doping
range. \textbf{The dependence of }$E_{p}$\textbf{\ on }$x$\textbf{\ }is
particularly interesting, since it \textbf{closely reflects that of }$\log
Q_{\text{max}}^{-1}$, as demonstrated in Fig. \ref{fig peak S}. Also the
pre-exponential factor $\tau_{0}$\ has a general decreasing trend with
increasing doping, passing from $1.4\times10^{-13}$~s to $10^{-10}$~s between
$x=0.015$ and $x=0.08$; for this reason the temperature of the peak measured
around 1~kHz remains close to 80~K, but it is understood that the peak
temperature depends on frequency. Values of $\tau_{0}$ as low as $10^{-10}$~s
are certainly too long for a polaronic relaxation, but may be understood in
terms of an extended defect like the stripe, with a strong lattice
contribution, as argued in what follows.

Before discussing the analogies between the doping dependences of relaxation
strength and activation energy, the present results are compared with the
predictions of the model of Morais Smith \textit{et al.} for the dynamics of
the pinned stripes \cite{MDH98}. From the balance between the electrostatic
interaction $\varepsilon_{c}$ of the hole stripe with the Sr$^{2+}$ dopants,
which tends to pin the stripe into a curved configuration, and the elastic
energy $\varepsilon_{l}$ for bending the stripe, which tends to keep it
straight, both the \textbf{collective length }$L_{c}$ for the stripe
reconformations and the \textbf{pinning barrier }$E_{p}$ for such
reconformations to occur are estimated. It turns out that
\begin{equation}
L_{c}=ax^{-5/6}\left(  \varepsilon_{l}/\varepsilon_{c}\right)  ^{2/3}
\label{Lc}%
\end{equation}
where $\varepsilon_{c}=e^{2}/\left(  \varepsilon_{0}a\right)  \sim0.1$~eV is
the Coulomb energy scale with an isotropic dielectric constant $\varepsilon
_{0}\simeq30\,$and $a\ $the lattice parameter, while $\varepsilon_{l}%
=J\simeq0.1$~eV is the elastic energy of the stripe determined by the AF
exchange constant $J$. Setting $\varepsilon_{l}=J$ tacitly assumes that the
hole stripe fluctuates in a static AF background of spins; this is appropriate
when the spin fluctuations of the Cu$^{2+}$ atoms are frozen below the
characteristic stripe fluctuation frequency under study, as in the CSG state,
but certainly not at 80~K for $x>x_{c}$. I suppose that the spin fluctuations
would lower the elastic energy $\varepsilon_{l}$ of the stripe, and therefore
also $L_{c}$, possibly making a crossover to the strong pinning regime with
$L_{c}<l\simeq a/\sqrt{x}$ , where $l$ is the mean distance between pinning
centers. Otherwise, with the above estimate of $\varepsilon_{l}\simeq
\varepsilon_{c}$ the collective pinning length $L_{c}\simeq ax^{-5/6}\ $is
always larger than $l$, resulting in the \textbf{weak pinning regime}. The
collective pinning energy is estimated as \cite{MDH98}
\begin{equation}
E_{p}=\left(  \varepsilon_{l}\varepsilon_{c}^{2}\right)  ^{1/3}x^{-1/6}%
\sim\left(  1300~\text{K}\right)  \,\,x^{-1/6}\,. \label{Ep}%
\end{equation}
In view of the rough nature of the estimate, especially of $\varepsilon_{c}$,
it is reasonable to expect that the $x^{-1/6}$ dependence of Eq. (\ref{Ep})
can be adopted, but with the constant factor as an adjusting parameter. The
blue line in Fig. \ref{fig peak S} is a fit of $E_{p}\left(  x\right)  $ of
peak S, excluding the $x\sim\frac{1}{8}$ region, with Eq. (\ref{Ep}), and the
constant is found to be $\left(  \varepsilon_{l}\varepsilon_{c}^{2}\right)
^{1/3}/k_{\text{B}}=840$~K. It can be concluded that \textbf{the agreement
between the general trend of }$E_{p}\left(  x\right)  $\textbf{\ and the
predictions for the collective pinning energy barrier of stripes pinned by the
Sr dopants of Ref. \cite{MDH98} is surprisingly good}, and provides further
support to the interpretation of peak S in terms of stripe depinning. It
should be noted, however, that the treatment of Ref. \cite{MDH98} does not
include the interaction with the lattice, which is at least as important as
the electrostatic in interaction with the dopants, as argued in the following Section.

\subsection{Hole-lattice coupling and octahedral
tilts\label{sect hole-lattice}}

Regarding the mechanism producing peak S, it would be tempting to simply
identify the hole stripe with a dislocation or domain wall, and apply the
models of anelastic relaxation from these linear defects, as proposed for the
relaxation in the CSG state in Sec. \ref{sect CSG}. The problem with peak S is
that at 80~K the hole stripes do not separate domains with different
orientations of the anisotropic strain, as is the CSG state, since at this
temperature the spin fluctuate very fast \cite{RBC98}. Therefore, it is not
possible to associate an anelastic strain $\varepsilon^{\text{an}}$
proportional to the area swept by the stripe (the continuum equivalent of the
elastic dipole of a point defect), as in the above models. Instead,
\textbf{different strain contributions }$\varepsilon^{\text{an}}%
$\textbf{\ should be associated with the different configurations that the
stripe may assume by thermal activation over the pinning barrier},
\textit{e.g. }straight or curved or kinked stripe segment. It is likely that
$\varepsilon^{\text{an}}$, like the distortion coupled to the presence of
holes \cite{Mar97}, is \textbf{mainly connected with the in-plane shear also
associated with the tilts of the octahedra}. Then, it is reasonable to assume
that $\varepsilon^{\text{an}}$ is an increasing function of the local degree
of tilting of the octahedra. This would partly contribute to the general
reduction of the relaxation strength with increasing doping, because both the
in-plane shear strain and tilt angles $\theta$ of the octahedra are decreasing
functions of doping (see Fig. \ref{fig LSCOpd}).

It has already pointed out how the sharp variations of $Q_{\text{max}}^{-1}$
with doping correspond to a variation in the degree of coupling between
lattice and stripe; it has to be explained why also the activation energy
$E_{p}$ presents similar features. An explanation can be found by observing
that the \textbf{octahedral tilts} (the most unstable lattice modes) are
responsible for the \textbf{stripe-strain interaction}, and therefore
determine the relaxation strength, but also cause \textbf{stripe pinning from
the lattice}, as demonstrated by the well known locking between LTT tilt
pattern and stripes. Therefore, large tilt amplitudes are associated with both
large relaxation strength and strong pinning from the lattice; in other words,
anomalies in the tilt-stripe coupling result in anomalies in both
$Q_{\text{max}}^{-1}$ and $E_{p}$.

\subsection{The picture of the slow stripe fluctuations after anelastic
spectroscopy\label{sect stripe pict}}

\begin{figure}[tbh]
\begin{center}
\includegraphics[
%natheight=281.375000pt,
%natwidth=801.750000pt,
%height=5.9902cm,
width=16.9645cm
]{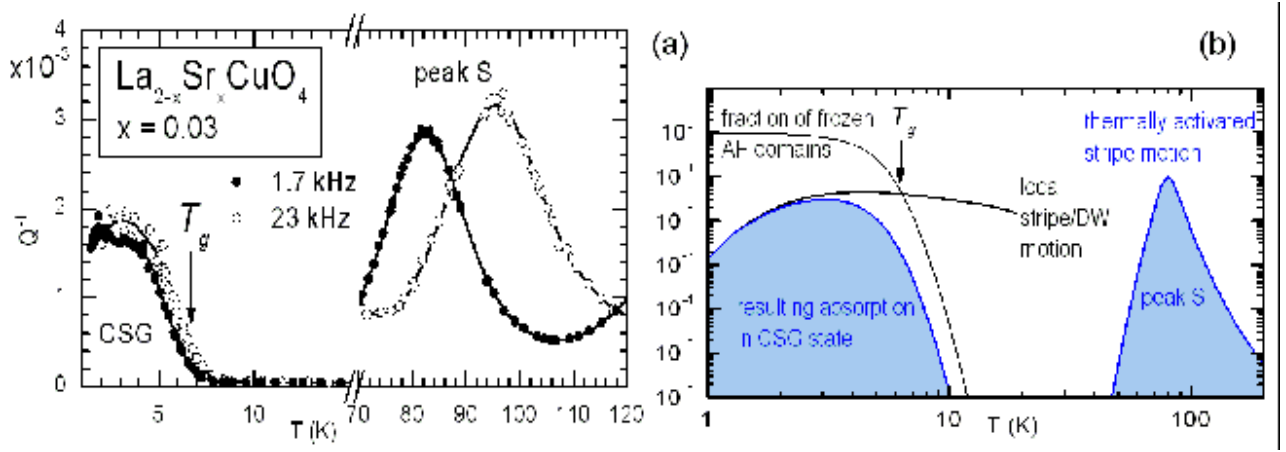}
\end{center}
\caption{(a) Features of the anelastic spectrum of La$_{1.97}$Sr$_{0.03}%
$CuO$_{4}$ associated with the dynamics of the charge stripes: freezing below
$T_{g}$ and thermally activated depinning above 80~K. (b) Sketch of how the
absorption below $T_{g}$ might be decomposed into the contribution from DW
fast local motion multiplied by the fraction of domains that are frozen.}%
\label{fig LSCO 03}%
\end{figure}

The two anelastic relaxation processes we attribute to the stripe motion are
again shown in Fig. \ref{fig LSCO 03}a in a sample with $x=0.03$. In a first
instance \cite{75,86,99} we assumed a strong pinning regime, where the CSG
absorption below $T_{g}$ is due to the motion between pinning points and the
80~K relaxation as the thermally activated depinning process. According to the
analysis of Morais Smith \textit{et al.} \cite{MDH98} the weak or collective
pinning regime would be verified also at very low doping, meaning that there
is no clear distinction between motion of stripe segments that are between or
at the Sr dopants; rather, there is a collective length $L_{c}\left(
x\right)  $ characteristic for the stripe displacements that minimize the free
energy, as mentioned in Sec. \ref{sect CSG} and \ref{sect depin}. If this type
of motion produces peak S around 80~K, its contribution becomes totally
negligible at $T_{g}<15$~K.

This is schematically illustrated in Fig. \ref{fig LSCO 03}b with
double logarithmic scale, where the extrapolation of peak S is clearly
negligible at $T_{g}$. The relaxation appearing below $T_{g}$ must therefore
originate from a much faster and probably more local type of stripe motion,
indicated as a (totally hypothetical)\ broad maximum. The fact that the
elastic energy loss is seen to grow only below $T_{g}$ indicates that, while
the dynamics may be associated to the stripes, the mechanism producing
anelasticity involves the frozen AF domains, as discussed in Sec.
\ref{sect CSG}. This is represented in Fig. \ref{fig LSCO 03}b with the
fraction of domains between stripes having a characteristic fluctuation
frequency $\tau^{-1}$ frozen below the measurement frequency $\omega$; the
resulting $Q^{-1}\left(  T\right)  $ curve is the product of this fraction
with the broad absorption that would be hypothetically measured if the AF
spins where already frozen at higher temperature, allowing the mechanism of
the stress induced domain wall motion to be operative. The fact that this
absorption is broad is indicated by the shape of $Q^{-1}\left(  T\right)  $
below $T_{g}$.

The mechanism governing the dynamics of the domain wall or stripe motion at
these temperatures is necessarily different from the thermally activated
overbarrier motion of peak S, and must be totally quantum mechanical. To my
knowledge there are no theoretical treatments of this problem; the stripe
relaxation time in the quantum limit $\tau\sim10^{-14}$~s estimated in Ref.
\cite{MDH98}, based on ohmic dissipation [$f_{\text{el}}\propto T$ in Eq.
(\ref{tau tunn})], is far too short for explaining the anelastic relaxation
below $T_{g}$ at our frequencies; in fact, the Debye factor $\omega
\tau/[1+\left(  \omega\tau\right)  ^{2}]$ would cut the absorption amplitude
by a factor of $10^{-10}$, making it unobservable. On the contrary, the broad
shape of the $Q^{-1}\left(  T\right)  $ curve below $T_{g}$ indicates that a
consistent weight of the distribution function of the relaxation frequencies
is in the kHz range.

\begin{figure}[tbh]
\begin{center}
\includegraphics[
%natheight=438.687500pt,
%natwidth=658.562500pt,
%height=6.2252cm,
width=9.3181cm
]{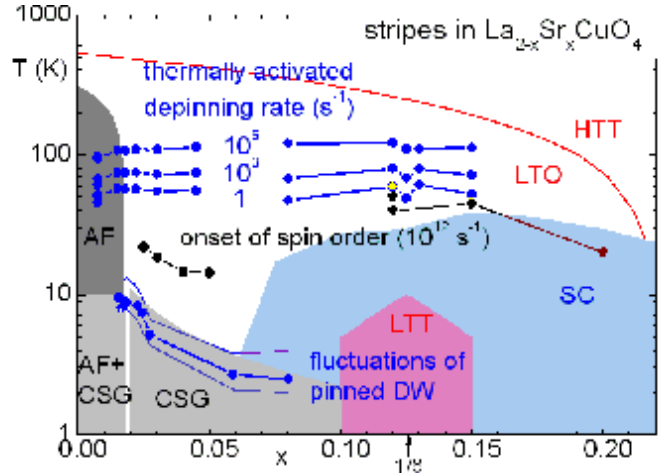}
\end{center}
\caption{Phase diagram of LSCO with the dynamic ranges of the stripe motions
deduced from the anelastic experiments.}%
\label{fig LSCO pd stripes}%
\end{figure}

Figure \ref{fig LSCO pd stripes} is the phase diagram of LSCO (see Fig.
\ref{fig LSCOpd}) with the dynamic ranges of the stripe motions deduced from
the anelastic experiments. The depinning rate curves have been calculated from
the relaxation time $\tau\left(  T\right)  \ $deduced from peak S; they
correspond to the mean relaxation rate for the stripes to overcome the pinning
potential from the Sr$^{2+}$ dopants, presumably in the weak pinning regime.
As explained in the previous Sections, the relaxation rate reflects the abrupt
changes of the effective pinning barrier in correspondence with changes in the
stripe-lattice coupling, like the change of the stripe orientation at
$x=0.055$ or the possible insurgence of LTT\ fluctuations near $x=\frac{1}{8}%
$. The black symbols are the temperatures for the onset of the incommensurate
spin order, deduced from the superlattice peaks in magnetic neutron
scattering; for $x<0.1$ these temperatures correspond to $T_{g}\left(
x,\omega\sim10^{12}~\text{s}^{-1}\right)  $, namely the onset of the CSG phase
on the neutron scattering time scale \cite{WSE99}. The CSG phase on the
(quasi)static time scale from magnetic susceptibility measurements is the gray
region. The points labeled ''fluctuations of pinned DW''\ and the two
accompanying upper and lower lines indicate the rise of the acoustic
absorption (10\%, inflection, 90\% of the step) in the CSG phase. The
absorption is due to the stress-induced motion of the AF clusters separated by
the hole stripes, and in these temperature region there is a substantial
spectral weight of stripe fluctuations with frequencies in the kHz region.

Before concluding I mention a recent ultrasonic experiment \cite{QLW05}, where
the influence of Nd codoping on LSCO has been studied. It is found that an
increase of the Nd content causes an increase of the stripe pinning barrier
deduced from peak S, and eventually the formation of LTT\ phase clamps a
fraction of stripes, as expected.

\section{Summary of the main results obtained in LSCO}

\subsection{Structural transformations}

The main feature of the anelastic spectrum of La$_{2-x}$Sr$_{x}$CuO$_{4}$ is
the \textbf{HTT/LTO structural transformation} at $T_{t}\left(  x\right)  $
(Sec. \ref{sect HTT/LTO}), which produces steps in both the real and imaginary
modulus, accompanied by some broad feature a little below $T_{t}$, due to the
motion of the twin walls. The analysis of the position and width of this
transformation has been useful for checking the actual doping level $x$ and
homogeneity, especially for $x<0.03$. The \textbf{LTO/LTT transformation} is
much more subtle, but can be identified as a rise of the modulus (see
\textit{e.g.} the upper panel of Fig. \ref{fig LSBCOTt}) occurring at the
expected $T_{d}$, as already observed in ultrasound experiments \cite{FNH90}.

\subsection{Absorption in the CSG state}

The other feature that has an evident relationship with other types of
experiments is the step-like rise of $Q^{-1}\left(  T\right)  $ below the
temperature $T_{g}$ at which the Cu spins are frozen into the \textbf{Cluster
Spin Glass} state (Sec. \ref{sect CSG}), as shown \textit{e.g.} in Fig.
\ref{fig LSCO CSG}. The occurrence of a clear $Q^{-1}\left(  T\right)  $ step
below $T_{g}$ convinced us that there must be sufficient coupling between spin
degrees of freedom and strain to produce visible effects in the anelastic
spectra. Considering the prevalent picture of the CSG state as due to
antiferromagnetically correlated spin clusters separated by hole-rich walls
(or stripes), it is obvious to assign the mechanism of elastic energy
dissipation to the \textbf{movement of the walls between spin clusters}. The
effect is analogous to the $Q^{-1}\left(  T\right)  $ step observed below
$T_{t}$ at the HTT/LTO transformation: in the case of the structural
transformation the fluctuating anelastic strain is provided by the different
orientations of the orthorhombic axes of different domains, while below
$T_{g}$ it is due to anisotropic magnetoelastic strain coupled with the
direction of the staggered magnetization in each spin cluster. Having
confidence in the correlation between $T_{g}$ and $Q^{-1}\left(  T\right)  $
step, and in the homogeneity of the samples evaluated from the narrow width of
their HTT/LTO transition, it has been possible to show that (Sec.
\ref{sect phasep}), contrary to the generally accepted opinion, below the
critical doping $x_{c}\simeq0.02$ the CSG phase does not disappear in favor of
a spin-glass state, but there is a phase separation between CSG and AF
hole-poor states, as also indicated by detailed neutron spectroscopy studies
\cite{MFY02}. The anelastic experiments are also in agreement with the
observation, mainly after neutron spectroscopy, that the hole stripes, or
equivalently the spin walls, are mostly parallel to each other and parallel to
the direction of the Cu-O bonds for $x>0.55,$so that they are clamped by the
lattice modulation of the LTT\ phase. In fact, the amplitude of the
$Q^{-1}\left(  T\right)  $ step in a sample transformed into the LTT\ phase
was reduced by almost 90\% with respect to those in LTO phase, consistent with
the hypothesis that \textbf{the majority of walls is clamped by the static LTT
lattice modulation} and therefore does not contribute to the dynamic compliance.

\subsection{Peak S}

Having found such close correlations between the amplitude and temperature of
the $Q^{-1}\left(  T\right)  $ step below $T_{g}$ as a function of doping and
the phenomenology of the hole stripes deduced from several types of
experiments, the next step was to proceed with the analogy between these
stripes and other linear defects and look for possible \textbf{thermally
activated depinning of the charge stripes from the Sr}$^{2+}$%
\textbf{\ dopants}. This kind of process has been theoretically studied but
never observed experimentally; this is understandable, since the dynamic
response of the single Cu spins and holes dominates most spectroscopies, like
magnetic, NMR, dielectric and optical. The anelastic response, instead, is
insensitive to the charge and spin dynamics, except for those excitations or
structures that are coupled to strain, as linear charge structures should be.
The theoretical estimate \cite{MDH98} for the (collective) pinning energy of
the stripes to the Sr dopants is $E_{p}/k_{\text{B}}\gtrsim1300$~K; therefore,
the possible signature of stripe depinning is a thermally activated $Q^{-1}$
peak with effective barrier $\sim E_{p}$. In addition, the intensity of this
peak should vanish at zero doping, where there are no holes, and for $x\ge
0.2$, where the holes are uniformly distributed in a metallic state. The
analysis of tens of anelastic spectra from samples with 13 different Sr doping
levels, allowed this process to be identified with \textbf{peak S} in Fig.
\ref{fig LSCOp80K}; besides the above requirements, the doping dependences of
its intensity and activation energy exhibit other outstanding features shown
in \ref{fig peak S}, like a drop at $x>0.55$, where the stripes have been
observed to change orientation from parallel to diagonal with respect to the
lattice modulation, and a local increase at $x\sim\frac{1}{8}$, where an
instability toward the LTT structure exists, whose modulation would be again
parallel to the prevalent stripe direction and commensurate with their mean
spacing. All these features have been explained in Sects. \ref{sect depin} and
\ref{sect hole-lattice} in terms of \textbf{coupling of the hole stripes to
the lattice through octahedral tilting}.

\subsection{Peaks O1 and O2}

The assignment of the other relaxation peaks found in LSCO is easily
accomplished after the identification of the peak due to the presence of
\textbf{interstitial oxygen} (O$_{\text{i}}$)\ in La$_{2}$CuO$_{4+\delta}$;
this follows straightaway from the dependence of the intensity of the pair of
peaks O1 and O2 in Fig. \ref{fig LCO O1O2T} on the content $\delta$ of excess
oxygen. Peak O1 is therefore assigned to hopping of isolated O$_{\text{i}}$
atoms, while peak O2, with slightly higher activation energy and prevailing at
higher $\delta$, to O$_{\text{i}}$ pairs. These peaks allow the presence of
O$_{\text{i}}$ well below $\delta\sim10^{-3}$ to be monitored, as estimated
from oxygen absorption and evolution experiments (Sec. \ref{sect pxT}); the
fact that peak O2 prevails over O1 even for $\delta\ll0.01$ is explained by
observing that the anisotropy of the elastic distortion due to O$_{\text{i}}$
(and therefore the intensity of peak O1)\ is proportional to the small
deviation of the orthorhombic structure of La$_{2}$CuO$_{4+\delta}$ from
tetragonal (Fig. \ref{fig O_int}), while stable O$_{\text{i}}$ pairs are
highly anisotropic defects (Sec. \ref{sect Oi str}). Incidentally, peaks O1
and O2 allow a very precise determination of the hopping rate of O$_{\text{i}%
}$ to be made and the existence of stable O$_{\text{i}}$ pairs to be assessed.

\subsection{Peak T}

Once established how to obtain La$_{2}$CuO$_{4+\delta}$ virtually free from
excess oxygen, we found that the anelastic spectrum of stoichiometric
defect-free La$_{2}$CuO$_{4}$ contains two extremely intense thermally
activated relaxation processes, causing strong absorption peaks and modulus
softenings up to 25\% above 150~K (at 1~kHz, \#4 in Fig. \ref{fig LCONCO_AS}
or peak T)\ and 10\% above 10~K (\#4 in Fig. \ref{fig LCONCO_AS} or peak LT).
These relaxation processes are hindered by doping and therefore must be due to
\textbf{intrinsic lattice mechanisms}, which must have to do with the most
unstable lattice modes of LSCO: the \textbf{octahedral tilting} giving rise to
the LTO and LTT structures (considering the low relaxation frequency involved,
$\tau^{-1}\sim\omega\sim10^{3}$~s$^{-1}$, by tilting I mean switching among
the several potential minima of the octahedra, as in Fig. \ref{fig TiltPot}b,
and not continuous tilting). In this respect, the comparison with Nd$_{2}%
$CuO$_{4}$ is particularly significant; in Nd$_{2}$CuO$_{4}$ the O atoms that
in La$_{2}$CuO$_{4}$ form the apices of the CuO$_{6}$ octahedra are shifted to
the interstitial positions (Fig. \ref{fig LCONCOstr}), so that there are no
octahedra, and in fact the anelastic spectrum of Nd$_{2}$CuO$_{4}$ is
completely flat. The evidence that octahedral tilts are involved comes from
the fact that peak T is observed also in the $^{139}$La NQR relaxation rate
(Fig. \ref{fig LCO ASNQR}), which is sensitive to fluctuations of the La-O
distances, and therefore to fluctuations of the octahedra. Additional clues to
the nature of these relaxation processes come from their dependence on doping
through substitution of La with Sr or through addition of interstitial O; the
latter type of doping is more effective in blocking the tilts of the
octahedra, and in fact $\delta<0.001$ is sufficient to completely suppress
peak T (Fig. \ref{fig LSCO block}b), while the effect of Sr is much weaker
(Figs. \ref{fig LSCO block}b and \ref{fig LSCOp80K}a). The fact that as little
as $\delta<0.001$ of O$_{\text{i}} $ completely blocks the octahedral motion
giving rise to peak T demonstrates that this a collective type of motion,
requiring the coordinated motion of hundreds of octahedra, and that can be
identified with the \textbf{solitonic tilt waves} predicted by Markiewicz
\cite{Mar93b}. It should be noted that this type of motion is possible only
for planar coordination of the octahedra with little correlation among
different planes, with the additional hypothesis that it is possible to start
from a ground state like the LTT tilt pattern that can be decomposed in weakly
correlated rows of octahedra (Fig. \ref{fig tiltwave}); in this manner
Markiewicz obtained a one-dimensional non linear equation of motion for the
tilts of the octahedra, which has solitonic solutions. On the contrary, with
rotation patterns like that in RuSr$_{2}$GdCu$_{2}$O$_{8}$ (Fig.
\ref{fig Ru octa}) the rotation of each octahedron is strongly coupled to
those of all the neighboring octahedra, so that no low energy excitations of
this type are possible.

\subsection{Peak LT}

The other \textbf{intrinsic lattice relaxation} mechanism in La$_{2-x}$%
Sr$_{x}$CuO$_{4}$ is peak LT below 10~K (Fig. \ref{fig LSCO block}b and
\ref{fig LSCO tunn1}). Contrary to peak T, it is only gradually suppressed by
the presence of O$_{\text{i}}$, indicating that the motions involved are of
local type, so that octahedra far from the O$_{\text{i}}$ atoms are not
affected; I imagine that even single O atoms in the CuO$_{2}$ planes may be
involved in such a fast relaxation. The most striking feature of peal LT is
its evolution with hole doping; in fact, the suppression of the intensities of
peaks T and LT is mainly due to steric effects, especially for O$_{\text{i}}$
($\delta\sim0.001$ of O$_{\text{i}}$ suppresses peak T but contributes very
little to hole doping). Figure \ref{fig LSCO tunn}a shows that hole doping
even enhances the intensity of peak LT, and above all greatly shifts to lower
temperature and narrows it, indicating an enhancement of the magnitude of the
fluctuation rate $\tau^{-1}$ and of its temperature derivative. This
phenomenology can be compared with the anelastic relaxation due to tunneling
of interstitial H\ in \textit{bcc} superconducting metals or due to two-level
systems in insulating and metallic glasses \cite{CCC98}. In such cases the
anelastic relaxation is due to transitions of the tunnel systems promoted by
the interaction with phonons and conduction electrons, whose contributions to
$\tau^{-1}\left(  T\right)  $ are well known; it is also established that the
direct interaction of the tunneling atoms with the conduction electrons is the
main responsible for such transitions. It is therefore clear that peak LT is
due to \textbf{single octahedra or O atoms tunneling between nearly equivalent
potential minima}, and that the transition rate within such tunnel systems
(the relaxation rate $\tau^{-1}$) is enhanced by the \textbf{interaction with
the holes}. Simple plotting of peak LT in double logarithmic scale shows that
$\tau^{-1}\propto T^{n}$ with $n\simeq5.4$ (Fig. \ref{fig LSCO tunn}b), which
is completely different from the $\tau^{-1}\left(  T\right)  $ laws known for
metallic systems; this is reasonable, since the hole excitations in these
cuprates are certainly different from those in metals, and the analysis of
peak LT would provide information on their excitation spectrum.

\chapter{YBCO}

The family of the YBa$_{2}$Cu$_{3}$O$_{6+x}$ superconductors (often
abbreviated in YBCO or Y-123 from the cation stoichiometry) is extensively
studied because of the particularly high superconducting temperature
$T_{\text{c}}\sim92$~K which make various applications possible, including
large scale electric devices and wires \cite{LGF01}. On the other hand, YBCO
is, among the cuprate superconductors, the one where nonstoichiometric oxygen
plays the most fundamental role, since it is the only responsible for hole
doping, but also has a wide stoichiometric range, $0<x<1$, and high mobility.
The diffusive oxygen\ jumps give rise to reorientation of the associated
elastic dipole and therefore the anelastic spectroscopy is particularly useful
in studying these cuprates.

\section{Structure and phase diagram}

\begin{figure}[tbh]
\begin{center}
\includegraphics[
%natheight=424.437500pt,
%natwidth=510.500000pt,
%height=235.875pt,
width=283.1875pt
]{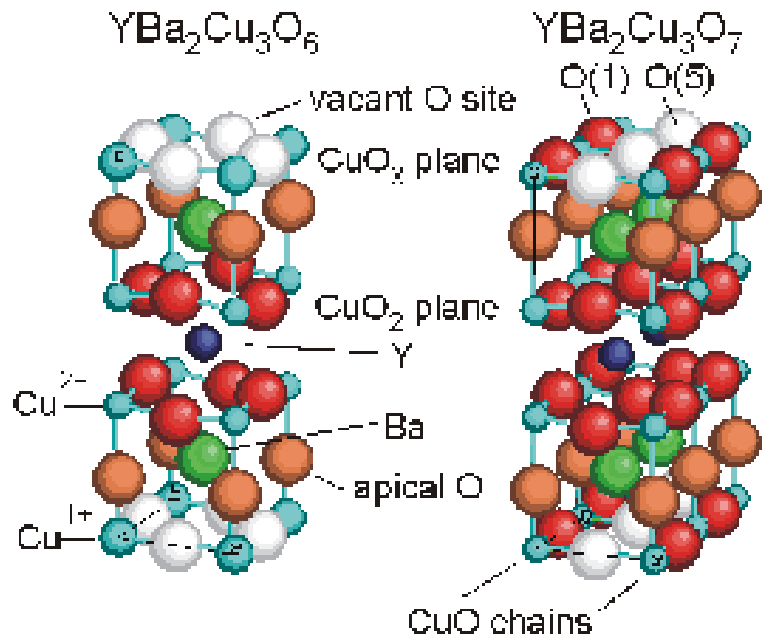}
\end{center}
\caption{Structure of YBa$_{2}$Cu$_{3} $O$_{6+x}$. Left: one cell of
tetragonal YBa$_{2}$Cu$_{3}$O$_{6}$; right: two cell of orthorhombic YBa$_{2}%
$Cu$_{3}$O$_{7}$. The vacant O\ sites in the CuO$_{x}$ planes are indicated as
white atoms.}%
\label{fig YBCO6_7}%
\end{figure}

\begin{figure}[tbh]
\begin{center}
\includegraphics[
%natheight=292.000000pt,
%natwidth=361.437500pt,
%height=2.6515in,
width=3.2776in
]{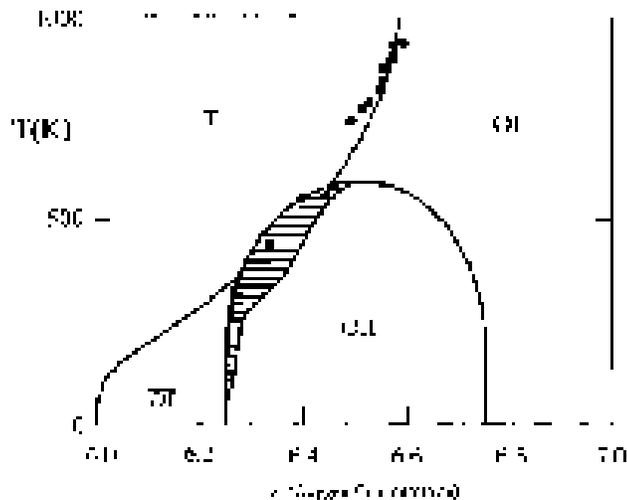}
\end{center}
\caption{Simplified phase diagram of YBCO (from Ref. \cite{MFC92}).}%
\label{fig YBCO_pd}%
\end{figure}

The structure of YBa$_{2}$Cu$_{3}$O$_{6+x}$ is shown in Fig. \ref{fig YBCO6_7}%
. The ideal YBa$_{2}$Cu$_{3}$O$_{6}$ has planes of Cu$^{1+}$ and the CuO$_{2}$
planes in the Cu$^{2+}$ oxidation state; there are no mobile charges, the
material is semiconducting and the Cu$^{2+}$ spins order antiferromagnetically
below $T_{\text{N}}>320$~K. The CuO$_{x}$ planes may accommodate O atoms up to
$x\le1$ and the stoichiometry can be controlled through temperature and
O$_{2}$ partial pressure. The hole doping in the CuO$_{2}$ planes is due to
the oxidation from these nonstoichiometric O atoms in the CuO$_{x}$ planes and
depends also on their ordering. In fact, the O atoms tend to form parallel
Cu-O chains. The right hand side of Fig. \ref{fig YBCO6_7} shows two cells of
the ideal orthorhombic structure of YBa$_{2} $Cu$_{3}$O$_{7}$ with the excess
O atoms perfectly ordered in Cu-O chains along the $b$ axis. In practice only
intermediate stoichiometries $0.05<x<1$ are achievable, and the $x-T$ phase
diagram (Fig. \ref{fig YBCO_pd}) contains three main phases: \textit{i)} the
\textbf{orthorhombic-I} (O-I) phase with parallel Cu-O chains along the $b$
axis when $x\simeq1$; \textit{ii)} the \textbf{orthorhombic-II} (O-II) phase
with Cu-O chains alternately filled and empty when $x\sim0.5$; \textit{iii)}
the \textbf{tetragonal} phase at $x<0.3$ or high temperature, where the oxygen
chain fragments are very short and oriented along both the $a$ and $b$
directions. Additional phases, \textit{e.g.} O-III with a pattern of
filled-filled-empty chains, may be obtained by carefully equilibrating the
sample in controlled O$_{2}$ atmosphere.

The structural phase diagram has been mainly determined by neutron diffraction
investigations \cite{JVP90,FCA90}, and is shown in Fig. \ref{fig YBCO_pd} in a
schematic version. Numerous studies have also appeared where the phase diagram
is reproduced by Monte Carlo simulations of models with different interaction
energies of the O atoms up to at least the next nearest neighbors. The most
successful model is the ASYNNNI (asymmetric next nearest neighbors
interactions) \cite{FCA90,UHS97,MLA01}. Such models have also been extended to
explain the hole doping in the CuO$_{2}$ planes, which depends on oxygen
ordering \cite{MFC92} as well as content in the CuO$_{x}$ planes, and are too
numerous to be reviewed here. It will be sufficient to mention that they
assume at least three interaction energies $V_{1}$, $V_{2}$ and $V_{3}$ for
nearest, 2nd and 3rd neighboring O atoms, with $V_{1}>0$ and $V_{2}<0$ in
order to reproduce the tendency to form parallel Cu-O chains, and $V_{3}>0$ in
order to reproduce the O-II phase with alternately filled and empty chains.

\section{Oxygen ordering and charge transfer between chains and
planes\label{sect YBCO CT}}

The mean length of the Cu-O chains may be estimated from the NQR $^{63}$Cu
spectra, where different peaks are found for Cu atoms which have zero, one or
two neighboring O atom in the CuO$_{x}$ plane, and therefore are 2, 3 or
4-fold coordinated, including the apical O atoms \cite{LSM96}; a quantitative
analysis of the spectra therefore provides the concentrations $n_{0}$, $n_{1}$
and $n_{2}$ of such atoms. These are related with the mean chain length and
degree of oxygen ordering; in fact, for infinitely long chains one has
$n_{2}=x$, $n_{1}=0$, $n_{0}=1-x$, while in the limit of a small concentration
of isolated O atoms one has $n_{2}=0$, $n_{1}=2x$, $n_{0}=1-2x$. The
concentration and nature of the holes doped by the nonstoichiometric O$^{2-}$
ions may be studied by X-ray absorption spectroscopies \cite{THM88,TBF92}, and
it turns out that for $x\rightarrow0$ Cu in the empty chains is in the
Cu$^{+}$ state, while in the CuO$_{2}$ planes it is in the Cu$^{2+}$ state, so
that the ideal undoped condition is
\begin{equation}
\text{YBa}_{2}\text{Cu}_{3}\text{O}_{6}=\text{Y}^{3+}\text{Ba}_{2}%
^{2+}\text{O}_{2}^{2-}\left(  \text{Cu}^{2+}\text{O}_{2}^{2-}\right)
_{2}\text{Cu}^{1+}\,;
\end{equation}
the addition of isolated oxygen\ in the CuO$_{x}$ planes oxidizes all the
pairs of neighboring Cu atoms into the Cu$^{2+}$ state, but chain fragments
with $n>1$ consecutive O\ atoms would require the oxidation of $n-1$ Cu atoms
to the Cu$^{3+}$ state, which is not observed \cite{BCD87,LSM96} (see also
Fig. \ref{fig chainfrags}). Therefore it may be supposed that in a first step
there are $n-1$ O$^{1-}$ atoms, or $n-1$ holes on the O$^{2-}$ atoms in the
chain; such holes may hop among the O atoms of the chain, providing a minor
contribution to the electrical conductivity. This behavior may be rationalized
in terms of electrostatic repulsion between the holes \cite{SHC94}:\ the
repulsion between two holes on the same atom (Cu$^{3+}=$ Cu$^{1+}+2h^{\bullet
}$) is too strong and they distribute over one O$^{2-}$ and one Cu$^{1+}$.
With increasing the chain length $n$, the electrostatic repulsion between the
holes in the oxygen chain atoms drives part of them, $m $, into the CuO$_{2}$
planes where they delocalize giving rise to most of the electric conduction
and superconduction. A model for the charge transfer from the chains to the
superconducting planes requires the knowledge of how many $m$ holes are
transferred from a chain fragment of length $n$ and has been developed by
Uimin \cite{UR92,UHS97}, based also on the experimental observation that for
long chains it is $m/n\sim0.7$ \cite{KMK92}. For short chains of length
$n=2,3$ and $4$, it has been estimated that $m=0,1,2$ holes are transferred to
the planes, and the transfer proceeds with increasing $n$ up to the optimal
value $m=0.7n$. For the interpretation of the anelastic spectra, it should be
noted that the migration of an isolated O atom in the CuO$_{x}$ plane does not
change the oxidation state of the crystal, while already the formation of an
oxygen pair causes the oxidation of the two neighboring Cu atoms, and the
joining to longer chains involves charge transfer with the CuO$_{2}$.
Therefore, the jumps of isolated O\ atoms are expected to be much easier than
those involving the joining to or coming out of a chain fragment, since the
latter involve substantial energy changes of the electronic system.

\section{Diffusive dynamics of oxygen in the CuO$_{x}$ planes
\label{sect YBCO diffus}}

While the features of the phase diagram have been extensively studied both
experimentally and theoretically, the knowledge of the diffusive and ordering
dynamics of oxygen is less detailed. The macroscopic diffusion has been
studied by tracer methods as recently summarized in \cite{Con01}, and by in-
or out-diffusion in a gas atmosphere in connection with thermogravimetry or
electrical resistance measurements to monitor the time dependent oxygen
content \cite{RR94}. The first type of experiments provides an average
macroscopic diffusion coefficient, which may depend on the microstructure and
possibly on inhomogeneous oxygen ordering, while the latter type of
experiments is also heavily affected by the kinetics of the oxygen exchange
with the gas phase and requires a model for the dependence of the electrical
resistivity on the filling of the CuO$_{x}$ planes. All these factors heavily
influence the apparent diffusion coefficient: for example, the apparent
activation energy for chemical diffusion may vary from 0.4~eV to 1~eV within
the same study, depending on microstructure \cite{LP93}, and even more from
experiment to experiment; an important role is attributed to the formation of
an oxygen rich shell in each grain, that obstructs further oxygen in-diffusion
during absorption experiments\cite{LP93}; a surface barrier of 1.7~eV has been
estimated for out-diffusion \cite{TTP88b}. The spread of the results on oxygen
diffusion in YBCO in the literature is even more impressive (several orders of
magnitude) in terms of the diffusion coefficient or hopping rate at a fixed
temperature. The anelastic results have the great advantage that, once
identified the elastic energy loss peak due to oxygen hopping, from the
condition $\omega\tau\simeq1$ at the peak temperature, Eq. (\ref{wt=1}), a
relaxation time $\tau$ very close to the oxygen hopping time in the bulk can
be measured, even when interpreted by different microscopic mechanisms such as
vacancy-assisted \cite{XCW89}, Zener pair relaxation \cite{CS91}, interacting
elastic dipoles \cite{BSW94}; in fact, different mechanisms may change the
ratio between relaxation and hopping time by less than one order of magnitude.
Instead, the evaluation of the local barrier for hopping depends on the type
of model that is assumed, ranging from 0.68~eV for a KWW relaxation
\cite{WGR93} to $1-1.4$~eV for Curie-Weiss interactions \cite{BFW97}.

\begin{figure}[tbh]
\begin{center}
\includegraphics[
%natheight=293.937500pt,
%natwidth=576.187500pt,
%height=7.8046cm,
width=15.2468cm
]{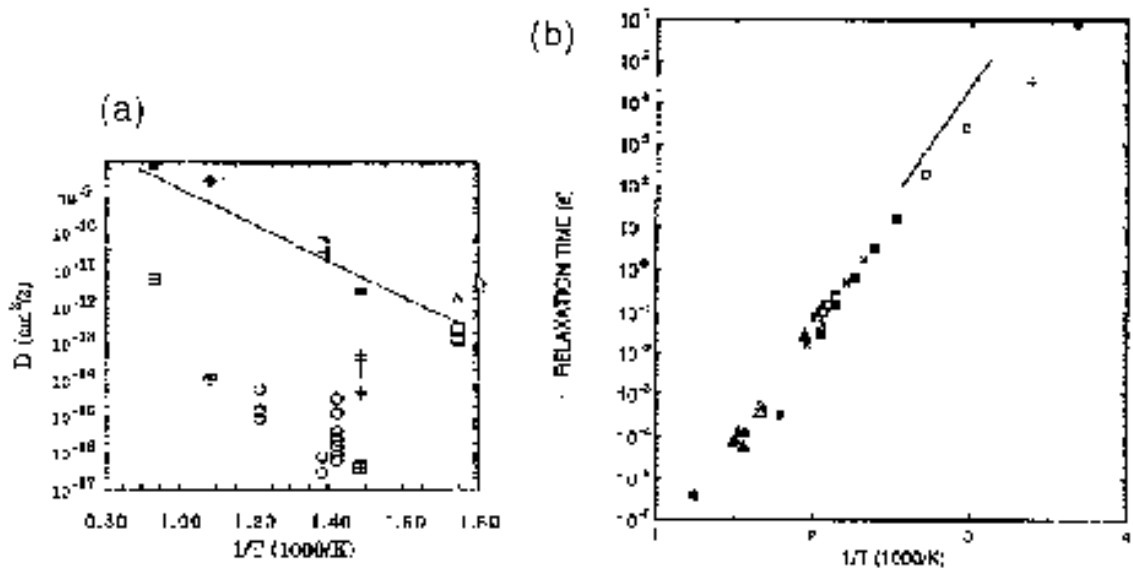}
\end{center}
\caption{Comparison between the chemical diffusion coefficient of O in YBCO
from various permeation experiments (a) and the hopping time of O deduced from
several anelastic experiments (b). The figures are from Ref. (\cite{RR94}).}%
\label{fig O dif}%
\end{figure}

\section{Anelastic measurements of the oxygen diffusive
jumps\label{sect YBCO AR O diff}}

The anelastic spectroscopy is certainly the best method to study in detail the
dynamics of the mobile O atoms in the CuO$_{x}$ planes. In fact, as explained
in detail in Sec. \ref{sect intr s}, an anisotropic elastic dipole is
associated to each O atom, which reorients by 90$^{\mathrm{o}}$ after a jump;
it is then possible to selectively probe different hopping processes in
different environments, through the analysis of the distinct maxima in the
$Q^{-1}\left(  T\right)  $ curve, occurring when the condition $\omega
\tau\left(  T\right)  =1$ is met for each process. Unfortunately, the present
Thesis reports only qualitative results of the high temperature peaks, because
for technical reasons the measurements were made in high vacuum, and oxygen
loss occurred during the measurement of the anelastic spectra above 500~K.
Therefore, the peak shapes were affected by the oxygen loss and could not be
reliably analyzed in order to extract precise information on the oxygen
dynamics. Still, the results clearly show the presence of two completely
distinct hopping regimes: \textit{i)} one that involves barriers of $\sim1$~eV
and has been studied by several authors with various techniques, \textit{ii)}
a much faster hopping over a barrier about 10 times smaller, that is observed
only at the lowest oxygen\ contents.

\subsection{Anelastic spectra at different oxygen
concentrations\label{sect YBCO HT}}

\begin{figure}[tbh]
\begin{center}
\includegraphics[
%natheight=582.687500pt,
%natwidth=907.125000pt,
%height=9.9573cm,
width=15.4686cm
]{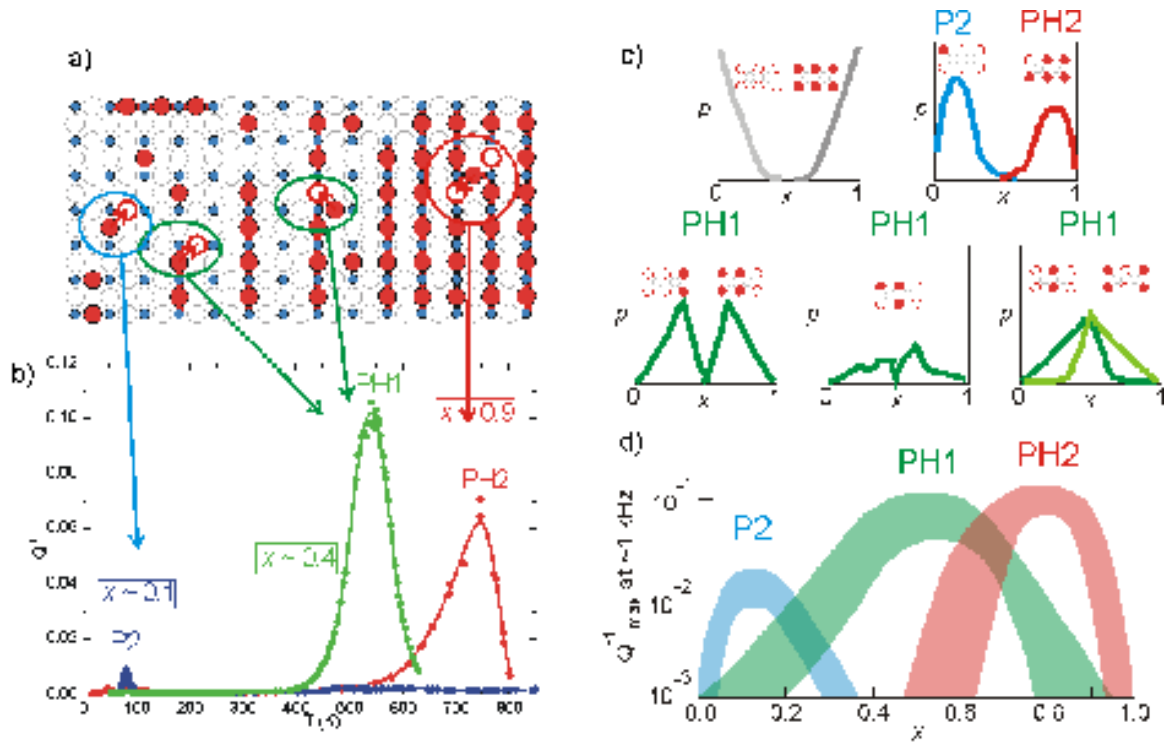}
\end{center}
\caption{(a) CuO$_{x}$ plane with $x$ increasing from left to right; the O
jumps causing the three main anelastic relaxation processes P2, PH1 and PH2
plotted in (b) are put in evidence. The intensities of these peaks are plotted
in (d) as a function of $x$, and are strongly correlated with the
probabilities of the corresponding clusters plotted in (c).}%
\label{fig ARCuOx}%
\end{figure}

The anelastic spectra of YBCO change completely with varying $x$ between 1 and
$\sim0$. Figure \ref{fig ARCuOx}b presents the anelastic spectra of YBCO
measured at $\sim1$~kHz at three representative values of $x$: 0.1, 0.5 and
0.9. Each of them contains a different peak; for compatibility with the
labeling originally given \cite{14,22,27,31}, they are called respectively P2
(at $\sim80$~K), PH1 (at $\sim550$~K) and PH2 (at $\sim750 $~K). All of them
are thermally activated, P2 with an activation energy of 0.11~eV, and PH1 and
PH2 with $E\simeq1$~eV. The upper panel represents the CuO$_{x}$ plane with
$x$ increasing from left to right; the oxygen jumps causing the three main
anelastic relaxation processes are put in evidence.

Starting from the superconducting phase with $x\sim0.9$, one finds peak PH2,
which disappears on lowering $x$ and, according to the pendulum experiment of
Xie \textit{et al.} \cite{XCW89}, also when $x\rightarrow1$. For this reason
those authors suggest a vacancy mechanism for the oxygen diffusion in the O-I
phase, as represented in Fig. \ref{fig ARCuOx}a, and associate peak PH2 with
such a hopping mechanism. For $0.2<$ $x<0.8$ peak PH1 is observed, with
intensity strongly dependent on thermal history; as shown later in Sec.
\ref{sect PH1}, peak PH1 cannot be attributed to hopping of isolated O atoms
and it must therefore be associated with jumps of oxygen\ aggregated into
chain fragments with the exclusion of those in the ordered O-I\ phase, which
produce peak PH2. Finally, at the lowest values of $x$ attainable, only peak
P2 is observed, which is therefore associated with the jumps of isolated O
atoms. Additional support to these assignments comes from the comparison
between the intensities $Q_{\text{max}}^{-1}\left(  x,j\right)  $ of the three
elastic energy loss peaks (Fig. \ref{fig ARCuOx}d) and the probabilities
$p\left(  x,j\right)  $ of the various oxygen clusters of type $j=$ P2, PH1,
PH2 or nothing (Fig. \ref{fig ARCuOx}c). These cluster probabilities have been
reproduced from Ref. \cite{MFC92}, where they are calculated with Monte Carlo
simulations on the ASYNNNI model assuming $T=300$~K. The clusters are
represented as insets in the various plots of $p\left(  x,j\right)  $ and
labeled according to the relaxation process $j$ that they cause according to
our interpretation; for PH1 the 5 clusters in the three plots have to be
summed together. The $Q_{\text{max}}^{-1}\left(  x,j\right)  $ curves have
been plotted from a large number of measurements on different samples,
including EuBa$_{2}$Cu$_{3}$O$_{6+x}$, and have a considerable error both in
the intensity (notice the logarithmic vertical scale) and in $x,$ since the
three peaks are not always as clearly distinguishable as in Fig.
\ref{fig ARCuOx}b and due to uncertainties in the determination of $x$; also,
the ends of the curves for $x<0.1$ and $x>0.9$ are a guess. Nevertheless, the
correlation between $Q_{\text{max}}^{-1}\left(  x,j\right)  $ and $p\left(
x,j\right)  $ from Ref. \cite{MFC92} is excellent, for all three relaxation
processes; a discrepancy can be found in the fact that $Q_{\text{max}}%
^{-1}\left(  x,\text{P2}\right)  $ is about an order of magnitude smaller than
$Q_{\text{max}}^{-1}\left(  x,\text{PH2}\right)  $, while $p\left(
x,\text{P2}\right)  $ is slightly larger than $p\left(  x,\text{PH2}\right)
$. A possible explanation might be that the $p\left(  x,j\right)  $ have been
calculated for $T=300$~K, while peak P2 is measured at lower temperature,
where the probability of isolated O atoms should be reduced in favor of chain
fragments; on the other hand, in the light of discussions in the next
Sections, I think that it is too much to look for a quantitative agreement
between $Q_{\text{max}}^{-1}\left(  x,j\right)  $ and these $p\left(
x,j\right)  $, and the comparison should remain on the qualitative level.

Before discussing in more detail the information obtainable from the three
relaxation processes P2, PH1 and PH2, I will discuss what kind of anelastic
spectra one would expect from the information presented in Sec.
\ref{sect anel}.

\subsection{Elastic dipole of oxygen in the CuO$_{x}$
plane\label{sect YBCO eldip}}

It is possible to estimate the elastic dipole associated with an O atom in the
CuO$_{x}$ plane by the $x$ dependence of the cell parameters, and it has been
noted that the elastic dipole is almost independent of the oxygen content and
ordering \cite{BFW97}. The estimate can be made by calculating the lattice
parameters $a_{\text{T}}$ and $a_{\text{O}}$, $b_{\text{O}}$ of the tetragonal
and orthorhombic cells in terms of the lattice parameter $a$ of the ideal
oxygen-free tetragonal cell and elastic dipoles $\lambda^{\left(  1\right)  }$
and $\lambda^{\left(  2\right)  }$ of O in the sites of type 1 and 2. These
sites are generally labeled as O(1) and O(5), depending whether the nearest
neighbor Cu atoms are in the $y$ or $x$ direction respectively (with
$x\parallel a$ and $y\parallel b$), but I\ will use the label 2 instead of 5.
Let us indicate the components of the elastic dipole associated with oxygen as
$\lambda_{1}=$ $\lambda_{xx}^{(2)}=$ $\lambda_{yy}^{(1)}$, $\lambda_{2}=$
$\lambda_{xx}^{(1)}=$ $\lambda_{yy}^{(2)}>\lambda_{1}$. The remaining
component along $z$ is not interesting, since it does not change after a jump.
The strain due to the occupation of the $n_{1}+n_{2}=x$ sites is
\begin{equation}
\varepsilon_{ij}=n_{1}\lambda_{ij}^{(1)}+n_{2}\lambda_{ij}^{(2)}%
\end{equation}
and the cell parameters in the O and T phases, assuming $n_{1}^{\text{(O)}%
}=x_{\text{O}}$, $n_{2}^{\text{(O)}}=0$ and $n_{1}^{\text{(T)}}=n_{2}%
^{\text{(T)}}=x_{\text{T}}/2,$ are:
\begin{align}
a_{\text{T}}  &  =a\left[  1+x_{\text{T}}\frac{1}{2}\left(  \lambda
_{1}+\lambda_{2}\right)  \right] \label{a(l)}\\
a_{\text{O}}  &  =a\left[  1+x_{\text{O}}\lambda_{1}\right]  \,,\quad
b_{\text{O}}=a\left[  1+x_{\text{O}}\lambda_{2}\right]
\end{align}
from which one deduces
\begin{align}
a  &  =\frac{a_{\text{T}}x_{\text{O}}-\frac{1}{2}\left(  a_{\text{O}%
}+b_{\text{O}}\right)  x_{\text{T}}}{x_{\text{O}}-x_{\text{T}}}\label{l(a)}\\
\left(  \lambda_{2}-\lambda_{1}\right)   &  =\frac{b_{\text{O}}-a_{\text{O}}%
}{x_{\text{O}}a}\\
\left(  \lambda_{1}+\lambda_{2}\right)   &  =\frac{b_{\text{O}}+a_{\text{O}}%
}{x_{\text{O}}a}=\left(  \frac{a_{\text{T}}-a}{a}\right)  \frac{2}%
{x_{\text{T}}}%
\end{align}
In the following table I consider the cell parameters measured by neutron
diffraction in two tetragonal and two orthorhombic samples in Ref.
\cite{CSH96}, and two pairs of data taken from Fig. 5 of Ref. \cite{JVP90}.%

\[%
\begin{tabular}
[c]{|l|l|l|l|l|l|l|l|l|}\hline
$x_{\text{T}}$ & $x_{\text{O}}$ & $a_{\text{T}}$ (\AA ) & $a_{\text{O}}$
(\AA ) & $b_{\text{O}}$ (\AA ) & Ref. & $a$ (\AA ) & $\left(  \lambda
_{2}-\lambda_{1}\right)  $ & $\frac{1}{2}\left(  \lambda_{1}+\lambda
_{2}\right)  $\\\hline
0.18 & 0.96 (O-I) & 3.8587 & 3.8227 & 3.8872 & \cite{CSH96} & 3.8596 &
0.0174 & -0.0012\\\hline
0.18 & 0.78 (O-III) & 3.8587 & 3.8265 & 3.8875 & \cite{CSH96} & 3.8592 &
0.0203 & -0.00073\\\hline
0.25 & 0.96 (O-I) & 3.8586 & 3.8227 & 3.8872 & \cite{CSH96} & 3.8599 &
0.0174 & -0.0013\\\hline
0.25 & 0.78 (O-III) & 3.8586 & 3.8265 & 3.8875 & \cite{CSH96} & 3.8591 &
0.0203 & -0.00069\\\hline
0.095 & 0.93 & 3.8600 & 3.8227 & 3.8872 & \cite{JVP90} & 3.8606 & 0.0180 &
-0.0016\\\hline
0.28 & 0.93 & 3.862 & 3.8227 & 3.8872 & \cite{JVP90} & 3.8650 & 0.0179 &
-0.0028\\\hline
\end{tabular}
\]
The sample with $x_{\text{O}}=0.78$ was detwinned and showing O-III
superstructure (one filled chain every three) and the authors report
negligible occupancy of O(5) in the ortho phases; therefore it should be
meaningful to assume $n_{1}^{\text{(O)}}=0$ also for such a relatively low
value of $x$. From these data it can be concluded, in accordance with Ref.
\cite{BFW97}, that $\left(  \lambda_{2}-\lambda_{1}\right)  \simeq
0.019\pm0.0015$ and $\frac{1}{2}\left(  \lambda_{1}+\lambda_{2}\right)
\simeq-0.0018\pm0.0001$.

An important remark must be made on this type of derivation of the elastic
dipole of oxygen in the CuO$_{x}$ planes: it is \textbf{valid for oxygen
aggregated into chains} or chain fragments, but \textbf{not necessarily for
isolated O atoms}; in fact, in Sects. \ref{sect YBCO P2}-\ref{sect YBCO P2.3}
it will be shown that around room temperature and below, the concentration of
isolated O atoms is very small also in highly oxygen\ deficient samples. As
explained in those Sections and in Sec. \ref{sect YBCO CT}, the state of
aggregated and isolated O atoms is different, due to charge transfer effects
dependent on the chain length, and it is not obvious that the elastic dipole
of an isolated O atom should be the same as that of an oxygen belonging to a
chain. Therefore, it is not obvious that the value of the anisotropy
$\Delta\lambda=0.019$ found above is appropriate for evaluating the anelastic
relaxation from oxygen hopping.

Let us see if it is compatible with the peak intensities reported in Fig
\ref{fig ARCuOx}. In Appendix B is calculated an estimate of the relaxation
strength for the Young's modulus of an isotropic polycrystal; neglecting
porosity, it should be%

\begin{equation}
\Delta=\frac{1}{15}\frac{cv_{0}}{k_{\text{B}}T\,\left\langle E^{-1}%
\right\rangle }\left(  \lambda_{1}-\lambda_{2}\right)  ^{2}\,\,,
\end{equation}
where $c$ is the concentration of relaxing dipoles. The Young's modulus of
ceramic YBa$_{2}$Cu$_{3}$O$_{6+x}$ corrected for porosity has been estimated
as $E=90$~GPa and 125~GPa for low and high $x$, respectively
\cite{WSA95,SLN95}. Setting $E=100$~GPa, $\rho\simeq6.3$~g/cm$^{3}$,
$v_{0}=175\times10^{-24}$~cm$^{3}$ we get
\begin{equation}
\Delta\sim\frac{30~\text{K}}{T}c\,. \label{relstr estim}%
\end{equation}
If we want to compare with the height of peak PH2 we set $T\sim750$~K and
$c\sim0.1$, since according to the vacancy mechanism \cite{XCW89} the
intensity is proportional to the O vacancies in the O-I phase and not to
$x\simeq0.9$; it turns out $Q_{\text{max}}^{-1}=\frac{\Delta}{2}\sim
2\times10^{-3}$, instead of the observed $10^{-2}-10^{-1}$ (Fig.
\ref{fig ARCuOx}). The intensity however, might be enhanced by the proximity
to the ordering transition \cite{BFW97}, by a Curie-Weiss like factor; from
the YBCO\ phase diagram (Fig. \ref{fig YBCO_pd}) it appears that at $x\sim0.9$
the transformation to the tetragonal phase, if any, is at $T>1000$~K, and
therefore $T=750$~K in the denominator should be substituted with
$T-T_{\mathrm{C}}>250$~K, which does not change much the situation. \textbf{It
seems therefore that }$\Delta\lambda=0.019$\textbf{\ is too small for
explaining the large relaxation strength we observe for peak PH2}, but these
estimates are really rough. In any case, it is difficult to draw quantitative
conclusions on the elastic dipole of oxygen from the anelastic spectra,
because, when measuring peaks PH1 and PH2 at high temperature, one has oxygen
loss during the measurement, while for peak P2 one needs an independent
estimate of the concentration $c$ of free oxygen.

\subsection{Interactions among the oxygen atoms in the Bragg-Williams
approximation\label{sect YBCO interac}}

As explained in detail in Sec. \ref{sect interact}, Wipf and coworkers
performed aftereffect \cite{SHW92,BSW94,BFW97} experiments on YBCO and took
into account the interactions among elastic dipoles in the Bragg-Williams
approximation, analogous to the Curie-Weiss approximation above $T_{\mathrm{C}%
}$. They showed that the anelastic aftereffect (the time dependence of strain
after application of a constant stress) for $x\simeq0.4$ around 380~K presents
an enhancement of the relaxation strength and time by a factor $T/\left(
T-T_{\mathrm{C}}\right)  ^{-1}$, where $T_{\mathrm{C}}$ is the temperature
below which the elastic dipoles associated with the O atoms start ordering
themselves, namely the temperature of the T/O phase transformation. Around
380~K, a critical concentration $x_{c}\simeq0.45$ is found, where both the
relaxation strength $\Delta\left(  x\right)  $ and relaxation time
$\tau\left(  x\right)  $ are enhanced by at least a factor 10 over the values
found away from $x_{c}$ \cite{BFW97}. The value $x_{c}\simeq0.45$ is somewhat
higher than $x\sim0.3-0.4$ deduced from the O-T line of the experimental phase
diagram \cite{MFC92,TPP99}.

\begin{figure}[tbh]
\begin{center}
\includegraphics[
%natheight=346.125000pt,
%natwidth=446.562500pt,
%height=6.1396cm,
width=7.9035cm
]{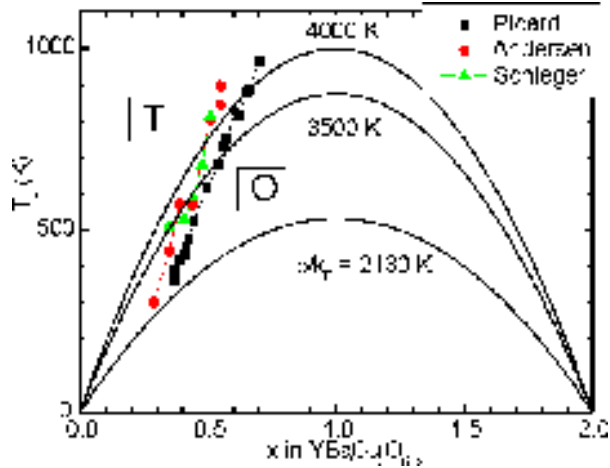}
\end{center}
\caption{Fit of the tetragonal/orthorhombic boundary with $T_{\mathrm{C}%
}\left(  x\right)  $.}%
\label{fig YBCO TO}%
\end{figure}

The elastic interaction alone cannot be the origin neither of the tetragonal
to orthorhombic transition nor of the several orthorhombic phases, which are
mainly determined by short range interactions of electronic origin, as
schematized in the ASYNNNI model. In fact, if one assumes that the ordering
temperature is given by $k_{\text{B}}T_{\mathrm{C}}=\alpha\frac{x}{2}\left(
1-\frac{x}{2}\right)  $[see eq. (\ref{TC})] and tries to fit the O-T boundary
of YBCO with such $T_{\mathrm{C}}\left(  x\right)  $, one obtains
$\alpha/k_{\text{B}}=2500-4000$~K, as shown in Fig. \ref{fig YBCO TO}, On the
other hand, Wipf \cite{Wip94} estimated the elastic contribution to $\alpha$
as $\alpha^{\text{el}}/k_{\text{B}}\sim290$~K, which is an order of magnitude
smaller than the value needed to reproduce the observed ordering temperatures.

Therefore, a treatment of the interaction among the O atoms in the
mean field approximation, as presented in Sec. \ref{sect interact}, is not
fully consistent if the main contribution to the interaction parameter
$\alpha$ is of local electronic origin; still, it is useful to reproduce
important effects in the relaxation strength and rate of the anelastic peaks
due to oxygen hopping, as shown by their critical enhancement observed in
after-effect experiments \cite{BFW97}.

\begin{figure}[tbh]
\begin{center}
\includegraphics[
%natheight=505.375000pt,
%natwidth=722.187500pt,
%height=10.7151cm,
width=15.2885cm
]{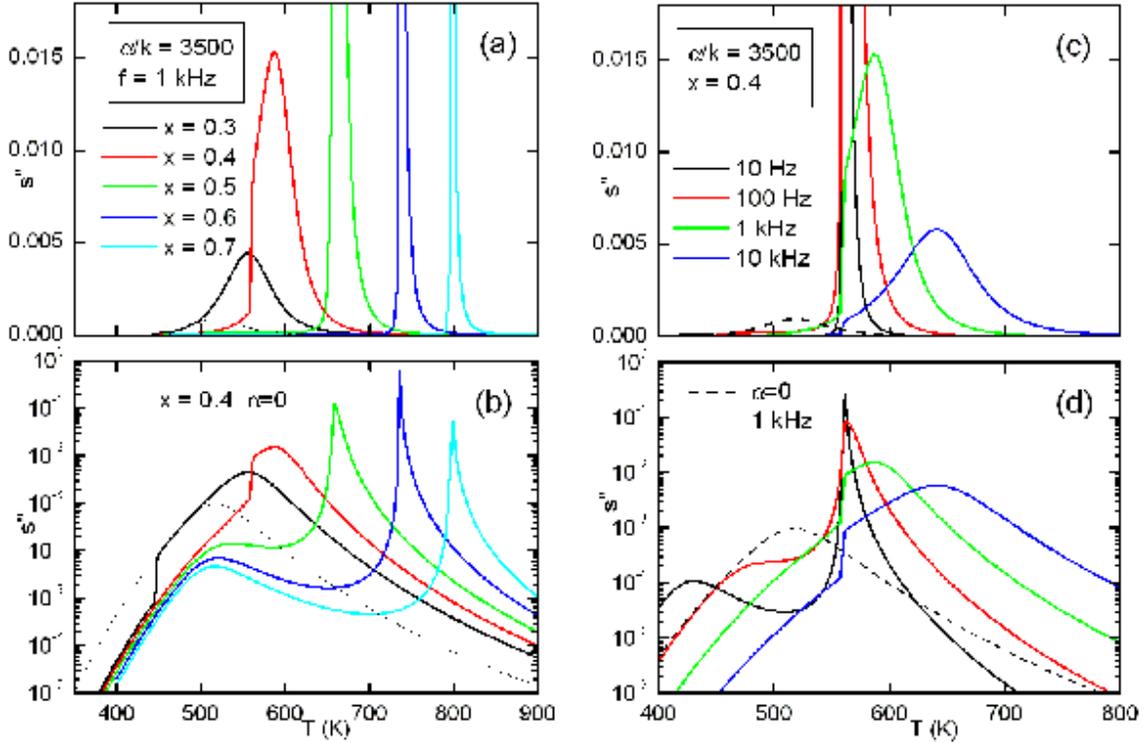}
\end{center}
\caption{$s^{\prime\prime}\left(  \omega,T\right)  $ curves in logarithmic
(upper) and linear (lower) scales calculated in the Bragg-Williams
approximation with $\tau_{0}=10^{-13}$~s, $E/k =11000$~K and $\omega/2\pi$,
$x$ and $\alpha$ as indicated in the legends.}%
\label{fig YBCO_BW}%
\end{figure}

According to the phase diagram of YBCO (Fig. \ref{fig YBCO_pd}), on varying
$x$ the temperature $T_{\mathrm{C}}\left(  x\right)  $ of the T/O-I\ ordering
transition crosses at $x\sim0.5$ the temperatures of peaks PH1 and PH2. Then,
it would be interesting to evaluate the effect of this ordering transition, at
least in the simple Bragg-Williams approximation presented in Sec.
\ref{sect interact}. Above $T_{\mathrm{C}}$, the simple Curie-Weiss-like
formula (\ref{CW}) can be used, but below $T_{\mathrm{C}}$ one has to
calculate the expressions (\ref{s BW}) and (\ref{tau_eff}), or equivalently
the order parameter $\xi=\frac{1}{2}\left(  c_{1}-c_{2}\right)  $. In Appendix
A numerical approximations are provided for $\xi\left(  c,T\right)  ,$which
might be useful for fitting purposes. As already noted, our data are generally
not susceptible to reliable analysis, since oxygen loss occurred during the
measurements, therefore I simply plot in Fig. \ref{fig YBCO_BW} some
$s^{\prime\prime}\left(  \omega,T\right)  $ curves calculated according to
Eqs. (\ref{s BW}), (\ref{tau_eff}) with the approximations (\ref{OPapprox}),
(\ref{OPmn}). The curves are plotted in linear (upper panels)\ and logarithmic
(lower panels) scales as a function of $x$ in (a) and (b) and as a function of
frequency in (c) and (d); the relaxation time in the absence of interaction
has been chosen as the typical hopping time of O, $\tau=10^{-13}$%
~s\thinspace$\exp\left(  11000/T\right)  $, the interaction parameter
$\alpha/k_{\text{B}}$ has been chosen as 3500~K, in order to have the
transition in the correct temperature range, and also plotted as a dashed line
is the reference peak with $\alpha=0$ (no interaction)\ at $T_{p}=520$~K (for
1~kHz). The ordering transition on cooling is clearly visible as a narrow
peak, when $T_{\mathrm{C}}>T_{p}$ and as a drop of intensity when
$T_{\mathrm{C}}<T_{p}$. Although the anelastic spectra we measured for $x<0.5$
contain some narrow peaks clearly due to oxygen ordering transitions
\cite{28}, there is no clear relationship with the curves calculated in Fig.
\ref{fig YBCO_BW} and it is evident that a \textbf{more complete treatment of
the short range interactions, like in the ASYNNNI model, is necessary to
describe the high-temperature anelastic spectra of YBCO}.

\subsection{Hopping in the O-I phase - peak PH2}

The information obtainable on PH2 from the present measurements is only
limited to the observation that it is a process occurring in the O-I\ phase,
distinct from PH1; in fact, the experiments are made in vacuum and rapid
oxygen loss occurs during the measurement, affecting both shape and intensity
of the peak. More reliable measurements of this process have been made by
other authors with the pendulum \cite{XCW89}, which allows some O$_{2}$
pressure to be maintained around the sample and moreover lowers the peak
temperature. The mechanism devised for PH2 in Fig. \ref{fig ARCuOx} is the one
requiring a neighboring vacancy, as suggested by Xie \textit{et al.}
\cite{XCW89}, on the basis of fact that its intensity vanishes for
$x\rightarrow1$. That analysis, however, is not fully consistent, since it
assumes an asymmetry between the energies in sites O(1) and O(5), $\Delta
E\left(  x\right)  =E_{5}-E_{1}$, which is maximum for $x=1$ and decreases to
0 when the sample becomes tetragonal. Such an asymmetry energy was taken into
account for reproducing the decrease of the apparent activation energy with
decreasing $x$, but it was not considered that it would also reduce the
intensity of the peak, due to the depopulation factor in Eq. (\ref{relstr}),
thereby accounting for a consistent part of the reduction of the peak
intensity when $x\rightarrow1$.

Among other measurements of peak PH2 \cite{CS91,CS93,WGR93}, I would like to
signal isothermal spectra measured with the pendulum \cite{MSB94} that
confirmed the presence of two peaks, the one at higher temperature is
associated with the orthorhombic phase and should therefore correspond to PH2,
while the other one to PH1. This confirmation is important, since it is based
on stationary and reproducible spectra.

\subsection{Other jumps of aggregated oxygen and oxygen ordering - peak
PH1\label{sect PH1}}

\begin{figure}[tbh]
\begin{center}
\includegraphics[
%natheight=316.750000pt,
%natwidth=579.625000pt,
%height=8.4065cm,
width=15.3346cm
]{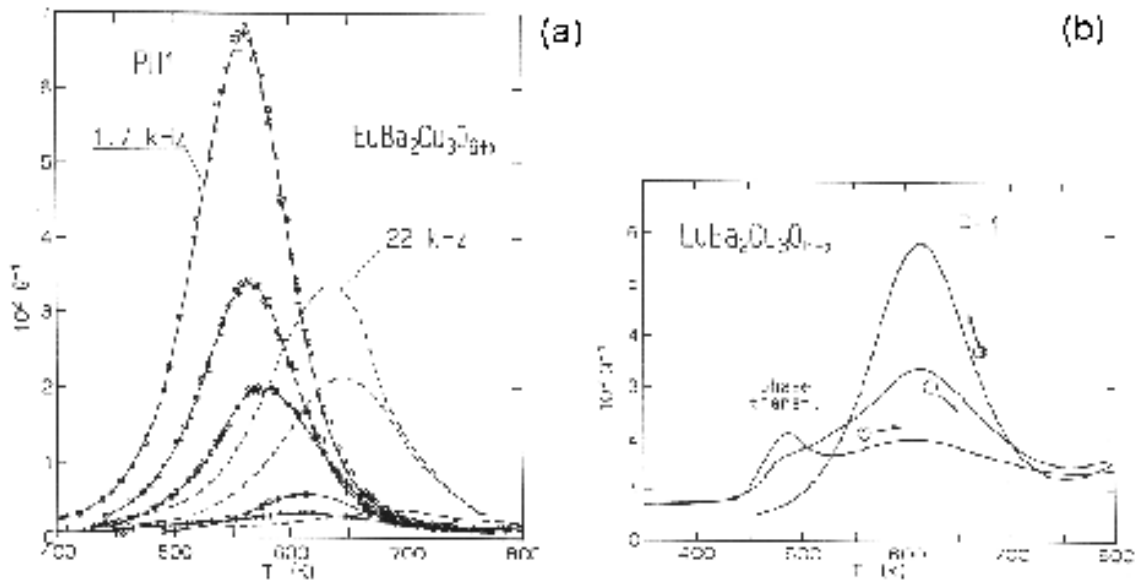}
\end{center}
\caption{(a) Peak PH1 measured in EuBa$_{2}$Cu$_{3}$O$_{6+x}$ starting from
$x\simeq0.5$; with decreasing $x,$the peak decreases in intesity and shifts to
higher temperature. (b) After quenching to room temperature from 990~K, peak
PH1 is strongly depressed, but reappears with aging at high temperature.}%
\label{fig YBCO PH1}%
\end{figure}

Figure \ref{fig YBCO PH1}a (from Ref. \cite{27}) shows peak PH1 measured in an
EuBa$_{2}$Cu$_{3}$O$_{6+x}$, which has anelastic spectra and all the physical
properties very similar to YBa$_{2}$Cu$_{3}$O$_{6+x}$. The peak with higher
intensity is for $x\simeq0.5$, and proceeding with the outgassing treatments
it decreases in intensity and shifts to higher temperature; the effective
activation energy, estimated from the shift of the peak at higher frequency,
also increases from 1.1 to 1.3~eV. While the decrease in intensity is easy to
understand after the considerations of Sec. \ref{sect YBCO HT}, the increase
in activation energy has no obvious explanation.

Figure \ref{fig YBCO PH1}b shows another interesting effect, which
\textbf{excludes the hopping of isolated O atoms as possible mechanism for
peak PH1}. Curve \#1 has been measured on heating after quenching from 990~K
to room temperature in few seconds, by pouring water on the quartz tube where
the sample was heated. After such a quenching the concentration of isolated O
atoms should be maximum, but peak PH1 is strongly depressed, and instead a
small peak appears near 480~K. The latter does not shift in temperature with
increasing frequency and is therefore signalling the occurrence of an
\textbf{ordering transition near 480~K}, like the frequency independent peak
at 560~K in the curves of Fig. \ref{fig YBCO_BW}d; it is difficult to say
whether the ordering occurring above 480~K is simply a lengthening of the Cu-O
chains or the formation of O-II domains. It is clear however, that peak PH1 is
due oxygen jumps within such long chains or domains and not to isolated O
atoms or even pairs or very short chains, which are promptly formed. In fact,
the peak develops with thermal cycling above 500~K (curves 2 and 3), and the
increase of intensity cannot be attributed to oxygen uptake, since the
measurements are made in high vacuum.

We also observed additional frequency independent anomalies up to 580~K
\cite{28}, which appear only on heating and not on cooling (as for curve 3 in
Fig. \ref{fig YBCO PH1}b), but still there is no better explanation than some
types of ordering transitions. This complex phenomenology demonstrates that
situation is even more complex than it appears from the diffraction
measurements of oxygen ordering.

\subsection{Isolated oxygen atoms - peak P2\label{sect YBCO P2}}

There is extensive evidence that the effective barrier for the diffusive jumps
of the O atoms in the CuO$_{x}$ planes is of the order of $1$~eV, as discussed
in the previous Sections. It should be noted, however, that none of the high
temperature $Q^{-1}\left(  T\right)  $ peaks presents the dependence on $x$
expected from the hopping of a concentration $c=x/2$ of an interstitial
species (if one regards oxygen in the CuO$_{x}$ plane as an interstitial
species, and takes into account that there are 2 sites available for each Cu
atom). As a matter of fact, none of the behaviors expected from an
interstitial solution is observed; e.g. the proportionality of the relaxation
strength to $c$ in the high dilution limit, or to $c\left(  1-c\right)  $ if
one takes into account the filling of a finite number of sites \cite{33}, or
to $c^{2}$ or some other power of $c$ if interstitial complexes contribute.
This is due to the strong interaction between the O atoms, and the difficulty
of probing the two limits $x\rightarrow0$ and $x\rightarrow1$. The interaction
among the O atoms is so strong to prevent the filling of the planes above
$c=0.5$ (or $x=1$). That concentration corresponds to the O-I phase, with
completely filled chains, and can be approached but probably never obtained.
We attempted at reaching the high dilution limit $x\rightarrow0$, but this
also cannot be done, due to the decomposition of YBCO in vacuum at high
temperature into the more stable oxides. Bormann and N\"{o}lting \cite{BN89}
found the decomposition limit $p\left(  T\right)  $ below which YBa$_{2}%
$Cu$_{3}$O$_{6+x}$ transforms into the more stable oxides Cu$_{2}$O,
BaCuO$_{2}$ and Y$_{2}$BaCuO$_{5}$ (the latter easily recognizable from the
green color), and found that the lowest attainable oxygen\ content is
YBa$_{2}$Cu$_{3}$O$_{6.05}$.

The first anelastic measurements on highly outgassed YBCO \cite{14,21} showed
a rather unexpected result: all the relaxation processes at high temperature
were completely suppressed \cite{21} and a new intense peak, labeled P2,
appeared around 60~K, close to a Debye relaxation with an \textbf{activation
energy of 0.11~eV} \cite{14} (see also Fig. \ref{fig ARCuOx}). Since
oxygen\ was certainly present in the CuO$_{x}$ planes at the level of several
molar percent also after outgassing, the most obvious conclusion was that P2
is due to hopping of isolated O\ atoms. This is striking, since \textbf{the
barrier for jumps of isolated O atoms would be 10 times lower than that for
jumps of oxygen within chains}, and also in any other known oxide; such a low
barrier is rather comparable that for interstitial H in \textit{bcc} metals.
In fact, the same relaxation process has been observed also in NQR experiments
and attributed to polaron hopping \cite{JBD94}. Even accepting that the O
atoms in the nearly empty CuO$_{x}$ planes overcome a barrier as little as
0.1~eV to perform a diffusive jump, some difficulties arise in conciliating
such a high mobility with the long time required for oxygen to reach an
equilibrium configuration; this led de Brion \textit{et al.} \cite{BHC90} to
assign P2 to hopping of isolated O atoms trapped by some defect, \textit{e.g.}
a vacancy in the apical oxygen sublattice. It has later been argued \cite{44}
how peak P2 is more likely to be due to oxygen rather than small polaron
hopping, considering that its activation energy is independent on
concentration, and how it is possible that the extremely fast mobility of
isolated O atoms may coexist with a slow kinetics for reaching equilibrium.

\begin{figure}[tbh]
\begin{center}
\includegraphics[
%natheight=310.312500pt,
%natwidth=577.687500pt,
%height=7.1478cm,
width=13.2544cm
]{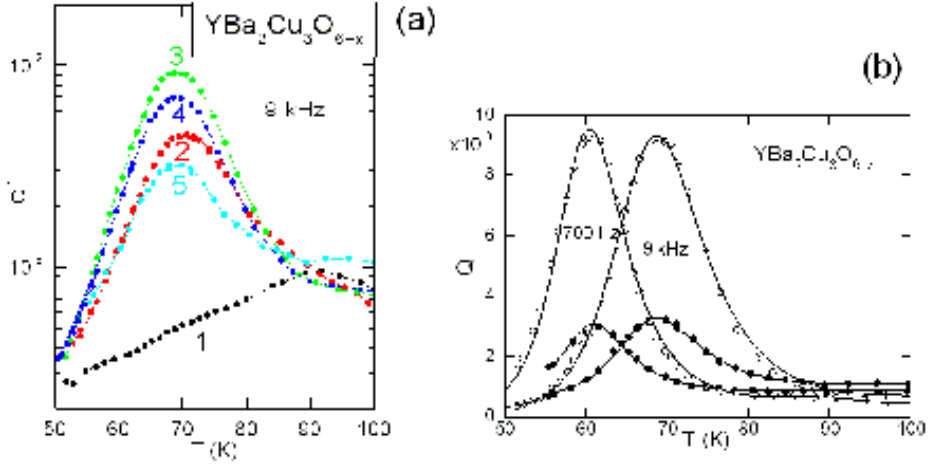}
\end{center}
\caption{(a) development of peak P2 with outgassing and aging at room
temperature; (b) fits of peak P2 measured at two frequencies in the states
corresponding to curves 2 and 5 in (a).}%
\label{fig YBCO_P2}%
\end{figure}

Part of the following results appeared in Ref. \cite{44}. We approached the
$x\rightarrow0$ limit by heating the samples in a vacuum of $\sim10^{-5}$~mbar
at progressively higher temperatures (up to 1000~K) for $\sim2$~h. The last
outgassing treatment at 1000~K was slightly beyond the decomposition limit,
and traces of superficial green phase had to be removed with emery paper. The
oxygen content after the final outgassing treatment was estimated from the $c$
lattice parameter by X-ray diffraction, and was found $x=0.05\pm0.05$, in
agreement with the results on the YBCO\ stability \cite{BN89}. Figure
\ref{fig YBCO_P2}a shows the evolution of peak P2, starting with $x\gtrsim0.4$
(curve 1); where the peak is practically absent, and peak P3 at 90~K (8~kHz)
is still visible; the peak appears after outgassing up to 810~K (curve 2),
with $x\sim0.3$, increases after 2~h in vacuum at 980~K (curve 3), and starts
decreasing with further outgassing (curve 4 after additional 2~h at 1000~K) or
after long aging at room temperature (curve 5 after 13 months).

The characteristics of peak P2 other than the intensity are practically
independent of $x$ or sample history, indicating relaxation of some simple
point defect. Figure \ref{fig YBCO_P2}b shows fits to the peak measured at two
vibration frequencies (0.7 and 9~kHz) in the states with the highest and
lowest intensity (curves 3 and 5 of Fig. \ref{fig YBCO_P2}a). For peak P2 the
Cole-Cole expression (\ref{pCC}) was found to provide a slightly better
interpolation than the Fuoss-Kirkwood one, and it has been allowed for
relaxation between slightly inequivalent sites [Eqs. (\ref{relstr})\ and
(\ref{tau asym}) with $\Delta E\neq0$], in order to reproduce the fact that
the peak intensity is slightly larger at higher temperature, instead of
exhibiting $1/T$ behavior (see the discussion at the end of Sec.
\ref{sect thermod}). Therefore peak P2 has been interpolated with
\begin{align}
Q^{-1}\left(  T\right)   &  =\frac{\Delta_{0}}{2T\cosh^{2}\left(  \frac{\Delta
E}{2k_{\text{B}}T}\right)  }\frac{\sin\left(  \frac{\pi}{2}\alpha\right)
}{\cosh\left[  \alpha\ln\left(  \omega\tau\right)  \right]  +\cos\left(
\frac{\pi}{2}\alpha\right)  }\label{eqn P2}\\
\tau &  =\tau_{0}e^{E/k_{\text{B}}T}/\cosh\left(  \frac{\Delta E}%
{2k_{\text{B}}T}\right)
\end{align}
plus a linear background and a small contribution from peak P3 around 90~K.
The peak may be fitted with $\tau_{0}=1.5\times10^{-13}$~s , $E/k_{\text{B}%
}=1290$~K, $\alpha=0.84$ and $\Delta E\simeq130$~K in all cases, except for
the final state (curve 5), where the slightly different values $\tau
_{0}=4\times10^{-14}$~s and $E/k_{\text{B}}=1370$~K give a better
interpolation. The slight shift of the peak temperature from curve 2 to curve
5 is due to changes of the resonance frequency of the sample. It can therefore
be concluded that \textbf{the relaxation parameters of peak P2 are independent
on }$x$\textbf{\ and sample conditions within experimental error}.

\subsection{Isolated and aggregated oxygen atoms\label{sect isol/aggr}}

The first question to answer is how is it possible that the oxygen jumps for
$x<0.3$ occur over a barrier about 10 times smaller than for $x>0.3$. The
proposed explanation is that there are two substantially different types of
jumps: fast jumps of isolated O atoms, and slow jumps, occurring at high
values of $x$, that involve breaking and formation of chemical bonds with a
neighboring O\ atom. Therefore, \textbf{the real barrier for oxygen\ hopping
in the CuO}$_{x}$\textbf{\ planes would be the one of peak P2, 0.11~eV, while
the additional 0.9-1.0~eV are due to the formation of chemical bonds and}, as
we shall see later, to the \textbf{electrostatic interaction between
neighboring O atoms}. This hypothesis is reasonable in view of the model for
the hole states and charge transfer from chains to planes presented in Sec.
\ref{sect YBCO CT}, where it has been pointed out that: \textit{i)} jumps of
isolated O\ atoms do not affect the electronic state; \textit{ii)} the
formation (or dissolution) of a pair of nearest neighbor O atoms involves the
change of a hole of Cu$^{3d}$ character into O$^{2p}$,\ with a higher energy
(but lower than having two holes on a same Cu$^{1+}$ atom); \textit{iii)}
jumps to or from longer chains involve charge transfer between chain and
CuO$_{2}$ plane.

\begin{figure}[tbh]
\begin{center}
\includegraphics[
%natheight=320.250000pt,
%natwidth=457.187500pt,
%height=2.1378in,
width=3.186in
]{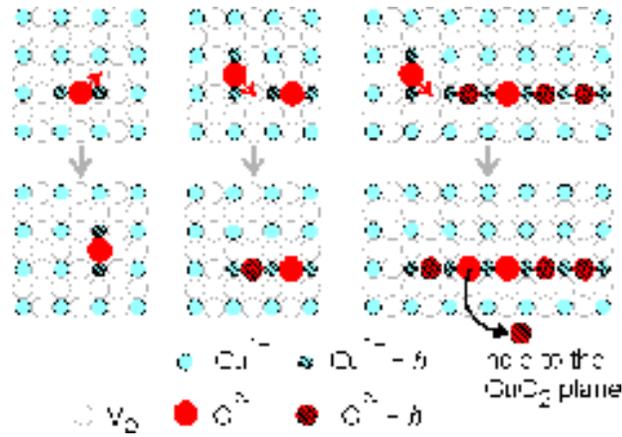}
\end{center}
\caption{Effect of different types of O jumps in the CuO$_{x}$ plane on the
holes. From left to right: 1) no change for an isolated O atoms; 2) a
Cu$^{3d}$ hole is converted into a O$^{2p}$ hole with higher energy; 3)\ after
the formation of a longer chain, a hole is pushed to the CuO$_{2}$ planes.}%
\label{fig chainfrags}%
\end{figure}

Figure \ref{fig chainfrags} presents a sketch of these processes. From these
considerations it appears that only the jumps of isolated O atoms involve only
a simple reorientation of the elastic dipole, as in the Snoek effect, while
all other types of jumps, besides a higher effective barrier, should also
present different initial and final energies, and possibly elastic dipoles.
The energies for creating a hole or changing its state are of the order of
several tenths of eV, which means thousands of kelvin in the temperature
scale. From Sec. \ref{sect thermod} it appears that in such cases the
depopulation factor makes the relaxation strength strongly increasing with
temperature, and this should be easily noted as a peak whose intensity is
higher when measured at higher frequency, instead of being smaller by the
$1/T\;$factor. The anelastic data at high temperature collected by us do not
evidence clearly such a behavior, even though the oxygen loss occurring during
the measurement in vacuum does not make this observation very reliable.

\begin{figure}[tbh]
\begin{center}
\includegraphics[
%natheight=158.187500pt,
%natwidth=158.187500pt,
%height=1.1173in,
width=1.1173in
]{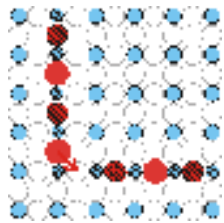}
\end{center}
\caption{Jumps of an O atom between terminals of two perpendicular chains of
similar length leaves the hole state unaffected.}%
\label{fig chainfrags2}%
\end{figure}A\ \textbf{possible explanation} for this fact would be that the
\textbf{anelastic relaxation observable at high temperature is due to those
jumps between Cu-O\ chains that do not change the overall hole state}, and
therefore whose initial and final electronic energies are the same. Figure
\ref{fig chainfrags2} presents the example of a jump between the ends of two
perpendicular chains. If both chains have $n<5$, no hole will be transferred
to or from the planes after the jump, at least according to Uimin's model
\cite{UR92,UHS97}; also, if their length is similar, then the total count of
the various types of holes will be the same before and after the jump. Jumps
of this type, although involve a high activation energy for the temporary
change of the electronic configuration, do have the same initial and final
energy, and therefore the related relaxation strength is proportional to $1/T
$, as usual. It should be noted that this type of jumps is probably involved
in the motion of twins. Considering that the hopping energy or bandwidth of
the holes is $\sim0.4$~eV \cite{SHC94}, it is likely that the difference
between initial and final energy of a jump that changes the electronic state
is of the order of tenths of eV; already setting $\Delta E=0.2$ eV in Eq.
(\ref{relstr}) the depopulation factor $\cosh^{-2}\left(  \Delta
E/2k_{\text{B}}T\right)  $ reduces the relaxation strength by 25 times at
500~K, a temperature where the peaks due to the chain rearrangements are
observed. These considerations have been overlooked up to now in all anelastic
studies of the oxygen jumps, which implicitly assume the equivalence between
the electronic energy before and after the oxygen jump. A reliable analysis of
such processes should take into account the energies involved in the
rearrangements of the chains and the statistical weights of the various chain
configurations; this is outside the scope of the present investigation, in
view of the impossibility of measuring a stable anelastic spectrum in vacuum
at high temperature.

\subsection{Slow achievement of equilibrium and expected concentration of
isolated oxygen atoms\label{sect YBCO P2.3}}

The mobility of the isolated O atoms deduced from peak P2 is extremely fast,
and corresponds to $10^{11}$ jumps per second at room temperature. With such
high a mobility one expects an almost instantaneous achievement of the
equilibrium configuration of the O atoms, contrary to observations. In fact,
oxygen\ ordering at room temperature may proceed for months both at high and
low values of $x$. A well known example is the fact that if a sample with
relatively low $x$ is quenched to room temperature, its superconducting
temperature is found to increase of over 10~K with aging at room temperature
over periods of days \cite{VPY90}; this is attributed to slow reordering of
oxygen at room temperature, with consequent change of the concentration of
holes injected in the CuO$_{2}$ planes and change of $T_{\mathrm{c}}$. Another
example is the slow decrease of the intensity of peak P2 with aging at room
temperature, which can be attributed to a slow decrease of the number of the
isolated O\ atoms at room temperature and below. The discrepancy between the
supposed fast hopping rate of the isolated O atoms and the slow reordering
kinetics has been put in evidence by de Brion \textit{et al.} \cite{BHC90}
with the following simple argument. In the low $x$ limit and assuming that
only O pairs are formed, the authors assumed that the time for reaching
equilibrium, $t_{\text{eq}}$, is the time for forming pairs starting from a
random distribution of isolated (free) O atoms; therefore they obtain
$t_{\text{eq}}\sim\tau_{f}/x^{2}$, where $\tau_{f}$ is the mean hopping time
of the free O atoms and $x$ their concentration. The extrapolation of $\tau$
from peak P2 to room temperature yields $\tau_{f}\sim10^{-11}$~s, which would
require $x\sim10^{-8}$ for obtaining $t_{\text{eq}}\sim1$~day and
$x\sim5\times10^{-10}$ for extending $t_{\text{eq}}$ to 1 year, as we
observed. Such concentrations are of course unrealistic, being $x>0.1$, and
this fact induced de Brion \textit{et al.} to attribute peak P2 to hopping of
oxygen\ trapped by some defect, or to the reorientation of O pairs, leaving
$\tau_{f}\sim\tau_{a}=\tau_{0}\exp\left(  E_{a}/k_{\text{B}}T\right)  $, the
usual hopping time for aggregated oxygen with a barrier of $E_{a}\sim1$~eV. A
flaw in this argument is the assumption that equilibrium is reached once all
the O atoms have had the opportunity of forming a pair or chain fragment for
the first time. This is not true, since the process of reaching equilibrium
will require further steps of dissociation and aggregation of the chain
fragments in the intermediate metastable configurations, and these require
jumps with the characteristic time $\tau_{a}$ of the aggregated oxygen; in
conclusion, the limit steps are those with $\tau_{a}$.

\begin{figure}[tbh]
\begin{center}
\includegraphics[
%natheight=232.812500pt,
%natwidth=812.437500pt,
%height=3.3323cm,
width=11.4839cm
]{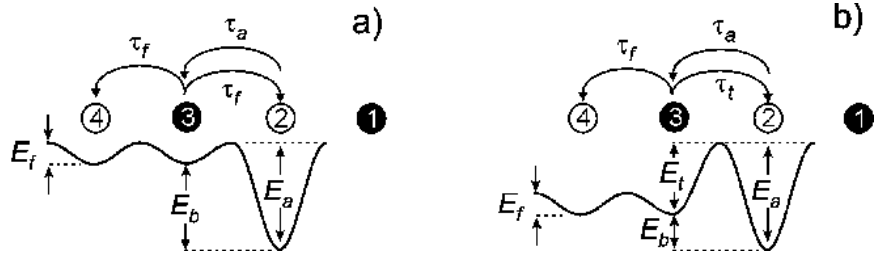}
\end{center}
\caption{Potential felt by an isolated O\ atom (3)\ on approaching another
O\ atoms (1) in site 2. In (a) only the binding energy $E_{b}$ of the pair is
taken into account; in (b) also the electrostatic repulsion is considered.}%
\label{fig YBCO pot}%
\end{figure}

The picture of only two hopping times, $\tau_{f}=\tau_{0}\exp\left(
E_{f}/k_{\text{B}}T\right)  $ with $E_{f}=0.11$~eV for hopping of isolated O
atoms and $\tau_{a}=\tau_{0}\exp\left(  E_{a}/k_{\text{B}}T\right)  $ with
$E_{a}\simeq1$~eV for leaving a chain fragment or diffusing between chains
(see Fig. \ref{fig YBCO pot}a), is nevertheless inadequate. In fact, the rate
equation for the equilibrium between free and aggregated O, in the simplest
form and neglecting any geometrical factor, would be
\begin{equation}
\frac{dc_{f}}{dt}\simeq\frac{c_{a}}{\tau_{a}}-\frac{c_{f}}{\tau_{f}}x\,,
\end{equation}
where $x$ is the probability that the jump of the free O atom joins another O
atom. By posing the condition of stationarity and recalling that $c_{a}%
+c_{f}=x$ and $\tau_{a}\gg\tau_{f}$, one obtains for the equilibrium
concentration of free oxygen
\begin{equation}
\overline{c}_{f}=\frac{x\tau_{f}}{x\tau_{a}+\tau_{f}}\simeq\frac{\tau_{f}%
}{\tau_{a}}=\exp\left(  \frac{E_{f}-E_{a}}{k_{\text{B}}T}\right)  =\exp\left(
\frac{-E_{b}}{k_{\text{B}}T}\right)  \,, \label{cf1}%
\end{equation}
which is of the order of $10^{-15}$ at room temperature with a binding energy
$E_{b}=E_{a}-E_{f}\simeq0.9$ eV. We find again that the concentration of free
O would be negligible at room temperature and below, while peak P2 requires
$c_{f}\sim10^{-4}$ below 100~K.

The explanation for this apparent inconsistency also explains why peak P2 is
not observed at $x>0.4$. In fact, it is proposed that \textbf{the
electrostatic repulsion between the O}$^{2-}$\textbf{\ ions, which is
effectively screened in the conducting phase by the mobile holes, becomes
strong in the semiconducting phase}, with $x<0.4$; the result is an increase
of the saddle point, as shown in Fig. \ref{fig YBCO pot}b. The electrostatic
energy of the two O atoms in the saddle point configuration between second and
first neighboring positions is $q^{2}\times0.6$~eV, where $q$ is the effective
oxygen\ charge, $q\simeq-2$, without taking into account any screening. It is
an energy 2.5 larger than $E_{a}$, and this demonstrates the importance of
screening in determining the short range interaction between the O atoms. If
the saddle point between sites 3 and 2 is larger than that between two free
sites 3 and 4, than we have to introduce a\ ''trapping'' activation energy
$E_{t}$ for the pair formation, and $E_{b}$ becomes $E_{a}-E_{t}$, where
$E_{t}$ is not known but can be a large fraction of $E_{a}$. The new picture
is that of a generally metastable concentration $c_{f}$ of free O\ atoms that
is small at room temperature and below, but larger than the equilibrium
$\overline{c}_{f}$, unless very prolonged aging over years occurs after the
last excursion to high temperature. These \textbf{free O atoms jump with an
extremely fast mean time }$\tau_{f}$\textbf{\ }and therefore approach
continuously other O\ atoms and chain fragments; the \textbf{mean time for
aggregating with other O, however, is the much longer} $\tau_{t}=\tau_{0}%
\exp\left(  E_{t}/k_{\text{B}}T\right)  $, \textbf{at least in the
semiconducting regions}. This fact maintains the concentration of excess free
O rather stable, making perfectly reasonable the observation of peak P2. In
this respect, it should also be noted that the concentration of free O atoms
at the peak temperature $T_{p}\simeq65$~K is not $\overline{c}_{f}\left(
T_{p}\right)  $, but much closer to the starting room temperature value
$c_{f}^{0}\left(  \text{RT}\right)  \ge\overline{c}_{f}\left(  \text{RT}%
\right)  $. In fact, assuming that the starting metastable room temperature
value $c_{f}^{0}\left(  \text{RT}\right)  $ is determined by the longer
hopping time $\tau_{a}$, during cooling, $\overline{c}_{f}\left(  T\right)  $
will be approached with the faster characteristic time $\tau_{t}$, but for a
measuring run with a cooling rate of the order of 1~K/min, $c_{f}$ will remain
frozen when $\tau_{t}$ reaches $\sim10$~min, therefore above $\overline{c}%
_{f}\left(  T^{*}\right)  $, where $\tau_{t}\left(  T^{*}\right)  =10$~min.
I\ do not know of any experimental indication of $E_{b}$, but a range of
values sufficient to freeze $c_{f}\sim10^{-4}$ necessary for observing peak P2
can be estimated setting $10^{-4}\simeq\overline{c}_{f}\left(  T^{*}\right)
\simeq\exp\left(  \frac{-E_{b}}{k_{\text{B}}T^{*}}\right)  $ and
$10$~min$\,=\tau_{t}\left(  T^{*}\right)  =\tau_{0}\exp\left(  \frac{-E_{t}%
}{k_{\text{B}}T^{*}}\right)  $ with the constraint $E_{b}+E_{t}=E_{a}\simeq
1$~eV. It results that for $0.2~$eV$\,<$ $E_{b}<$ $0.3$~eV one freezes
$10^{-5}<\overline{c}_{f}\left(  T^{*}\right)  <10^{-4}$ around 250~K$\,<T^{*}%
<$ 280~K.

At last, we note that in the proposed picture, at room temperature and below,
the rearrangement of the aggregated O\ atoms is extremely slow ($\tau
_{a}\left(  290~\text{K}\right)  \simeq6$~h) and a minor concentration of
highly mobile O atoms remains, that are bound to diffuse in the semiconducting
regions free of O chains. Since their mean hopping time $\tau_{f}$ is much
faster than the mean time for joining other O\ atoms, $\tau_{t}$, being
$\tau_{f}/\tau_{t}=\exp\left[  \left(  E_{f}-E_{t}\right)  /k_{\text{B}%
}T\right]  >10^{2}$, they can be considered as effectively free from the
strong interactions giving rise to the ordered phase and described in Sects.
\ref{sect interact} and \ref{sect YBCO interac}. In fact such interactions
involve changes of the great majority of aggregated O, which is frozen on the
time scale of $\tau_{f}$. It is therefore consistent to ignore any ordering
temperature $T_{\mathrm{C}}$ in eq. (\ref{eqn P2}). The introduction of a mean
asymmetry $\Delta E$ $\simeq130$~K between the free sites is justified by the
fact that the regions free of O pairs and chains, where the isolated O atoms
diffuse, are certainly small and their site energies are strongly perturbed by
strain due to the surrounding aggregated O. This effect is particularly
evident for defect relaxation with small activation energy, giving rise in
this case to $\Delta E/E=0.1$, but is unnoticed for the high temperature
relaxation with $\Delta E/E=0.01$. The magnitude of $\Delta E$ is typical of
the site energy shifts due to strain interactions, as discussed in the end of
Sec. \ref{sect thermod}.

\section{Motion of oxygen in off-center positions of the Cu-O chains}

\begin{figure}[tbh]
\begin{center}
\includegraphics[
%natheight=2.182800in,
%natwidth=3.170400in,
%height=2.1828in,
width=3.1704in
]{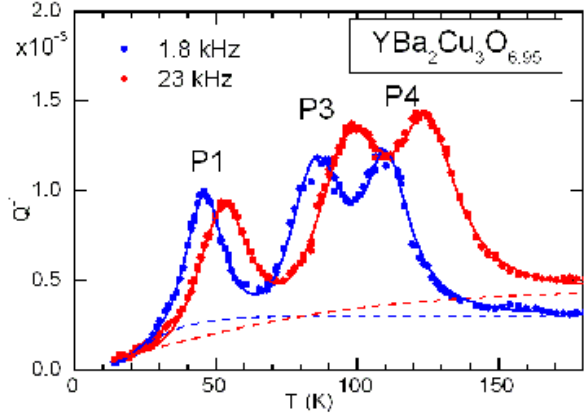}
\end{center}
\caption{Low temperature elastic energy loss of YBCO measured at 1.8 and 23
kHz. The continuous lines are a fit as explained in the text with the dashed
lines as background$.$}%
\label{fig P1P3P4}%
\end{figure}

The low temperature anelastic spectrum of well oxygenated YBCO is shown in
Fig. \ref{fig P1P3P4} and presents three peaks, which we label P1, P3 and P4,
with activation energies of 0.11, 0.16 and 0.19~eV respectively \cite{20}.
Peak P1 will be dealt with in the next section.

The continuous lines in Fig. \ref{fig P1P3P4} are a fit with the
Fuoss-Kirkwood expression (\ref{pFK}) for peaks P3 and P4 and (\ref{pFK2}) for
P1 plus a background of the form $\tanh\left[  \left(  T-T_{0}\right)
/w\right]  $ (dashed lines). Again, in order to reproduce the intensities of
the peaks at both frequencies with the same set of parameters it is necessary
to introduce the possibility that relaxation occurs between energetically
non-equivalent states. This requirement is normal in such a disordered system
like YBCO at temperatures below 100~K (see the end of Sec. \ref{sect thermod}%
). It is therefore necessary to introduce a correction of the type $\cosh
^{-2}\left(  \Delta E/2k_{\text{B}}T\right)  $ to the relaxation strength and
$\cosh^{-1}\left(  \Delta E/2k_{\text{B}}T\right)  $ to the relaxation time
(Eqs. (\ref{relstr}) and (\ref{tau asym})). The parameters used for the fit
are reported in the table.%

\[%
\begin{tabular}
[c]{|l|l|l|l|l|l|}\hline
peak & $E/k_{\text{B}}$ (K) & $\Delta E/k_{\text{B}}$ (K) & $\tau_{0}$ (s) &
$\alpha$ & $\beta$\\\hline
P1 & 908 & 61 & $5.9\times10^{-13}$ & 0.29 & .57\\\hline
P3 & 1730 & 186 & $1.3\times10^{-13}$ & 0.49 & \\\hline
P4 & 2450 & 211 & $2.1\times10^{-14}$ & 0.79 & \\\hline
\end{tabular}
\]

\begin{figure}[tbh]
\begin{center}
\includegraphics[
%natheight=297.375000pt,
%natwidth=361.437500pt,
%height=3.0787in,
width=3.7386in
]{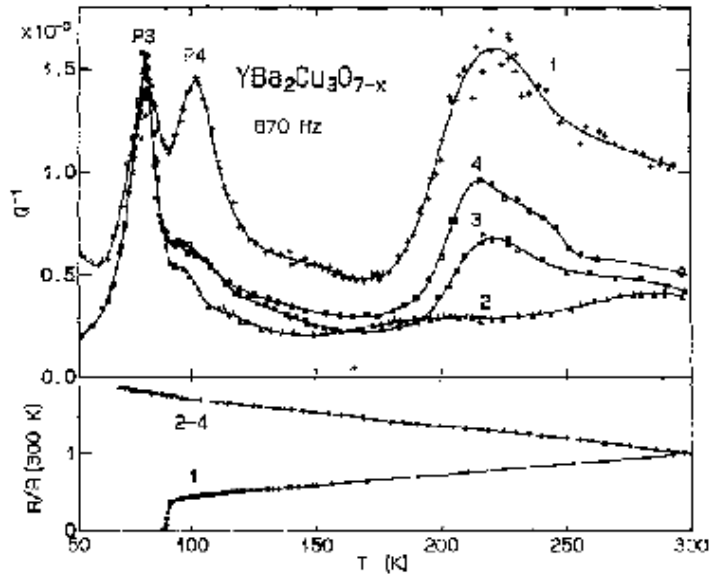}
\end{center}
\caption{Elastic energy loss and resistance normalized to room temperature of
an YBCO sample with $x\simeq0.95$ (curves 1) and after outgassing to
$x\sim0.5$ (curves 2-4) [from \cite{20}].}%
\label{fig P3P4}%
\end{figure}

The pre-exponential factors $\tau_{0}$ are all compatible with point defects,
while the activation energies of P3 and P4, 0.15 and 0.21~eV, are very close
to those deduced from the condition $\omega\tau=1$ at the maxima of the peaks
measured by several authors in a broad range of frequencies, 0.16 and 0.19
respectively \cite{20}. The peaks are also considerably broader than pure
Debye relaxations, and the asymmetry parameters $\Delta E$ are within 10\%\ of
the activation energy; they do not affect much the other parameters but
definitely improve the fit. The evolution of peaks P3 and P4 with doping is
shown in Fig. \ref{fig P3P4}: after an outgassing treatment to $x\sim0.5$, as
deduced from resistivity (absence of superconductivity down to 70~K and
increase of resistivity by 20 times at room temperature), peak P4 was almost
suppressed, while P3 slightly enhanced. The broad peak at 220~K seems to be
connected with some type of oxygen ordering, as briefly discussed in Sec.
\ref{sect 240K}.

The existence of relaxation processes with low activation energy and
disappearing when the Cu-O\ chains become shorter and more disordered is
naturally put in relation with the fact that the \textbf{O atoms in the Cu-O
chains actually occupy positions that are 0.15~\AA \ off-centre along the }%
$a$\textbf{\ axis}. This fact is deduced from the anomalously large
Debye-Waller factor of these atoms in the $a$ direction, from refinements of
neutron \cite{FJY88} and X-ray \cite{WGK90} diffraction spectra. Such
anomalous thermal factors are relatively common in perovskite-related
materials and are indication of off-centre occupation. In addition, a
M\"{o}ssbauer experiment indicated that such zig-zag configurations are
dynamic with a correlation time $>10^{-7}$~s \cite{NH89}. The doping
dependence in Fig. \ref{fig P3P4} indicates that P4 rather than P3 should be
associated with these off-center jumps, since the outgassing treatment at
least halved the number of the off-center O atoms in the chains, and P3 did
not decrease at all. On the other hand, the reduction of the intensity of P4
by much more than a factor $\frac{1}{2}$, might indicate that \textbf{only the
O atoms in relatively long and ordered chains are free to jump between
off-center position}; in fact, in a phase of disordered chain fragments, most
of the pairs of off-center sites would be highly inequivalent, and the O atoms
would remain in the sites of smaller energy.

\begin{figure}[tbh]
\begin{center}
\includegraphics[
%natheight=311.437500pt,
%natwidth=281.312500pt,
%height=2.1767in,
width=1.97in
]{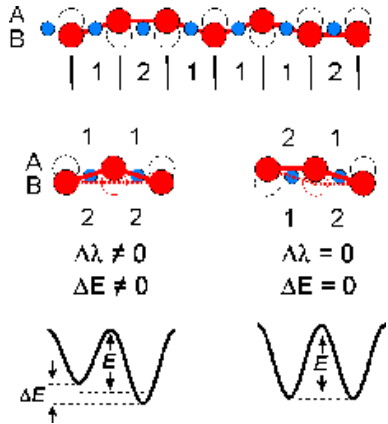}
\end{center}
\caption{Possible types of O jumps between the off-center positions A and B of
the zig-zag Cu-O chains.}%
\label{fig zigzag}%
\end{figure}

A\ model has therefore been developed of how the jumps of the O\ atoms between
the off-centre positions might produce anelastic relaxation \cite{20}. The
reasoning goes along the lines of Sec. \ref{sect thermod}-\ref{sect kin}, with
the additional observation that \textbf{the relaxing units are pairs of
neighboring O atoms}. In fact, an isolated O atom hopping between the two
off-center positions would not give rise to anelastic relaxation, since its
hopping would simply cause an inversion of the off-centre defect, and the
centrosymmetric elastic dipole would not change. Instead, if we consider pairs
of O atoms or chain segments, than we can distinguish between pairs of atoms
that are on a same side of the chain or on both sides. If we call A and B the
sites on either side of the chain, we can distinguish between chain segments
of type 1, in AA or BB sites and having elastic dipole $\lambda_{1}$, and
segments of type 2 in AB or BA sites with elastic dipole $\lambda_{2}$. An
oxygen jump like that on the left in Fig. \ref{fig zigzag} will change the
contribution to the elastic strain by $\Delta\varepsilon^{\text{an}}=2\left(
\lambda_{1}-\lambda_{2}\right)  =2\Delta\lambda$, while a jump like that on
the right hand will let $\varepsilon^{\text{an}}$ unaffected.

Also the profile of the potential energy is different in the two
cases, since in the first case relaxation occurs between states differing in
local strain $2\Delta\lambda$ and in energy by $2\Delta E\neq0$, while in the
second case also the energy must be the same by symmetry (simple inversion of
the configuration). Therefore, the origin of the consistent asymmetry $\Delta
E/k_{\text{B}}\simeq210$~K found for peak P4, may be indicative of the energy
inequivalence between the two types of chain segments, rather than disorder.

\subsection{Possible (anti)ferroelastic and (anti)ferroelectric ordering of
the off-center oxygen atoms\label{sect AFE}}

\begin{figure}[tbh]
\begin{center}
\includegraphics[
%natheight=361.437500pt,
%natwidth=303.312500pt,
%height=7.6772cm,
width=6.4515cm
]{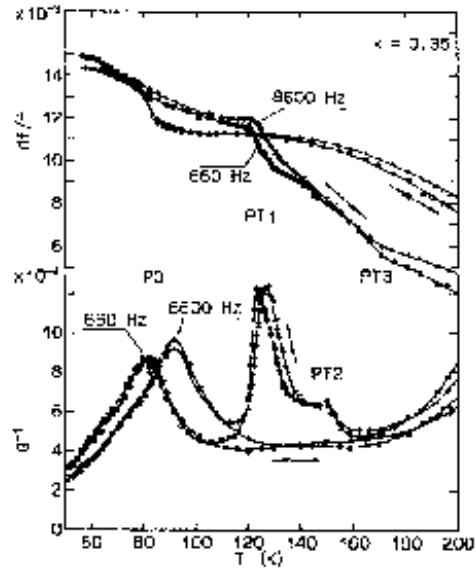}
\end{center}
\caption{Elastic energy loss and resonance frequency ( $\propto\sqrt{E}$) of
YBa$_{2}$Cu$_{3}$O$_{6.35}$ measured on cooling and subsequent heating after
20~h at 80~K. }%
\label{fig PT120K}%
\end{figure}

Finally, we point out that (anti)ferroelastic correlations may arise between
the off-centre O atoms, which also involve electric dipoles along the $a$
axis, associated with O$^{2-}$ atoms that are out-of-axis with respect to the
row of Cu$^{2+}$ atoms along $b$. Therefore, such correlations would also be
of (anti)ferroelectric nature, although the electric field from these dipoles
is effectively screened in the metallic or superconducting environment, and is
therefore hardly detectable in experiments. Nonetheless, interactions between
neighboring O\ atoms along the chain or even across adjacent chains cannot be
excluded and might even result in short-range (anti)ferroelectric and
(anti)ferroelastic domains. Indeed, the $Q^{-1}\left(  T\right)  $ curves
shown here are measured on cooling, but the measurements on heating after some
aging below 90~K exhibit clear phase transformations between 120 and 170~K
\cite{25}, indicated with PT1 and PT2 in Fig. \ref{fig PT120K}. In Ref.
\cite{25} we discussed how these anomalies in both dissipation and modulus
might arise from ferroelectric and/or antiferroelectric domains that are
formed during low temperature aging along the Cu-O chains. We also proposed a
possible evolution of peak P4 with decreasing $x$ toward a slower and more
correlated dynamics, as suggested by a shift to higher temperature and
broadening of the peak in some measurements of outgassed samples. Such an
increase of the correlation between the chain segments would be due to the
decreased electric screening when going into the semiconducting state.

\begin{figure}[tbh]
\begin{center}
\includegraphics[
%natheight=365.437500pt,
%natwidth=512.000000pt,
%height=6.4778cm,
width=9.0523cm
]{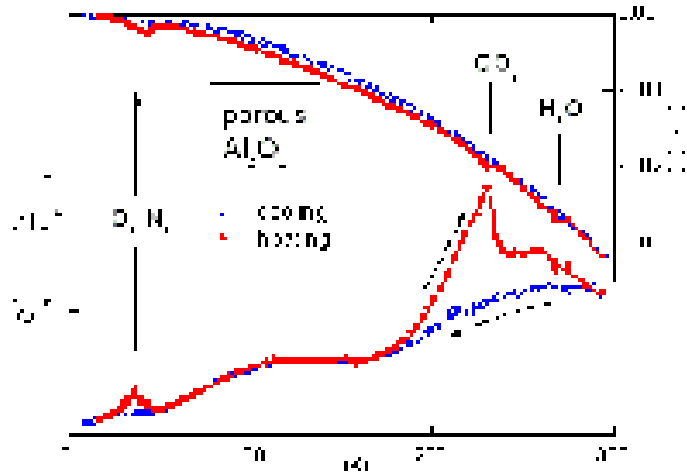}
\end{center}
\caption{Elastic energy loss coefficient and Young's modulus of porous alumina
measured on cooling and subsequent heating. The anomalies on heating
correspond the solid/liquid transitions of residual gases adsorbed by the
sample.}%
\label{fig Al2O3}%
\end{figure}

In general, one should be very cautious in considering phase transitions like
these, since during several hours of low temperature aging in an ordinary
vacuum system without particular precautions \textbf{a porous sample may
adsorb residual gases and then present their solid/liquid transitions on
heating}. This fact is exemplified by the measurement made on a sample of
alumina with $\sim50\%$ porosity (kindly supplied by E. Roncari at CNR-ISTEC,
Faenza, Italy) presented in Fig. \ref{fig Al2O3} without using the adsorption
pump during the experiment.

The measurement on heating present three anomalies around
$35-45$~K, 230~K and 268~K, attributable to O$_{2}$/N$_{2}$, CO$_{2}$ and
H$_{2}$O which were adsorbed onto the porous sample, especially at LHe
temperature, and which pass from the solid to the liquid state and finally
desorb on heating. It was also possible to monitor an increase of pressure in
the system in correspondence with these anomalies. The anomaly of Fig.
\ref{fig PT120K}, however, should be intrinsic of YBCO, since it occurs at a
temperature where nothing is observed in the alumina dummy sample.

\section{Bipolaron reorientation in the overdoped state}

The anelastic relaxation process with the lowest activation energy,
$E=0.11$~eV, is labelled P1 and is observed only in the overdoped state: it is
absent with $x<0.85$ and its intensity rises very fast on approaching full
oxygenation \cite{48}. This is a really noticeable feature, since most of the
physical properties of YBCO are practically constant in the range $0.85<x<1$,
including the superconducting temperature $T_{\mathrm{C}}\left(  x\right)  $,
which presents a maximum at the so-called optimal doping $x\simeq0.93$, but
whose variation remains within 4\% in that range, as shown in Fig.
\ref{fig TcP1vsx} from Ref. \cite{BSH95}.

\begin{figure}[tbh]
\begin{center}
\includegraphics[
%natheight=590.187500pt,
%natwidth=816.062500pt,
%height=7.317cm,
width=10.0935cm
]{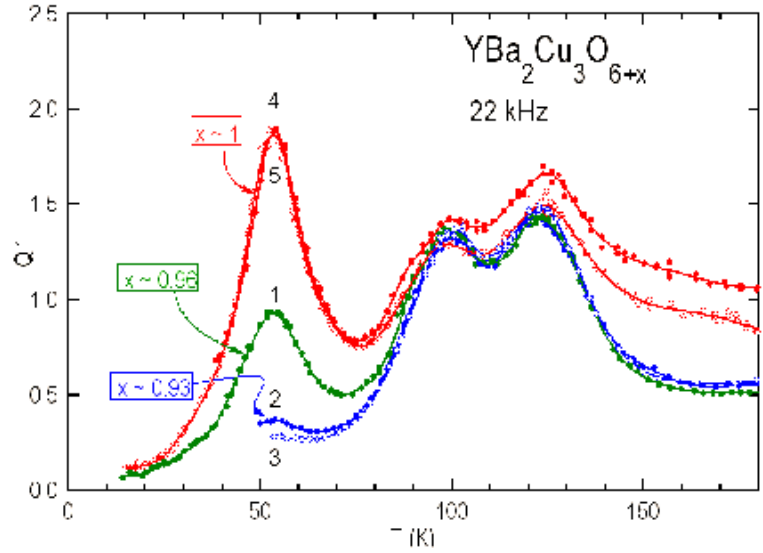}
\end{center}
\caption{Anelastic spectrum of YBa$_{2}$Cu$_{3}$O$_{6+x}$ at various values of
$x$ in the overdoped region, with peaks P1, P3 and P4, starting from low
temperature.}%
\label{fig YBCO P1}%
\end{figure}

Figure \ref{fig YBCO P1} presents the anelastic spectrum of YBCO below 180~K,
measured at 22~kHz on the same sample in several different states from
optimally doped to overdoped, obtained in different ways over a period of 6
years. The peaks at 100~K and 120~K (P3 and P4 discussed above) remain
remarkably stable, and only the intensity of P1 at 50~K changes by at least
one order of magnitude. Curve 1 is the as-prepared state, where $x$ was
estimated from the lattice parameters as $x_{1}\simeq0.93$ \cite{48}. The
curve is closely similar to that measured on an identical sample 6 years
earlier and therefore demonstrates the \textbf{stability of samples and
anelastic spectra}. Notice that, in view of the near constancy of the lattice
parameters for $x>0.9$, the above estimate of $x$ has great uncertainty, and
should be increased. In fact, after a mild annealing at 470~K for 20~h in a
vacuum better than $10^{-5}$~mbar, some oxygen\ loss certainly occurred, and
$T_{\mathrm{C}}$ increased from 90.3~K to 91.9~K, indicating that the sample
passed from the overdoped to the optimally doped state.

With the help of the $T_{\mathrm{C}}\left(  x\right)  $ curve from literature
\cite{BSH95}, we can set $x_{1}=0.96$ and $x_{2}=0.93$, as shown in Fig.
\ref{fig TcP1vsx}. After such a small decrease of the oxygen\ content, the
intensity of P1 decreased by 3-4 times (curve 2); after 15 days the intensity
was further reduced (curve 3), possibly due to some oxygen reordering. The
sample was then oxygenated as fully as possible in the UHV system described in
Sec. \ref{sect UHV}, by heating to 600~$^{\mathrm{o}}$C, introducing a static
atmosphere of 1250~mbar O$_{2}$ and slowly cooling to room temperature (where
the O$_{2}$ pressure became 740~mbar); the cooling rate was 0.2~K/min except
between 490 and 300~$^{\mathrm{o}}$C, where it was reduced to 0.1~K/min. After
this full oxygenation the intensity of P1 become about twice the original one
(curve 4), and remained stable during subsequent runs (curve 5). The sample
has been finally outgassed in the UHV system and equilibrated at
460~$^{\mathrm{o}}$C with a known amount of O$_{2} $, resulting in a final
pressure of 9.6~torr O$_{2}$, which should result in $x=0.85$ according to the
phase diagram of Ref. \cite{SHY91}. From the amount of gas absorbed it was
estimated $x=0.89\pm0.04$, and therefore it can be set $x_{6}=0.87$; at this
oxygen\ content no trace of peak P1 was present.

\begin{figure}[tbh]
\begin{center}
\includegraphics[
%natheight=425.875000pt,
%natwidth=483.250000pt,
%height=6.0451cm,
width=6.8535cm
]{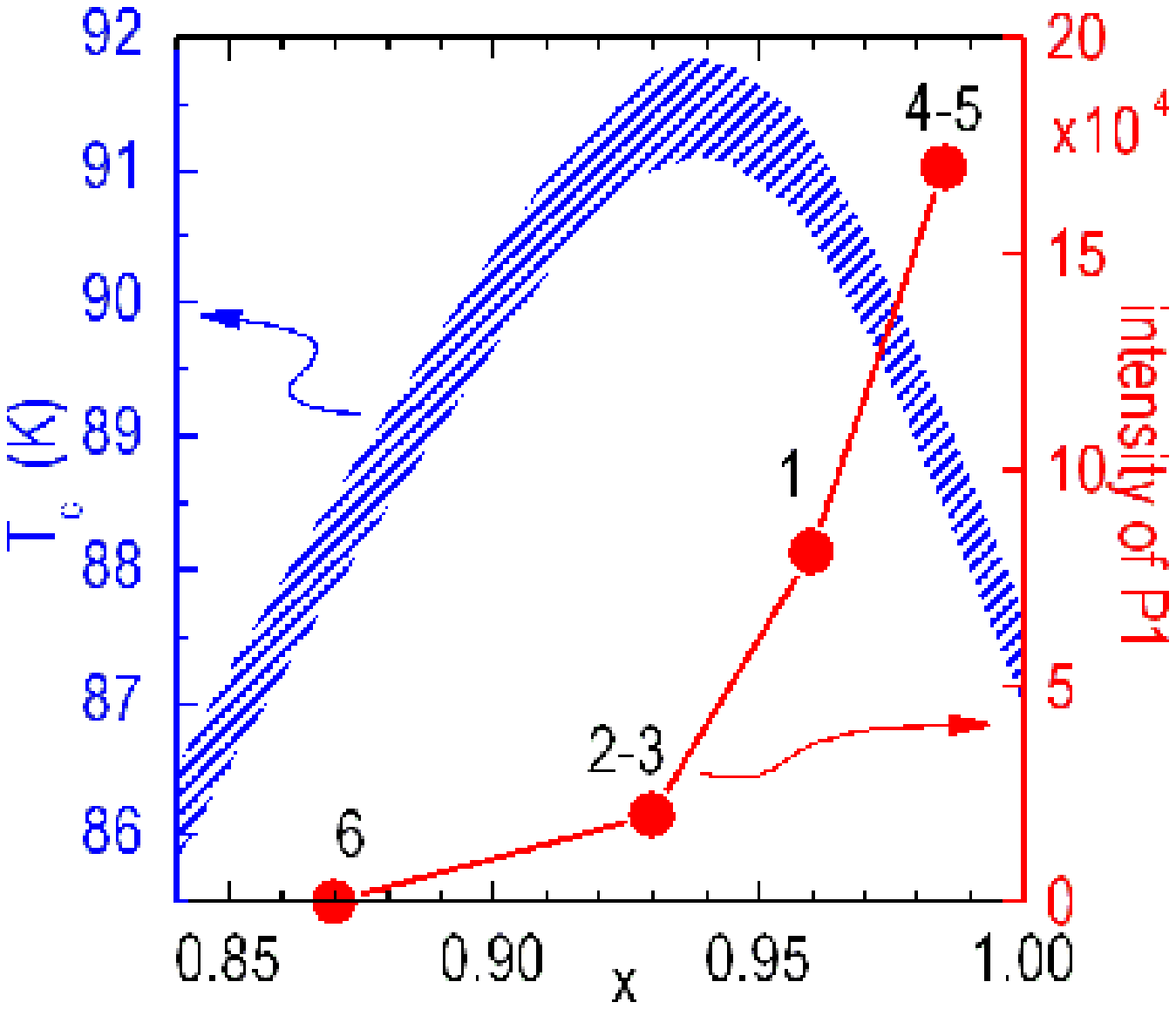}
\end{center}
\caption{Maximum of $T_{\mathrm{C}}\left(  x\right)  $ inYBa$_{2}$Cu$_{3}%
$O$_{6+x}$ (after \cite{BSH95}). Superimposed in red are the intensities of
peak P1 corresponding to the curves in Fig. \ref{fig YBCO P1} at values of $x$
estimated with the help of the $T_{\mathrm{C}}\left(  x\right)  $ curve.}%
\label{fig TcP1vsx}%
\end{figure}

\begin{figure}[tbh]
\begin{center}
\includegraphics[
%natheight=6.146100cm,
%natwidth=8.180200cm,
%height=6.1461cm,
width=8.1802cm
]{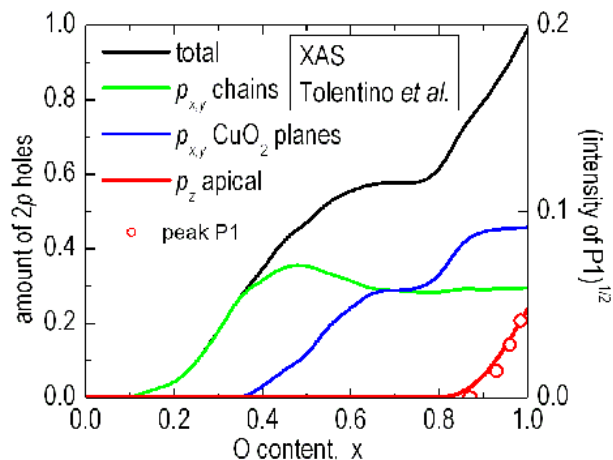}
\end{center}
\caption{Concentration of holes with prevalent O $2p$ character, measured by
XAS (from \cite{TBF92}). The concentration of $p_{z}$ holes is evidenced in
red, and is in good agreement with $\sqrt{Q_{max}^{-1}}$ of peak P1 (right
ordinate axis).}%
\label{fig YBCO holes}%
\end{figure}

Considering that \textbf{all the structural parameters remain practically
constant within the range of }$x$\textbf{\ in which P1 develops}, it is
impossible to associate its mechanism with structural defects, impurities,
defects in the order of the Cu-O chains, or off-center atoms, as also
discussed in detail in \cite{48}. The only quantity that to our knowledge
starts increasing in the same doping range is the concentration of
\textbf{holes of character }$p_{z}$, deduced from the peaks in the X-ray
absorption spectra (XAS) of the Cu-K\ edge \cite{TBF92}, and shown in Fig.
\ref{fig YBCO holes}.

\begin{figure}[tbh]
\begin{center}
\includegraphics[
%natheight=568.250000pt,
%natwidth=463.312500pt,
%height=7.2071cm,
width=5.8891cm
]{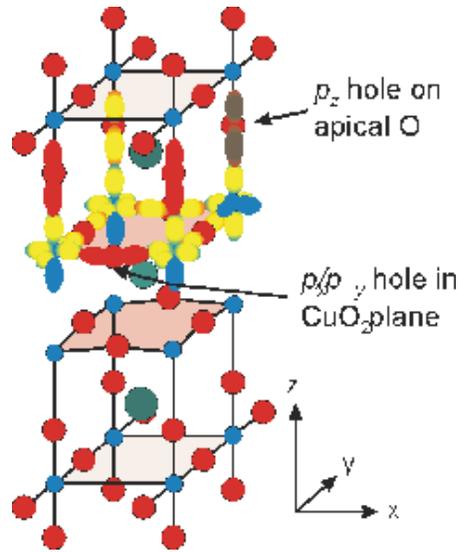}
\end{center}
\caption{$p_{x}$ hole in the CuO$_{2}$ plane and $p_{z}$ hole on an apical O
in YBa$_{2}$Cu$_{3}$O$_{7}.$}%
\label{fig YBCO orb}%
\end{figure}

These holes, having symmetry $p_{z}$, must reside on the apical O
atoms (see Fig. \ref{fig YBCO6_7}), which are the only atoms with consistent
hybridization of the $2p_{z}$ orbitals. The correspondence between the growth
of the concentration of these holes for $x>0.8$ and the growth of peak P1 is
demonstrated in Fig. \ref{fig YBCO holes}, where $\sqrt{Q_{\text{max}}^{-1}}$
of P1 is also plotted. In Fig. \ref{fig YBCO orb} the $p_{z}$ orbitals are
sketched together with those of O $2p_{x}$ and O $2p_{y}$ character,
hybridized with the Cu $3d$ orbitals in the CuO$_{2}$ planes. Regarding the
possible anelastic relaxation processes arising from hopping of the holes
among these orbitals, the hopping between $p_{x}$ and $p_{y}$ would of course
produce anelastic relaxation, but these holes are extremely mobile, in a band
giving rise to superconductivity. The relaxation rate $\tau^{-1}$
characterizing the equilibrium between the $p_{x}$ and $p_{y}$ populations is
certainly much faster than the thermally activated rate over a barrier of
0.11~eV found for peak P1. Also the process of change of a hole from $p_{x/y}
$ in the CuO$_{2}$ plane to apical $p_{z}$ can hardly explain peak P1, since,
as already discussed for peak P2 in Sec. \ref{sect isol/aggr}, different
electronic states are expected to differ in energy in the order of 0.1~eV or
more; this would lead to a depression of the relaxation strength by a factor
$\cosh^{2}\left(  \Delta E/2k_{\text{B}}T\right)  $ which, already for $\Delta
E=0.1$~eV would be $3\times10^{-9}$. This means that the relaxation process
corresponding to the $p_{x/y}\leftrightarrow p_{z}$ exchange is unobservable.

The hopping of a hole among the $p_{z}$ orbitals of different apical O atoms
might well have a correlation time like that of P1, since these orbitals are
too much separated to form a band, and two intermediate $p_{x/y}%
\leftrightarrow p_{z}$ exchanges are necessary. In this case initial and final
energy are the same and there is no problem of depression of the relaxation
strength; rather, anelastic relaxation is impossible because of the
equivalence of all the $p_{z}$ orbitals also under the application of any
stress. On the other hand, if two holes on neighboring apical O\ atoms form a
stable pair, a \textbf{bipolaron}, the resulting elastic dipole has
orthorhombic symmetry and \textbf{reorients itself together with the hole
pair}. It is therefore proposed that peak P1 is due to the reorientation of
pairs of holes, or small bipolarons, on neighboring apical O atoms.

\textbf{The intensity of P1 is then expected to be proportional to
the square of the concentration of }$p_{z}$\textbf{\ holes}, and this is
verified in Fig. \ref{fig YBCO holes}, where the square root of the intensity
of P1 (closed circles and right ordinate) corresponds very well with the
number of $p_{z}$ holes. The activation energy of 0.11~eV would be connected
wit the $p_{x/y}\leftrightarrow p_{z}$ exchange process, necessary for the
bipolaron reorientation.

\begin{figure}[tbh]
\begin{center}
\includegraphics[
%natheight=425.875000pt,
%natwidth=568.500000pt,
%height=6.0451cm,
width=8.0506cm
]{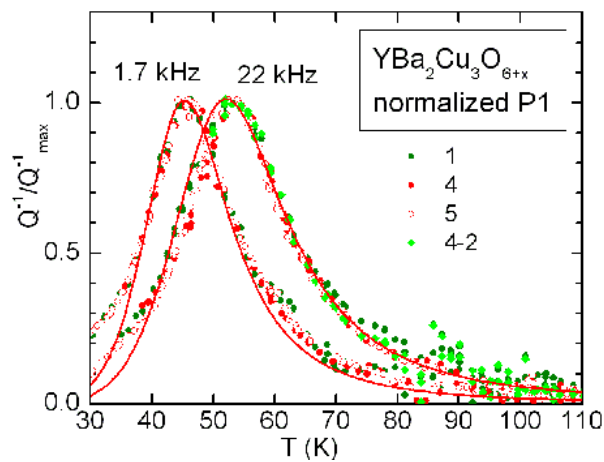}
\end{center}
\caption{Peak P1 after linear background subtraction and normalization to the
peak intensity; the symbols are the same of Fig. \ref{fig YBCO P1} . The
continuous line is the fit as explained in text.}%
\label{fig P1norm}%
\end{figure}

It is noticeable that the \textbf{position in temperature and shape of peak P1
do not change with doping within experimental error}. Figure \ref{fig P1norm}
presents the experimental data of Fig \ref{fig YBCO P1} after subtraction of a
linear background and normalization to the peak intensity, and all data fall
on the same curves. As an additional check of the reliability of the
background subtraction, also the difference between curves 4 and 2 is found to
fall on the same curves, and it does not depend on the choice of the background.

The continuous lines are a fit with the Jonscher expression, eq.
(\ref{Jonscher}), with $\tau_{1}=\tau_{2}=$ $\tau_{0}\exp\left(
E/k_{\text{B}}T\right)  $ $,$ $\tau_{0}=5\times10^{-13}$~s, $E/k_{\text{B}%
}=945$~K, $\alpha=0.33$ and $\beta=0.41$. The values of $\alpha$ and $\beta$
are definitely smaller than 1, indicating considerable broadening with respect
to the monodispersive Debye case. This fact is not unexpected, especially for
a low-temperature relaxation process in a rather complex and disordered system
like YBCO, where random energy shifts due to interaction with various defects,
including disorder in the Cu-O chains, may easily be of the order of $0.1-0.2$
times the activation energy. In the present case, however, the main source of
disorder are O\ vacancies in the Cu-O chains, and their concentration $1-x$
decreases of about 4 times from curve 1 to curve 4 without resulting in any
narrowing of the peak. Neither can the dynamic interaction among the $p_{z}$
holes be the source of broadening, since also their concentration changes by
more than 4 times. \textbf{The spectrum of relaxation times should therefore
be ascribed to the interaction with phonons and the other charge carriers},
which remain practically constant within the range $0.85<x<1$ (see also Fig.
\ref{fig YBCO holes}).

\section{The anomaly near 240~K\label{sect 240K}}

In YBCO there is a relaxation peak in the absorption around 240~K with the
characteristics of glassy dynamics, accompanied by a hysteresis in the Young's
modulus below the same temperature \cite{13,LF88,LK88}. This anomaly can
easily be masked by the freezing transition at $\sim220$~K of the pump oils
adsorbed in porous samples from the measurement vacuum system, as first
noticed by Gzowski \cite{KGG91}; in addition, the hysteresis in the elastic
moduli may be affected by the sample microstructure \cite{EGL87,BBG93}, since
the anisotropic thermal expansion may cause severe stresses from grain to
grain during thermal cycling. Nonetheless, an intrinsic effect exists, has a
strong dependence on the oxygen content, and has been confirmed by several
later works. Regarding the possible mechanism, there are different proposals,
like an oxygen ordering transition \cite{13,LF88,WSA95}, a ferroelectric or
ferromagnetic transition \cite{PPN92} or some polaronic mechanism
\cite{YW02,YHW04}.

\section{Anomalies at the superconducting temperature}

There are various reports of elastic anomalies at the superconducting
transition \cite{KKA90,Dom93,LLH94}, also by means of non traditional
techniques \cite{KMG94} and we also found that sometimes anomalies in the
Young's modulus appear near $T_{\text{c}}$. I\ will not try to analyze such
anomalies, considered the multitude of effects affecting the elastic moduli of
ceramic samples (see also the previous Section), and, as observed by
Mizubayashi \textit{et al.} \cite{MTO88}, the presence of the modulus defects
associated with peaks P3 and P4 in the same temperature range (see Fig.
\ref{fig P1P3P4}).

\section{Summary of the main results obtained in YBCO}

The interpretation of the anelastic spectra of YBCO starts from the
realization that they must be dominated by the hopping with consequent
reorientation of the anisotropic elastic dipole of a concentration $0<x<1$ of
nonstoichiometric O\ atoms in the CuO$_{x}$ planes. One can convince himself
or herself of this fact, by noting that in LSCO a concentration $\delta<0.01$
of interstitial O atoms dominates the $Q^{-1}\left(  T\right)  $ curves (Fig.
\ref{fig LCO O1O2T}b), even though these hopping defects are in a first
approximation isotropic (in a tetragonal cell). The barrier for the jumps of O
in the CuO$_{x}$ planes of YBCO has been determined as $\sim1$~eV from several
anelastic relaxation and other types of experiments, but the inspection of
spectra at $x\sim0.1$, 0.4 and 0.9 (Fig. \ref{fig ARCuOx}b) demonstrates that
the situation is actually more complex. The complexity should arise from the
fact that when an O atom jumps into or out of a chain, not only causes a
reorientation of its elastic dipole, but also a change of the electronic
energy, possibly by tenths of eV, as discussed in Sec. \ref{sect isol/aggr}.

\subsection{Peak PH2}

The peak that is observed at the highest values of $x$ is obviously due to
hopping of O atoms in the O-I phase, where the O\ atoms are ordered into
parallel chains. From the $Q^{-1}\left(  T\right)  $ curves presented here it
is impossible to extract any quantitative information, due to the loss of
oxygen during the measurements in vacuum, and therefore I accept the picture
based on low frequency anelastic experiments \cite{XCW89} that oxygen hopping
requires the presence of a vacancy in the neighboring chain. In Sec.
\ref{sect isol/aggr}, however, it is pointed out that the existing analyses
overlook the change in electronic energy that occurs when an O atom leaves or
joins a chain; this fact might drastically reduce the number of jumps that
actually contribute to the dynamic compliance (due to the depopulation factor
Eq. (\ref{relstr})) to only those that leave the electronic energy almost
unchanged, like the jumps occurring in the twin walls (Fig.
\ref{fig chainfrags2}).

\subsection{Peak PH1}

Also in a broad range $0.3<x<0.8$ there is a peak with an activation energy
$\sim1$~eV, but its temperature is definitively lower than that of PH2, and it
is labeled here as PH1 (Fig. \ref{fig LCO O1O2T}b and \ref{fig YBCO PH1}a);
the fact that PH1 and PH2 are actually distinct peaks is reliability confirmed
by low frequency isothermal measurements \cite{MSB94}. The difference between
PH1 and PH2 apparently resides in the fact that the latter occurs in an
environment of full Cu-O chains, while PH1 might involve all the other jumps,
as depicted in Fig. \ref{fig ARCuOx}a (save the possible condition that only
jumps of the type of Fig. \ref{fig chainfrags2} are observable).

\subsection{Peak P2}

The unexpected consequence of reducing $x$ below $0.3$ is that peak PH1
disappears in favor of peak P2 at much lower temperature (Fig.
\ref{fig LCO O1O2T}b and \ref{fig YBCO_P2}). This is the only relaxation
process that may be assigned to the hopping of isolated O atoms in the
CuO$_{x}$ plane; in fact, quenching experiments exclude that PH1 may contain
contributions from isolated O atoms (Fig. \ref{fig YBCO PH1}b). Considering
that hopping of isolated O atoms must produce a $Q^{-1}\left(  T\right)  $
well visible down to concentrations well below 0.01, it must be concluded that
this peak is P2, with an activation energy of only 0.11~eV. To my knowledge,
this is the lowest barrier for oxygen\ hopping ever reported, comparable to
that for hopping of interstitial H in metals. The main difference between
jumps of isolated O atoms and those causing peaks PH1 and PH2 is that the
latter involve changes in the chemical or electronic state, as discussed in
Sec. \ref{sect isol/aggr}.

In Sec. \ref{sect YBCO P2.3} it is also discussed how the unscreened
electrostatic repulsion between O atoms in the oxygen-poor semiconducting
phase should prevent isolated O atoms to form pairs or join existing chain
fragments, so explaining why a small but finite concentration of isolated O
atoms survives at the low temperature where peak P2 is observed (otherwise the
free O atoms would immediately aggregate into the energetically favorable
chains). This also explains the long times (years) necessary for oxygen to
reach an equilibrium configuration at room temperature.

\subsection{Peak P4}

At relatively high oxygen contents three $Q^{-1}\left(  T\right)  $ peaks are
observed, with activation energies of $0.08-0.21$~eV (Fig. \ref{fig P1P3P4}).
Of these, peak P4 presents a doping dependence compatible with a process
occurring in the O-I phase, and it has therefore been associated with short
jumps of the O atoms between the sites of the Cu-O chains, which have been
shown to be off-centre with respect to the chain axis by diffraction
experiments \cite{FJY88,WGK90}.

\subsection{Peak P1}

The evolution of peak P1 with doping, instead, is very peculiar (Figs.
\ref{fig YBCO P1} and \ref{fig TcP1vsx}); in fact, it appears only for
$x\gtrsim0.87$ and then grows more than linearly with $x$ in a doping region
where virtually all the physical properties, including the superconducting
critical temperature, remain almost constant. The only physical quantity that
has been reported to display a similar doping dependence is the amount of
holes in the $p_{z}$ orbitals of the apical O atoms \cite{TBF92}. Such
orbitals have negligible overlap with each other, so that the $p_{z}$ holes
cannot form a band but have to pass to orbitals in the CuO$_{2}$ planes or
CuO$_{x}$ chains in order to move; this process may well require an activation
energy like that of P1, 0.08~eV, and therefore the mechanism proposed for P1
is the reorientation of pairs of $p_{z}$ holes (bipolarons); in fact, the
motion of single holes, having the same symmetry of the lattice, would not
cause any anelastic relaxation. A confirmation that pairs of $p_{z}$ holes are
involved is the proportionality of the intensity of P1 to the square of the
concentration of these holes (Fig. \ref{fig YBCO holes}).

\chapter{Ru-1212}

\section{Structure\label{sect Ru struct}}

\begin{figure}[tbh]
\begin{center}
\includegraphics[
%natheight=243.437500pt,
%natwidth=305.062500pt,
%height=6.0451cm,
width=7.563cm
]{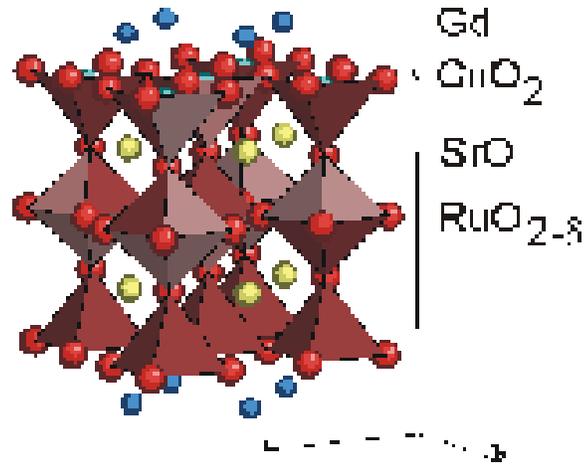}
\end{center}
\caption{Structure of Ru-1212 with the RuO$_{6}$ and CuO$_{5}$ polyhedra put
in evidence.}%
\label{fig Ru poly}%
\end{figure}

The ruthenocuprates are a relatively new class of HTS,\cite{BWB95} which has
attracted much interest for the apparent coexistence of ferromagnetism and
superconductivity \cite{FWB99}. RuSr$_{2}$GdCu$_{2}$O$_{8}$ (Ru-1212) has a
cell similar to that of YBCO, with Gd instead of Y, SrO instead of BaO layers,
and RuO$_{2}$ instead of CuO$_{x}$ planes. Figure \ref{fig Ru poly} puts in
evidence the CuO$_{5}$ pyramids, found also in YBCO, and RuO$_{6}$ octahedra,
absent in YBa$_{2}$Cu$_{3}$O$_{6+x}$ where at most CuO$_{4}$ squares in the
$bc$ planes exist in correspondence with the chains. From the structural
analogy with YBCO, if O vacancies may be introduced in Ru-1212, they are
expected to be in the RuO$_{2}$ planes. The RuO$_{6}$ octahedra are slightly
rotated about the $c$ axis in an antiferrodistortive pattern (see Sec.
\ref{sect Ru rot} below), and the rotation angles decreases from
14.4$^{\mathrm{o}}$ at $T=0$~K to 13.9$^{\mathrm{o}}$ at room temperature
\cite{CJS00}; a transformation to the symmetric unrotated structure is
expected at higher temperature, but not yet observed.

Doping in Ru-1212 is supposed to be due to the mixed valence of Ru$^{4+/5+}$,
which, from X-ray absorption spectroscopy \cite{MZA99} and bond valence sums
\cite{LJH01}, is found to be $40-50\%$ Ru$^{4+}$ and the remaining fraction
Ru$^{5+}$. The mixed valence of Ru is reflected in mixed valence of Cu, with
each Ru$^{4+}$ considered to inject a hole in one of the CuO$_{2}$ planes
above or below therefore providing the holes in the usual superconducting
CuO$_{2}$ planes. However, the deduced concentration of holes $p=$ $\frac
{1}{2}\left[  \text{Ru}^{4+}\right]  =$ $0.2-0.25$ in the CuO$_{2}$ planes
contrasts with the transport properties, which are typical of underdoped
cuprates, therefore with $p\simeq0.1$. A possible explanation for this
incongruence is that most of the holes are trapped by defects or by magnetic
order and do not contribute to conductivity \cite{MZA99}. Of course, the
presence of O vacancies would affect the charge balance and doping, and
therefore also in Ru-1212, as in the other cuprates, it is of general interest
to study their formation, dynamics and possible ordering.

In spite of extensive experimental investigations, nominally identical samples
may be superconducting or nonsuperconducting, and there is no consensus yet,
about the influence of the oxygen stoichiometry and sample microstructure.
Some studies find that annealing at high temperature in vacuum or inert
atmosphere causes considerable oxygen deficiency \cite{BWB95,MAJ03} while
others find no influence at all \cite{PTW99,CJS00} even up to
800~$^{\mathrm{o}}$C \cite{HFA00}. It has then proposed that the prolonged
annealings affect the cation ordering \cite{KDM00}, the grain boundaries
\cite{LMC01c}, or the microstructure \cite{TLW00}.

The $c$ lattice parameter is almost perfectly 3 times greater than $a$, so
that there is little mismatch between domains oriented perpendicularly with
each other; as a consequence, there is little driving force for the domain
growth, and the domain size is very small, unless prolonged aging is made. For
this reason, it has been proposed that the reason why superconductivity often
presents character of granularity is the high number boundaries where the
CuO$_{2}$ planes meet at an angle of 90$^{\mathrm{o}}$; prolonged anneals
therefore would favor superconductivity due to the growth of these
microdomains and consequent reduction of granularity, rather than
stoichiometry changes \cite{MZA99}.

Although interesting effects are found also in the anelastic spectrum of
Ru-1212 below room temperature \cite{111}, in the present Thesis only the high
temperature \cite{104} results will be considered.

\section{Oxygen vacancies\label{sect Ru V_O}}

\begin{figure}[tbh]
\begin{center}
\includegraphics[
%natheight=397.000000pt,
%natwidth=397.000000pt,
%height=8.4285cm,
width=8.4285cm
]{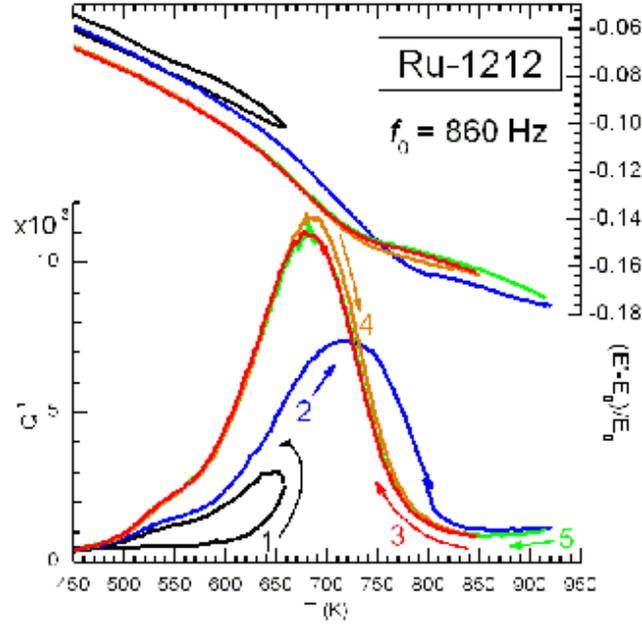}
\end{center}
\caption{Evolution of the anelastic spectrum of Ru-1212 measured in vacuum at
progressively higher temperatures.}%
\label{fig Ru spectra}%
\end{figure}

Figure \ref{fig Ru spectra} presents a series of measurements in vacuum on a
same sample at heating/cooling rates of $1.5-3$~K/min, reaching increasing
maximum temperatures. The elastic energy loss coefficient $Q^{-1}$ is shown in
the lower part, while the upper part is the change of the Young's modulus
relative to the initial value at $T=0$~K. A dissipation peak starts developing
above 600~K, and it is due to O\ vacancies (V$_{\text{O}}$); in fact, the
oxygen\ loss when heating above that temperature was confirmed by the increase
of the oxygen partial pressure during heating ramps in the UHV system. The
peak, which is centered at 690~K at 860~Hz and we label as P1, is completely
developed after heating up to 850~K and remains stable during further
heating/cooling runs up to 930~K. The $Q^{-1}\left(  T\right)  $ and $E\left(
T\right)  $ curves measured concomitantly at 9.8~kHz (not shown in the figure)
where shifted to higher temperature, indicating that the peak is thermally
activated, as expected from jumps of V$_{\text{O}}$. The relatively fast
achievement of an equilibrium spectrum with an intense thermally activated
peak, without all the complex phenomenology found in YBCO, indicates that
\textbf{only a small concentration }$\delta$\textbf{\ of V}$_{\text{O}}%
$\textbf{\ can be introduced in RuSr}$_{2}$\textbf{GdCu}$_{2}$\textbf{O}%
$_{8-\delta}$, in accordance with other results in the literature. The oxygen
loss estimated from the mass loss or gain of a sample subjected to outgassing
or oxygenation in the UHV system is $\delta\sim0.02-0.03$.

\begin{figure}[tbh]
\begin{center}
\includegraphics[
%natheight=216.312500pt,
%natwidth=577.687500pt,
%height=6.0385cm,
width=16.3011cm
]{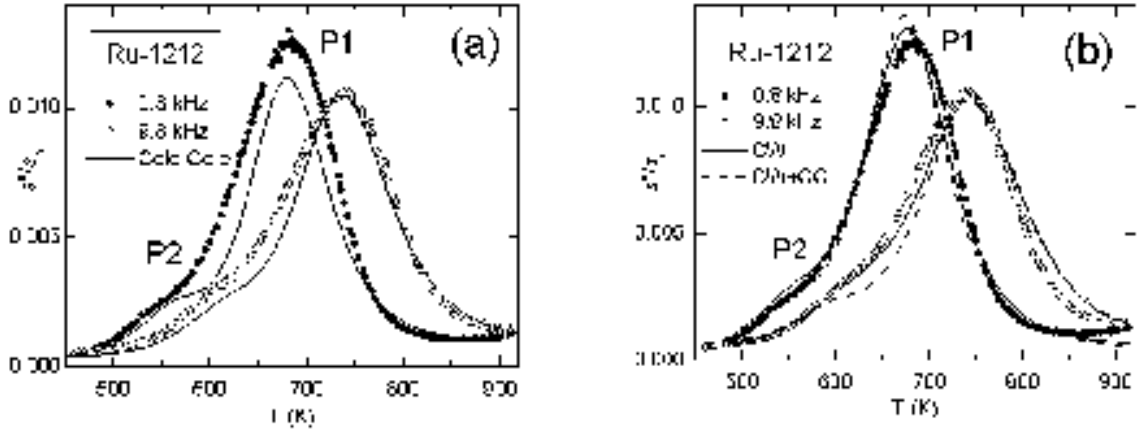}
\end{center}
\caption{Imaginary part of the compliance of Ru-1212 after stabilization of
the two peaks. The continuous lines are fits with Cole-Cole peaks (a)\ and
including the Curie-Weiss corrections (b).}%
\label{fig Ru fit}%
\end{figure}

Since the final spectrum (curves 3 to 5) is stable, it is meaningful to
analyze it to make deductions on the dynamics of the O vacancies. In addition,
in view of the good quality of the data and rather high values of $Q^{-1}$,
instead of $Q^{-1}\left(  \omega_{i},T\right)  $ I chose to fit the imaginary
part of the compliance $s^{\prime\prime}$ referred to the value $s_{0}$
extrapolated to $T=0$~K (the low temperature measurements are not reported
here):
\begin{equation}
\frac{s^{\prime\prime}\left[  \omega_{i}\left(  T\right)  ,T\right]  }{s_{0}%
}=Q^{-1}\left[  \omega_{i}\left(  T\right)  ,T\right]  \,\frac{s^{\prime
}\left[  \omega_{i}\left(  T\right)  ,T\right]  }{s_{0}}=Q^{-1}\left[
\omega_{i}\left(  T\right)  ,T\right]  \,\left[  \frac{\omega_{i}\left(
0\right)  }{\omega_{i}\left(  T\right)  }\right]  ^{2}%
\end{equation}
so avoiding any approximation of $Q^{-1}\ll1$; I also took into account the
variation of the resonance frequencies $\omega_{i}\left(  T\right)  $ by 6\%
and therefore $s^{\prime}$ by 12\% within the broad temperature range spanned
by the peak. In fact, formulas like eq. (\ref{CW}) are valid for both
$s^{\prime\prime}$ and $s^{\prime}$ due to the relaxing defect, whereas
$s^{\prime}$ generally contains an important temperature dependent
contribution $s_{\text{el}}^{\prime}$ from the elastic constants. The
resulting $s^{\prime\prime}$ curves are plotted in Fig. \ref{fig Ru fit} for
both the 1st and 5th vibration modes. There are in fact two peaks: the main
one, P1, at 680~K and a smaller one, P2, at 550~K at 0.8~kHz. Figure
\ref{fig Ru fit}a presents a fit with two peaks of the Cole-Cole type, eq.
(\ref{pCC}). It is apparent that such a model of relaxation is inadequate for
reproducing the main peak; in fact, \textit{i)}\ the intensity of the main
peak decreases faster than $1/T$; \textit{ii)} the parameters for the
relaxation time are $\tau_{0}=2.7\times10^{-18}$~s and $E/k_{\text{B}}%
=21800$~K; \textit{iii)} even with these unphysical values of $\tau_{0}$ and
$E$ it is impossible to reproduce the asymmetrical broadening of the peak. All
these features indicate an \textbf{anomalous enhancement of both compliance
and relaxation time on cooling, as expected from interacting O vacancies in
the Curie-Weiss-like approach} of Sec. \ref{sect interact}.

Therefore, the fitting function is (see Eq. (\ref{CW}))%

\begin{align}
s^{\prime\prime}\left(  \omega,T\right)   &  =\Delta\tilde{s}\frac
{\omega\tilde{\tau}}{1+\left(  \omega\tilde{\tau}\right)  ^{2}}\\
\Delta\tilde{s}  &  =\frac{\Delta s}{T-T_{\mathrm{C}}},\quad\tilde{\tau}%
=\frac{\tau_{0}\exp\left(  E/k_{\text{B}}T\right)  }{1-T_{\mathrm{C}}/T}%
\end{align}
for P1 and again the Cole-Cole formula (\ref{pCC}) for P2. The resulting fit
is the dashed curve in Fig. \ref{fig Ru fit}b, with the following parameters:
$\tau_{0}=1\times10^{-13}~$s, $E=1.33~$eV, $T_{\mathrm{C}}=460$~K for P1 and
$\tau_{0}=2\times10^{-17}$~s, $E=1.4$~eV, $\alpha=0.63$ for P2. The fit to P1
is definitely better, even without any broadening parameter, and with a
perfectly reasonable value of $\tau_{0}$; this demonstrates that \textbf{a
mean-field scheme of interaction between the V}$_{\text{O}}$\textbf{\ elastic
dipoles is able to reproduce the salient properties of P1}.

In order to further improve the agreement with the experiment, I\ introduced a
peak broadening according to the Cole-Cole expression, obtaining the
continuous lines with the parameters $\tau_{0}=1.1\times10^{-15}~$s,
$E=1.46~$eV, $T_{\mathrm{C}}=470$~K, $\alpha=0.82$ for P1 and $\tau
_{0}=2\times10^{-14}$~s, $E=1.1$~eV, $\alpha=0.7$ for P2. This broadening is
not excessive and may be justified by the high density of boundaries between
the three possible orientations of the $c$ axis (see Sec. \ref{sect Ru struct}%
) and of antiphase boundaries between domains with opposite directions of the
rotation angle of the octahedra about the $c$ axis; the parameters of P1,
including the onset temperature $T_{\mathrm{C}}$ of orientational ordering of
the elastic dipoles are little affected, but those of P2 assume more physical
values. The fact that the fit is not perfect may be ascribed to an inadequacy
of the description of the dipole interaction in the mean-field approximation,
but also to a variation of the various physical parameters in the broad
temperature range from 500 to 900~K. A slow variation in the electronic and/or
structural parameters may result in a temperature dependent activation energy
$E$. For example, the rotation angle of the RuO$_{6}$ octahedra is a linearly
decreasing function of temperature within the $0-300$~K range \cite{CJS00} and
probably continues decreasing also above room temperature, certainly affecting
the hopping parameters of the O\ vacancies. Incidentally, one might expect a
\textbf{structural transformation to non-rotated octahedra} above room
temperature, similarly to the case of LSCO\ and many perovskites, but
\textbf{the anelastic spectra do not show any trace of such a transformation
up to 920~K}.

Regarding the minor peak P2, no clear indication of $T_{\mathrm{C}}>0$ can be
found, also due to the fact that the peak hardly emerges from the tail of P1,
but it must also be related with V$_{\text{O}}$. Two possible origins of P2
are:\ \textit{i)} the presence of V$_{\text{O}}$ also in the CuO$_{2}$ planes,
although with a much smaller concentration than in the RuO$_{2}$
ones;\ \textit{ii)} V$_{\text{O}}$ trapped by some defect in the RuO$_{2}$
planes, like Cu substituting Ru.

\subsection{Rotations of the RuO$_{6}$ octahedra\label{sect Ru rot}}

\begin{figure}[h]
\begin{center}
\includegraphics[
%natheight=149.875000pt,
%natwidth=535.562500pt,
%height=3.218cm,
width=11.35cm
]{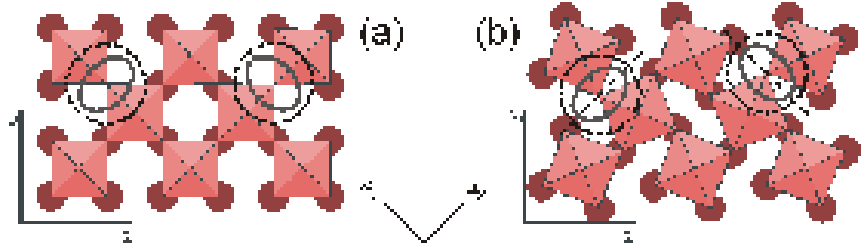}
\end{center}
\caption{RuO$_{6}$ octahedra and elastic dipoles (ellipsoids) of two O
vacancies neglecting (a) and considerin the rotations of the octahedra about
the $c$ axis (b). The dashed unit circles represent the absence of strain.}%
\label{fig Ru octa}%
\end{figure}

The fact that the RuO$_{6}$ octahedra are rotated about the $c$ axes (by about
14$^{\text{o}}$ at room temperature)\ complicates only slightly the treatment
of the anelasticity due to oxygen hopping within the RuO$_{2}$ planes. Figure
\ref{fig Ru octa}a shows the case of non-rotated octahedra, and the $x$ and
$y$ axes are chosen parallel to the Ru-O bonds (at 45$^{\mathrm{o}}$ with the
standard setting of the $a$ and $b$ cell parameters). The elastic dipoles
$\lambda$ associated with V$_{\text{O}}$ will therefore have major and minor
axes parallel to $x$ or $y$; the ellipsoids represent the principal values of
$\lambda$ at two O vacancies, compared with the unit circles of no strain at
regular O sites (the deviation from unity is exaggerated, and actually an
anisotropic expansion instead of contraction might occur, as shown for YBCO in
Sec. \ref{sect YBCO eldip}). Figure \ref{fig Ru octa}b includes the rotations
of the octahedra, again exaggerated, and it can be seen that the strain
ellipsoids at the O vacancies remain oriented along $x$ or $y$, which are
still symmetry planes of the structure. In conclusion, \textbf{it is possible
to neglect the rotation of the octahedra in analyzing the anelasticity from
the hopping of O vacancies}.

One might ask \textbf{why collective dynamic rotations of the octahedra of the
type observed in La}$_{2}$\textbf{CuO}$_{4}$\textbf{\ are not found in
Ru-1212}, even though it has planes of octahedra able to rotate without
constraints from outside the plane. The reason is that soliton-like rotation
waves cannot develop for rotations about the $c$ axis of octahedra coupled in
the $ab$ plane; in fact, it is not possible to decouple rotations along
different rows as for the LTT\ pattern of La$_{2}$CuO$_{4}$ and therefore it
is not possible to derive a one-dimensional equation of motion with solitonic
solutions. This can be readily checked in Fig. \ref{fig Ru octa}, where it
appears that the rotation of a single octahedron determines the rotations of
all the octahedra in the same plane.

\subsection{Elastic dipole of the oxygen vacancy and comparison with YBCO}

In view of the low concentration $\delta$ of V$_{\text{O}}$ in the
RuO$_{2-\delta}$ planes, it is natural to think in terms of relaxation of
V$_{\text{O}}$ possessing an elastic dipole $\lambda^{\text{V}}$, instead of
jumps of O atoms with elastic dipole $\lambda^{\text{O}}$, but the two
treatments are otherwise equivalent, with $\lambda^{\text{O}}=-\lambda
^{\text{V}}$. This is apparent, for example, also from the symmetry in $c$ and
$1-c$ in the equations describing the relaxation of interacting dipoles in
Sec. \ref{sect interact}, where $c=\frac{\delta}{2}$ for RuO$_{2-\delta}$ or
$c=\frac{x}{2}$ for CuO$_{x}$. From the intensity of peak P1 and the estimated
concentration $\delta=0.02-0.03$ of V$_{\text{O}}$ one can evaluate the
anisotropic part $\left|  \lambda_{2}-\lambda_{1}\right|  $of the elastic
dipole of V$_{\text{O}}$. The estimate is very rough, due to the
polycrystalline and highly porous ($\sim50\%$) nature of the samples. As
already done in Sec. \ref{sect YBCO eldip} and shown in detail in Appendix B,
neglecting porosity (which is of the order of 50\% in the present samples),
the relaxation strength for V$_{\text{O}}$ hopping is
\begin{equation}
\Delta=\frac{1}{15}\frac{cv_{0}}{k_{\text{B}}\left(  T-T_{\mathrm{C}}\right)
\,\left\langle E^{-1}\right\rangle }\left(  \Delta\lambda\right)  ^{2}\,\,,
\end{equation}
where the Curie-Weiss correction has been included. The Young's modulus of our
sample deduced from Eq. (\ref{f flex}) was $E=40$~GPa, probably due to the
large porosity. Inserting $v_{0}=170\times10^{-24}$~cm$^{3}$, $c=0.025$,
$T_{\mathrm{C}}=470$~K, $\Delta=$ $2Q_{\text{max}}^{-1}$ $=0.025$ at 680~K one
obtains $\Delta\lambda\sim0.4$; the error in this estimate is large, due to
sample porosity and to uncertainties in the determination of $c $, but is
about 20 times larger than the value estimated in Sec. \ref{sect YBCO eldip}
for YBCO from the lattice parameters. A possible justification for the large
difference is that the assumption that $\lambda^{\text{O}}=-\lambda^{\text{V}%
}$ would be valid in the absence of electronic effects, which instead are important.

\subsection{Possible roles of the oxygen vacancies in determining the
superconducting properties}

The samples used in the present investigation had been annealed for 1 week at
1070~$^{\mathrm{o}}$C in flowing O$_{2}$, in order to achieve
superconductivity, but no treatments at high O$_{2}$ pressure were done.
Still, the superconducting transition was in two steps: a first step at 45~K,
indicative of intragrain superconductivity, and a second one at 19~K, when the
supercurrents flow also between different grains. In these conditions, the
anelastic spectrum above room temperature is flat until the oxygen loss starts
above 600~K; looking at the $Q^{-1}\left(  T\right)  $ curve 1 just below
600~K it appears that the initial concentration of O vacancies is at least one
order of magnitude smaller than after outgassing, \textit{i.e.} RuSr$_{2}%
$GdCu$_{2}$O$_{8-\delta}$ with $\delta<0.002$. This means that even without
high pressure oxygenation the concentration of O\ vacancies in Ru-1212 is very
small, and, also after outgassing, it does not exceed $\delta\sim0.02-0.03$
(see Sec. \ref{sect Ru V_O}). These values of $\delta$ cannot account for the
deficit of conducting holes with respect to the concentration of Ru$^{4+}$
ions, as discussed in Sec. \ref{sect Ru V_O}, and therefore the present study
supports the view \cite{MZA99} that the scarce reproducibility of the
superconducting properties from sample to sample are due to the microstructure
rather than to oxygen stoichiometry.

\chapter{Conclusions}

The main findings described in the present Thesis from the analysis of a large
number of anelastic spectra of HTS\ cuprates can be divided into three types:
\textit{i)}\ information on the diffusive motion of nonstoichiometric oxygen
and its ordering; \textit{ii)} local and collective motion of oxygen between
off-center positions; \textit{iii)} low frequency fluctuations of the hole
stripes pinned by the dopants.

\subsection{Oxygen diffusion}

The most relevant results on the oxygen mobility have been obtained in
YBa$_{2}$Cu$_{3}$O$_{6+x}$, where stoichiometry and ordering of oxygen\ play a
fundamental role in determining hole doping. It has been demonstrated that,
although it is generally considered that there is a barrier of $\sim1$~eV for
oxygen hopping in the CuO$_{x}$ planes, there are actually three distinct
types of jumps: in the almost fully oxygenated O-I phase, possibly requiring
the presence of an O vacancy in the otherwise filled Cu-O chains; jumps
involving more sparse Cu-O\ chain fragments; jumps of isolated O atoms in the
semiconducting state over a barrier of only 0.11~eV. The latter process
implies an extraordinarily high mobility of the O$^{2-}$ ions in the nearly
empty CuO$_{x}$ planes, but is not easily identifiable with other experimental
techniques, since the O\ atoms have a strong tendency to form chains, and
their concentration as isolated ions is very small already at room
temperature. A discussion of the apparent paradoxes posed by such a short
hopping time ($10^{-11}$~s at room temperature) and the long characteristic
times for reordering (years) has been provided, based on the hypothesis that
the electrostatic repulsion between O$^{2-}$ ions is scarcely screened in the
semiconducting states, and rises the barrier for oxygen clustering. An
analysis of the oxygen\ jumps involving changes in the electronic state, also
due to charge transfer between Cu-O chains and CuO$_{2}$ planes, demonstrates
that the situation is more complex than generally supposed in other anelastic
spectroscopy studies.

The ruthenocuprate RuSr$_{2}$GdCu$_{2}$O$_{8}$ is isostructural with YBa$_{2}
$Cu$_{3}$O$_{6+x}$, except for having RuO$_{2-\delta}$ planes completely
filled with oxygen, instead of the CuO$_{x}$ chains at most half-filled, and
it is of interest to ascertain whether the O vacancies are responsible for the
difficulty in obtaining reproducible superconducting states. The high
temperature anelastic spectra of RuSr$_{2}$GdCu$_{2}$O$_{8-\delta}$ show that
their concentration remains low even after long annealings in vacuum. It is
also shown that the O\ vacancies jump over a barrier of 1.4~eV, create an
anisotropic distortion (elastic dipole) much larger than that of oxygen\ in
the CuO$_{x}$ planes of YBCO, and it is possible to describe the
high-temperature anelastic spectrum of Ru-1212 taking into account the long
range elastic interaction among these elastic dipoles.

In La$_{2-x}$Sr$_{x}$CuO$_{4}$ it is possible to introduce excess oxygen in
interstitial positions between layers of CuO$_{6}$ octahedra. It has been
shown that, although in the ideal tetragonal structure oxygen jumps between
these interstitial positions do not cause reorientation of the elastic dipole,
this occurs in the slightly distorted orthorhombic structure, and the elastic
energy loss peaks associated with jumps of isolated and paired interstitial O
atoms have been identified. The formation of stable pairs confirms the
existence of peroxyde species lowering the hole transfer to the conducting
CuO$_{2}$ planes, but it is also shown that such pairs form at much lower
concentrations than formerly believed.

\subsection{Off-centre positions and tilts of the octahedra}

In addition to the diffusive motion of nonstoichiometric oxygen, the anelastic
spectra demonstrate that oxygen may perform smaller jumps between off-center
positions both in YBCO and LSCO. In the first case, an anelastic relaxation
process has been identified with the hopping of oxygen between off-center
positions on either side of the Cu-O chains (the zig-zag nature of the chains
is suggested by diffraction experiments showing anomalously high oxygen
thermal factors); it is also suggested that an ordering transition of the
chain O atoms in (anti)ferroelastic and (anti)ferroelectric fashion might be
responsible for anomalies observed between 120 and 170~K on heating.

In LSCO\ off-center hopping is connected with the unstable tilts of the oxygen
octahedra, responsible for the structural transformations; the anelastic
spectra presented here demonstrate that at temperatures much lower than that
of the tetragonal-to-orthorhombic transformation and where also the twin
motion is frozen, the dynamic modulus (1~kHz) of stoichiometric and
defect-free La$_{2}$CuO$_{4}$ presents still two hardenings of $\sim25\%$ on
cooling below 150~K and of 10\% below 10~K; these magnitudes are huge, if one
considers that they are not associated to any structural transformation. The
first is thermally activated and has been identified as tilt waves of
solitonic type within the planes of octahedra, while the second has been
identified as faster and local motion of single octahedra or O atoms, with the
dynamics determined by quantum tunneling. The fact that octahedral tilts are
involved has also been confirmed by NQR measurements, but a consistent picture
of these lattice motions has been possible thanks to a complete analysis of
the effect of doping through excess oxygen and substitutional Sr. It has
therefore been possible to clearly distinguish between collective and local
motion and to determine that the local tunneling is strongly coupled to hole
excitations, with as little as 3\% doping making its dynamics so fast to bring
the maximum of the relaxation process below 1~K for frequencies of kHz. This
finding suggests that at higher doping the anharmonic potential for the O
atoms might be of the type that has been proposed to enhance the
electron-phonon coupling to levels necessary for explaining the
high-$T_{\mathrm{c}}$\ superconductivity \cite{BBB91}.\emph{\ }It has also
been shown that the spectrum of the hole excitations responsible for the
tilt-hole coupling is different from the known cases of electron-lattice
interaction governing atomic tunnel systems in other metals or superconductors.

\subsection{Hole stripes and antiferromagnetic clusters}

It has been a surprise to discover that the anelastic spectra of La$_{2-x}%
$Sr$_{x}$CuO$_{4}$ contain clear signatures of the freezing of the Cu$^{2+}$
spins into the cluster spin glass state at $T_{g}\left(  x\right)  $:\ a steep
rise of the acoustic absorption at $T_{g}$. This absorption has been
interpreted in terms of stress induced motion of the walls between the
antiferromagnetic clusters with uncorrelated directions of the staggered
magnetization, and presenting an anisotropic strain thanks to magnetoelastic
coupling. A neutron scattering study \cite{MFY02} is also confirmed, where, at
variance with the prevalent opinion, it is proposed that an electronic phase
separation occurs for $x<0.02$, with cluster spin glass domains coexisting
with long range antiferromagnetic order.

The interest in this finding is enhanced by the fact that, according to the
prevalent interpretation, these domain walls coincide with the hole stripes,
whose role in high-$T_{\text{c}}$ superconductivity is much discussed, and
whose fluctuation dynamics has not been probed by other experiments. Carrying
on the analogy between the hole stripes interacting with the Sr$^{2+}$ dopants
and other linear defect pinned by impurities, a relaxation process around 80~K
at 1~kHz has been identified as due to stripes overcoming the collective
pinning barrier by thermal activation, with the measured barrier in very good
agreement with a theoretical estimate \cite{MDH98}. The assignment of such a
relaxation process to the depinning stripe mechanism has been done after the
analysis of a large number of anelastic spectra with Sr doping ranging from
$x=0.008$ to 0.20. In this manner, it has also been possible to evidence the
effect of stripe-lattice coupling through the tilts of the octahedra, with
clear changes when the prevalent stripe orientation passes from parallel to
diagonal with respect to the lattice modulation, and near the doping
$x=\frac{1}{8}$, where the stripe spacing becomes commensurate with the
lattice modulation. The locking of the stripes with domains of low-temperature
tetragonal phase near $x=\frac{1}{8}$ has also been seen as a strong
depression of the absorption in the cluster spin glass phase in samples
containing Ba instead of Sr, due to the reduction of the fraction of mobile
stripes. I would like to point out that such phenomena are hardly detectable
by other spectroscopies, which are dominated by the response of single charges
and spins (dielectric and magnetic susceptibilities, NMR, $\mu$SR) or may
probe much shorter characteristic times (light and neutron spectroscopies).
The anelastic spectroscopy is instead insensitive to individual charge and
spin excitations, and probes only those that are coupled to strain.

Even though it is commonly believed that anelastic experiments are too much
affected by grain boundaries, ''defects'' and other vagaries, to be able to
detect more fundamental excitations or processes, they are as reliable as any
other spectroscopy, and the anelastic spectra are perfectly repeatable after
years if the samples are good. I hope I showed that anelastic spectroscopy is
able to provide reliable and often unique information not only on diffusive,
reorientational and tunneling atomic motions, including collective modes like
tilts of oxygen octahedra, but also on elusive charge and spin structures like
the hole stripes in high-$T_{\text{c}}$ cuprates.

%\QTP{bibsection}

\chapter{Appendices}

\section{Appendix A - analytical approximation of the order parameter in the
Bragg-Williams model of the CuO$_{x}$ planes}

It is convenient to introduce the order parameter $\xi$, such that
\begin{equation}
c_{1,2}=\frac{c}{2}\left(  1\pm\xi\right)  \label{OP}%
\end{equation}
with $\xi=0$ and 1 in the completely disordered and ordered states,
respectively. Equation (\ref{c1})\ becomes
\begin{equation}
\ln\left(  \frac{\left(  1+\xi\right)  \left(  2-c+c\xi\right)  }{\left(
1-\xi\right)  \left(  2-c-c\xi\right)  }\right)  =\frac{\alpha c}{k_{\text{B}%
}T}\xi=\frac{4}{\left(  2-c\right)  }\frac{T_{\mathrm{C}}}{T}\xi\label{eqOP}%
\end{equation}

\begin{figure}[tbh]
\begin{center}
\includegraphics[
%natheight=358.250000pt,
%natwidth=449.625000pt,
%height=6.3548cm,
width=7.9606cm
]{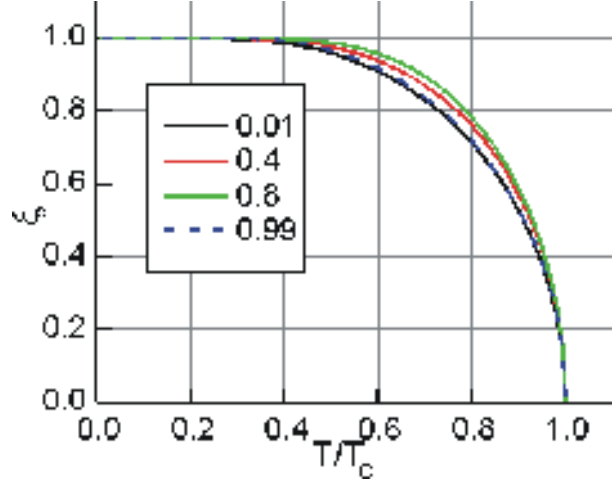}
\end{center}
\caption{Order parameter vs reduced temperature for various concentrations.}%
\label{fig OP}%
\end{figure}

A solution $\xi\neq0$ is possible when the slope of the logarithm in the left
hand at small $\xi$, $\frac{4}{2-c}$, is larger than the slope of the linear
term, $\frac{\alpha c}{k_{\text{B}}T}$, in the right hand equation; and this
determines the critical temperature
\begin{equation}
k_{\text{B}}T_{\mathrm{C}}=\alpha\frac{c}{2}\left(  1-\frac{c}{2}\right)  \;
\label{TC}%
\end{equation}
Figure \ref{fig OP} shows the order parameter $\xi\left(  t\right)  $,
solution of eq. (\ref{eqOP}) computed with Mathematica, versus the reduced
temperature $t=T/T_{\mathrm{C}}$ for some values of the concentration of
dipoles $c$. For $t>0.2$, the $\xi\left(  t\right)  $ curves become steeper
with $c$ increasing up to 0.7, and then acquire again a shape similar to the
low-$c$ case. Note that for the case of oxygen in the CuO$_{x}$ planes of
YBCO, there are two oxygen sites per Cu atom, so that in principle $0<x<2$ and
the concentration has to be defined as $c=\frac{x}{2}$; however, without
doping in the Y or Ba sublattices, it is $0<x<1$, or $c\leq0.5$. Instead, the
RuO$_{x}$ planes of Ru-1212 have $c\simeq1$.

The limit $t\rightarrow0$ or $\xi\rightarrow1$ can be solved analytically by
setting $\xi=1-y$ and obtaining, to first order in $y$,
\begin{equation}
\ln\left(  \frac{2}{y\left(  1-c\right)  }\right)  \simeq\frac{4}{\left(
2-c\right)  t}\rightarrow y=\frac{2}{\left(  1-c\right)  }\exp\left(
-\frac{4}{\left(  2-c\right)  t}\right)  \label{OPlowt}%
\end{equation}

\begin{figure}[tbh]
\begin{center}
\includegraphics[
%natheight=512.000000pt,
%natwidth=512.000000pt,
%height=6.3548cm,
width=6.3548cm
]{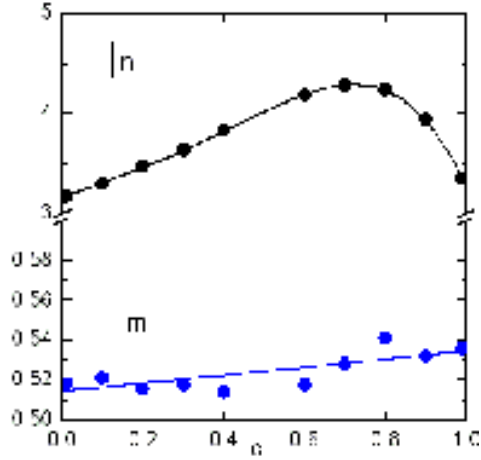}
\end{center}
\caption{Values of the $m$ and $n$ parameters in eq. (\ref{OPapprox}) for
fitting $\xi\left(  c,t\right)  $.}%
\label{fig mn}%
\end{figure}

For fitting purposes, it is convenient to have an analytical approximation to
$\xi\left(  c,t\right)  $. A tractable expression, valid for $0<c<0.8$ and
$\xi>10^{-2}$, is the following:
\begin{equation}
\xi\left(  c,t\right)  =\left(  1-t^{n}\right)  ^{m} \label{OPapprox}%
\end{equation}
with $n$ and $m$ depending on concentration as (see also Fig. \ref{fig mn})
\begin{align}
n  &  =3.153+1.785c-2.681c^{2}+9.312c^{3}-8.293c^{4}\label{OPmn}\\
m  &  =0.51+0.02c
\end{align}

For the case $c>0.8$, relevant for Ru-1212, a simple expression like
(\ref{OPapprox}) is inadequate, but luckily the anelastic relaxation due to
oxygen hopping in Ru-1212 at frequencies above $10^{2}$~Hz is well above
$T_{\mathrm{C}}$ and there is no need of computing $\xi\left(  c,t\right)  $.

\section{Appendix B - relaxation strength in a tetragonal polycrystal}

The Young's modulus of a tetragonal crystal along the direction $\mathbf{\hat
{n}}$ is (see \textit{e.g. }\cite{SS82})%

\begin{align}
E^{-1}\left(  \mathbf{\hat{n}}\right)   &  =S_{11}\left(  n_{1}^{4}+n_{2}%
^{4}\right)  +S_{33}n_{3}^{4}+\left(  S_{44}+2S_{13}\right)  \left(  n_{1}%
^{2}+n_{2}^{2}\right)  n_{3}^{2}+\label{E tetr}\\
&  +\left(  S_{66}+2S_{12}\right)  n_{1}^{2}n_{2}^{2}+2S_{16}n_{1}n_{2}\left(
n_{1}^{2}-n_{2}^{2}\right) \nonumber
\end{align}
where $S_{16}=0$ for various classes, including $4/mmm$, which is appropriate
for Ru-1212 \cite{CJS00} and tetragonal YBCO; in this case eq. (\ref{E tetr})
coincides with eq. (6.2-3) of \cite{NB72} and eq. (A5) of \cite{NH65}. The O
atom or equivalently the O vacancy in the CuO$_{x}$ or RuO$_{2-\delta}$ planes
can be considered as a $\left\langle 100\right\rangle $\ orthorhombic defect,
which causes relaxation of the elastic constant \cite{NH65,NB72}
\begin{equation}
\delta\left(  S_{11}-S_{12}\right)  =\frac{cv_{0}}{kT}\frac{1}{2}\left(
\lambda_{1}-\lambda_{2}\right)  ^{2}\,=\delta S^{B_{1}}.
\end{equation}
Adopting the group theoretical compliances of a tetragonal crystal from Eq.
(3.5), p. 125 and Table 5 of \cite{NH65}, the elastic strain $\varepsilon
_{i}=s_{ij}\sigma_{j}$ in matrix form may be written as
\begin{align}
&  \quad\quad\,\,\,\,\,\,\,\,\,\,\,\,\,\,\,
\begin{array}
[c]{cccccc}%
A_{1},1\,\, & A_{2},2\,\,\,\,\,\,\,\,\,\,\,\, & B_{1}\quad\quad\,\, &
B_{2}\,\,\,\,\,\, & E\quad\,\, & E\quad
\end{array}
\nonumber\\
&
\begin{array}
[c]{c}%
A_{1},1\\
A_{2},2\\
B_{1}\\
B_{2}\\
E\\
E
\end{array}
\left[
\begin{array}
[c]{cccccc}%
s_{33} & \sqrt{2}s_{13} & 0 & 0 & 0 & 0\\
\sqrt{2}s_{13} & s_{11}+s_{12} & 0 & 0 & 0 & 0\\
0 & 0 & s_{11}-s_{12} & 0 & 0 & 0\\
0 & 0 & 0 & \frac{1}{2}s_{66} & 0 & 0\\
0 & 0 & 0 & 0 & \frac{1}{2}s_{66} & 0\\
0 & 0 & 0 & 0 & 0 & \frac{1}{2}s_{66}%
\end{array}
\right]  \left[
\begin{array}
[c]{c}%
\sigma_{3}\\
\frac{1}{\sqrt{2}}\left(  \sigma_{1}+\sigma_{2}\right) \\
\frac{1}{\sqrt{2}}\left(  \sigma_{1}-\sigma_{2}\right) \\
\sigma_{6}\\
\sigma_{4}\\
\sigma_{5}%
\end{array}
\right]
\end{align}
In terms of these compliances eq. (\ref{E tetr}) can be written as
\begin{align}
E^{-1}\left(  \mathbf{\hat{n}}\right)   &  =\overset{S_{11}}{\overbrace
{\frac{1}{2}\left(  S^{A,22}+S^{B_{1}}\right)  }}\left(  n_{1}^{4}+n_{2}%
^{4}\right)  +\overset{S_{33}}{\overbrace{S^{A,11}}}n_{3}^{4}+\overset
{S_{44}+2S_{13}}{\overbrace{\left(  \sqrt{2}S^{A,12}+2S^{E}\right)  }}\left(
n_{1}^{2}+n_{2}^{2}\right)  n_{3}^{2}+\nonumber\\
&  +\,\overset{S_{66}+2S_{12}}{\overbrace{\left(  S^{A,22}+2S^{B_{2}}%
-S^{B_{1}}\right)  }}n_{1}^{2}n_{2}^{2} \label{E tetr2}%
\end{align}
In the hypothesis that stress is uniform over the polycrystalline sample
(Reuss approximation), in order obtain the average $\left\langle
\,E^{-1}\right\rangle $ one has to average eq. (\ref{E tetr2}) over the
angles, with $\left\langle \,n_{i}^{4}\right\rangle =\frac{1}{5}$,
$\left\langle n_{i}^{2}n_{j}^{2}\right\rangle =\frac{1}{15}$:
\begin{align}
\left\langle E^{-1}\right\rangle  &  =\frac{1}{5}S^{A,11}+\frac{4}{15}%
S^{A,22}+\frac{2\sqrt{2}}{15}S^{A,12}+\frac{2}{15}S^{B_{1}}+\frac{2}%
{15}S^{B_{2}}+\frac{4}{15}S^{E}=\\
&  =\frac{1}{5}\left(  2S_{11}+S_{33}\right)  +\frac{1}{15}\left(
2S_{44}+S_{66}+4S_{13}+2S_{12}\right) \nonumber
\end{align}
The relaxation of the compliance due to jumps of the orthorhombic defect comes
only from $\delta S^{B_{1}}\neq0$ and therefore
\begin{equation}
\delta\left\langle E^{-1}\right\rangle =\frac{2}{15}\delta S^{B_{1}}=\frac
{2}{15}\delta\left(  S_{11}-S_{12}\right)
\end{equation}
and the relaxation strength is
\begin{equation}
\Delta=\frac{\delta\left\langle E^{-1}\right\rangle }{\left\langle
E^{-1}\right\rangle }=\frac{2}{15}\frac{\delta\left(  S_{11}-S_{12}\right)
}{\left\langle E^{-1}\right\rangle }=\frac{1}{15}\frac{cv_{0}}{k_{\text{B}%
}T\,\left\langle E^{-1}\right\rangle }\left(  \lambda_{1}-\lambda_{2}\right)
^{2} \label{relstr OinYBCO}%
\end{equation}

Porosity strongly affects the elastic moduli \cite{LLH94} and also the
relaxation strength in a non-trivial manner, and I simply propose the naive
argument that a fraction $p$ of void due to porosity should decrease both $E$
and $\delta E^{-1}$ by $1-p$ (neglecting the effects of the shape of the pores
on $E$); therefore it should be
\begin{equation}
\Delta=\frac{\left(  \delta E^{-1}\right)  _{\text{eff}}}{\left(
E^{-1}\right)  _{\text{eff}}}=\left(  1-p\right)  ^{2}\frac{\delta E^{-1}%
}{E^{-1}}\,.
\end{equation}

%\bibliographystyle{abbrv}
%\bibliography{biblio}

\begin{thebibliography}{999}                                                                                              %


\bibitem {LGF01}D. Larbalestier, A. Gurevich, D.M. Feldmann and A. Polyanskii,
Nature \textbf{414}, 368 (2001).

\bibitem {KBF03}S.A.Kivelson, I.P.Bindloss, E.Fradkin, V.Oganesyan,
J.M.Tranquada, A. Kapitulnik and C. Howald, Rev. Mod. Phys. \textbf{75}, 1201 (2003).

\bibitem {Ric97}T.M. Rice, Physica C \textbf{282-287}, xix (1997).

\bibitem {BSR96}A. Bianconi, N.L. Saini, T. Rossetti, A. Lanzara, A. Perali,
M. Missori, H. Oyanagi, H. Yamaguchi, Y. Nishihara and D.H. Ha, Phys. Rev. B
\textbf{54}, 12018 (1996).

\bibitem {KFE98}S.A. Kivelson, E. Fradkin and V.J. Emery, Nature \textbf{393},
550 (1998).

\bibitem {Zaa00}J. Zaanen, Nature \textbf{404}, 714 (2000).

\bibitem {Zen48}C. Zener, \textit{Elasticity and Anelasticity of Metals}.
(Univ. of Chicago Press, Chicago, 1948).

\bibitem {NB72}A.S. Nowick and B.S. Berry, \textit{Anelastic Relaxation in
Crystalline Solids}. (Academic Press, New York, 1972).

\bibitem {SFG01}A.T. Savici, Y. Fudamoto, I.M. Gat, T. Ito, M.I. Larkin, Y.J.
Uemura, G.M. Luke, K.M. Kojima, Y.S. Lee, M.A. Kastner, R.J. Birgeneau and K.
Yamada, \textit{Mechanical Spectroscopy Q$^{-1}$ 2001: with Applications to
Materials Science}. ed. by R. Schaller, G. Fantozzi and G. Gremaud, (Trans
Tech Publications, Totton, UK, 2001).

\bibitem {Nye57}J.F. Nye, \textit{Physical Properties of Crystals; their
Representation by Tensors and Matrices}. (Oxford University Press, Oxford, UK, 1957).

\bibitem {LB1}G. Leibfried and N. Breuer, \textit{Point Defects in Metals I}.
(Springer, Berlin, 1978).

\bibitem {SS82}Yu.I. Sirotin and M.P. Shaskolskaya, \textit{Fundamentals of
crystal physics}. (Mir Publishers, Moscow, 1982).

\bibitem {33}F. Cordero, Phys. Rev. B \textbf{47}, 7674 (1993). In this paper
there are some errors without consequences on any of the conclusions: Eq. (4)
should be $\Delta=$ $...=$ $M\frac{\partial\varepsilon^{\text{an}}}%
{\partial\sigma}$ and the line below ''$\varepsilon=\sigma/M$''; in Eqs. (9,
11, 19, 20, 30, 39, 40, 48, 49, 62) $1/v_{0}$ must be substituted with $v_{0}$
or $1/\left(  Mv_{0}\right)  $ with $Mv_{0}$.

\bibitem {WN84}H. Wipf and K. Neumaier, Phys. Rev. Lett. \textbf{52}, 1308 (1984).

\bibitem {LL5}L.D. Landau and E.M. Lifshitz, \textit{Statistical Physics}.
(Pergamon, London, 1959).

\bibitem {WK96}H. Wipf and B. Kappesser, J. Phys.: Condens. Matter \textbf{8},
7233 (1996).

\bibitem {RBC98}A. Rigamonti, F. Borsa and P. Carretta, Rep. Prog. Phys.
\textbf{61}, 1367 (1998).

\bibitem {KBP00}S. Kamba, V. Bovtun, J. Petzelt, I. Rychetsky, R. Mizaras, A.
Brilingas, J. Banys, J. Grigas and M. Kosec, J. Phys.: Condens. Matter
\textbf{12}, 497 (2000).

\bibitem {116}F. Cordero, M. Corti, F. Craciun, C. Galassi, D. Piazza and F.
Tabak, Phys. Rev. B \textbf{71}, 94112 (2005).

\bibitem {FK41}R.M. Fuoss and J.G. Kirkwood, J. Am. Chem. Soc. \textbf{63},
385 (1941).

\bibitem {48}G. Cannelli, R. Cantelli, F. Cordero, F. Trequattrini and M.
Ferretti, Phys. Rev. B \textbf{54}, 15537 (1996).

\bibitem {Jon75}A.K. Jonscher, Nature \textbf{256}, 673 (1975).

\bibitem {FCA90}D. de Fontaine, G. Ceder and M. Asta, Nature \textbf{343}, 544 (1990).

\bibitem {AP95}M. Ausloos and A. Pekalski, Phys. Rev. B \textbf{52}, 4577 (1995).

\bibitem {SPA99}K. Skwarek, A. Pekalski and M. Ausloos, Eur. Phys. J. B
\textbf{11}, 369 (1999).

\bibitem {BSW94}F. Brenscheidt, D. Seidel and H. Wipf, J. Alloys and Compounds
\textbf{211/212}, 264 (1994).

\bibitem {Dat82}S. Dattagupta, J. Phys. F: Metal Phys. \textbf{12}, 1363 (1982).

\bibitem {DBR82}S. Dattagupta, R. Balakrishnan and R. Ranganathan, J. Phys. F:
Met. Phys. \textbf{12}, 1345 (1982).

\bibitem {HWD92}G. Haneczok, M. Weller and J. Diehl, phys. stat. sol. (b)
\textbf{172}, 557 (1992).

\bibitem {ACC02}C. Artini, M.M. Carnasciali, G.A. Costa, M. Ferretti, M.R.
Cimberle, M. Putti and R. Masini, Physica C \textbf{377}, 431 (2002).

\bibitem {KGG91}B. Kusz, M. Gazda, O. Gzowski, I. Davoli and S. Stizza,
\textit{IV National Conference on High Transition Temperature
Superconductivity - SATT4}{, Parma, Italy}, 13 (1991).

\bibitem {74}S. Braccini, C. Casciano, F. Cordero, F. Corvasce, M. De Sanctis,
R. Franco, F. Frasconi, E. Majorana, G. Paparo, R. Passaquieti, P. Rapagnani,
F. Ricci, D. Righetti, A. Solina and R. Valentini, Meas. Sci. Technol.
\textbf{11}, 467 (2000).

\bibitem {97}F. Cordero, R. Cantelli and M. Ferretti, J. Appl. Phys.
\textbf{92}, 7206 (2002).

\bibitem {63}F. Cordero and R. Cantelli, Physica C \textbf{312}, 213 (1999).

\bibitem {CCJ92}F.C. Chou, J.H. Cho and D.C. Johnston, Physica C \textbf{197},
303 (1992).

\bibitem {RJS93}P.G. Radaelli, J.D. Jorgensen, A.J. Schultz, B.A. Hunter, J.L.
Wagner, F.C. Chou and D.C. Johnston, Phys. Rev. B \textbf{48}, 499 (1993).

\bibitem {ZCG94}J.-S. Zhou, H. Chen and J.B. Goodenough, Phys. Rev. B
\textbf{50}, 4168 (1994).

\bibitem {BBK99}E.S. Bozin, S.J.L. Billinge, G.H. Kwei and H. Takagi, Phys.
Rev. B \textbf{59}, 4445 (1999).

\bibitem {RMA95}C. Rial, E. Mor{\`{a}}n, M.A. Alario-Franco, U. Amador and
N.H. Andersen, Physica C \textbf{254}, 233 (1995).

\bibitem {AMH89}J.D. Axe, A.H. Moudden, D. Hohlwein, D.E. Cox, K.M. Mohanty,
A.R. Moodenbaugh and Y. Xu, Phys. Rev. Lett. \textbf{62}, 2751 (1989).

\bibitem {CHC91}M.K. Crawford, R.L. Harlow, E.M. McCarron, W.E. Farneth, J.D.
Axe, H. Chou and Q. Huang, Phys. Rev. B \textbf{44}, 7749 (1991).

\bibitem {PCK91}W.E. Pickett, R.E. Cohen and H. Krakauer, Phys. Rev. Lett.
\textbf{67}, 228 (1991).

\bibitem {NSM93}M. Nohara, T. Suzuki, Y. Maeno, T. Fujita, I. Tanaka and H.
Kojima, Phys. Rev. Lett. \textbf{70}, 3447 (1993).

\bibitem {KWI95}Y. Koyama, Y. Wakabayashi, K. Ito and Y. Inoue, Phys. Rev. B
\textbf{51}, 9045 (1995).

\bibitem {MWZ98}A.R. Moodenbaugh, L. Wu, Y. Zhu, L.H. Lewis and D.E. Cox,
Phys. Rev. B \textbf{58}, 9549 (1998).

\bibitem {TR91}E. Takayama-Muromachi and D.E. Rice, Physica C \textbf{177},
195 (1991).

\bibitem {FNH90}T. Fukase, T. Nomoto, T. Hanaguri, T. Goto and Y. Koike,
Physica B \textbf{165/6}, 1289 (1990).

\bibitem {BMF91}A. Bussmann-Holder, A. Migliori, Z. Fisk, J.L. Sarrao, R.G.
Leisure and S.-W. Cheong, Phys. Rev. Lett. \textbf{67}, 512 (1991).

\bibitem {TII89}H. Takagi, T. Ido, S. Ishibashi, M. Uota, S. Uchida and Y.
Tokura, Phys. Rev. B \textbf{40}, 2254 (1989).

\bibitem {CBC93}F.C. Chou, F. Borsa, J.H. Cho, D.C. Johnston, A. Lascialfari,
D.R. Torgeson and J. Ziolo, Phys. Rev. Lett. \textbf{71}, 2323 (1993).

\bibitem {LG98b}E. Lai and R.J. Gooding, Phys. Rev. B \textbf{57}, 1498 (1998).

\bibitem {LAK01}A.N. Lavrov, Y. Ando, S. Komiya and I. Tsukada, Phys. Rev.
Lett. \textbf{87}, 017007 (2001).

\bibitem {GSB97}R.J. Gooding, N.M. Salem, R.J. Birgeneau and F.C. Chou, Phys.
Rev. B \textbf{55}, 6360 (1997).

\bibitem {NBB98}Ch. Niedermayer, C. Bernhard, T. Blasius, A. Golnik, A.
Moodenbaugh and J.I. Budnick, Phys. Rev. Lett. \textbf{80}, 3843 (1998).

\bibitem {MFY02}M. Matsuda, M. Fujita, K. Yamada, R.J. Birgeneau, Y. Endoh and
G. Shirane, Phys. Rev. B \textbf{65}, 134515 (2002).

\bibitem {TSA95}J.M. Tranquada, B.J. Sternlieb, J.D. Axe, Y. Nakamura and S.
Uchida, Nature \textbf{375}, 561 (1995).

\bibitem {TIU99}J.M. Tranquada, N. Ichikawa and S. Uchida, Phys. Rev. B
\textbf{59}, 14712 (1999).

\bibitem {WSE99}S. Wakimoto, G. Shirane, Y. Endoh, K. Hirota, S. Ueki, K.
Yamada, R.J. Birgeneau, M.A. Kastner, Y.S. Lee, P.M. Gehring and S.H. Lee,
Phys. Rev. B \textbf{60}, 769 (1999).

\bibitem {WBK00}S. Wakimoto, R.J. Birgeneau, M.A. Kastner, Y.S. Lee, R. Erwin,
P.M. Gehring, S.H. Lee, M. Fujita, K. Yamada, Y. Endoh, K. Hirota and G.
Shirane, Phys. Rev. B \textbf{61}, 3699 (2000).

\bibitem {MDH98}C. Morais Smith, Yu.A. Dimashko, N. Hasselmann and A.O.
Caldeira, Phys. Rev. B \textbf{58}, 453 (1998).

\bibitem {86}F. Cordero, A. Paolone, R. Cantelli and M. Ferretti, Phys. Rev. B
\textbf{64}, 132501 (2001).

\bibitem {88}R.S. Markiewicz, F. Cordero, A. Paolone and R. Cantelli, Phys.
Rev. B \textbf{64}, 54409 (2001).

\bibitem {93}A. Paolone, F. Cordero, R. Cantelli and M. Ferretti, Phys. Rev. B
\textbf{66}, 94503 (2002).

\bibitem {94}A. Paolone, R. Cantelli, F. Cordero, M. Corti, A. Rigamonti and
M. Ferretti, Int. J. Mod. Phys. B \textbf{17}, 512 (2003).

\bibitem {99}F. Cordero, A. Paolone, R. Cantelli and M. Ferretti, Phys. Rev. B
\textbf{67}, 104508 (2003).

\bibitem {CCJ93}J.H. Cho, F.C. Chou and D.C. Johnston, Phys. Rev. Lett.
\textbf{70}, 222 (1993).

\bibitem {Mar97}R.S. Markiewicz, J. Phys. Chem. Solids \textbf{58}, 1179 (1997).

\bibitem {BD02}S.J.L. Billinge and P.M. Duxbury, Phys. Rev. B \textbf{66},
64529 (2002).

\bibitem {YLE97}K. Yamada, C.H. Lee, Y. Endoh, G. Shirane, R.J. Birgeneau and
M.A. Kastner, Physica C \textbf{282-287}, 85 (1997).

\bibitem {FYH02}M. Fujita, K. Yamada, H. Hiraka, P.M. Gehring, S.H. Lee, S.
Wakimoto and G. Shirane, Phys. Rev. B \textbf{65}, 64505 (2002).

\bibitem {TTU89}Y. Tokura, H. Takagi and S. Uchida, Nature \textbf{337}, 345 (1989).

\bibitem {LLN90}W.-K. Lee, M. Lew and A.S. Nowick, Phys. Rev. B \textbf{41},
149 (1990).

\bibitem {SMM94}J.L. Sarrao, D. Mandrus, A. Migliori, Z. Fisk, I. Tanaka, H.
Kojima, P.C. Canfield and P.D. Kodali, Phys. Rev. B \textbf{50}, 13125 (1994).

\bibitem {Joh97}D.C. Johnston, \textit{Handbook of Magnetic Materials}. ed. by
K.H.J. Buschow, p. 1 (North Holland, 1997).

\bibitem {PR91}J.C. Phillips and K.M. Rabe, Phys. Rev. B \textbf{44}, 2863 (1991).

\bibitem {MC00}A.R. Moodenbaugh and D.E. Cox, Physica C \textbf{341-348}, 1775 (2000).

\bibitem {BBC98}P. Blakeslee, R.J. Birgeneau, F.C. Chou, R. Christianson, M.A.
Kastner, Y.S. Lee and B.O. Wells, Phys. Rev. B \textbf{57}, 13915 (1998).

\bibitem {FLH95}H.H. Feng, Z.G. Li, P. Hor, S. Bhavaraju, J.F. DiCarlo and
A.J. Jacobson, Phys. Rev. B \textbf{51}, 16499 (1995).

\bibitem {59}F. Cordero, C.R. Grandini, G. Cannelli, R. Cantelli, F.
Trequattrini and M. Ferretti, Phys. Rev. B \textbf{57}, 8580 (1998).

\bibitem {61}F. Cordero, C.R. Grandini and R. Cantelli, Physica C
\textbf{305}, 251 (1998).

\bibitem {CJ96}F.C. Chou and D.C. Johnston, Phys. Rev. B \textbf{54}, 572 (1996).

\bibitem {KWK96}B. Kappesser, H. Wipf and R.K. Kremer, J. Low Temp. Phys.
\textbf{105}, 1481 (1996).

\bibitem {JDP89}J.D. Jorgensen, B. Dabrowski, S. Pei, D.R. Richards and D.G.
Hinks, Phys. Rev. B \textbf{40}, 2187 (1989).

\bibitem {CCF89}C. Chaillout, S.W. Cheong, Z. Fisk, M.S. Lehmnann, M. Marezio,
B. Morosin and J.E. Schirber, Physica C \textbf{158}, 183 (1989).

\bibitem {HSH96}D. Haskel, E.A. Stern, D.G. Hinks, A.W. Mitchell, J.D.
Jorgensen and J.I. Budnick, Phys. Rev. Lett. \textbf{76}, 439 (1996).

\bibitem {CHC95}M.K. Crawford, R.L. Harlow, E.M. McCarron, W.E. Farneth, J.D.
Axe, H. Chou and Q. Huang, J. Phys. Chem. Solids \textbf{56}, 1459 (1995).

\bibitem {MC97}C.R. Michel and N. Casan-Pastor, Physica C \textbf{278}, 149 (1997).

\bibitem {62}F. Cordero, M. Corti, M. Campana, A. Rigamonti, R. Cantelli and
M. Ferretti, Int. J. Mod. Phys. B \textbf{13}, 1079 (1999).

\bibitem {KIA90}T. Kamiyiama, F. Izumi, H. Asano, H. Takagi, S. Uchida, Y.
Tokura, E. Takayama- Muromachi, M. Matsuda, K. Yamada, Y. Endoh and Y. Hidaka,
Physica C \textbf{172}, 120 (1990).

\bibitem {Mar93b}R.S. Markiewicz, Physica C \textbf{210}, 264 (1993).

\bibitem {64}F. Cordero, R. Cantelli, M. Corti, M. Campana and A. Rigamonti,
Phys. Rev. B \textbf{59}, 12078 (1999).

\bibitem {Aub75}S. Aubry, J. Chem. Phys. \textbf{62}, 3217 (1975).

\bibitem {TR87}S. Torre and A. Rigamonti, Phys. Rev. B \textbf{36}, 8274 (1987).

\bibitem {BMN97}M. Braden, A.H. Moudden, S. Nishizaki, Y. Maeno and T. Fujita,
Physica C \textbf{273}, 248 (1997).

\bibitem {BRN98}M. Braden, W. Reichardt, S. Nishizaki, Y. Mori and Y. Maeno,
Phys. Rev. B \textbf{57}, 1236 (1998).

\bibitem {MYY98}H. Matsui, M. Yamaguchi, Y. Yoshida, A. Mukai, R. Settai, Y.
Onuki, H. Takei and N. Toyota, J. Phys. Soc. Jpn. \textbf{67}, 3687 (1998).

\bibitem {111}A. Paolone, F. Cordero, R. Cantelli, G.A. Costa, C. Artini, A.
Vecchione and M. Gombos, J. Magn. Magn. Mater. \textbf{272-276}, 2106 (2004).

\bibitem {72}F. Cordero, R. Cantelli and M. Ferretti, Phys. Rev. B
\textbf{61}, 9775 (2000).

\bibitem {CCC98}G. Cannelli, R. Cantelli, F. Cordero and F. Trequattrini,
\textit{Tunneling Systems in Amorphous and Crystalline Solids}. ed. by P.
Esquinazi, p. 389 (Springer, Berlin, 1998).

\bibitem {BBB91}A. Bussmann-Holder, A.R. Bishop and I. Batistic, Phys. Rev. B
\textbf{43}, 13728 (1991).

\bibitem {WUE00}S. Wakimoto, S. Ueki, Y. Endoh and K. Yamada, Phys. Rev. B
\textbf{62}, 3547 (2000).

\bibitem {JCR01}M.-H. Julien, A. Campana, A. Rigamonti, P. Carretta, F. Borsa,
P. Kuhns, A.P. Reyes, W.G. Moulton, M. Horvatic, C. Berthier, A. Vietkin and
A. Revcolevschi, Phys. Rev. B \textbf{63}, 144508 (2001).

\bibitem {75}F. Cordero, A. Paolone, R. Cantelli and M. Ferretti, Phys. Rev. B
\textbf{62}, 5309 (2000).

\bibitem {HCM99}N. Hasselmann, A.H. Castro Neto, C. Morais Smith and Y.
Dimashko, Phys. Rev. Lett. \textbf{82}, 2135 (1999).

\bibitem {BS01}S. Bogner and S. Scheidl, Phys. Rev. B \textbf{64}, 54517 (2001).

\bibitem {BSZ01}M. Bosch, W. van Saarloos and J. Zaanen, Phys. Rev. B
\textbf{63}, 92501 (2001).

\bibitem {NSV90}T. Nattermann, Y. Shapir and I. Vilfan, Phys. Rev.B
\textbf{42}, 8577 (1990).

\bibitem {QLW05}J.F. Qu, Y. Liu, F. Wang, X.Q. Xu and X.G. Li, Phys. Rev. B
\textbf{71}, 94503 (2005).

\bibitem {MFC92}R. McCormack, D. de Fontaine and G. Ceder, Phys. Rev. B
\textbf{45}, 12976 (1992).

\bibitem {JVP90}J.D. Jorgensen, B.W. Veal, A.P. Paulikas, L.J. Nowicki, G.W.
Crabtree, H. Claus and W.K. Kwok, Phys. Rev. B \textbf{41}, 1863 (1990).

\bibitem {UHS97}G. Uimin, H. Haugerud and W. Selke, Physica C \textbf{275}, 93 (1997).

\bibitem {MLA01}D. M{\o }nster, P.-A. Lindg{\aa }rd and N.H. Andersen, Phys.
Rev. B \textbf{64}, 224520 (2001).

\bibitem {LSM96}H. L{\"{u}}tgemeier, S. Schmenn, P. Meuffels, O. Storz, R.
Sch{\"{o}}llhorn, Ch. Niedermayer, I. Heinmaa and Yu. Baikov, Physica C
\textbf{267}, 191 (1996).

\bibitem {THM88}J.M. Tranquada, S.M. Heald, A.R. Moodenbaugh and ouwen Xu,
Phys. Rev. B \textbf{38}, 8893 (1988).

\bibitem {TBF92}H. Tolentino, F. Baudelet, A. Fontaine, T. Gourieux, G. Krill,
J.Y. Henry and J. Rossat-Mignod, Physica C \textbf{192}, 115 (1992).

\bibitem {BCD87}A. Bianconi, A.C. Castellano, M. Desantis, P. Rudolf, P.
Lagarde, A.M. Flank and A. Marcelli, Solid State Commun. \textbf{63}, 1009 (1987).

\bibitem {SHC94}P. Schleger, W.N. Hardy and H. Casalta, Phys. Rev. B
\textbf{49}, 514 (1994).

\bibitem {UR92}G. Uimin and J. Rossat-Mignod, Physica C \textbf{199}, 251 (1992).

\bibitem {KMK92}A. Krol, Z.H. Ming, Y.H. Kao and N. N, Phys. Rev. B
\textbf{45}, 2581 (1992).

\bibitem {Con01}K. Conder, Mater. Sci. Engin.: R \textbf{32}, 41 (2001).

\bibitem {RR94}J.L. Routbort and S.J. Rothman, J. Appl. Phys. \textbf{76},
5615 (1994).

\bibitem {LP93}J.R. LaGraff and D.A. Payne, Phys. Rev. B \textbf{47}, 3380 (1993).

\bibitem {TTP88b}K.N.Tu,C.C.Tsuei,S.I. Park, and A. Levi, Phys. Rev. B
\textbf{38}, 772 (1988).

\bibitem {XCW89}X.M. Xie, T.G. Chen and Z.L. Wu, Phys. Rev. B \textbf{40},
4549 (1989).

\bibitem {CS91}J.R. Cost and J.T. Stanley, J. Mater. Res. \textbf{6}, 232 (1991).

\bibitem {WGR93}J. Woirgard, P. Gadaud, A. Riviere and B. Kaya, Mat. Sci.
Forum \textbf{119-121}, 719 (1993).

\bibitem {BFW97}F. Brenscheidt, K. Foos and H. Wipf, Europhys. Lett.
\textbf{39}, 275 (1997).

\bibitem {14}G. Cannelli, R. Cantelli and F. Cordero, Phys. Rev. B
\textbf{38}, 7200 (1988).

\bibitem {22}M. Canali, G. Cannelli, R. Cantelli, F. Cordero, M. Ferretti and
F. Trequattrini, Physica C \textbf{185-189}, 897 (1991).

\bibitem {27}G. Cannelli, R. Cantelli, F. Cordero, F. Trequattrini and M.
Ferretti, Solid State Commun. \textbf{82}, 433 (1992).

\bibitem {31}G. Cannelli, R. Cantelli, F. Cordero and F. Trequattrini,
Supercond. Sci. Tech. \textbf{5}, 247 (1992).

\bibitem {CSH96}H. Casalta, P. Schleger, P. Harris, B. Lebech, N.H. Andersen,
R. Liang, P. Dosanjh and W.N. Hardy, Physica C \textbf{258}, 321 (1996).

\bibitem {WSA95}Q. Wang, G.A. Saunders, D.P. Almond, M. Cankurtaran and K.C.
Goretta, Phys. Rev. B \textbf{52}, 3711 (1995).

\bibitem {SLN95}Y. Shindo, H. Ledbetter and H. Nozaki, J. Mater. Res.
\textbf{10}, 7 (1995).

\bibitem {SHW92}D. Seidel, A. Hornes and H. Wipf, Europhys. Lett. \textbf{18},
307 (1992).

\bibitem {TPP99}R. T{\'{e}}tot, V. Pagot and C. Picard, Phys. Rev. B
\textbf{59}, 14748 (1999).

\bibitem {Wip94}H. Wipf, Solid State Commun. \textbf{91}, 713 (1994).

\bibitem {28}G. Cannelli, R. Cantelli, F. Cordero, F. Trequattrini, S. Ferraro
and M. Ferretti, Solid State Commun. \textbf{80}, 715 (1991).

\bibitem {CS93}J.R. Cost and J.T. Stanley, Scripta metall. mater. \textbf{28},
773 (1993).

\bibitem {MSB94}Y. Mi, R. Schaller and W. Benoit, J. Alloys Comp.
\textbf{211/212}, 283 (1994).

\bibitem {BN89}R. Bormann and J. N{\"{o}}lting, Appl. Phys. Lett. \textbf{54},
2148 (1989).

\bibitem {21}G. Cannelli, R. Cantelli, F. Cordero, M. Ferretti and F.
Trequattrini, Solid State Commun. \textbf{77}, 429 (1991).

\bibitem {JBD94}G. Jang, C. Bucci, R. De Renzi, G. Guidi, M. Varotto, C. Serge
and P. Radaelli, Physica C \textbf{226}, 301 (1994).

\bibitem {BHC90}S. de Brion, J.Y. Henry, R. Calemczuk and E. Bonjour,
Europhys. Lett. \textbf{12}, 281 (1990).

\bibitem {44}G. Cannelli, R. Cantelli, F. Cordero, N. Piraccini, F.
Trequattrini and M. Ferretti, Phys. Rev. B \textbf{50}, 16679 (1994).

\bibitem {VPY90}B.W. Veal, A.P. Paulikas, H. You, H. Shi, Y. Fang and J.W.
Downey, Phys. Rev. B \textbf{42}, 6305 (1990).

\bibitem {20}G. Cannelli, R. Cantelli, F. Cordero, M. Ferretti and L. Verdini,
Phys. Rev. B \textbf{42}, 7925 (1990).

\bibitem {FJY88}M. Francois, A. Junod, K. Yvon, A. W. Hewat, J. J. Capponi, P.
Strobel, M. Marezio and P. Fischer, Solid State Commun. \textbf{66}, 1117 (1988).

\bibitem {WGK90}W. Wong-Ng, F. W. Gayle, D. L. Kaiser, S. F. Watkins and F. R.
Fronczek, Phys. Rev. B \textbf{41}, 4220 (1990).

\bibitem {NH89}A. Nath and Z. Homonnay, Physica C \textbf{161}, 205 (1989).

\bibitem {25}G. Cannelli, M. Canali, R. Cantelli, F. Cordero, S. Ferraro, M.
Ferretti and F. Trequattrini, Phys. Rev. B \textbf{45}, 931 (1992).

\bibitem {BSH95}V. Breit, P. Schweiss, R. Hauff, H. W{\"{u}}hl, H. Claus, H.
Rietschel, A. Erb and G. M{\"{u}}ller-Vogt, Phys. Rev. B \textbf{52}, 15727 (1995).

\bibitem {SHY91}P. Schleger, W.N. Hardy and B.X. Yang, Physica C \textbf{176},
261 (1995).

\bibitem {13}G. Cannelli, R. Cantelli, F. Cordero, G.A. Costa, M. Ferretti and
G.L. Olcese, Europhys. Lett. \textbf{6}, 271 (1988).

\bibitem {LF88}T. Laegreid and K. Fossheim, Europhys. Lett. \textbf{6}, 81 (1988).

\bibitem {LK88}H.M. Ledbetter and S.A. Kim, Phys. Rev. B \textbf{38}, 11857 (1988).

\bibitem {EGL87}S. Ewert, S. Guo, P. Lemmens, F. Stellmach, J. Wynants, G.
Arlt, D. Bonnenberg, H. Kliem, A. Comberg and H. Passing, Solid State Commun.
\textbf{64}, 1153 (1987).

\bibitem {BBG93}E. Biagi, E. Borchi, R. Garre, S. Degennaro, L. Masi and L.
Masotti, phys. stat. sol. (a) \textbf{138}, 249 (1993).

\bibitem {PPN92}L.N. Pal-Val, P.P. Pal-Val, V.D. Natsik and V.I. Dotsenko,
Solid State Commun. \textbf{81}, 761 (1992).

\bibitem {YW02}X.N. Ying and Y.N. Wang, Solid State Commun. \textbf{123}, 511 (2002).

\bibitem {YHW04}X.N. Ying, Y.N. Huang and Y.N. Wang, Supercond. Sci. Technol.
\textbf{17}, 347 (2004).

\bibitem {KKA90}T.J. Kim, J. Kowalewski, W. Assmus and W. Grill, Z. Physik B
\textbf{78}, 207 (1990).

\bibitem {Dom93}J. Dominec, Supercond. Sci. Technol. \textbf{6}, 153 (1993).

\bibitem {LLH94}H. Ledbetter, M. Lei, A. Hermann and Z. Sheng, Physica C
\textbf{225}, 397 (1994).

\bibitem {KMG94}T.J. Kim, E. Mohler and W. Grill, J. Alloys and Compounds
\textbf{212}, 318 (1994).

\bibitem {MTO88}H.Mizubayashi,K. Takita, and S. Okuda, Phys. Rev. B
\textbf{37}, 9777 (1988).

\bibitem {BWB95}L. Bauernfeind, W. Widder and H.F. Braun, Physica C
\textbf{254}, 151 (1995).

\bibitem {FWB99}A. Fainstein, E. Winkler, A. Butera and J. Tallon, Phys. Rev.
B \textbf{60}, R12597 (1999).

\bibitem {CJS00}O. Chmaissem, J.D. Jorgensen, H. Shaked, P. Dollar and J.L.
Tallon, Phys. Rev. B \textbf{61}, 6401 (2000).

\bibitem {MZA99}A.C. McLaughlin, W. Zhou, J.P. Attfield, A.N. Fitch and J.L.
Tallon, Phys. Rev. B \textbf{60}, 7512 (1999).

\bibitem {LJH01}R.S. Liu, L.-Y. Jang, H.-H. Hung and J.L. Tallon, Phys. Rev. B
\textbf{63}, 212507 (2001).

\bibitem {MAJ03}M. Matvejeff, V.P.S. Awana, L.-Y. Jang, R.S. Liu, H. Yamauchi
and M. Karppinen, Physica C \textbf{392-396}, 87 (2003).

\bibitem {PTW99}D.J. Pringle, J.L. Tallon, B.G. Walker and H.J. Trodahl, Phys.
Rev. B \textbf{59}, R11679 (1999).

\bibitem {HFA00}R.W. Henn, H. Friedrich, V.P.S. Awana and E. Gmelin, Physica C
\textbf{341-348}, 457 (2000).

\bibitem {KDM00}P.W. Klamut, B. Dabrowski, M. Maxwell, J. Mais, O. Chmaissem,
R. Kruk, R. Kmiec and C.W. Kimball, Physica C \textbf{341-348}, 455 (2000).

\bibitem {LMC01c}B. Lorenz, R.L. Meng, J. Cmaidalka, Y.S. Wang, J. Lenzi, Y.Y.
Xue and C.W. Chu, Physica C \textbf{363}, 251 (2001).

\bibitem {TLW00}J.L. Tallon, J.W. Loram, G.V.M. Williams and C. Bernhard,
Phys. Rev. B \textbf{61}, R6471 (2000).

\bibitem {104}F. Cordero, M. Ferretti, M.R. Cimberle and R. Masini, Phys. Rev.
B \textbf{67}, 144519 (2003).

\bibitem {HST99}A.W. Hunt, P.M. Singer, K.R. Thurber and T. Imai, Phys. Rev.
Lett. \textbf{82}, 4300 (1999).

\bibitem {NH65}A.S. Nowick and W.R. Heller, Adv. Phys. \textbf{14}, 101 (1965).
\end{thebibliography}

\chapter*{List of acronyms and symbols}

\section*{Acronyms}

\quad

AF = antiferromagnetic

CSG = Cluster Spin Glass

DW = Domain Wall

EXAFS = Extended X-ray Absorption Fine Structure

HTS = high-temperature superconductor/superconductivity

HTT = High-Temperature Tetragonal

LBCO = La$_{2-x}$Ba$_{x}$CuO$_{4}$

LSCO = La$_{2-x}$Sr$_{x}$CuO$_{4}$

LTO = Low-Temperature Orthorhombic

LTT = Low-Temperature Tetragonal

$\mu$SR = muon spin relaxation

NCO = Nd$_{2}$CuO$_{4+\delta}$

NMR = Nuclear Magnetic Relaxation

NQR = Nuclear Quadrupolar Relaxation

Ru-1212 = RuSr$_{2}$GdCu$_{2}$O$_{8}$

SG = Spin Glass

TS = Tunnel System

YBCO = YBa$_{2}$Cu$_{3}$O$_{6+x}$\newpage

\section*{Symbols}

\quad

$c=$ molar concentration

$c_{ij}=$ elastic stiffness (matrix notation)

$\Delta=$ relaxation strength

$\Delta E=$ energy difference between two states in a relaxation process

$E=$ Young's modulus, activation energy

$E_{p}=$ pinning barrier for the stripes

$\varepsilon_{ij}=$ strain tensor ($\varepsilon_{i}$ in matrix notation)

$J=$ spectral density (Eq. (\ref{FDT}))

$L_{c}=$ collective pinning length of the stripes

$\lambda_{ij}=$ elastic dipole tensor ($\lambda_{i}$ in matrix notation)

O = oxygen

O$_{\text{i}}=$ interstitial O atom

$p=$ hole density

$Q^{-1}=\frac{s^{\prime\prime}}{s^{\prime}}=\frac{c^{\prime\prime}}{c^{\prime
}}=$ elastic energy loss coefficient

$s_{ij}=$ elastic compliance (matrix notation)

$\sigma_{ij}=$ stress tensor ($\sigma_{i}$ in matrix notation)

$\tau=$ relaxation time

$T_{\text{c}}=$ superconducting temperature

$T_{\mathrm{C}}=$ Curie-Weiss-like ordering temperature

$T_{d}$ $=$ temperature of the LTO/LTT transition

$T_{g}=$ temperature for freezing into the CSG phase

$T_{\text{N}}=$ N\'{e}el temperature for long range AF ordering

$T_{t}$ $=$ temperature of the HTT/LTO transition

V$_{\text{O}}=$ oxygen vacancy

\chapter*{Acknowledgments}

Most of the results presented here are due to a long lasting collaboration
with Professors Rosario Cantelli and Gaetano Cannelli, which I thank for
having taught me the anelastic relaxation technique.

Immediately after the announcenments of the new high-$T_{\text{c}}$
superconductor YBCO, Cantelli understood that it was worth making anelastic
relaxation measurements on this material, and thanks to his promptness and
that of Maurizio Ferretti, who immediately started preparing YBCO samples, we
were able to produce new results from the beginning of the intense research
activity that would have involved so many groups all over the world.

Thanks are due therefore to Prof. Maurizio Ferretti, with whom a really
productive collaboration started, and still goes on also on other materials.
Without his commitment to preparing and studying a really large number of high
quality samples we would not have been able to reveal so many interesting
effects in the HTS cuprates.

I should thank also Prof. Attilio Rigamonti, with his deep and vast knowledge
on phase transitions, magnetism, superconductors, ...; every time I talk with
him I try to learn something new. The collaboration with Rigamonti and
Maurizio Corti not only allowed us to confirm our interesting results on the
tilt waves of the octahedra in LSCO, but also to comprehend better their
nature and later to study in a really productive way the spin and charge
inhomogeneities in LSCO.

To several anelastic experiments have contributed Francesco Trequattrini and
Annalisa Paolone; Annalisa had also the courage of starting to interpret our
low temperature anelastic results on LSCO in terms of freezing into the
cluster spin glass phase, which was not so obvious on the basis of the first data.

There is also a collaboration with Prof. Carlos Roberto Grandini, who stayed
in my laboratory for almost one year, participating to the experiments on the
Nd cuprates, and who gave me a warm hospitality in his Department at UNESP in
Brasil. Some students of the Degree in Physics have also contributed to
measurements on the YBCO materials, Marco Canali and Nadia Piraccini.

Thanks are due, of course, to Prof. Hiroshi Mizubayashi, who proposed me to
write this Thesis, and so generously provided us in Rome with the technique
for making anelasticity measurements on thin films. It was gratifying to
discuss the Thesis with the other members of the Committee, Professors K.
Kadowaki, K. Takita, S. Kojima and H. Tanimoto, whose acute comments allowed
improvements to the Thesis to be made; I thank them and Prof. Mizubayashi also
for their extremely careful reading of the manuscript.

\end{document}